\documentclass{article}
\usepackage[utf8]{inputenc}
\usepackage{amssymb}
\usepackage{amsbsy}
\usepackage{latexsym}
\usepackage{amsmath}
\allowdisplaybreaks
\usepackage{array}
\usepackage{multirow}
\usepackage{graphicx}
\usepackage{caption}
\usepackage{subcaption}
\usepackage{algorithmic}
\usepackage{float}
\usepackage{hyperref}
\usepackage{url} 
\usepackage{algorithm}
\usepackage{amsthm}
\usepackage{verbatim}
\usepackage{color}
\usepackage{geometry}  
\usepackage{dsfont}
\usepackage[mathscr]{eucal}
\usepackage{booktabs}
\usepackage{cleveref}
\usepackage{appendix}

\theoremstyle{definition}

\newtheorem{definition}{Definition}

\newcommand{\modL}[1]{\left|#1\right|}
\newcommand{\paraL}[1]{\left(#1\right)}
\newcommand{\sumintK}[1]{\sum_{\sigma \in \mathcal{E}(K) \bigcap \mathcal{E}_{\text{int}}} \frac{|\sigma|}{|K|}}
\newcommand{\sumintKK}[1]{\sum_{\sigma \in \mathcal{E}(K)} \frac{|\sigma|}{|K|}}
\newcommand{\sumK}[1]{\sum_{K \in \mathcal{T}} |K|}
\newcommand{\sumintall}[1]{\sum_{\sigma \in \mathcal{E}} |\sigma|}
\newcommand{\norm}[1]{\left\lVert#1\right\rVert}
\newcommand{\dx}[1]{\ \mathrm{d}x}
\newcommand{\dy}[1]{\ \mathrm{d}y}
\newcommand{\dt}[1]{\ \mathrm{d}t}
\newcommand{\dbx}[1]{\ \mathrm{d}\mathbf{x}}
\newcommand{\divx}[1]{\text{div}_x}
\newcommand{\divh}[1]{\text{div}_h}

\usepackage{bm}

\newcommand*\xbar[1]{%
	\hbox{%
		\vbox{%
			\hrule height 0.5pt 
			\kern0.4ex
			\hbox{%
				\kern-0.05em
				\ensuremath{#1}%
				\kern-0.00em
			}%
		}%
	}%
}

\newcommand{\mU}{\bm{U}}

\newcommand{\delx}{\Delta x}
\newcommand{\dely}{\Delta y}

\newcommand{\jph}{{j+\frac{1}{2}}}
\newcommand{\jmh}{{j-\frac{1}{2}}}

\newcommand{\kph}{{k+\frac{1}{2}}}
\newcommand{\kmh}{{k-\frac{1}{2}}}

\def\softd{{\leavevmode\setbox1=\hbox{d}%
		\hbox to 1.05\wd1{d\kern-0.4ex{\char039}\hss}}}

\title{Selection criteria for the compressible Euler equations: numerical study\footnote{\textbf{Funding:} This work was supported by the DAAD DST (German-India) Project based personnel exchange programme: Development and analysis of higher-order
structure-preserving numerical methods for hyperbolic balance laws;  Project
 525853336; SPP 2410 “Hyperbolic Balance Laws: Complexity,
Scales and Randomness. M. L. gratefully acknowledges support of the Gutenberg Research Fellowship.}}
\author{Megala Anandan$^{1}$, Bahulayan Chalil Aravind$^{2}$, M\'{a}ria Luk\'{a}\v{c}ov\'{a}-Medvid'ov\'{a}$^{1,3}$, \\ S. V. Raghurama Rao$^{2}$, Kedar Shridhar Wagh$^{2}$, Changsheng Yu$^{1}$, Jiahui Zhang$^{1}$}
\date{}

\begin{document}

\maketitle

\begin{center}
\small
$^{1}$Johannes Gutenberg University of Mainz, Germany\\
$^{2}$Indian Institute of Science Bangalore, India\\
$^{3}$RMU Co-Affiliate, Technical University of Darmstadt, Germany\\[0.5em]
E-mail: manandan@uni-mainz.de
\end{center}

\abstract{%
We numerically investigate selection criteria for dissipative weak solutions of the compressible Euler equations. Using the Kelvin-Helmholtz problem as a benchmark, we compare four standard numerical schemes - the viscous finite volume (VFV) method, the discrete velocity Boltzmann finite volume (DVBFV) method, the Runge-Kutta discontinuous Galerkin (RKDG) method, and the discontinuous Galerkin spectral element method (DGSEM), and demonstrate that each may converge to a different dissipative weak solution. We evaluate these solutions with respect to several selection criteria based on entropy production, total energy, and energy defect, and examine the sensitivity of selection functionals to the numerical diffusion inherent in each scheme.
} 

\section{Introduction}

\label{Sec: Introduction}
The Euler system describes the time evolution of the density $\varrho = \varrho(t,\mathbf{x})$, the momentum $\mathbf{m} = \mathbf{m}(t,\mathbf{x})$, and the energy $E=E(t,\mathbf{x})$ of a compressible inviscid fluid confined to a spatial domain $\Omega \subset \mathbb{R}^d$, $d=1,2,3$. It expresses the conservation of mass, momentum and energy as follows:
\begin{eqnarray}
    \partial_t \varrho + \divx{} \mathbf{m} = 0, \nonumber \\
    \partial_t \mathbf{m} + \divx{} \paraL{\frac{\mathbf{m} \otimes \mathbf{m}}{\varrho}} + \nabla_x p = \mathbf{0}, \label{Euler} \\
    \partial_t E + \divx{} \paraL{(E+p)\frac{\mathbf{m}}{\varrho}} = 0. \nonumber
\end{eqnarray}
Here, $p = p(\varrho,\mathbf{m},E)$ is the pressure determined by a suitable equation of state. For the sake of simplicity, we consider the periodic conditions or the impermeability conditions $\mathbf{m}\cdot\mathbf{n}\textbar_{\partial\Omega}=0$ on the boundary of $\Omega$, and the initial state of the system is defined through the initial conditions, $\varrho(0,\cdot)=\varrho_0(\cdot)$, $\mathbf{m}(0,\cdot)=\mathbf{m}_0(\cdot)$, $E(0,\cdot)=E_0(\cdot)$. The entropy admissibility condition based on the second law of thermodynamics is given by
\begin{equation}
\label{entropy}
    \partial_t S + \divx{} \paraL{S \frac{\mathbf{m}}{\varrho}} \geq 0,
\end{equation}
where $S = \frac{\varrho}{\gamma-1}\ln(\frac{p}{\varrho^\gamma})$ is the total entropy of the system. 
\par Since the solutions of \eqref{Euler}-\eqref{entropy} are known to develop discontinuities in finite time even for smooth initial data, system \eqref{Euler}-\eqref{entropy} is considered in the weak (or distributional) sense. However, system \eqref{Euler}-\eqref{entropy} is ill-posed in the class of weak solutions. The first numerical evidence of this was presented by Elling \cite{Elling2006}. It was later proved rigorously, applying convex integration in the pioneering work of De Lellis, Sz\'{e}kelyhidi and others \cite{Chiodaroli2014,Chiodaroli2015,ChiodaroliKreml2014,Chiodaroli2021,deLellis2010,Feireisl2020}. Indeed, the multi-dimensional Euler system \eqref{Euler}-\eqref{entropy} admits infinitely many physically admissible weak solutions for a generic class of initial data. Hence, the question of an appropriate selection criterion to obtain a unique physically reasonable solution for the multi-dimensional Euler equations remains a challenging open problem.  
\par The main goal of this paper is to discuss possible selection criteria in the class of generalised, dissipative weak solutions and to illustrate the behaviour of some standard numerical methods for compressible Euler equations with respect to these selections. 

\subsection{Dissipative weak solutions}
\par In light of the ill-posedness of the multi-dimensional compressible Euler equations in the class of weak solutions, the question of suitable selection criteria to recover well-posedness of the compressible Euler equations is crucial. To this end, we propose to enlarge the class of functions in which the unique selection is sought to guarantee its convexity, compactness, and boundedness in a suitable topology. In \cite{feireisllukacova2025,BreitFeireislHofmanova2020}, Feireisl~et~al.\ proposed the set of dissipative weak (DW) solutions as a suitable set for the formulation of physically reasonable selections. Moreover, DW solutions are also suitable to analyse the convergence of numerical schemes. It has been shown in \cite{Brezina2018,Breit2020,FeireislHofmanova2020,FeireislLukacovaMizerova2020,FeireislLukacovaMizerova2020_2} that consistent (satisfying the weak form of Euler system modulo a local truncation error) and stable (satisfying uniform energy bounds) approximations of the compressible Euler system either converge strongly to a weak solution, or weakly to a generalised (dissipative measure-valued) solution. It is convenient to work with the DW solutions, instead of the dissipative measure-valued solutions. DW solutions may be interpreted as the expected values with respect to parametrised Young measures \cite{FeireislLukacova_book2021}. In the recent literature, several efficient and robust numerical schemes that are known to have positivity preservation and entropy stability have been shown to converge to the DW solutions \cite{LukacovaYuan2023,Abgrall2023,LukacovaOffner2023,KuzminLukáčováMedvidováÖffner2025}. 
\par In what follows, we recall the definition of DW solutions for the Euler system \eqref{Euler}-\eqref{entropy}. 
\begin{definition}[Dissipative weak (DW) solution]
Let the initial data satisfy 
\begin{equation*}
    \varrho_0 \in L^1(\Omega), \mathbf{m}_0 \in L^1(\Omega;\mathbb{R}^d), E_0 = E(\varrho_0,\mathbf{m}_0,S_0) \text{ and } \int_{\Omega} E(\varrho_0,\mathbf{m}_0,S_0) \dbx{} < \infty. 
\end{equation*}
We say that $[\varrho,\mathbf{m},S]$ is a DW solution to the Euler system in $[0,T) \times \Omega$, $0<T< \infty$, if the following holds:
\begin{itemize}
    \item \textbf{Regularity.} $(\varrho,\mathbf{m},S)$ belongs to the class
    \begin{gather*}
        \varrho \in C_{\text{weak,loc}}\paraL{[0,T);L^{\gamma}(\Omega)}, \mathbf{m} \in C_{\text{weak,loc}}\paraL{[0,T);L^{\frac{2\gamma}{\gamma+1}}(\Omega;\mathbb{R}^d)}, \\
        S \in L^{\infty}\paraL{0,T;L^{\gamma}(\Omega)} \ \cap \ BV_{\text{weak}}\paraL{[0,T);L^{\gamma}(\Omega)}, \\
        \int_{\Omega} E(\varrho,\mathbf{m},S) (t,\cdot) \dbx{} \leq \int_{\Omega} E(\varrho_0,\mathbf{m}_0,S_0) \dbx{} \text{ for any } 0\leq t < T. 
    \end{gather*}
    \item \textbf{Continuity equation.} The integral identity
    \begin{equation*}
        \int_0^T \int_{\Omega} \left[ \varrho \partial_t \varphi + \mathbf{m} \cdot \nabla_x \varphi \right] \dbx{} \dt{} = - \int_{\Omega} \varrho_0 \varphi(0,\cdot) \dbx{}
    \end{equation*}
    holds for any test function $\varphi \in C_c^1\paraL{[0,T) \times \Omega}$.
    \item \textbf{Momentum equation.} The integral identity
    \begin{multline*}
        \int_0^T \int_{\Omega} \left[ \mathbf{m} \cdot \partial_t \boldsymbol{\varphi} + \mathds{1}_{\varrho>0} \frac{\mathbf{m}\otimes\mathbf{m}}{\varrho} : \nabla_x \boldsymbol{\varphi} + p(\varrho,S) \divx{} \boldsymbol{\varphi} \right] \dbx{} \dt{} \\ 
        = - \int_0^T \int_{\Omega} \nabla_x \boldsymbol{\varphi} : \text{d}\mathfrak{R}(t) \dbx{} - \int_{\Omega} \mathbf{m}_0 \cdot \boldsymbol{\varphi}(0,\cdot) \dbx{}
    \end{multline*}
    holds for any test function $\boldsymbol{\varphi} \in C_c^1\paraL{[0,T) \times \Omega;\mathbb{R}^d}$. Here, $\mathfrak{R}$ is the so-called Reynolds stress defect, such that
    \begin{equation*}
        \mathfrak{R} \in L^{\infty}\paraL{(0,T);\mathcal{M}^+\paraL{\Omega;\mathbb{R}_{\text{sym}}^{d\times d}}}. 
    \end{equation*}
    \item \textbf{Entropy inequality.} The inequality
    \begin{gather*}
        \left[\int_{\Omega} S \varphi \dx{}\right]_{t=\tau_1-}^{t=\tau_2+} \geq \int_{\tau_1}^{\tau_2} \int_{\Omega} \left[ S \partial_t \varphi + \left\langle \mathcal{V}_{t,x};\mathrm{1}_{\widetilde{\varrho}>0} \paraL{\widetilde{S}\frac{\widetilde{\mathbf{m}}}{\widetilde{\varrho}}} \right\rangle \cdot \nabla_x \varphi \right] \dbx{} \dt{}, \\ 
        S(0-,\cdot) = S_0
    \end{gather*}
    holds for any $0 \leq \tau_1 \leq \tau_2 < T$, any test function $\varphi \in C_c^1\paraL{[0,T) \times \Omega}$, $\varphi \geq 0$. Here, $\left\{ \mathcal{V}_{t,x} \right\}_{(t,x) \in (0,T) \times \Omega}$ is a parametrized probability (Young) measure,
    \begin{gather*}
        \mathcal{V}_{t,x} \in L^{\infty}\paraL{(0,T) \times \Omega;\mathcal{P}\paraL{\mathbb{R}^{d+2}}}, \mathbb{R}^{d+2} = \left\{ \widetilde{\varrho} \in \mathbb{R}, \widetilde{\mathbf{m}} \in \mathbb{R}^d, \widetilde{S} \in \mathbb{R}  \right\}; \\
        \left\langle \mathcal{V}; \widetilde{\varrho} \right\rangle = \varrho, \left\langle \mathcal{V}; \widetilde{\mathbf{m}} \right\rangle = \mathbf{m}, \left\langle \mathcal{V}; \widetilde{S} \right\rangle = S. 
    \end{gather*}
    \item \textbf{Compatibility of the energy and Reynolds stress defects.} There exists a non-increasing function $\mathcal{E}:[0,T) \to [0,\infty)$ such that
    \begin{eqnarray*}
        \mathcal{E}(0-) &=& \int_{\Omega} E(\varrho_0,\mathbf{m}_0,S_0) \dbx{}, \\
        \mathcal{E}(\tau+) &=& \int_{\Omega} E(\varrho,\mathbf{m},S)(\tau,\cdot) \dbx{} + \mathfrak{E} \text{ for any } 0 \leq \tau < T.
    \end{eqnarray*}
    Here $\mathfrak{E} \in L^{\infty}\paraL{(0,T;\mathcal{M}^+\paraL{\Omega}}$ is the energy defect satisfying
    \begin{equation*}
        \min \{ 2, d (\gamma - 1) \} \mathfrak{E} \leq \text{trace} [\mathfrak{R}] \leq \max \{ 2, d (\gamma - 1) \} \mathfrak{E}. 
    \end{equation*}
\end{itemize}
\end{definition}

Numerical solutions obtained by structure-preserving numerical schemes may, in general, not converge strongly \cite{FeireislHofmanova2020}. This typically arises in turbulent flows, such as the Kelvin-Helmholtz problem. The Reynolds stress defect plays an important role in such oscillatory flows. It is known \cite{FeireislLukacova_book2021} that the Reynolds stress vanishes and the limit is a weak solution if and only if the convergence is strong. Thus, the turbulent solutions, which are the (weak$-*$) limits of consistent approximations, are DW solutions and not weak solutions of the Euler system. Even if the convergence to DW solutions is not strong, it is desirable to recover strong convergence at least for some observables. The concept of $\mathcal{K}-$ convergence, see Koml\'{o}s \cite{Komlos1967}, which utilises the \textit{Ces\`{a}ro} averages that are obtained by averaging the numerical realisations over different mesh resolutions, allows to obtain the desired strong convergence, see Feireisl, Luk\'{a}\v{c}ov\'{a}~et.~al.\ \cite{FeireislLukaocvaMizerova2019,Feireisletal2021,LukacovaSheYuan2025}. This averaging procedure mimics the Strong Law of Large Numbers in probability, and $\mathcal{K}-$ convergence also yields strong convergence in space and time for the associated Young measures. 

\subsection{Selection criteria}

\par Clearly, the Euler system is also non-unique in the class of DW solutions. Therefore, we are motivated to identify additional criteria that select physically relevant solutions. B\v{r}ezina and Feireisl introduced the concept of admissible measure-valued solutions for the Euler system in \cite{BrezinaFeireisl2017}, and established their existence for any suitable initial data. These admissible solutions maximise entropy production among all solutions emanating from given initial data. 

The dynamical systems approach, originating from the Markov selection in stochastic analysis, focuses on solution paths of bounded variation in time that take values in the phase space. This method was first studied by Breit, Feireisl, and Hofmanová \cite{BreitFeireislHofmanova2020} for the isentropic Euler system, and later extended to the full Euler system in \cite{BreitFeireislHofmanova2020b}. The authors proved the existence of a selection procedure such that the chosen solutions are maximal in the sense of maximal entropy production, and satisfy the semi-group property with respect to time.
The selection is implemented by maximising certain functionals.
Specifically, the DW solutions $[\varrho,\mathbf{m},S]$ with initial data $[\varrho_0,\mathbf{m}_0,S_0]$ are selected if they satisfy
\begin{multline}
\label{functionals}
    \int_0^\infty \exp (-\lambda t)\beta (\varrho(t,\cdot),\mathbf{m}(t,\cdot),S(t,\cdot)) \dt{}  \\  
    \leq \int_0^\infty \exp (-\lambda t) \beta (\tilde{\varrho}(t,\cdot),\tilde{\mathbf{m}}(t,\cdot),\tilde{S}(t,\cdot)) \dt{}, \ \forall [\tilde{\varrho},\tilde{\mathbf{m}},\tilde{S}] \in \mathcal{U} [\varrho_0,\mathbf{m}_0,S_0].
\end{multline}
Here $\beta$ is a bounded continuous function, $\lambda>0$, and $\mathcal{U} [\varrho_0,\mathbf{m}_0,S_0]$ denotes the set of all DW solutions emanating from the initial data $[\varrho_0,\mathbf{m}_0,S_0]$. By assigning special functions $\beta$, the admissible DW solutions can be selected. Nevertheless, different choices of $\beta$ and $\lambda$ may lead to different solution semiflows, and there is no further a priori guidance on how to select $\lambda$ and $\beta$. Thus, this semiflow selection cf.~\cite{BreitFeireislHofmanova2020,BreitFeireislHofmanova2020b} yields infinitely many selection steps.

In \cite{FeireislJungelLukacova2025}, Feireisl, J\"{u}ngel, Luk\'{a}\v{c}ov\'{a} proposed a two-step selection process in the class of DW solutions for the isentropic Euler system. This consists of minimising the following functionals. First, the energy functional 
\begin{equation*}
    \int_0^\infty \exp (-t) \mathcal{E}(t) \dt{}
\end{equation*}
is minimised. Here, $\mathcal{E}$ is a scalar, time-dependent function, the so-called total energy. In the convergence analysis of structure-preserving numerical methods, $\mathcal{E}$ is obtained as a weak$^*$ limit of an approximate sequence of energies $\int_\Omega E_n \dx{} = \int_\Omega \left(\frac{1}{2}\frac{\modL{\mathbf{m}_n}^2}{\varrho_n} + \frac{1}{\gamma-1}\varrho_n^\gamma\right)\dx{}$, for $n \to \infty$. We point out that the total energy $\mathcal{E}$ is, in general, not strictly convex. Note, however, that for admissible weak solutions the total energy reduces to the mean energy, i.e. $\mathcal{E}(t) = \int_\Omega E(\varrho,\mathbf{m})\dx{} \ (t)$ a.e. $t \in (0,T)$. Due to the thermodynamic stability imposed on the equations of state $E(\varrho,\mathbf{m})$ is strictly convex.  

As shown in \cite{FeireislJungelLukacova2025}, if a unique solution is selected in the first step, it is an admissible weak solution. Otherwise, we proceed with the second selection step and select a truly measure-valued solution by minimising a suitable strictly convex functional on the set of DW solutions selected in the first step. In \cite{FeireislJungelLukacova2025}, it was suggested to minimise
\begin{equation*}
    \int_0^\infty \exp (-t) \left[ \norm{\varrho}_{L^q(\Omega)}^q (t) + \norm{\mathbf{m}}_{L^q(\Omega)}^q (t) + \modL{\mathcal{E}(t)}^q \right] \dt ,
\end{equation*}
where $1<q\leq \frac{2\gamma}{\gamma +1}$. 

For the full Euler system, the following two-step selection was proposed by Feireisl, Luk\'{a}\v{c}ov\'{a} in \cite{feireisllukacova2025}
\begin{gather*}
    \int_0^\infty \exp (-t) \paraL{ \int_\Omega -S(t,\cdot) \ \dx{} } \dt{}, \\
    \int_0^\infty \exp (-t) \paraL{\int_\Omega E(\varrho,\mathbf{m}, S) (t,\cdot) \ \dx{}} \dt{}. 
\end{gather*}
By combining the above two functionals, the following one-step selection, i.e., the minimisation of 
\begin{equation*}
    \int_0^{\infty} \exp (-t) \int_{\Omega} \left( E\left( \widetilde{\varrho},\widetilde{\mathbf{m}},\widetilde{S} \right) - \overline{\vartheta} \widetilde{S} \right) \, \dx{} \ \dt{} 
\end{equation*}
was also proposed in \cite{feireisllukacova2025}. It represents the Bregman distance between $(\widetilde{\varrho}, \widetilde{\mathbf{m}}, \widetilde{S})$ and the equilibrium solution $(\overline{\varrho},0,\overline{\theta})$ that maximizes the entropy production, $\overline{\varrho} = \frac{1}{\modL{\Omega}} M_0$. Here $\overline{\vartheta} = \frac{\mathcal{E}_0}{c_v M_0}, \mathcal{E}_0 = \int_{\Omega} \widetilde{E}(t=0,\mathbf{x}) \dx{}, M_0 = \int_{\Omega} \widetilde{\varrho}(t=0,\mathbf{x}) \dx{}, c_v = 2.5$. We refer to \cite{feireisllukacova2025} for further details on the derivation.  
Recently, another approach aimed at selecting “maximally turbulent” solutions was proposed by Klingenberg, Markfelder, and Wiedemann \cite{KlingenbergMarkfelderWiedemann2025}. 

The goal of this paper is to discuss numerical possibilities of the selection criterion. For the Kelvin-Helmholtz problem, numerical solutions obtained from different schemes have been observed to converge strongly to different DW solutions via the $\mathcal{K}-$convergence mentioned earlier, see \cite{chu2025numerical}. Here, we compare four different numerical schemes, each with different polynomial degrees ($N=0,1,2,4,6,8$): the viscous finite volume (VFV) method, the discrete velocity Boltzmann finite volume (DVBFV) method, the Runge-Kutta discontinuous Galerkin (RKDG) method, and the discontinuous Galerkin spectral element method (DGSEM) with subcell resolution. We present the mesh-averaged densities computed by each method for the Kelvin–Helmholtz problem, each approximating a different and “truly” dissipative measure-valued solution. The latter is illustrated by computing the defects, which are nonzero. Next, guided by the selection criteria discussed above, we compute the entropy production, total energy, and energy defect. We also examine the behaviour of these solutions with respect to functionals involving different choices of $\beta$ and $\lambda$. In the present paper, we are interested in the cases where $\beta= S$ and $\beta = E (\varrho,\mathbf{m},S)$. In particular, we consider two values of $\lambda$: $\lambda=0$ and $\lambda=1.0$. Further details will be provided in \Cref{Sec: criteria}. It is worth noting that if a solution is selected for all $\lambda >0$, then it is maximal in the sense of Dafermos' \emph{local} entropy production principle, see \cite{BreitFeireislHofmanova2020b}. However, we have found that such maximal solutions must be weak solutions, as shown in \cite{FeireislJungelLukacova2025,FeireislLukacovaYu2025}, and they generally do not align with oscillatory solutions that were observed, for example, in our numerical simulation for the Kelvin-Helmholtz problem. Moreover, we will demonstrate that the values of the selection functionals are highly sensitive to the numerical diffusion inherent in each numerical method.

\par The rest of the paper is organised as follows: In \Cref{Sec: Plots and Convergence}, we describe the Kelvin-Helmholtz problem and present the plots of numerical densities and their \textit{Ces\`{a}ro} averages obtained by the studied numerical schemes. We also show the experimental order of convergence in the $L_1$ error norm for both density and its \textit{Ces\`{a}ro} average. Further, in \Cref{Sec: criteria}, we present numerical realisation of the selection functionals, and compare the different numerical methods (detailed in Appendix~\ref{Sec: Num schemes}) with respect to these functionals. 

\section{Numerical studies of the Kelvin-Helmholtz problem} 
\label{Sec: Plots and Convergence}
In the following numerical test, we choose the computational domain to be $\Omega = [0, 1]^2$ with periodic boundary conditions. The initial values are given by
\begin{equation*}
    \left(\varrho,\frac{\mathbf{m}}{\varrho},p\right) = \begin{cases}
         (2,-0.5, 0, 2.5), &\quad \text{if}\quad I_1 \leq y \leq I_2\\
         (1, 0.5, 0, 2.5), &\quad \text{otherwise}
    \end{cases},
\end{equation*}
where the interface profiles, 
\begin{equation*}
    I_j = I_j (x) = J_j + \epsilon Y_j (x), \ j = 1, 2,
\end{equation*}
are chosen to be small perturbations around the lower $y = J_1 = 0.25$ and the upper $y = J_2 = 0.75$ interfaces, respectively. The function $Y_j$ is given as,
\begin{equation*}
    Y_j(x) = \sum_{k=1}^m a_j^k cos\left(b_j^k + 2\pi k x\right), \quad j=1,2,
\end{equation*}
where $a_j^k \in [0, 1]$ and $b_j^k$ , $j = 1, 2$, $k = 1, \dots , m$ are arbitrary but fixed numbers given in \Cref{tab:init-data}. The coefficients $a_j^k$ have been normalized such that $\sum_{k=1}^m a_j^k=1$, in order to guarantee that $\modL{I_j(x) -  J_j} \leq \epsilon$ for $j = 1, 2$. Here, the parameters are set as $m = 10$, $\epsilon = 0.01$, and the value of the ratio of specific heats  $\gamma$ is $1.4$.
\par
\begin{figure}[!ht]
    \centering
    \subfloat[$t=1$]{\label{fig:VFV_N1_t1}%
        \includegraphics[width=.333\linewidth]{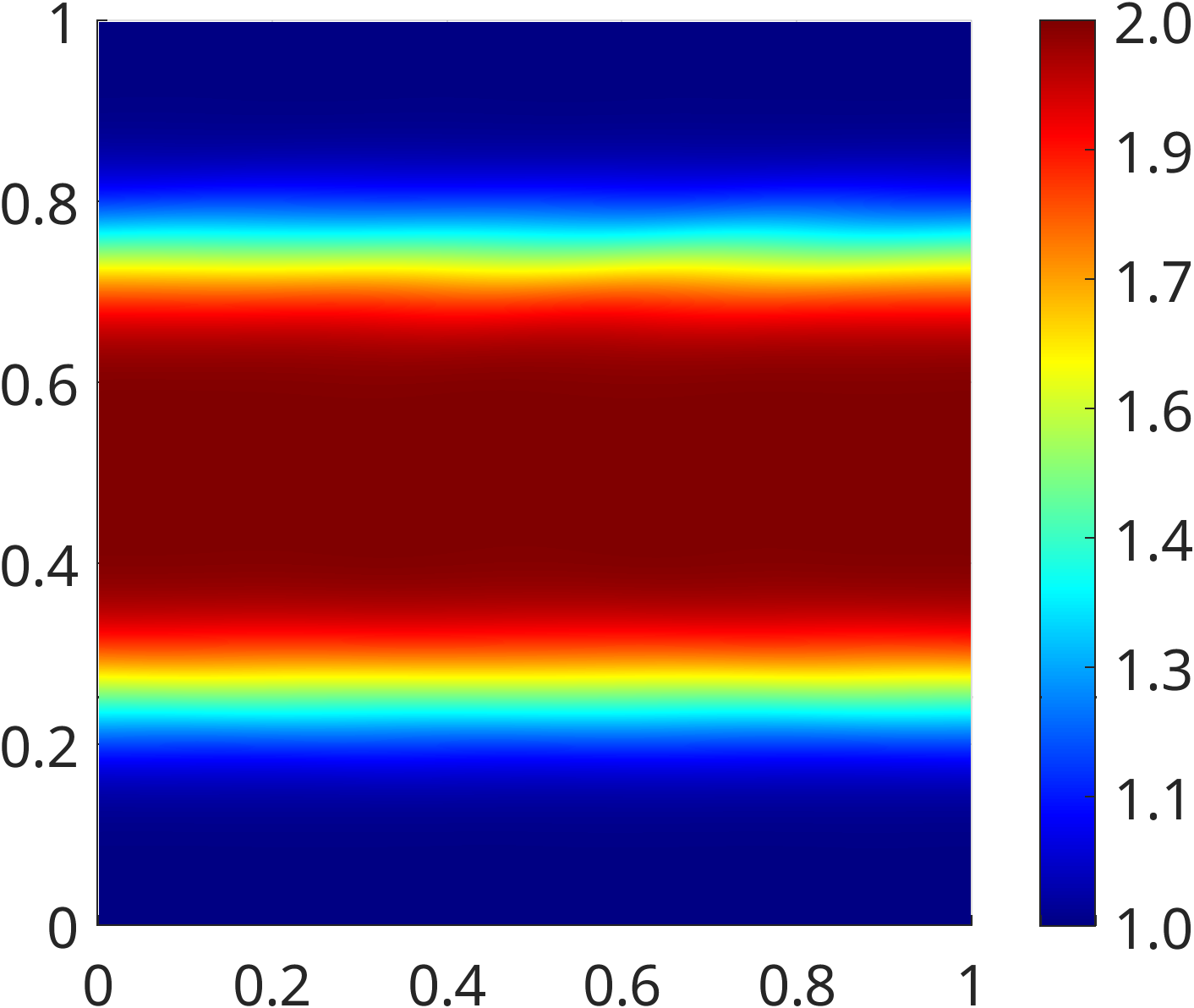}}\hfill
    \subfloat[$t=2$]{\label{fig:VFV_N1_t2}%
        \includegraphics[width=.333\linewidth]{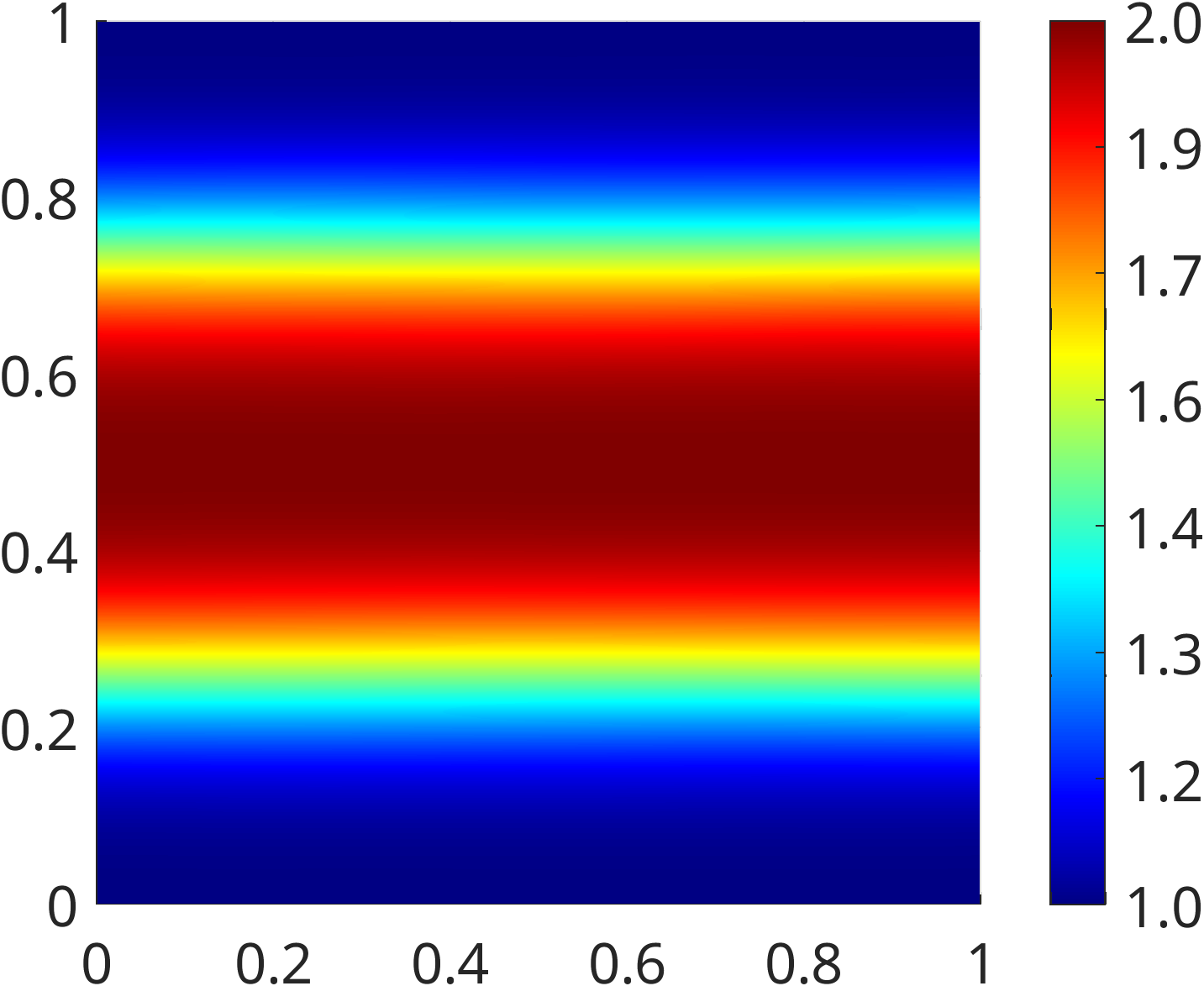}}\hfill
    \subfloat[$t=10$]{\label{fig:VFV_N1_t10}%
        \includegraphics[width=.333\linewidth]{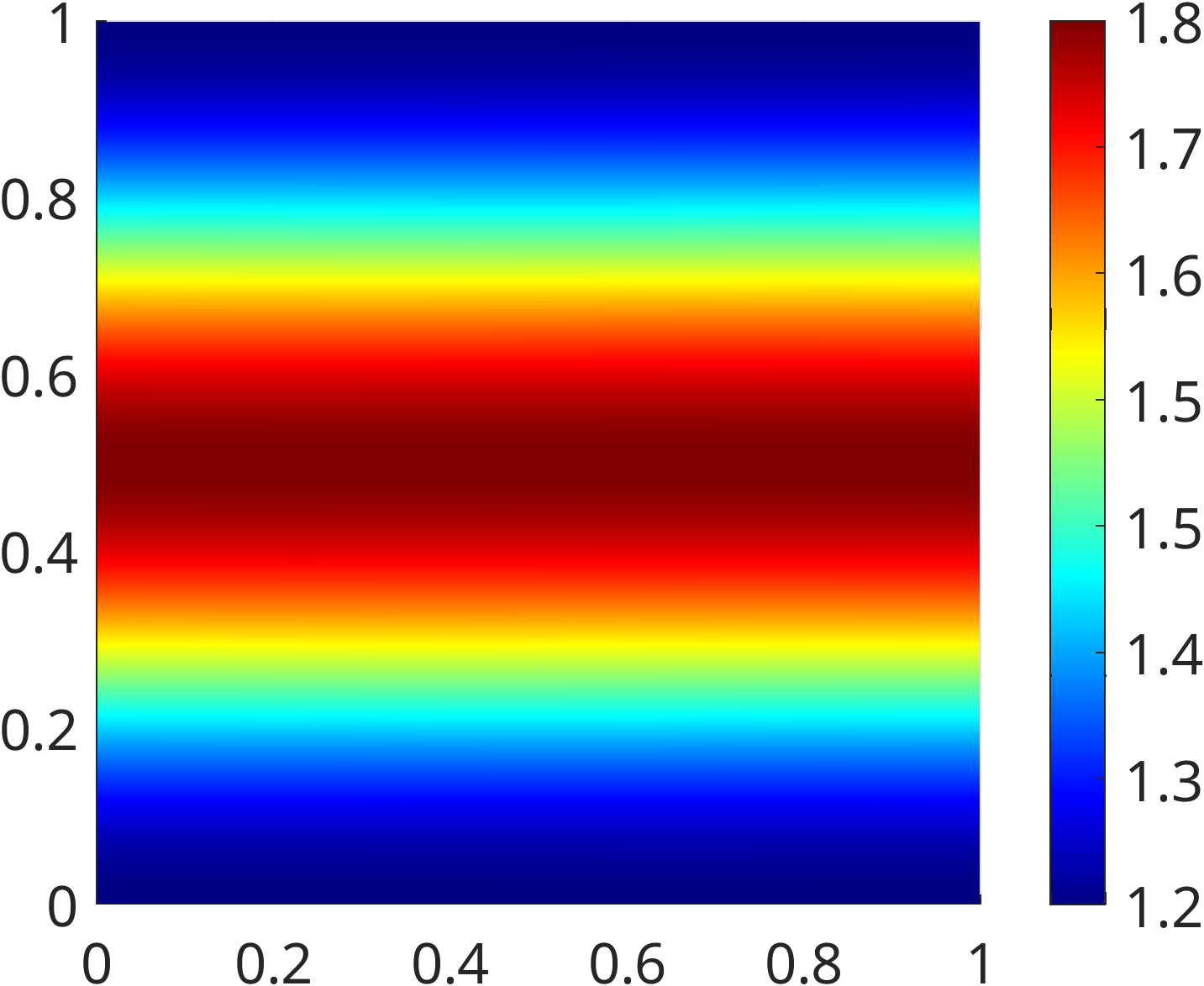}}
    \vspace{-8pt}
    \subfloat[$t=1$]{\label{fig:VFV_N1_avg_t1}%
        \includegraphics[width=.333\linewidth]{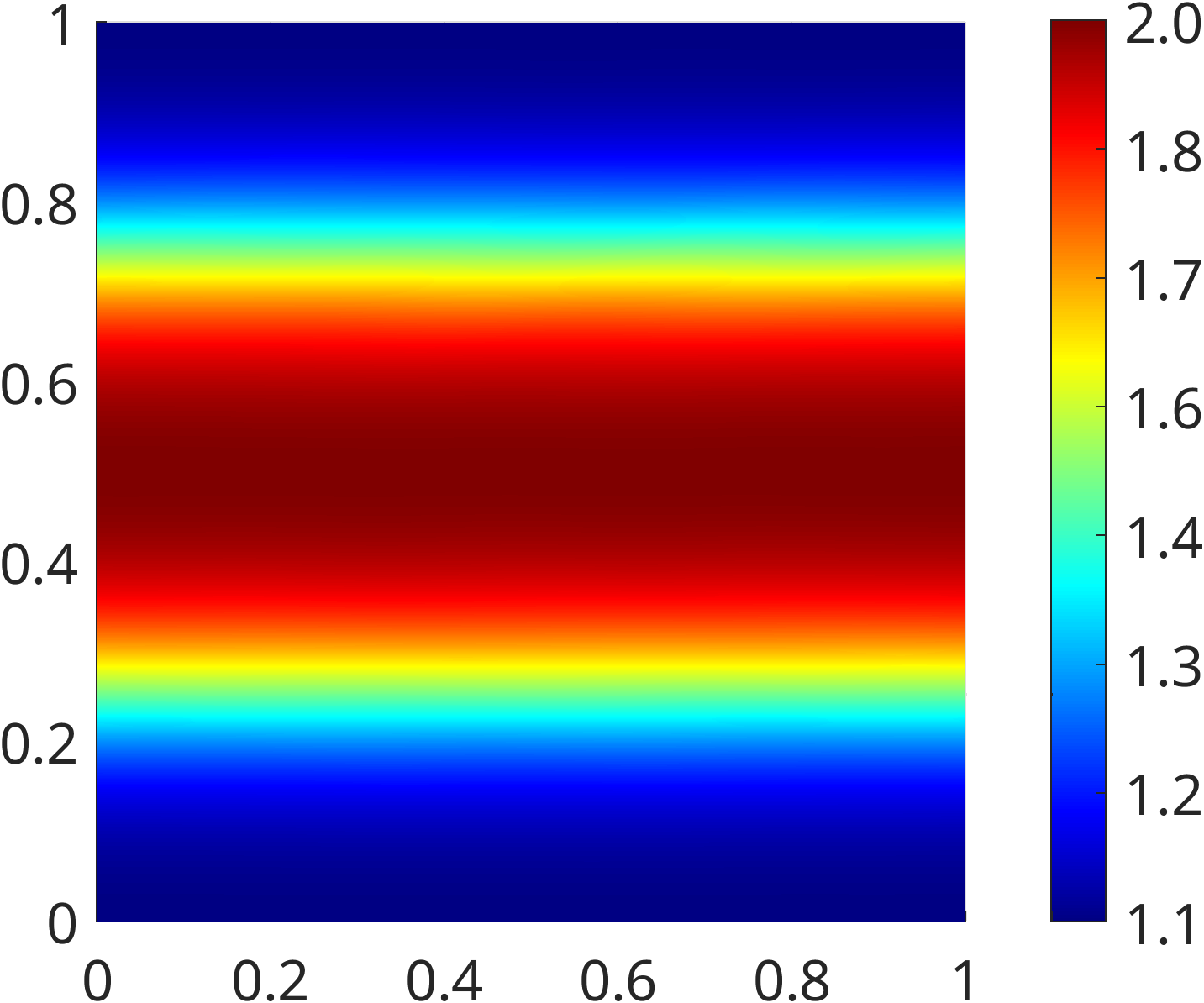}}\hfill
    \subfloat[$t=2$]{\label{fig:VFV_N1_avg_t2}%
        \includegraphics[width=.333\linewidth]{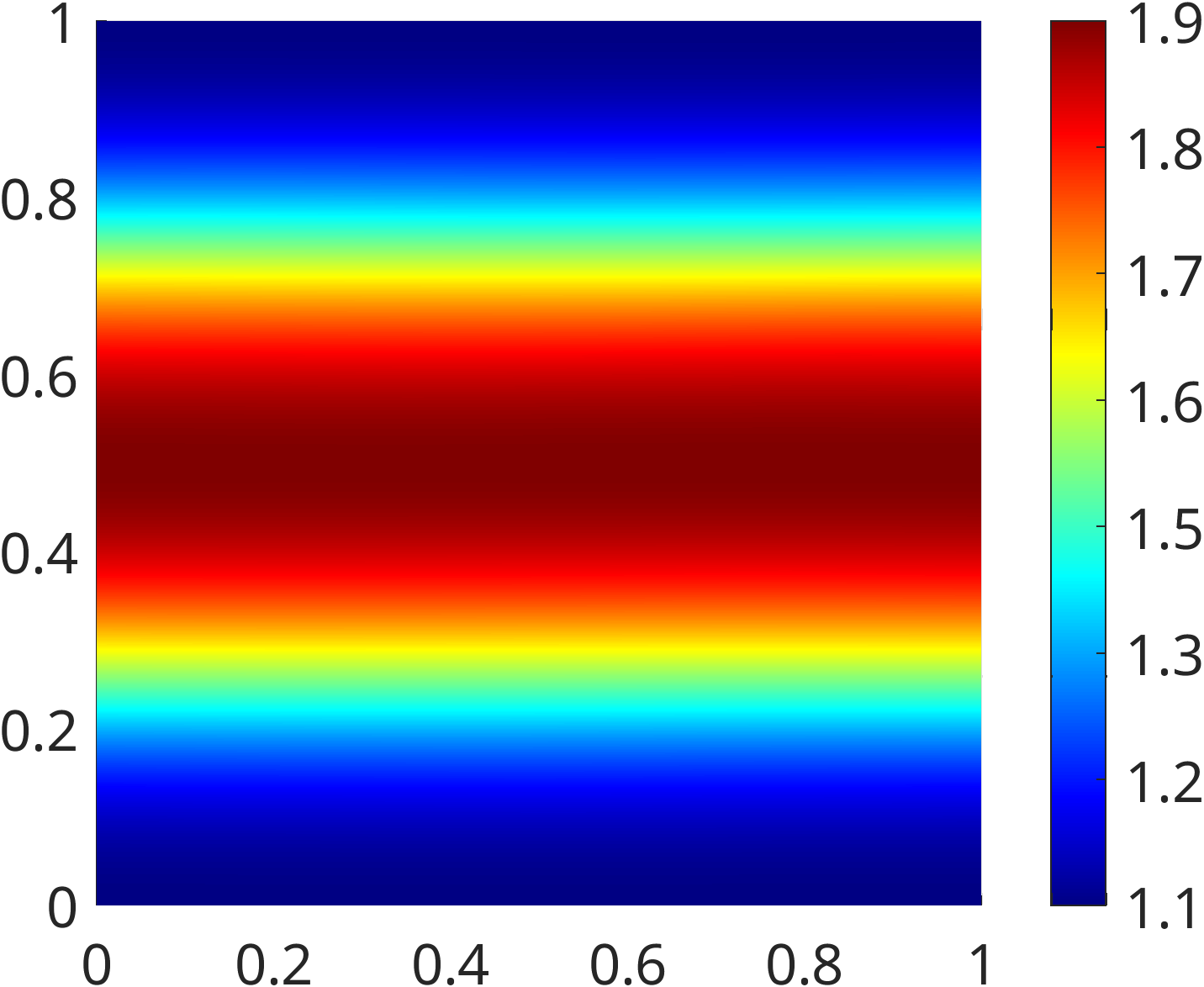}}\hfill
    \subfloat[$t=10$]{\label{fig:VFV_N1_avg_t10}%
        \includegraphics[width=.333\linewidth]{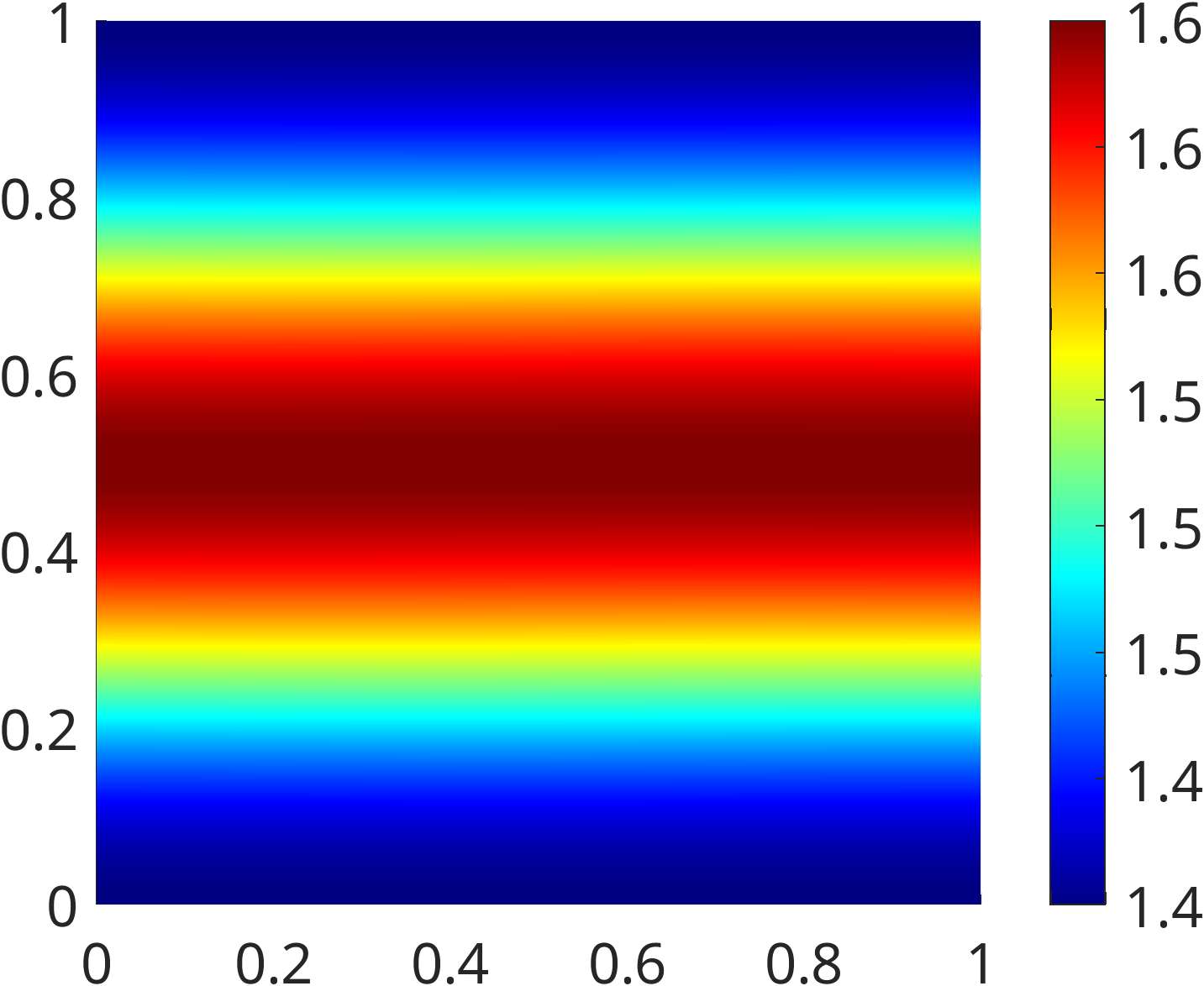}}\par
    \caption{Density (top row) and its \textit{Ces\`{a}ro} average (bottom row) for VFV method with $N=0$ at different times.}
    \label{fig:VFV_N1}
\end{figure}

\begin{figure}[ht]
    \centering
    \subfloat[$t=1$]{\label{fig:VFV_N2_t1}%
        \includegraphics[width=.333\linewidth]{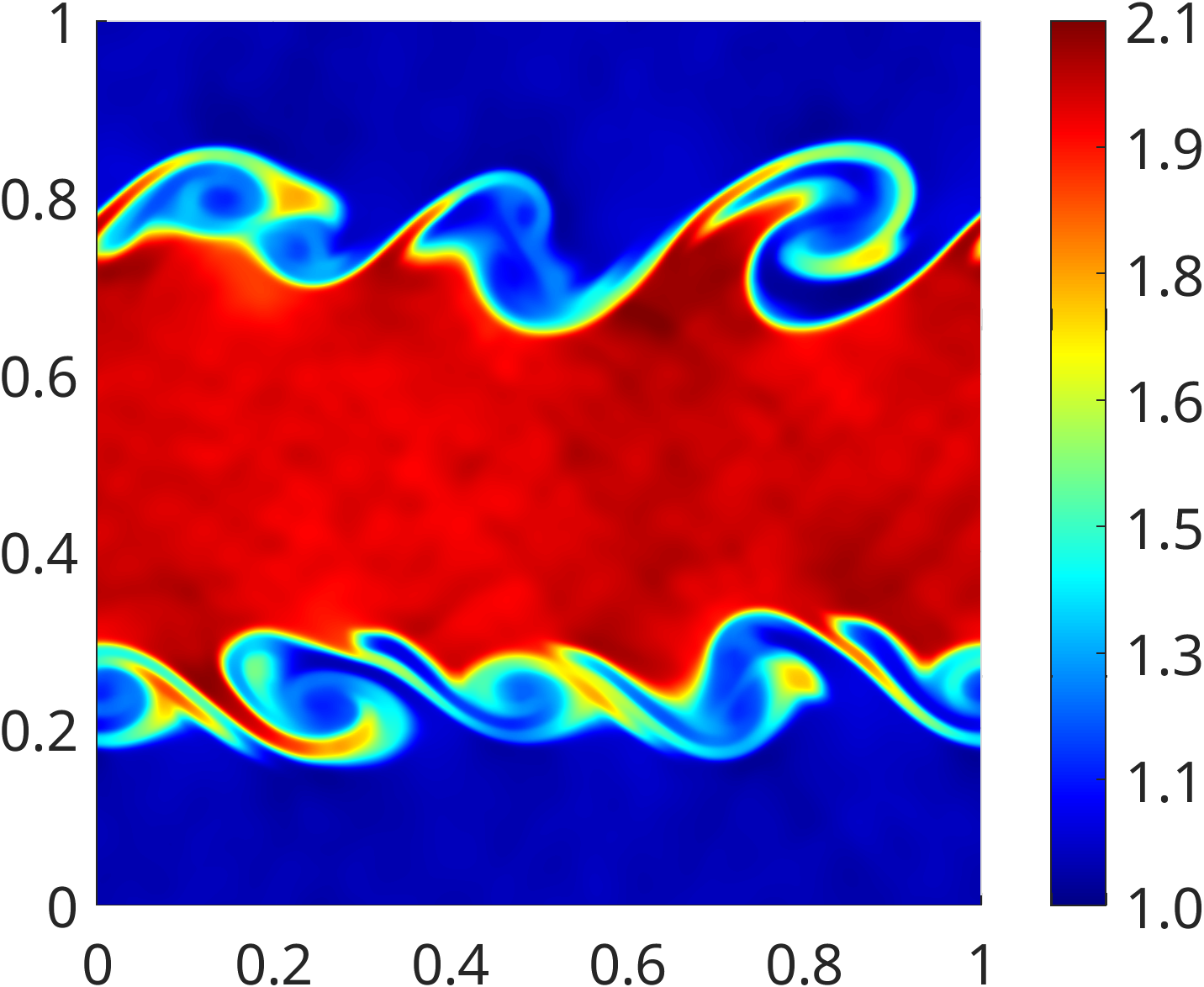}}\hfill
    \subfloat[$t=2$]{\label{fig:VFV_N2_t2}%
        \includegraphics[width=.333\linewidth]{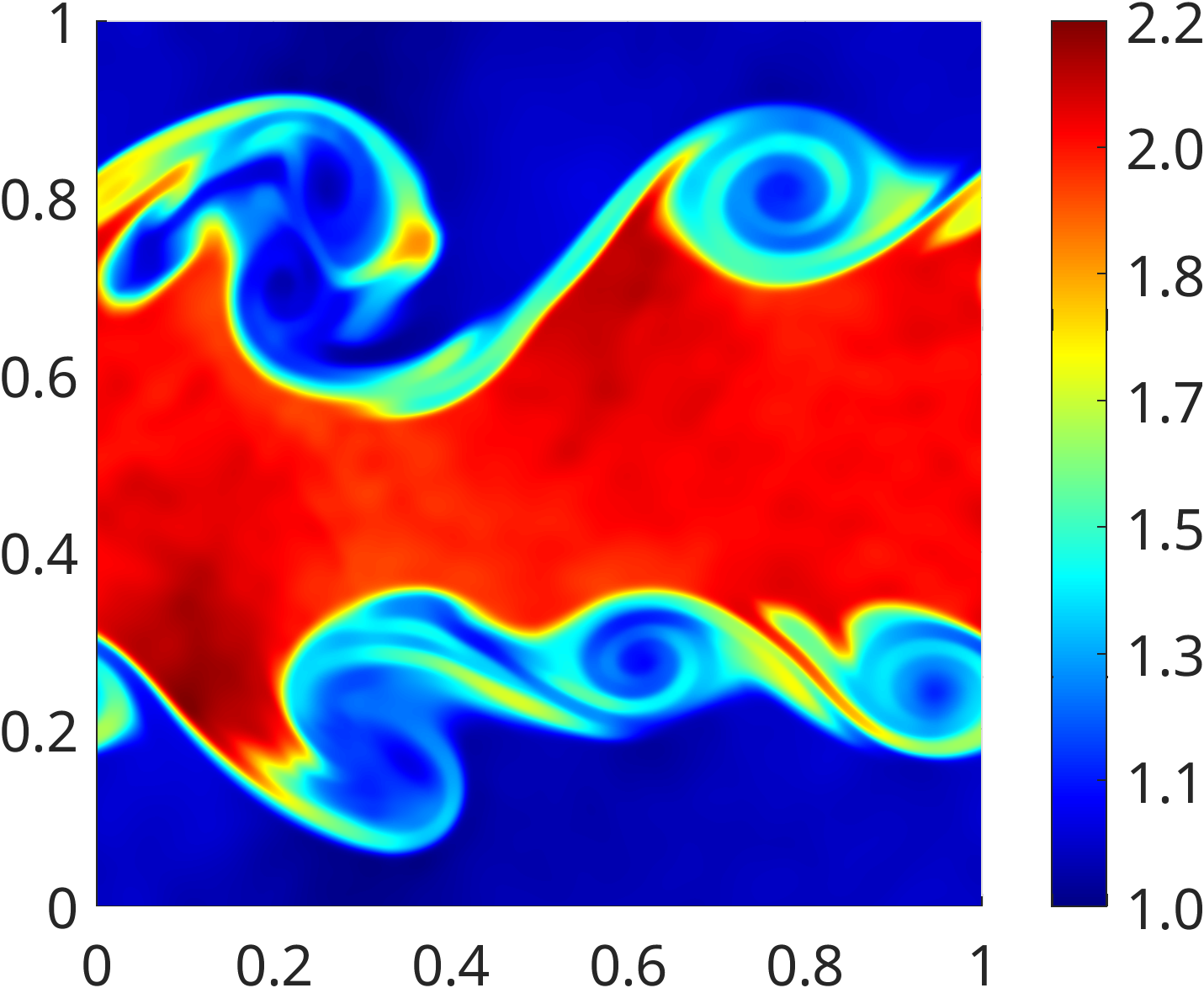}}\hfill
    \subfloat[$t=10$]{\label{fig:VFV_N2_t10}%
        \includegraphics[width=.333\linewidth]{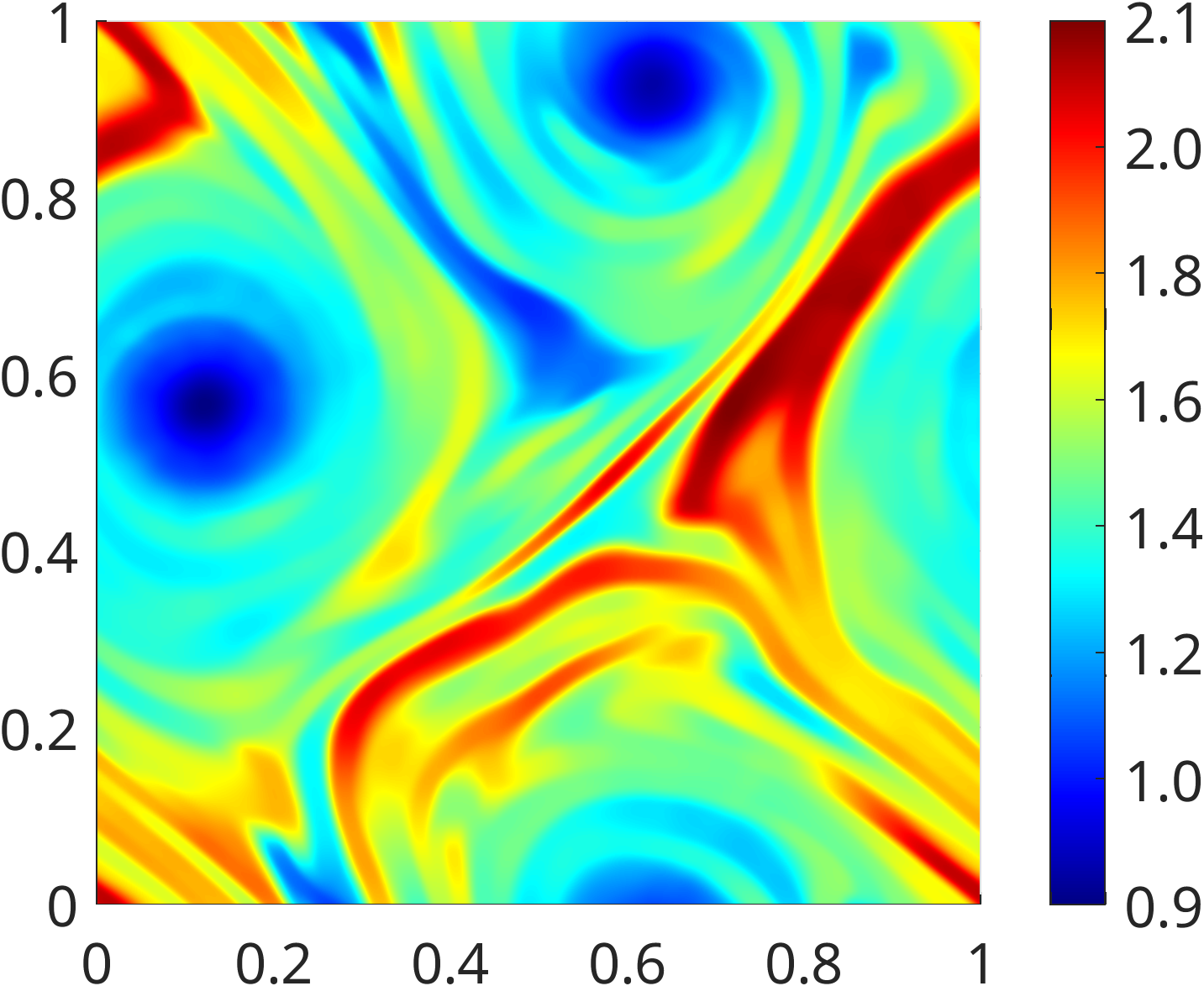}}
    \vspace{-8pt}
    \subfloat[$t=1$]{\label{fig:VFV_N2_avg_t1}%
        \includegraphics[width=.333\linewidth]{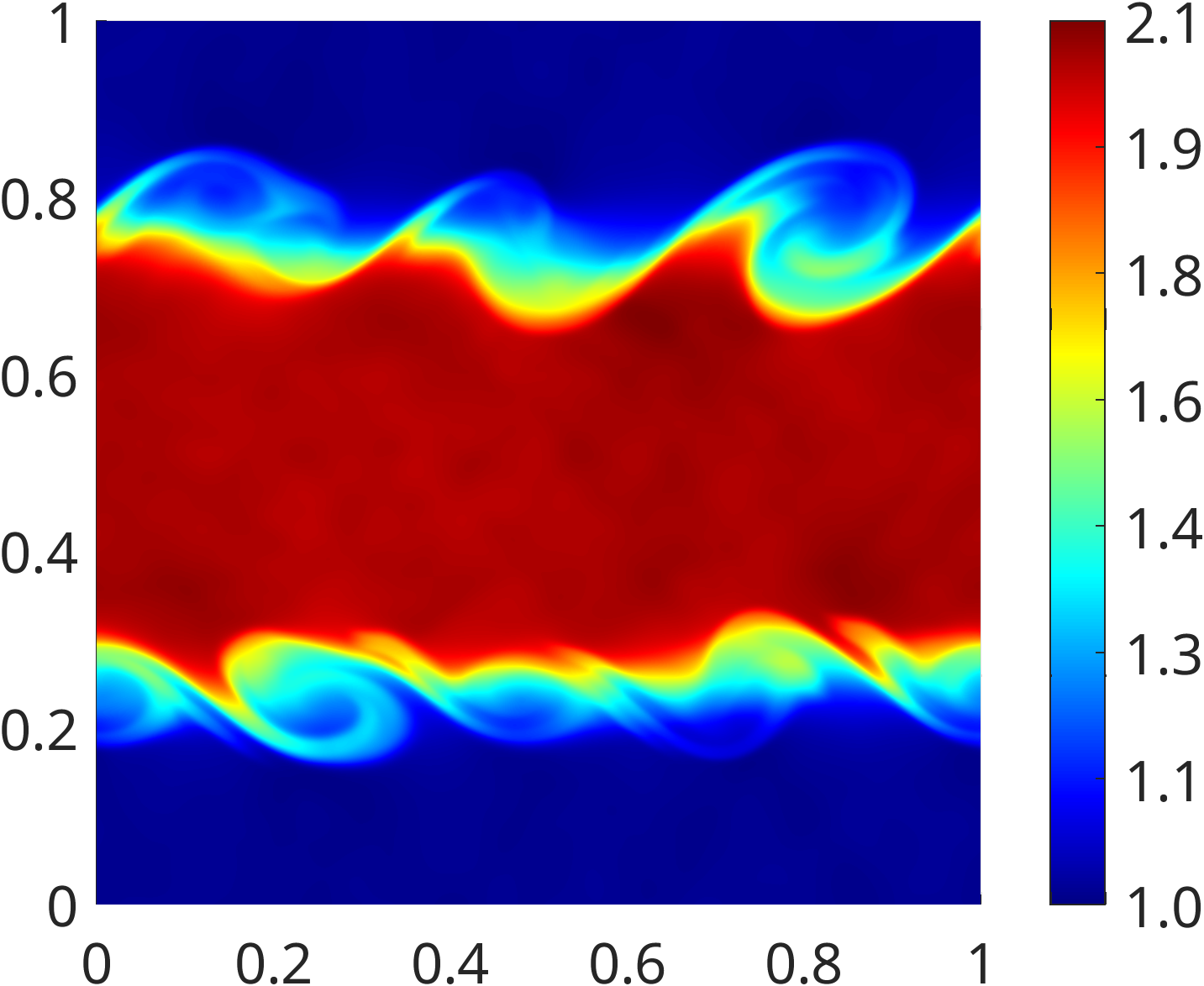}}\hfill
    \subfloat[$t=2$]{\label{fig:VFV_N2_avg_t2}%
        \includegraphics[width=.333\linewidth]{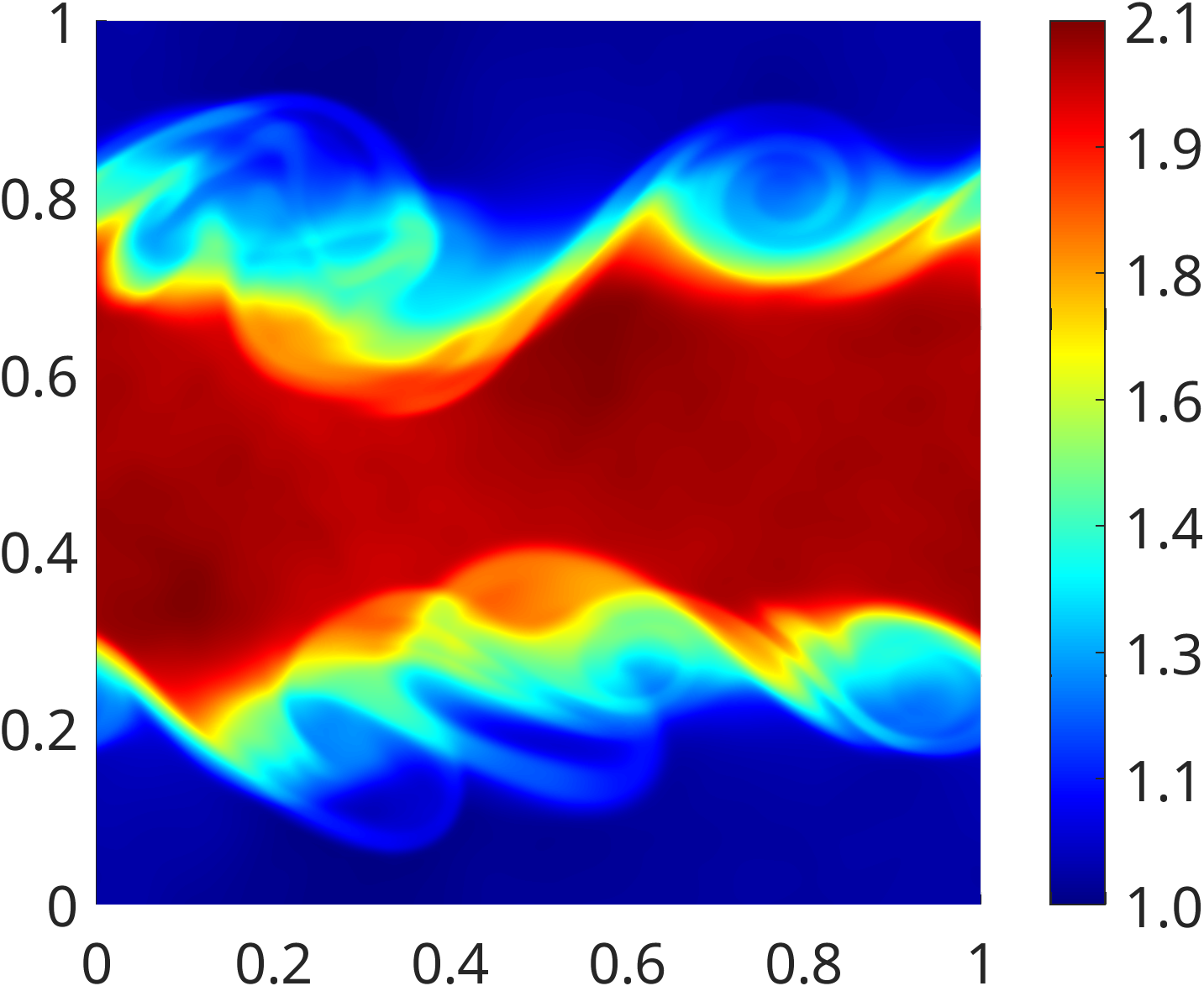}}\hfill
    \subfloat[$t=10$]{\label{fig:VFV_N2_avg_t10}%
        \includegraphics[width=.333\linewidth]{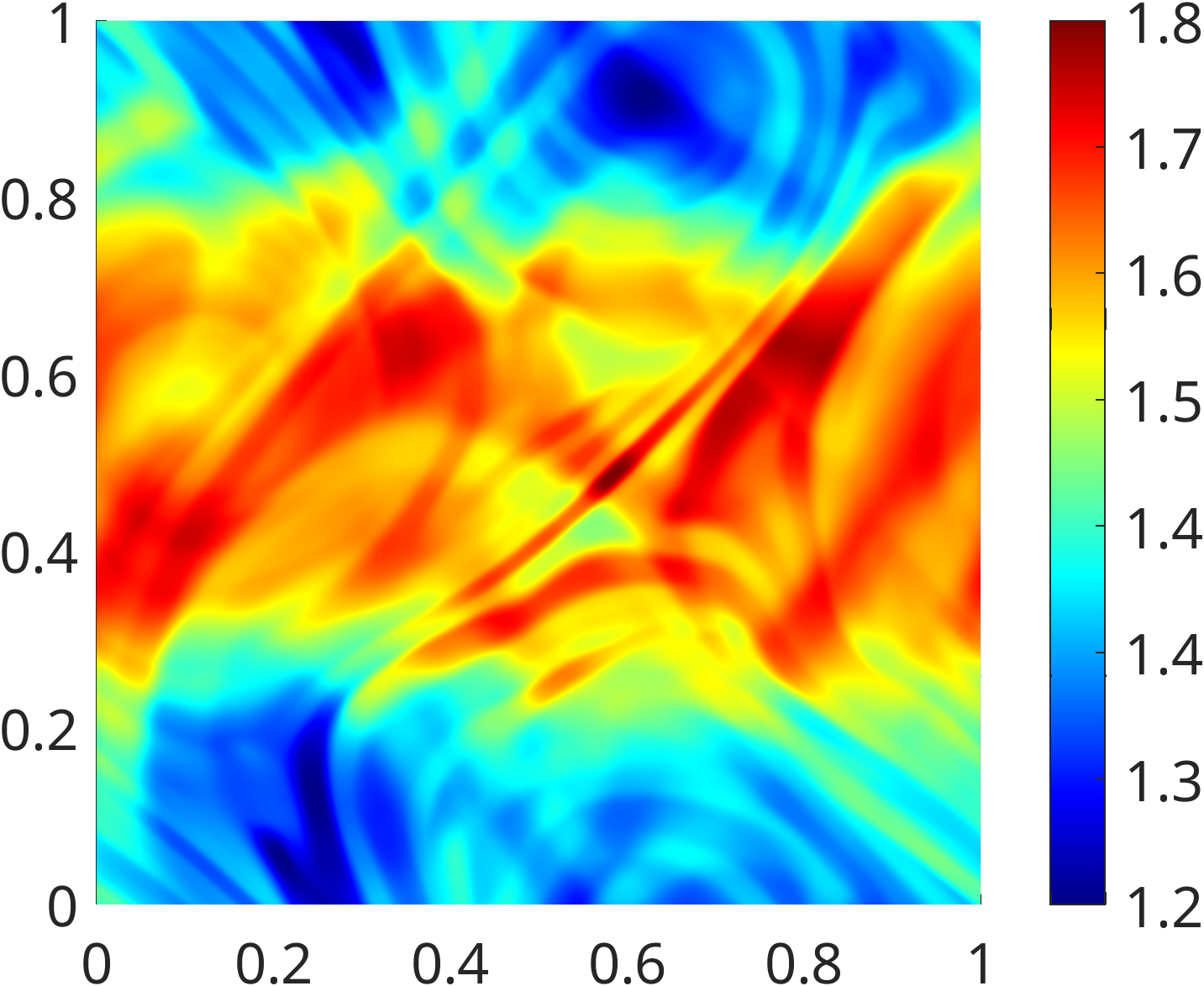}}\par
    \caption{Density (top row) and its \textit{Ces\`{a}ro} average (bottom row) for VFV method with $N=1$ at different times.}
    \label{fig:VFV_N2}
\end{figure}

\begin{figure}[ht]
    \centering
    \subfloat[$t=1$]{\label{fig:VFV_N3_t1}%
        \includegraphics[width=.333\linewidth]{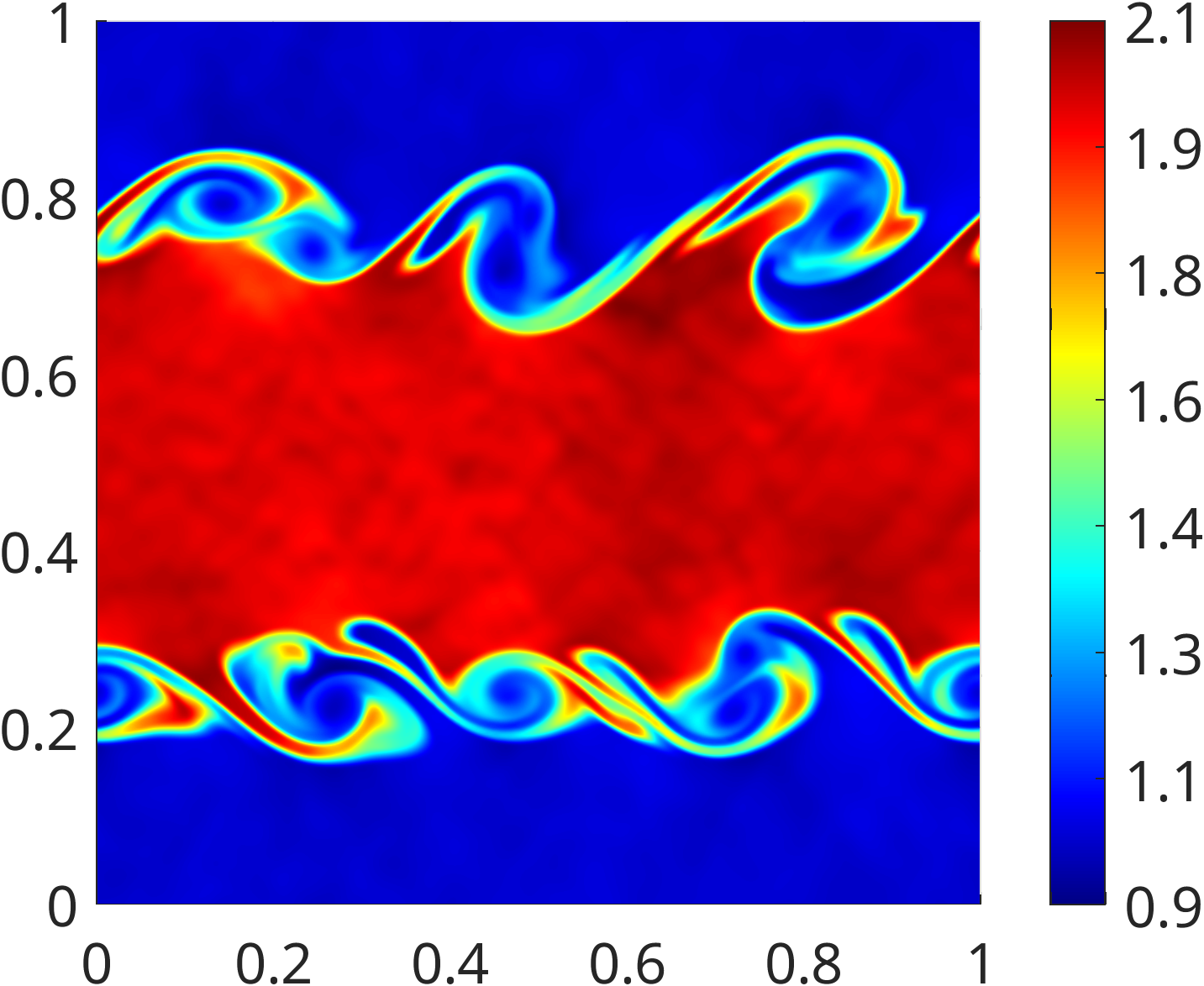}}\hfill
    \subfloat[$t=2$]{\label{fig:VFV_N3_t2}%
        \includegraphics[width=.333\linewidth]{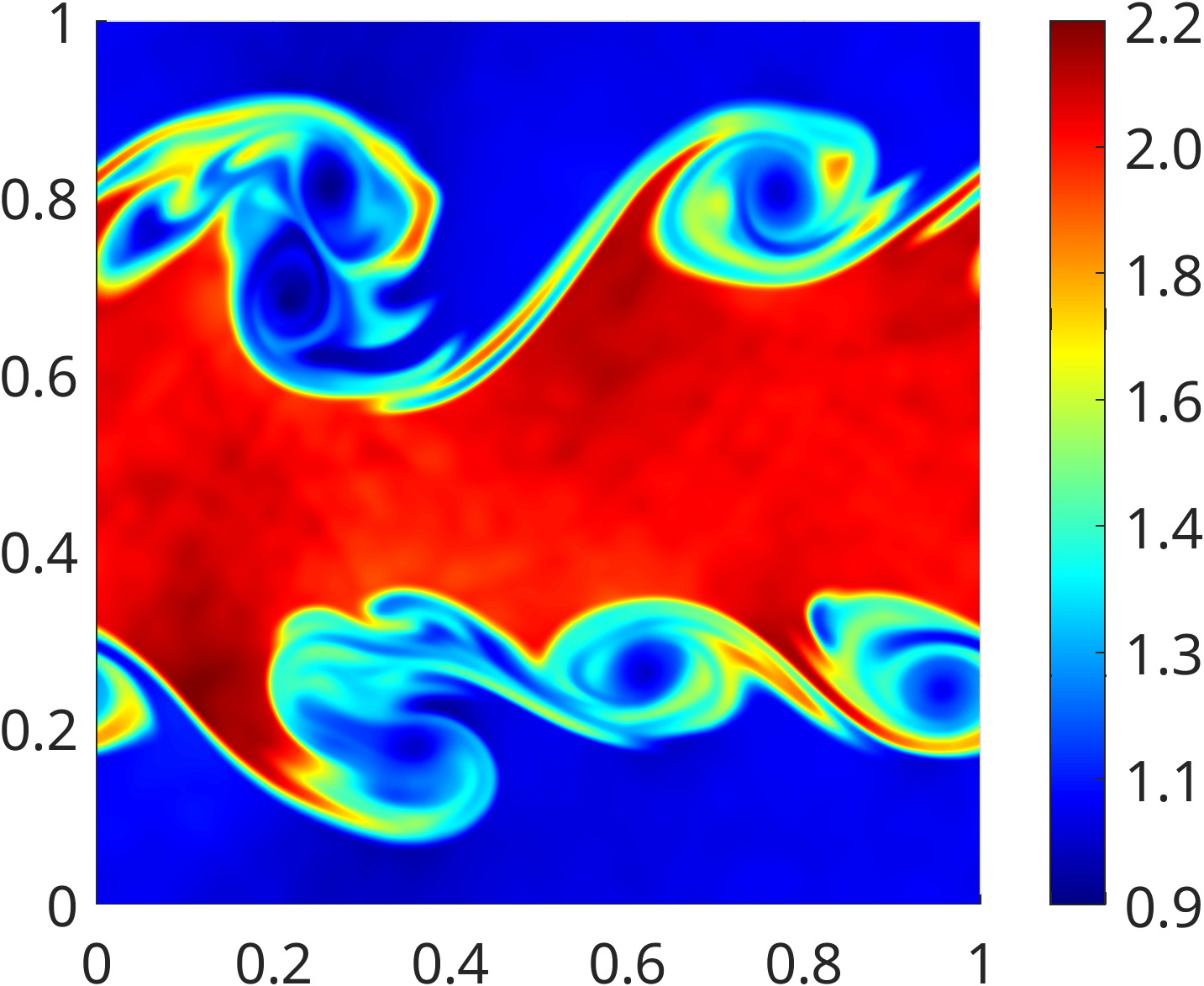}}\hfill
    \subfloat[$t=10$]{\label{fig:VFV_N3_t10}%
        \includegraphics[width=.333\linewidth]{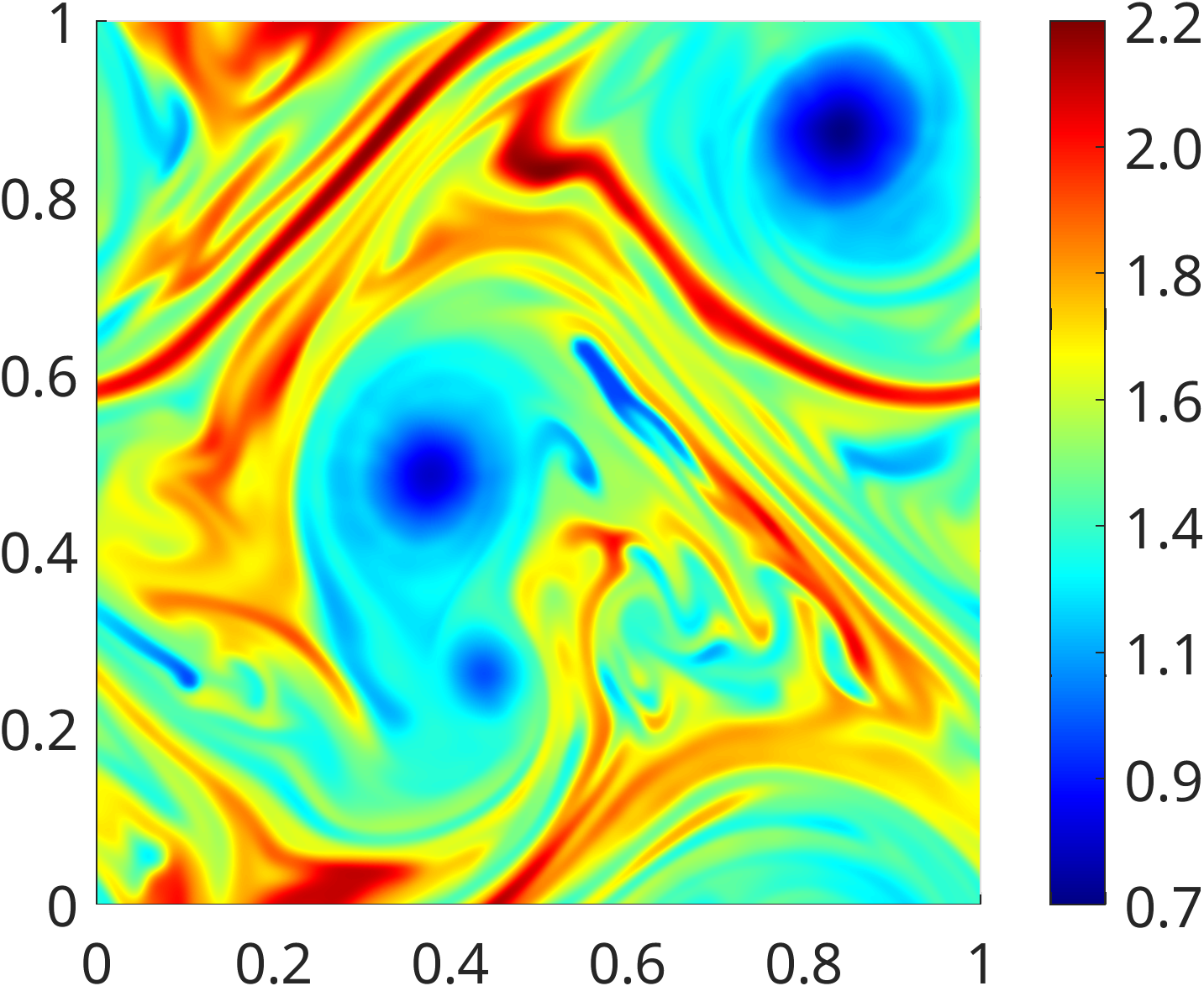}}
    \vspace{-8pt}
    \subfloat[$t=1$]{\label{fig:VFV_N3_avg_t1}%
        \includegraphics[width=.333\linewidth]{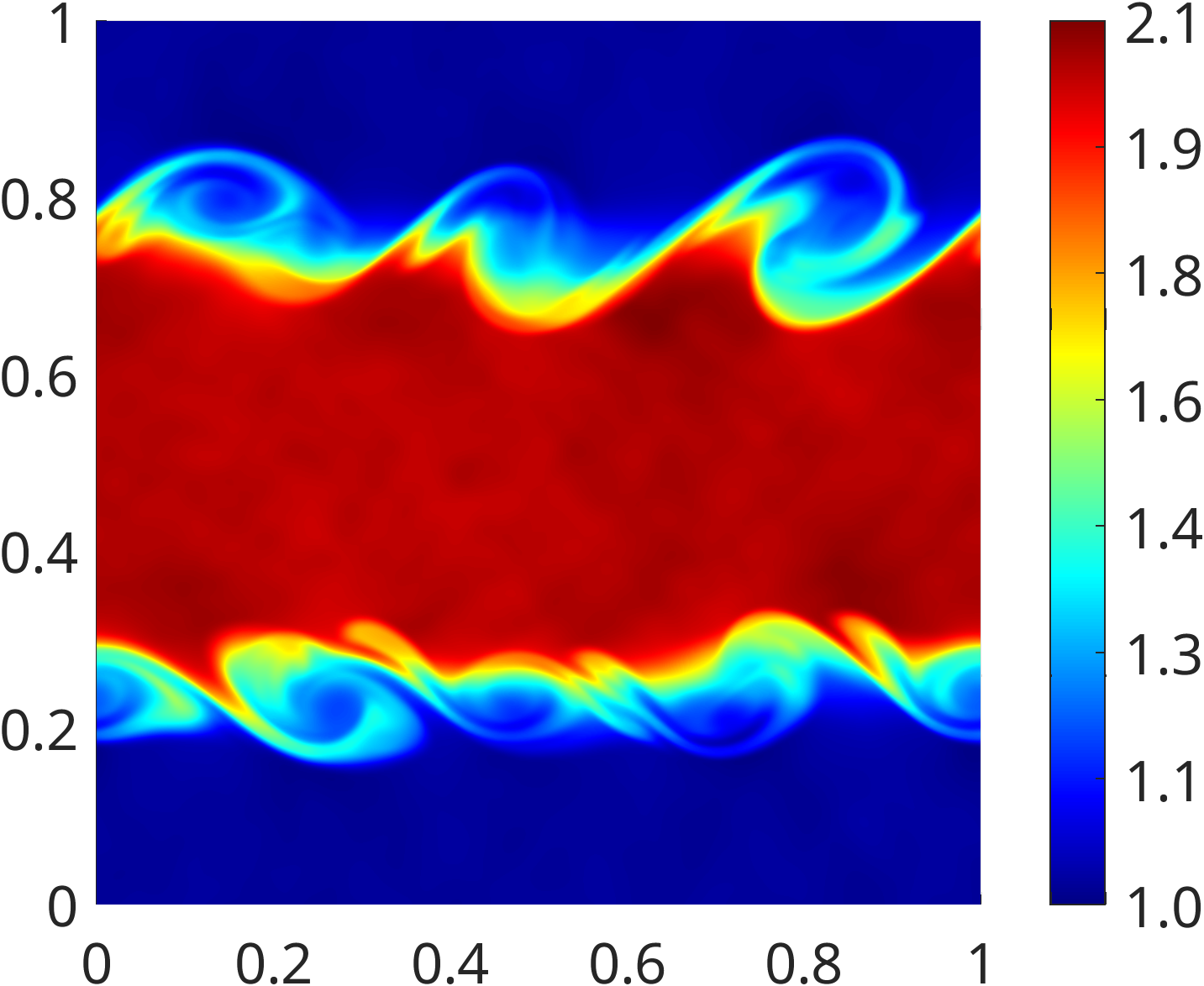}}\hfill
    \subfloat[$t=2$]{\label{fig:VFV_N3_avg_t2}%
        \includegraphics[width=.333\linewidth]{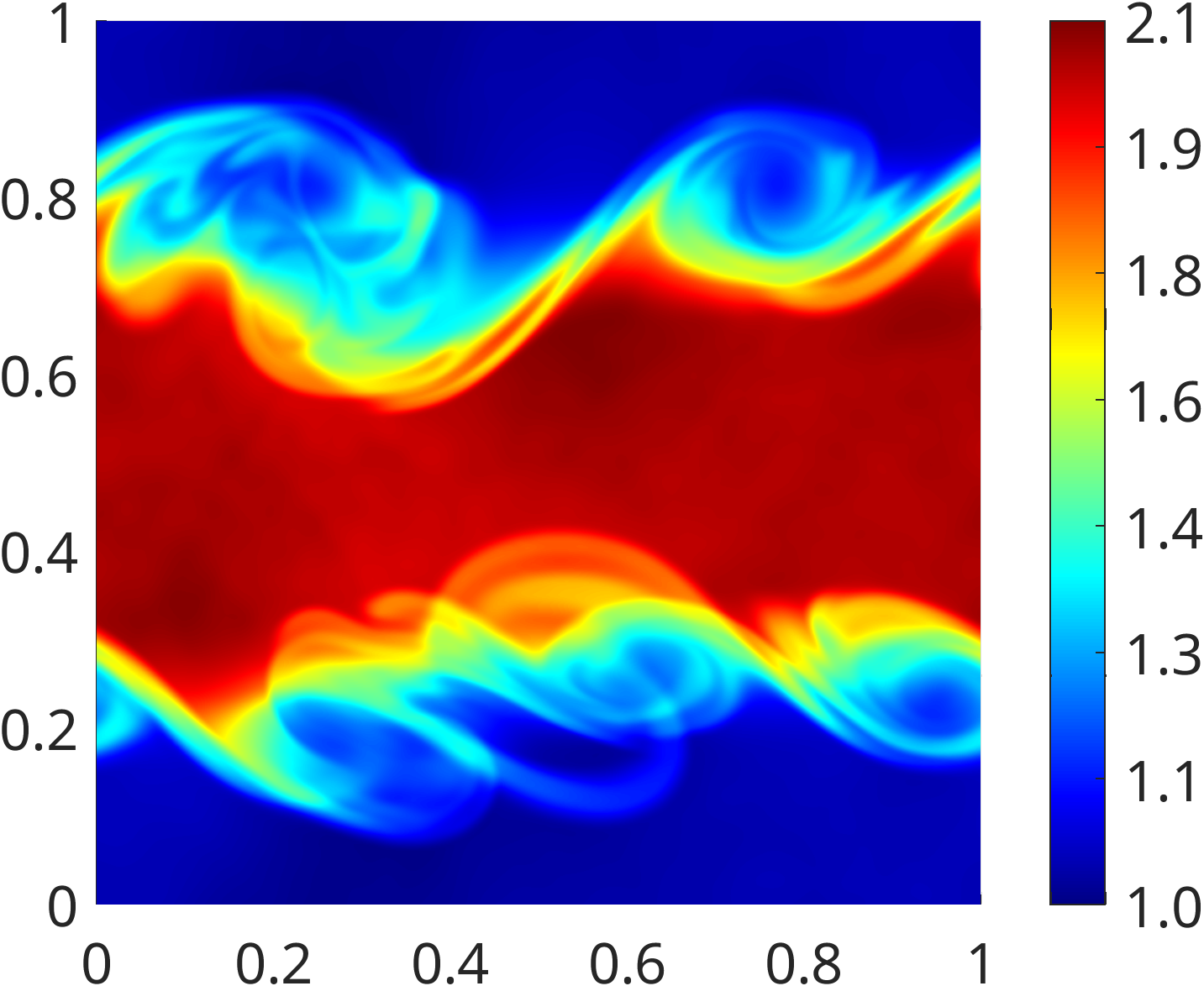}}\hfill
    \subfloat[$t=10$]{\label{fig:VFV_N3_avg_t10}%
        \includegraphics[width=.333\linewidth]{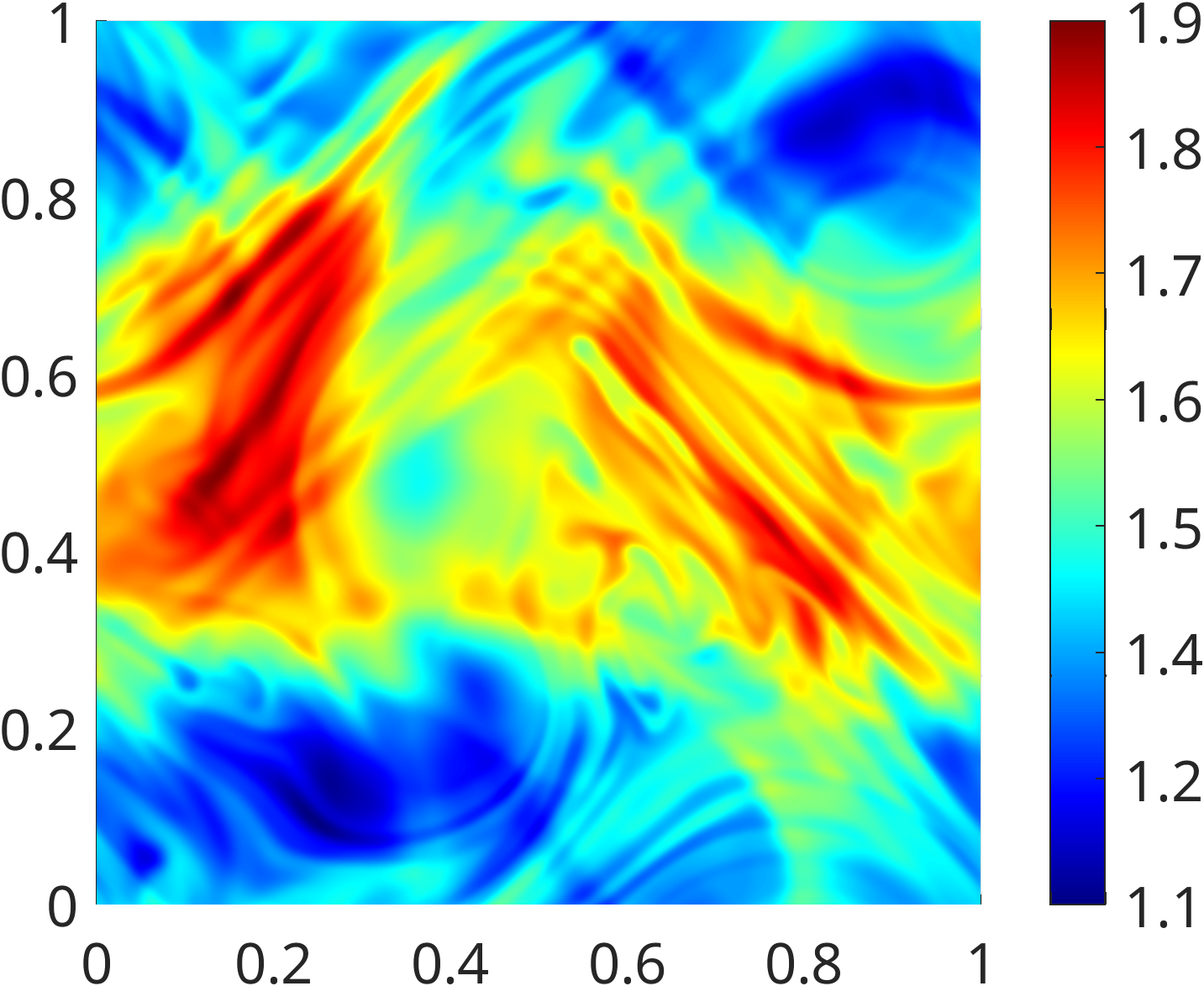}}\par
    \caption{Density (top row) and its \textit{Ces\`{a}ro} average (bottom row) for VFV method with $N=2$ at different times.}
    \label{fig:VFV_N3}
\end{figure}

\begin{figure}[ht]
    \centering
    \subfloat[$t=1$]{\label{fig:VFV_N5_t1}%
        \includegraphics[width=.333\linewidth]{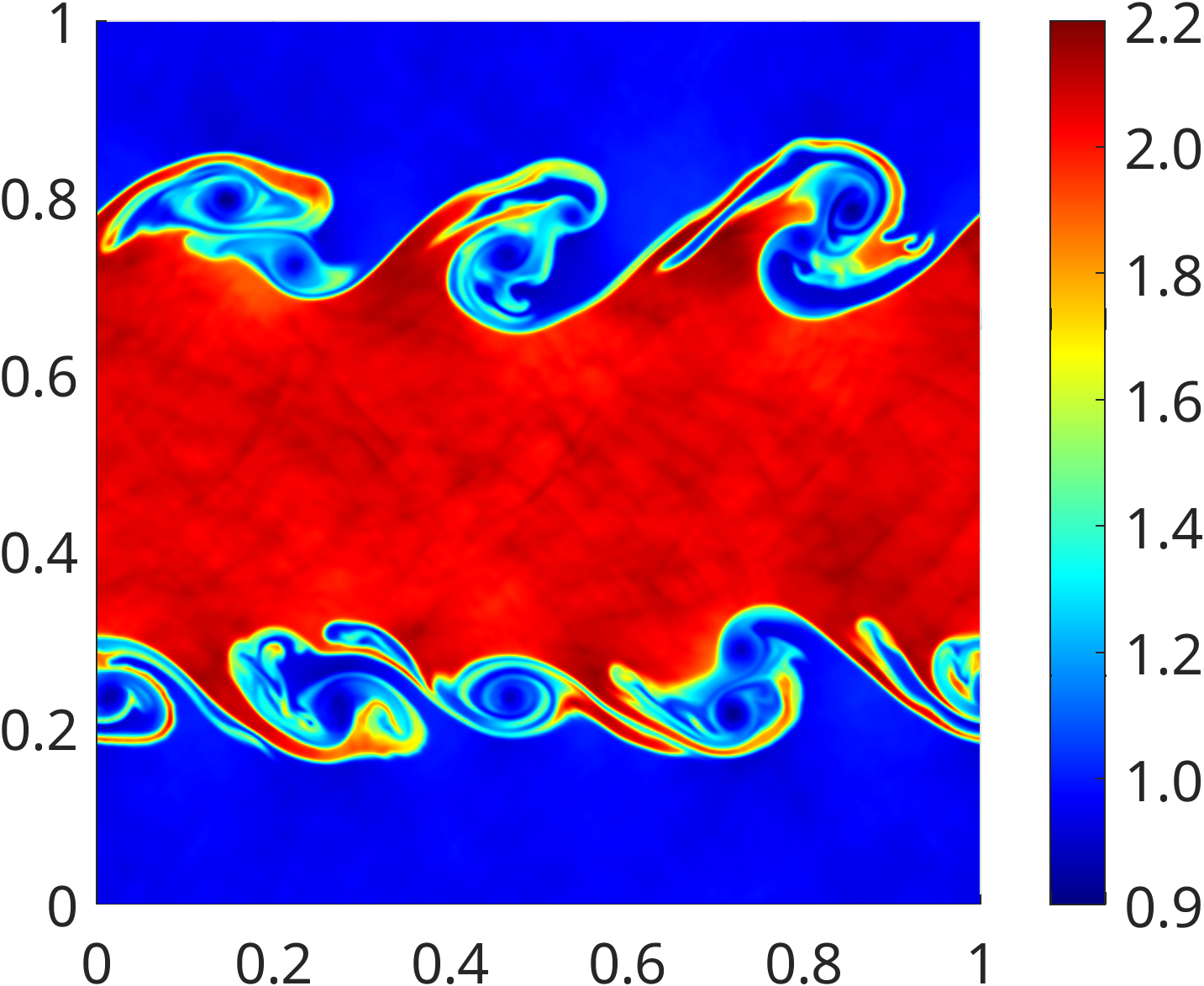}}\hfill
    \subfloat[$t=2$]{\label{fig:VFV_N5_t2}%
        \includegraphics[width=.333\linewidth]{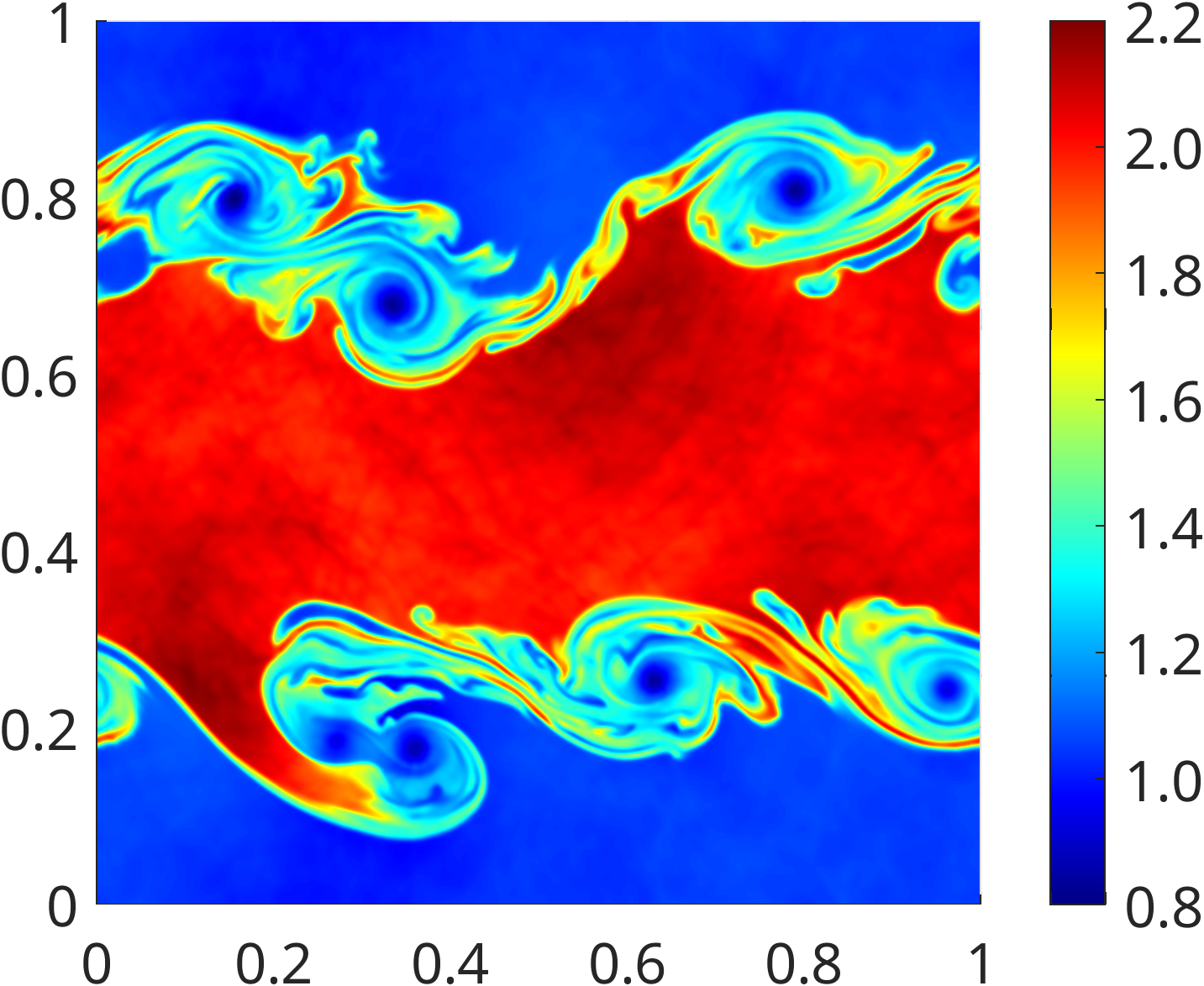}}\hfill
    \subfloat[$t=10$]{\label{fig:VFV_N5_t10}%
        \includegraphics[width=.333\linewidth]{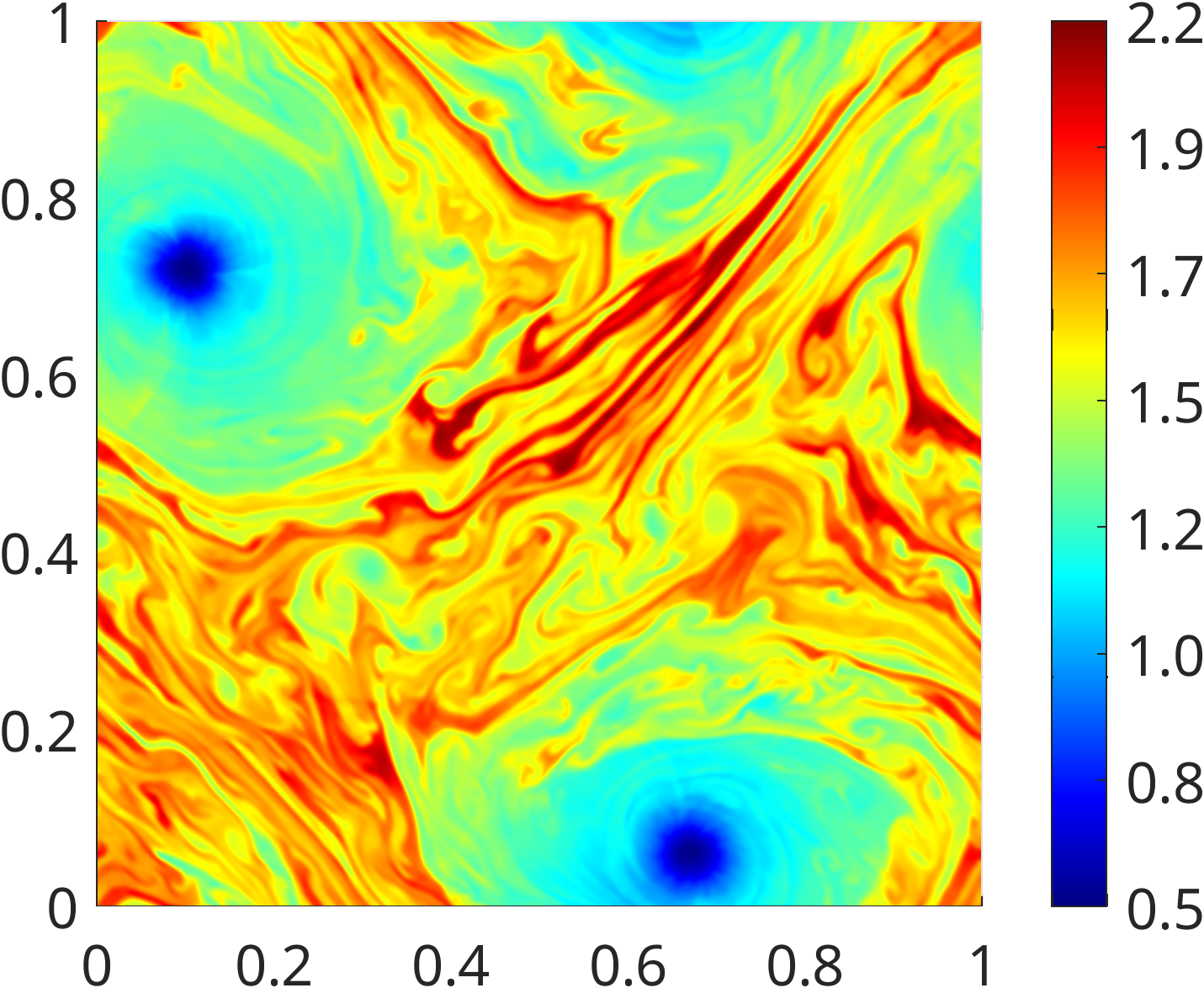}}
    \vspace{-8pt}
    \subfloat[$t=1$]{\label{fig:VFV_N5_avg_t1}%
        \includegraphics[width=.333\linewidth]{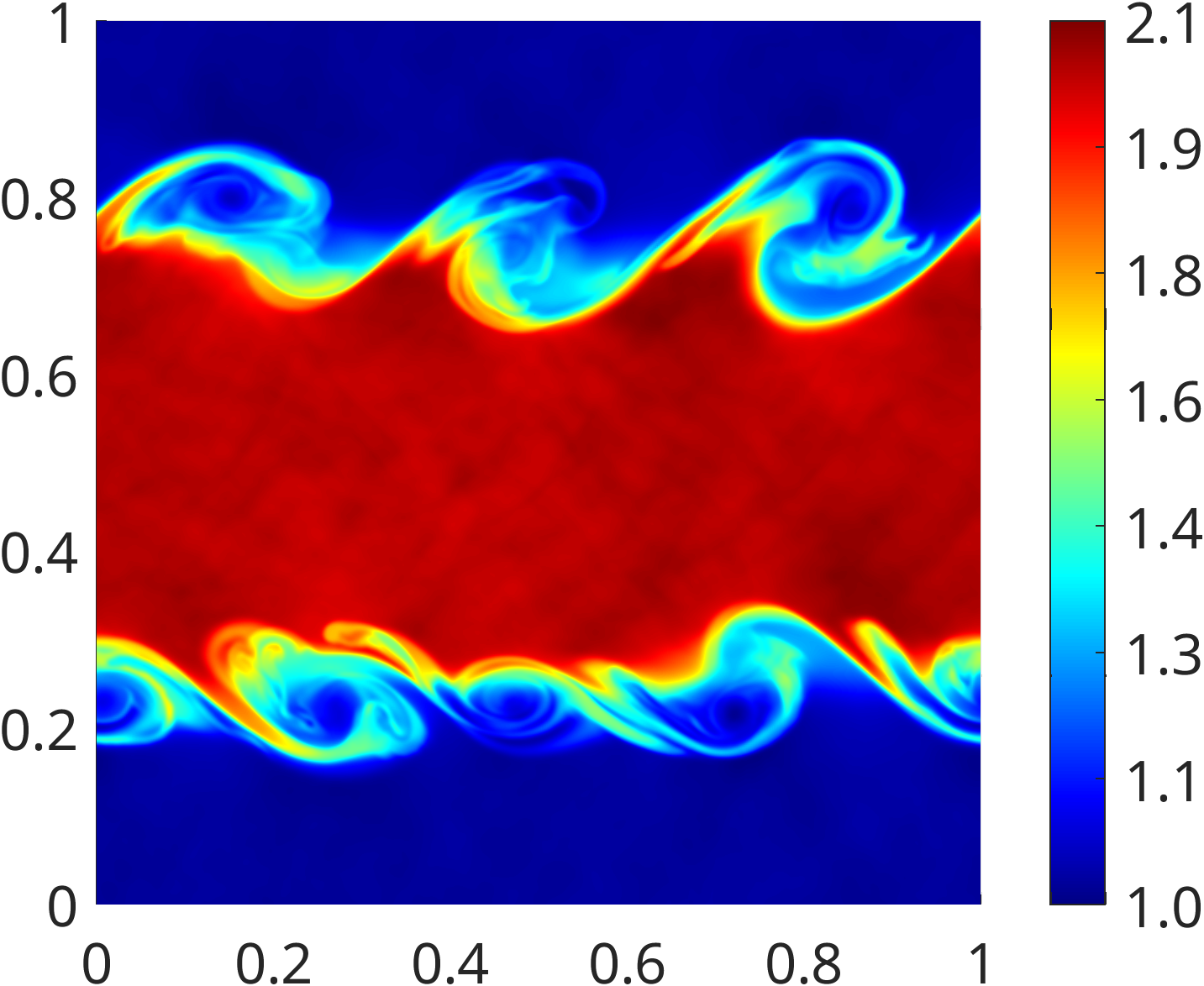}}\hfill
    \subfloat[$t=2$]{\label{fig:VFV_N5_avg_t2}%
        \includegraphics[width=.333\linewidth]{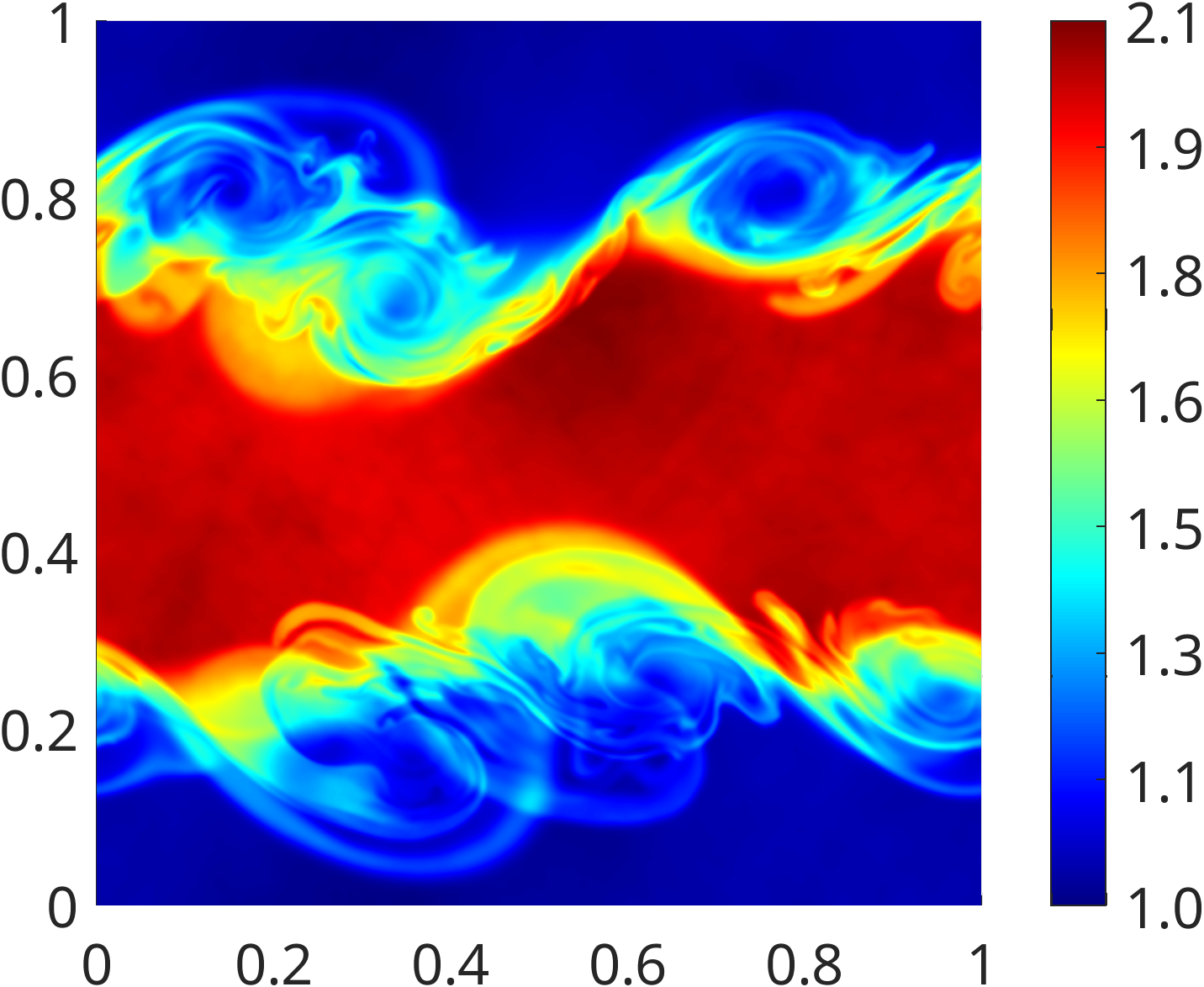}}\hfill
    \subfloat[$t=10$]{\label{fig:VFV_N5_avg_t10}%
        \includegraphics[width=.333\linewidth]{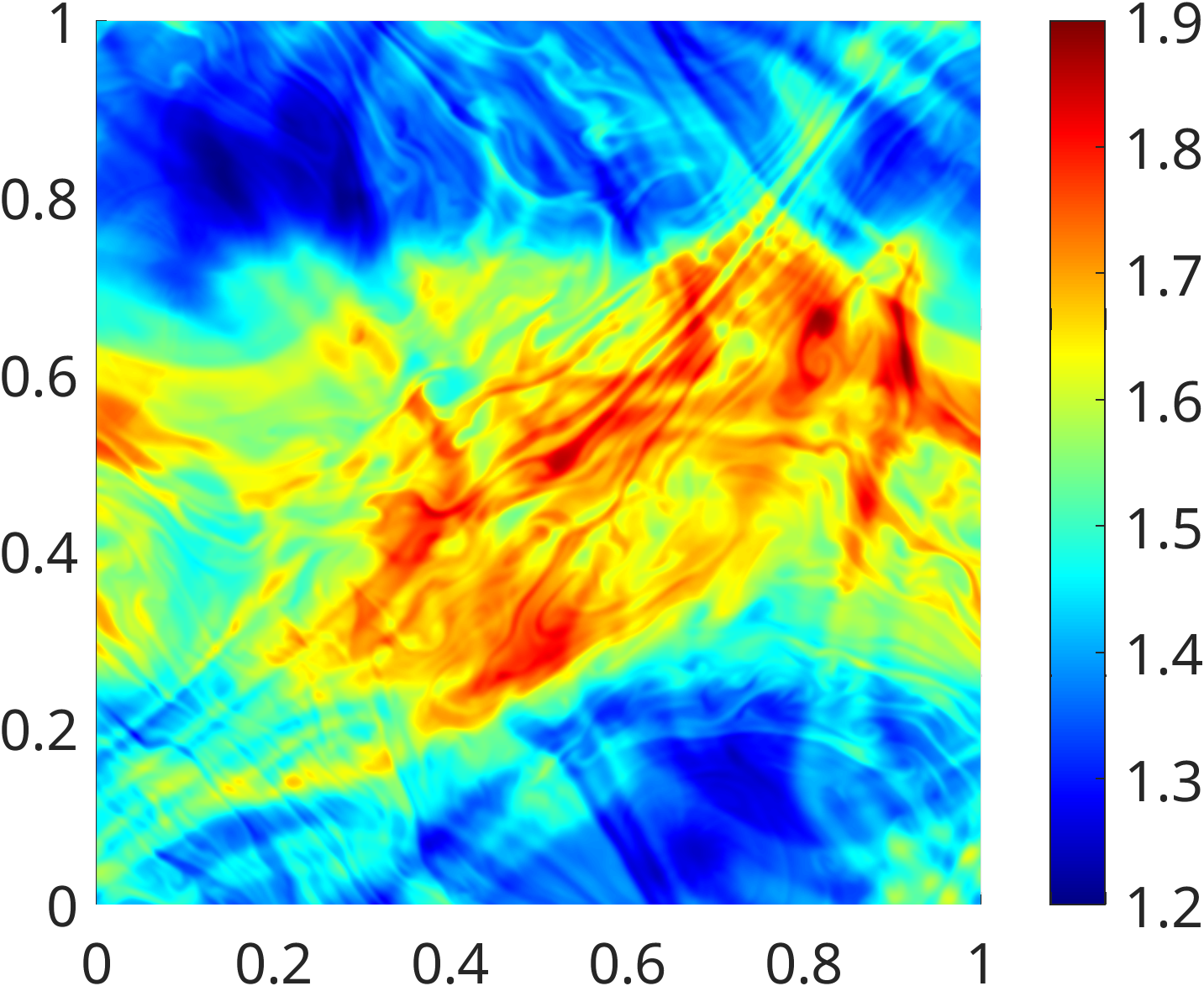}}\par
    \caption{Density (top row) and its \textit{Ces\`{a}ro} average (bottom row) for VFV method with $N=4$ at different times.}
    \label{fig:VFV_N5}
\end{figure}

\begin{figure}[ht]
    \centering
    \subfloat[$t=1$]{\label{fig:VFV_N7_t1}%
        \includegraphics[width=.333\linewidth]{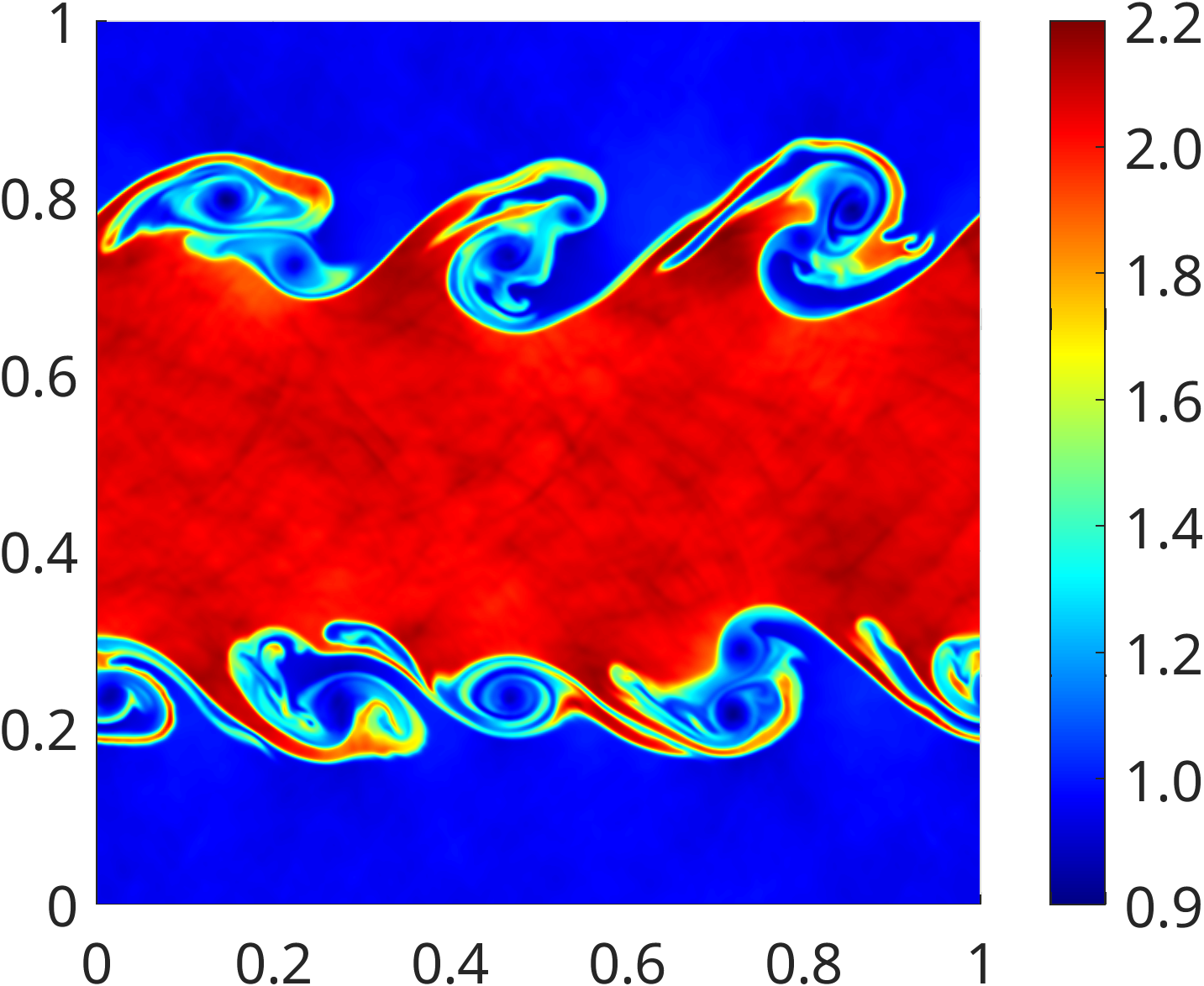}}\hfill
    \subfloat[$t=2$]{\label{fig:VFV_N7_t2}%
        \includegraphics[width=.333\linewidth]{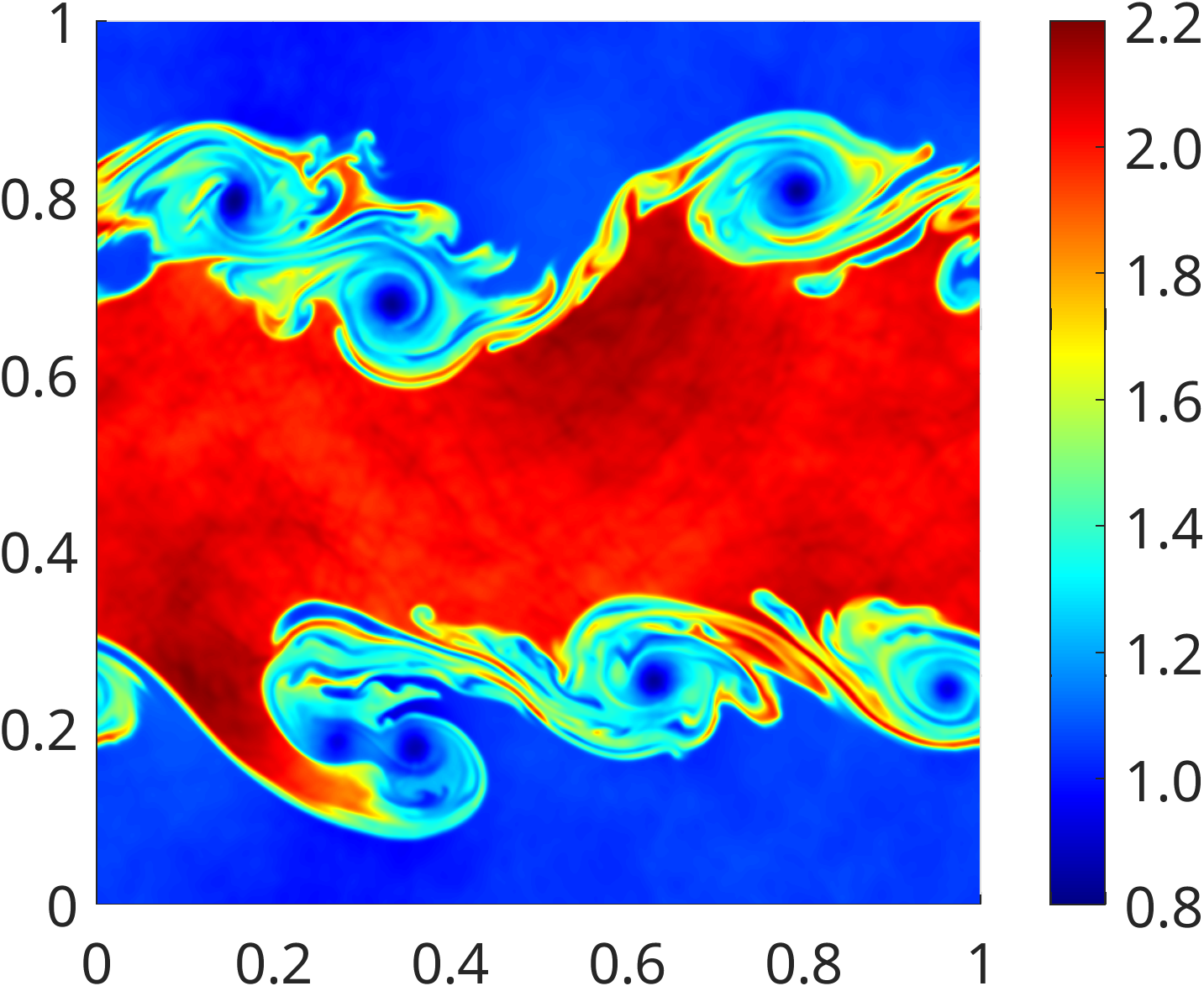}}\hfill
    \subfloat[$t=10$]{\label{fig:VFV_N7_t10}%
        \includegraphics[width=.333\linewidth]{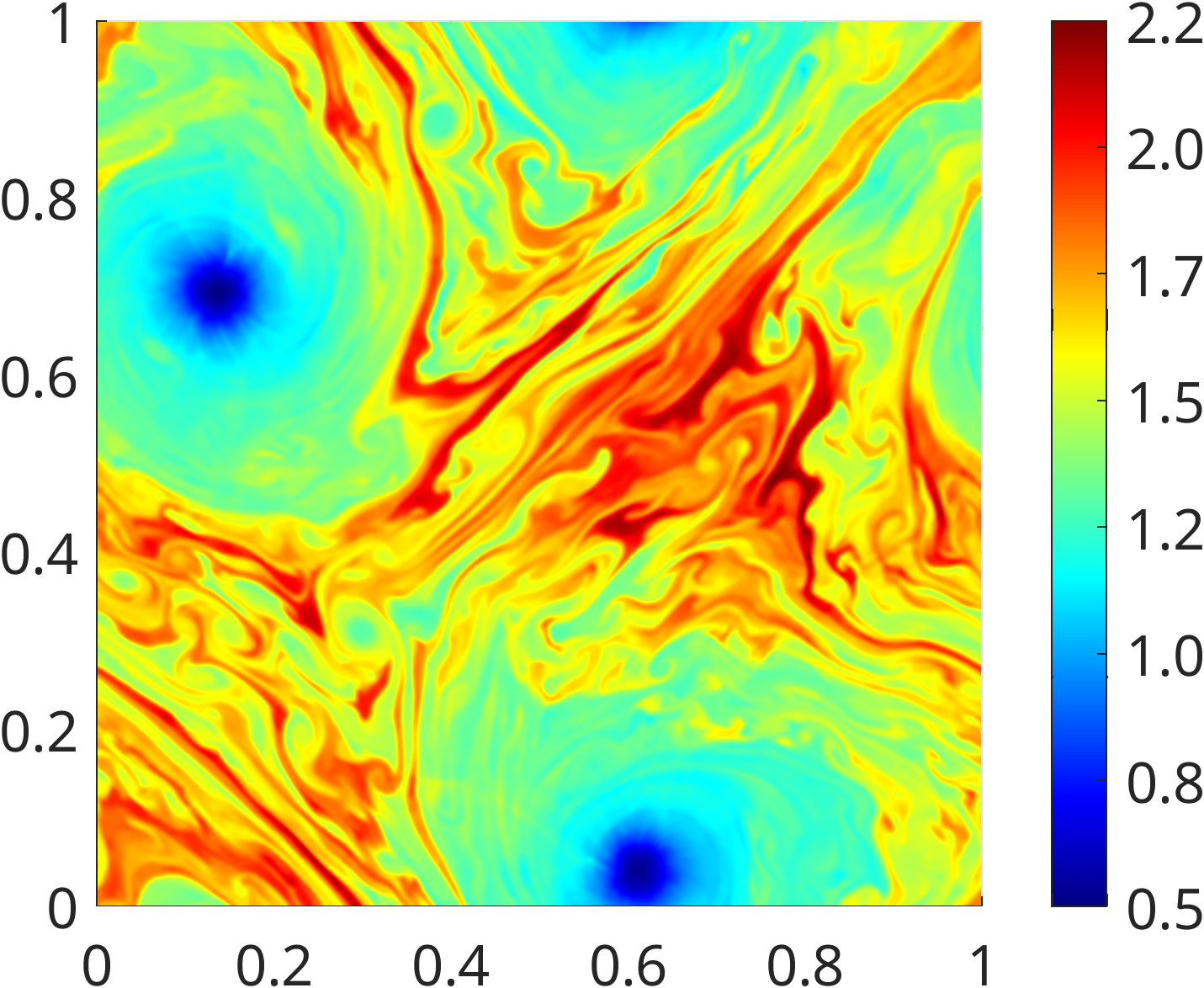}}
    \vspace{-8pt}
    \subfloat[$t=1$]{\label{fig:VFV_N7_avg_t1}%
        \includegraphics[width=.333\linewidth]{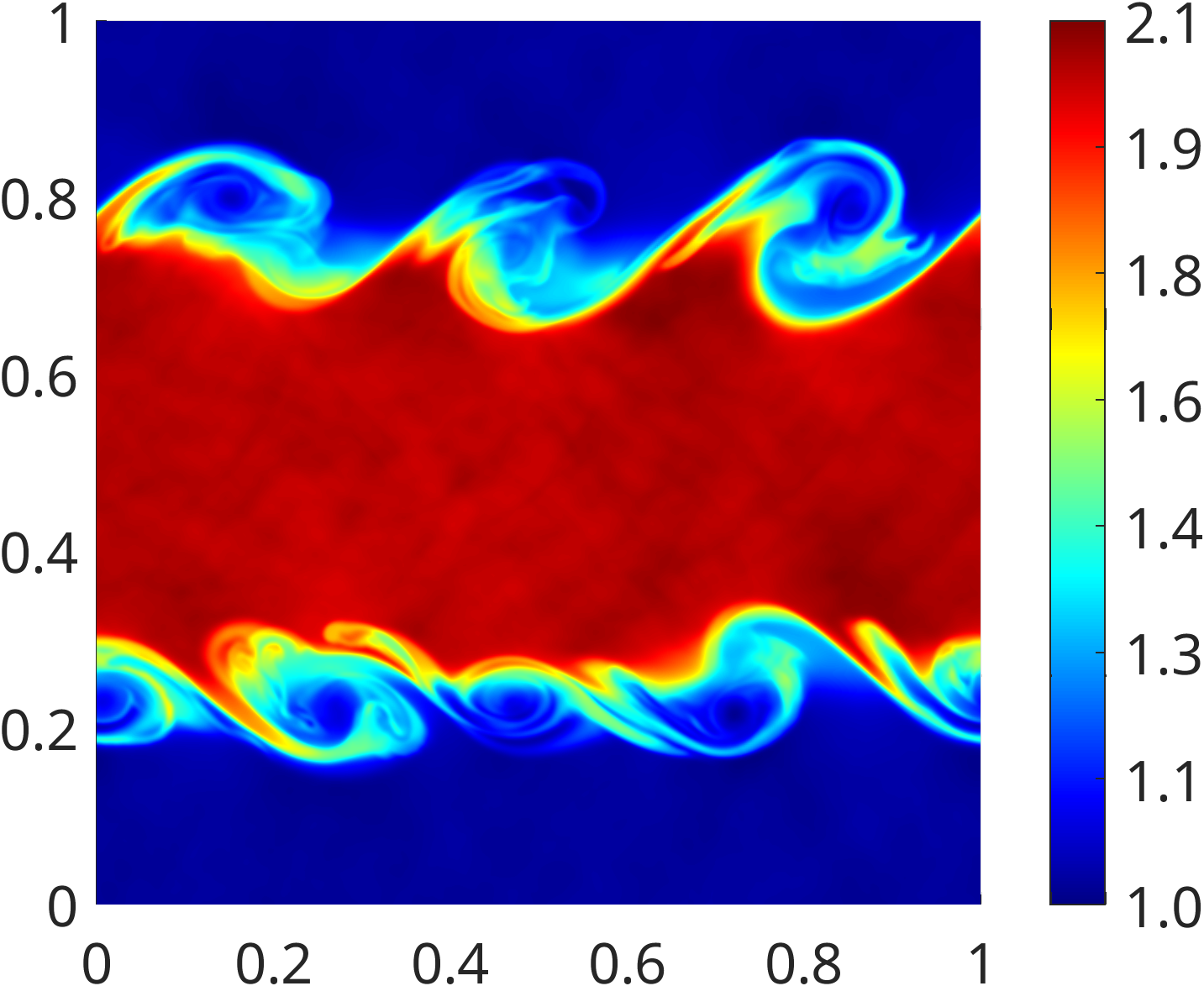}}\hfill
    \subfloat[$t=2$]{\label{fig:VFV_N7_avg_t2}%
        \includegraphics[width=.333\linewidth]{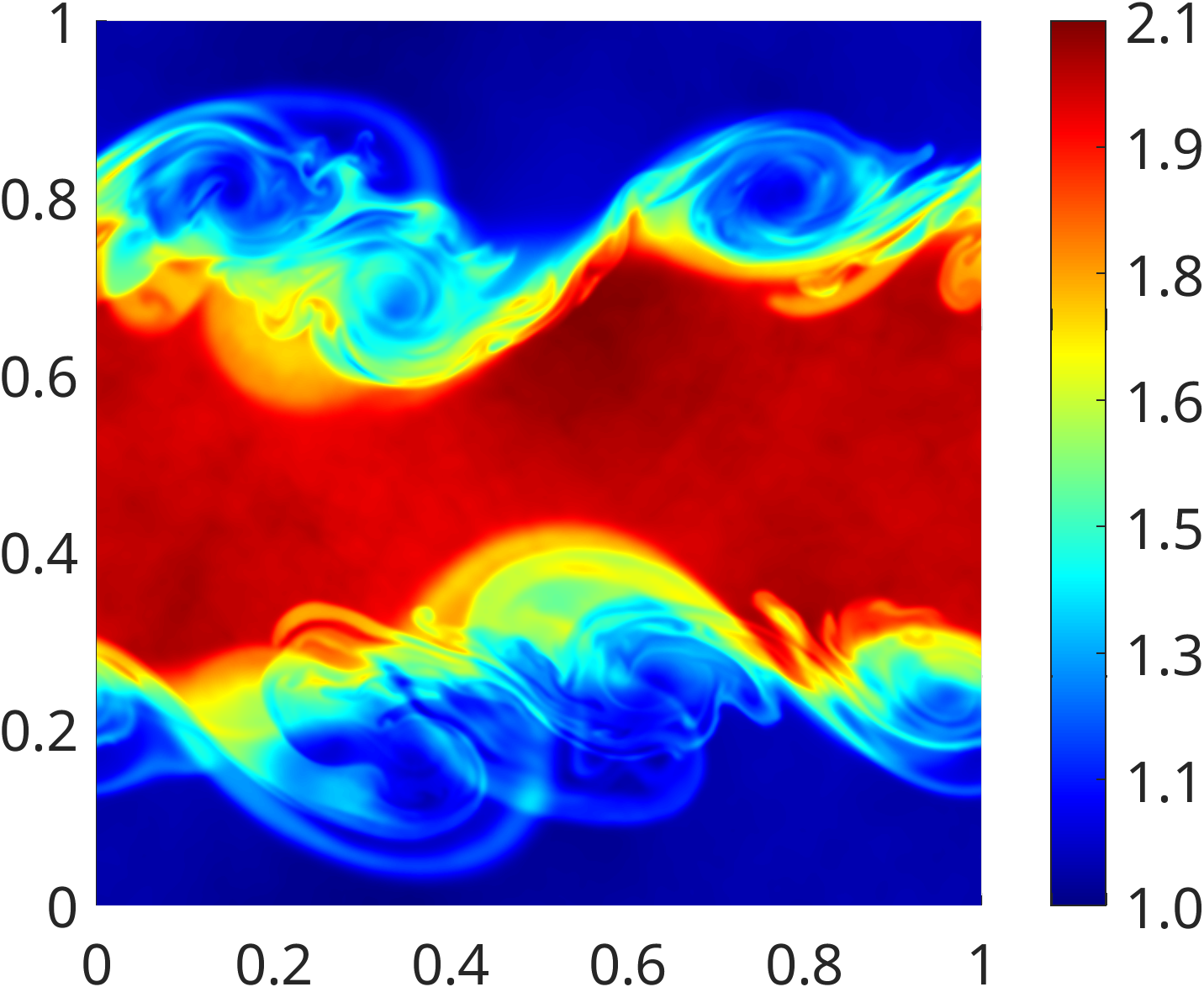}}\hfill
    \subfloat[$t=10$]{\label{fig:VFV_N7_avg_t10}%
        \includegraphics[width=.333\linewidth]{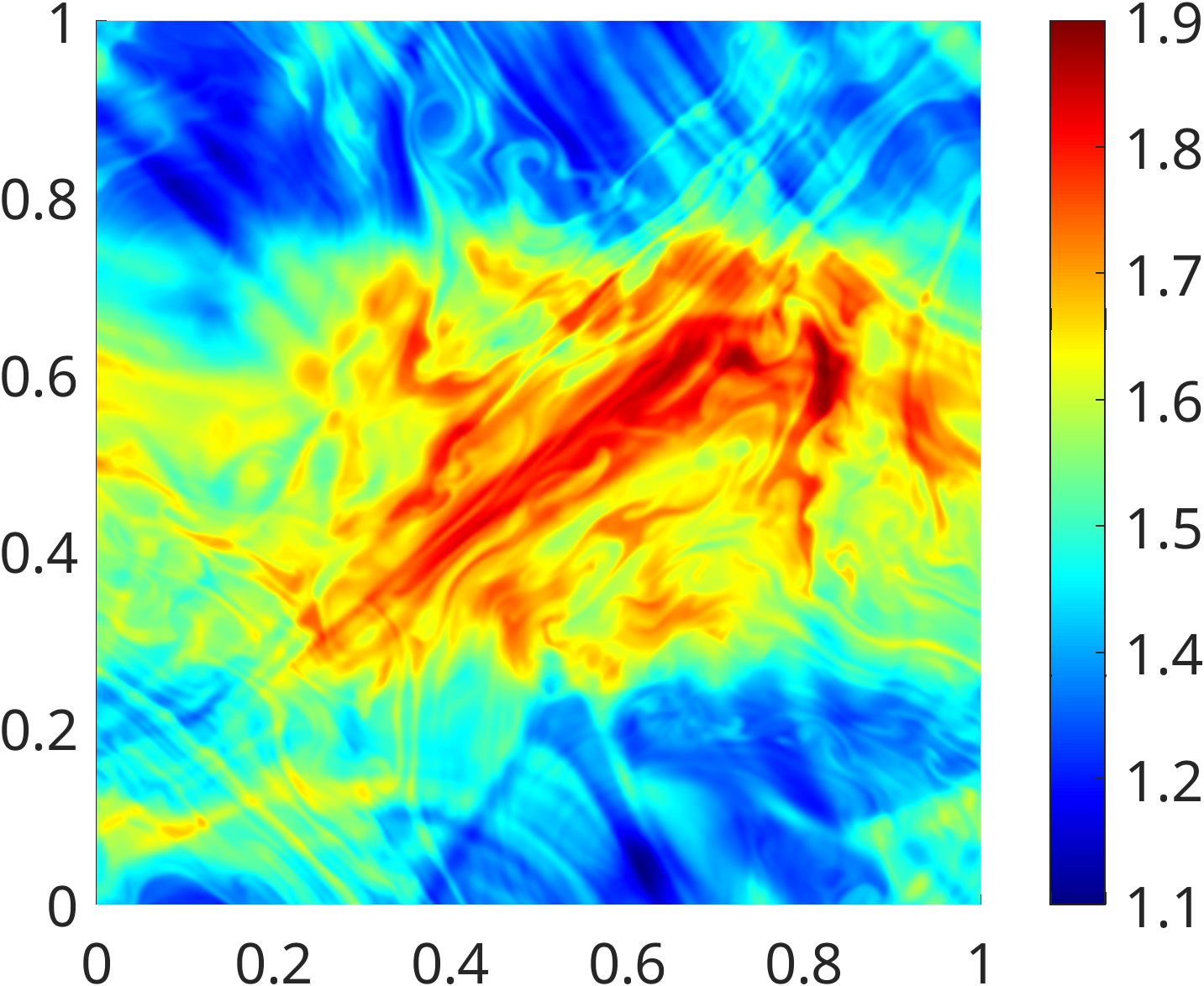}}\par
    \caption{Density (top row) and its \textit{Ces\`{a}ro} average (bottom row) for VFV method with $N=6$ at different times.}
    \label{fig:VFV_N7}
\end{figure}

\begin{figure}[ht]
    \centering
    \subfloat[$t=1$]{\label{fig:VFV_N9_t1}%
        \includegraphics[width=.333\linewidth]{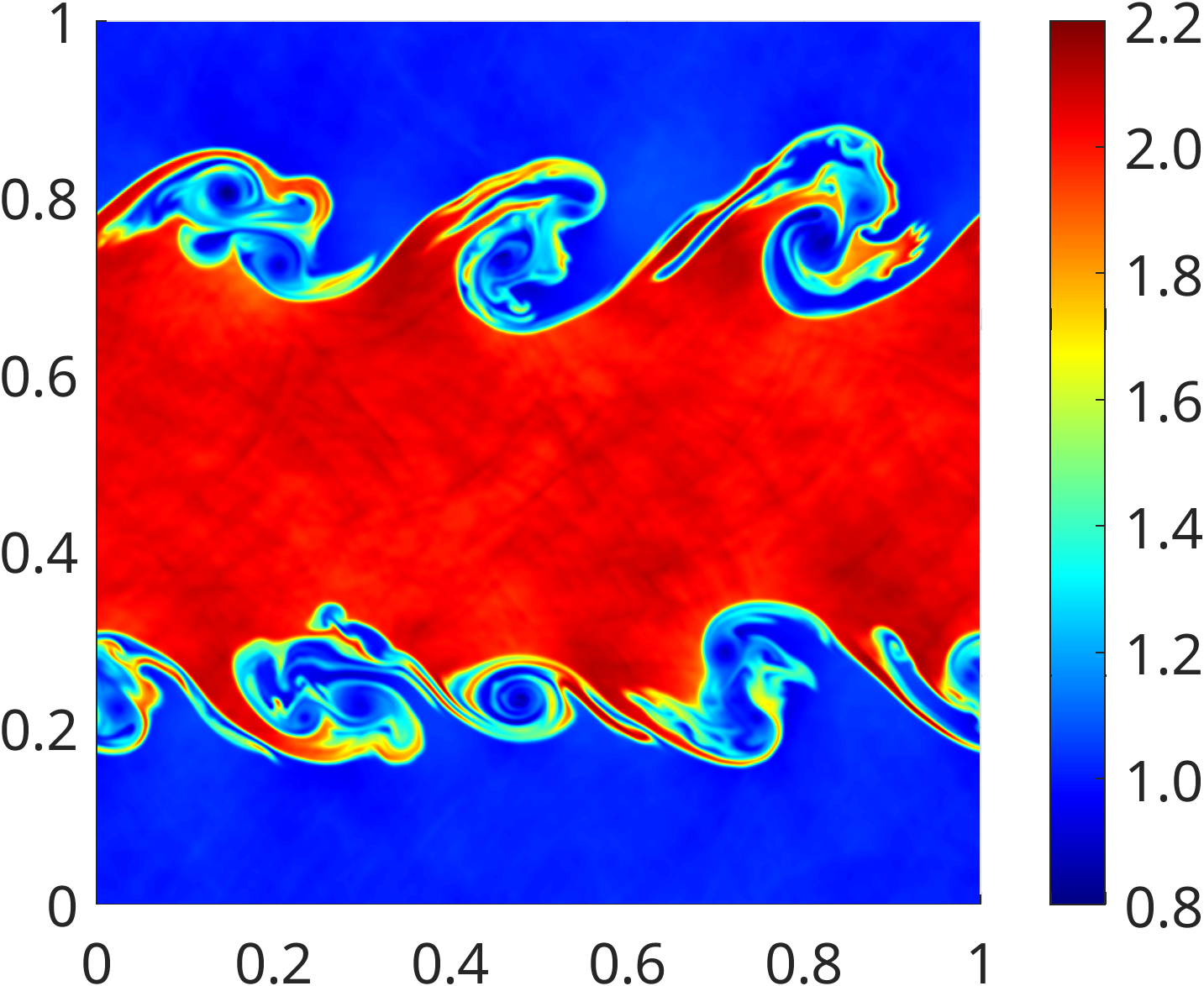}}\hfill
    \subfloat[$t=2$]{\label{fig:VFV_N9_t2}%
        \includegraphics[width=.333\linewidth]{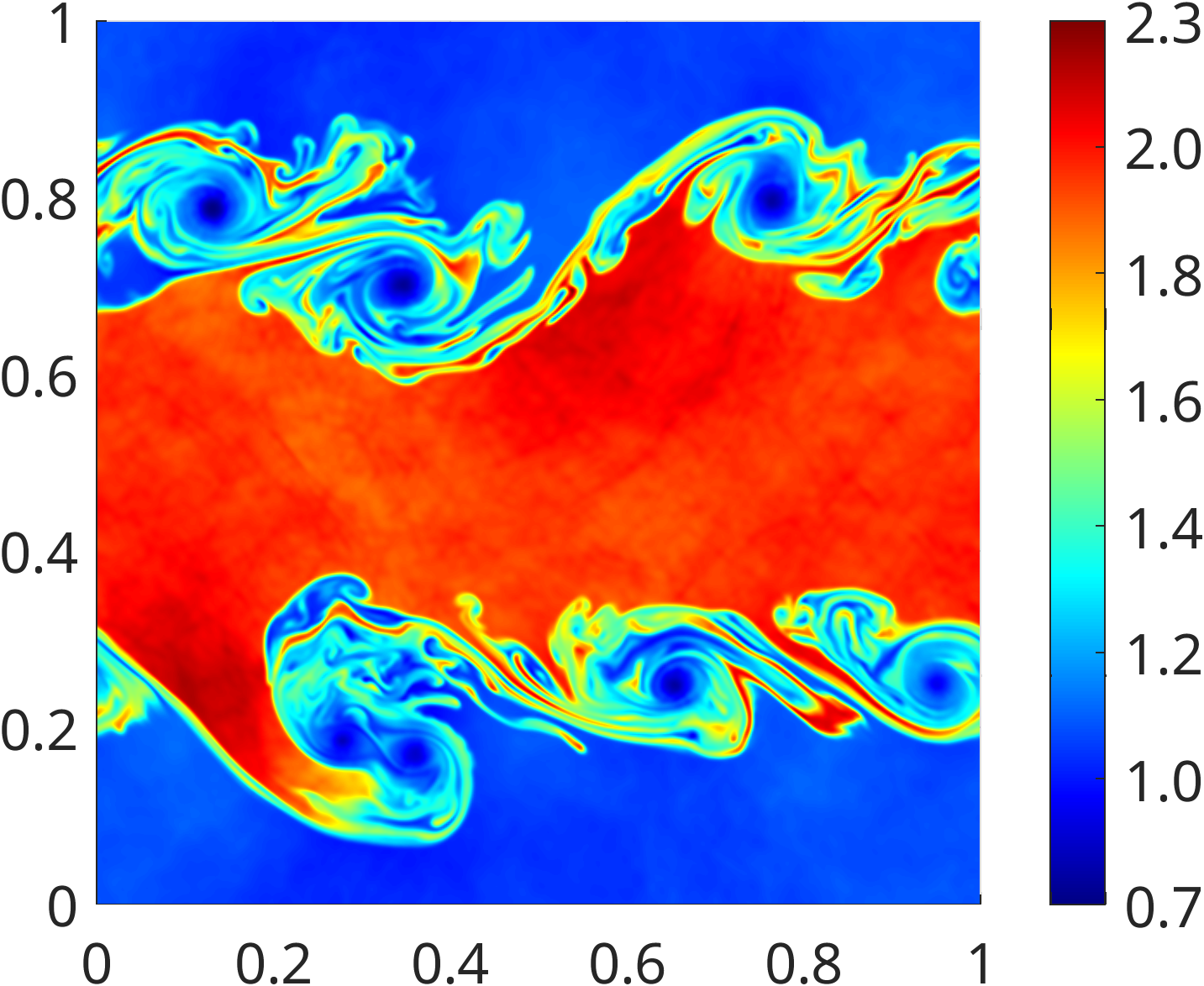}}\hfill
    \subfloat[$t=10$]{\label{fig:VFV_N9_t10}%
        \includegraphics[width=.333\linewidth]{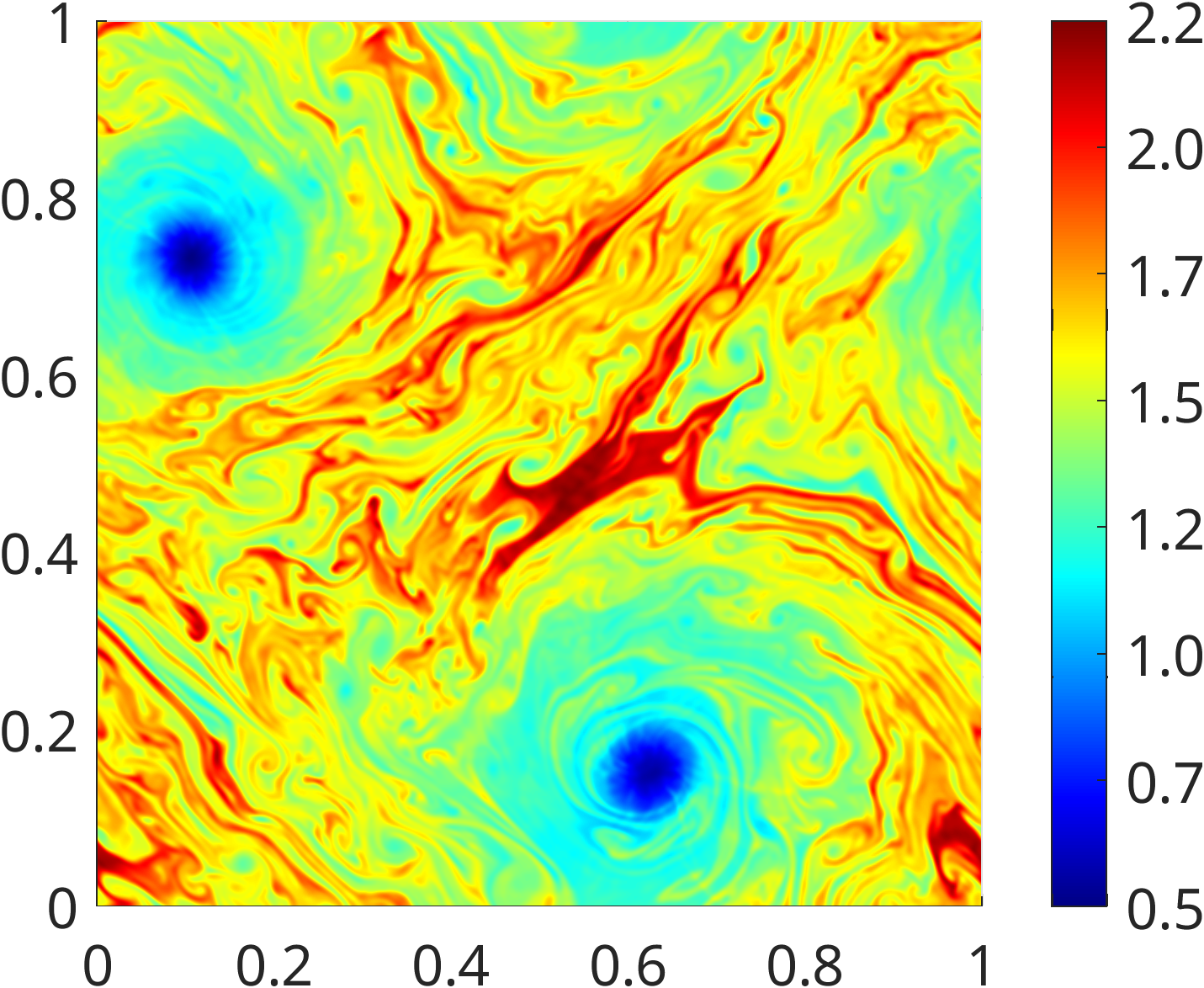}}
    \vspace{-8pt}
    \subfloat[$t=1$]{\label{fig:VFV_N9_avg_t1}%
        \includegraphics[width=.333\linewidth]{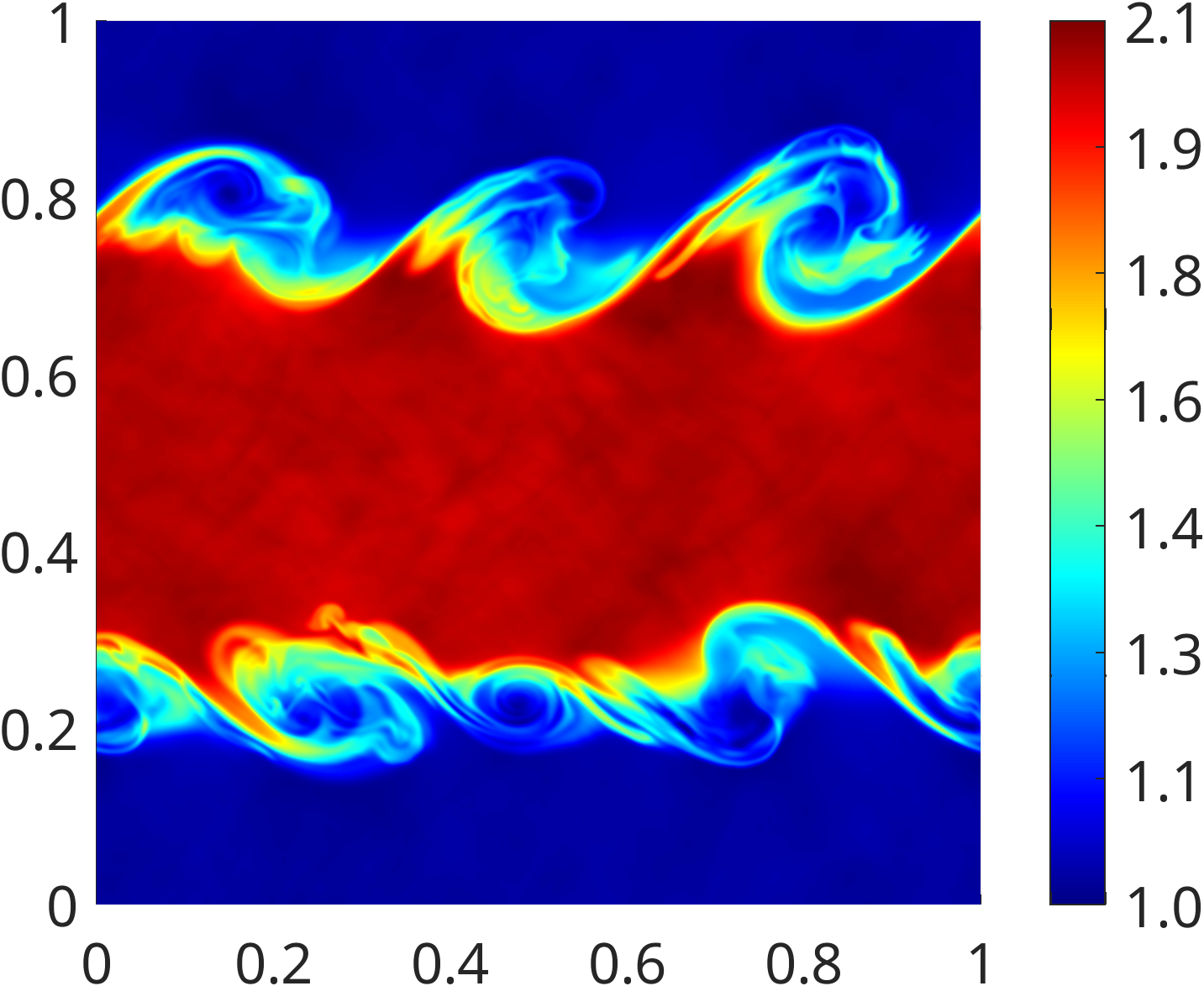}}\hfill
    \subfloat[$t=2$]{\label{fig:VFV_N9_avg_t2}%
        \includegraphics[width=.333\linewidth]{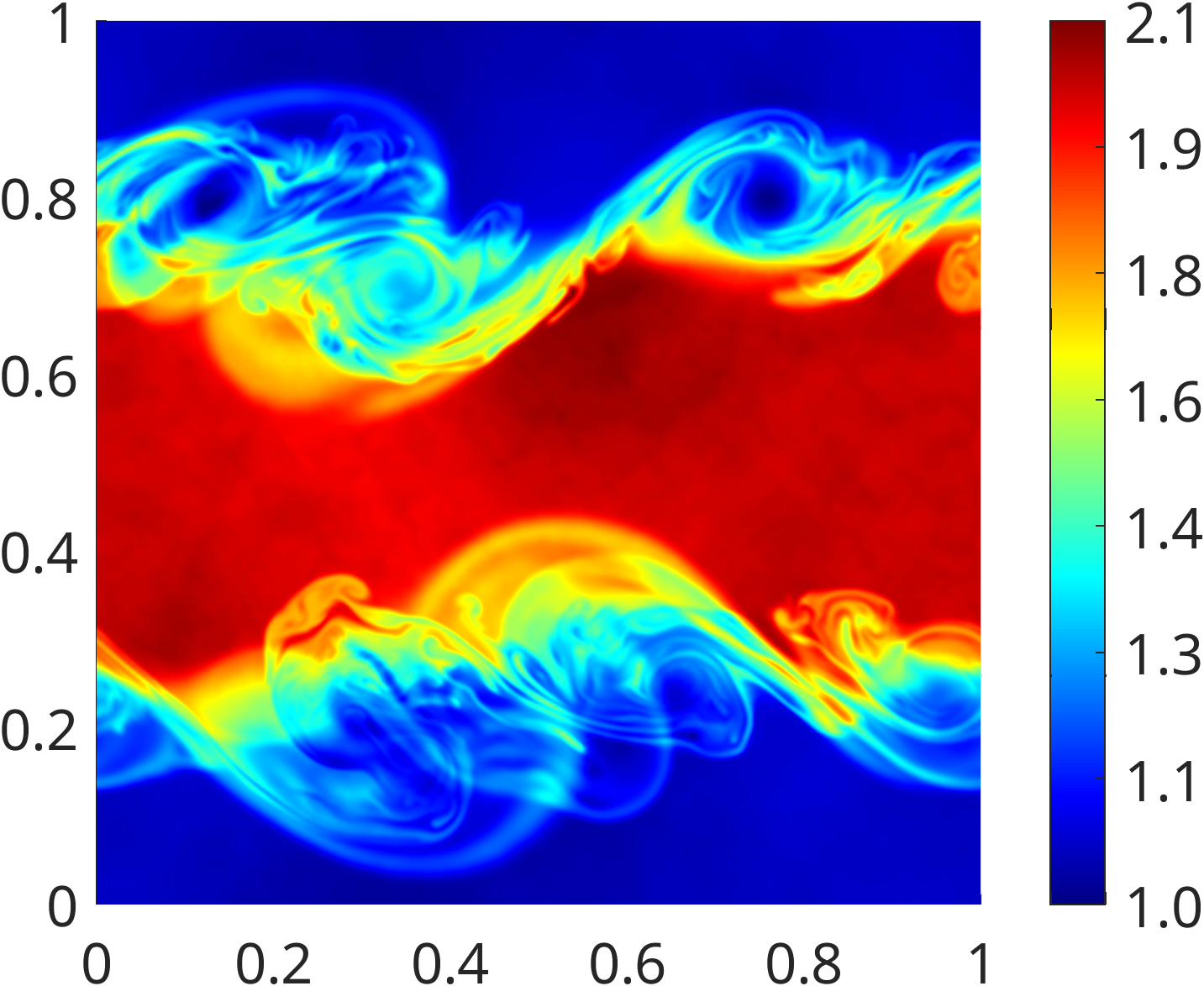}}\hfill
    \subfloat[$t=10$]{\label{fig:VFV_N9_avg_t10}%
        \includegraphics[width=.333\linewidth]{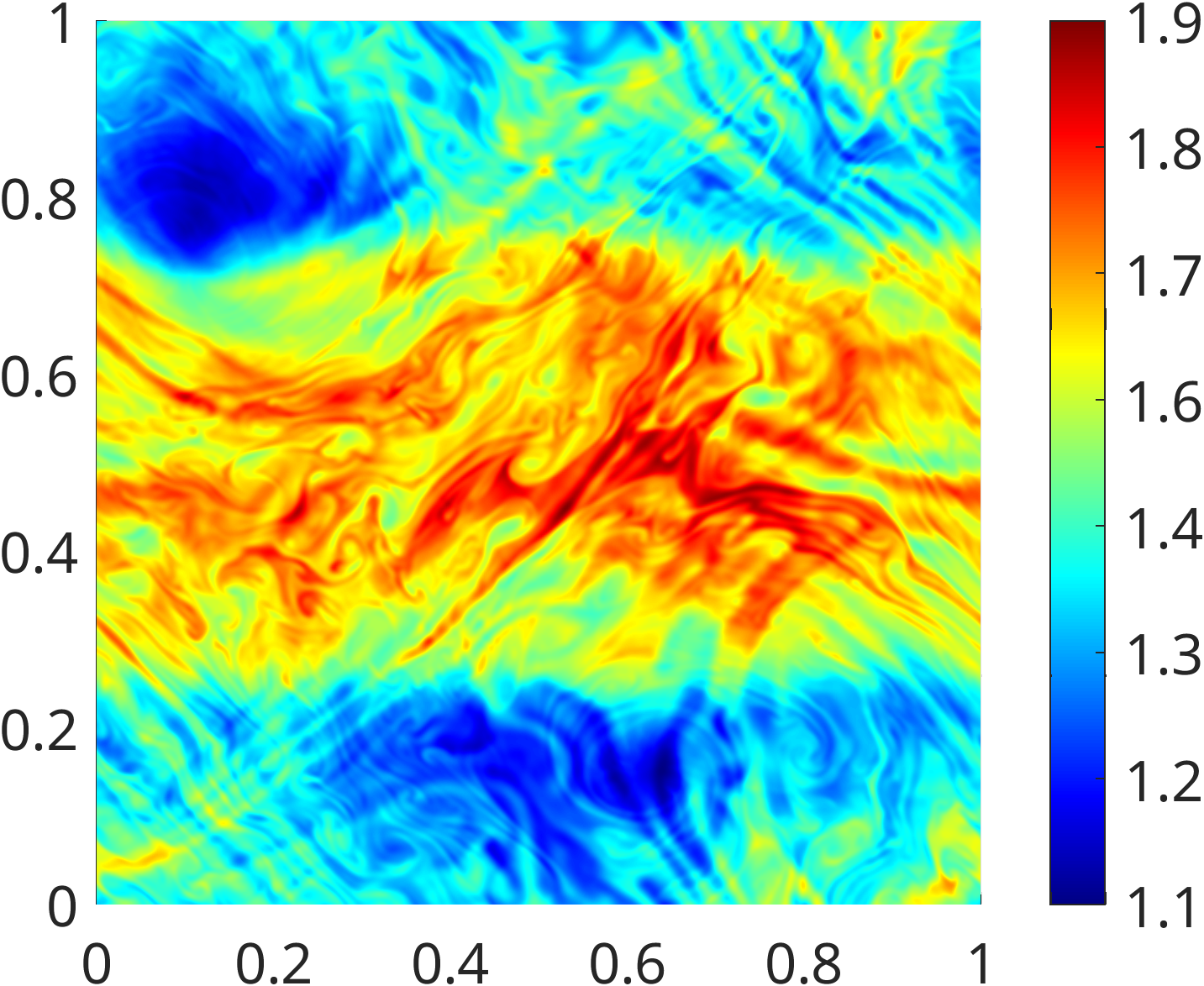}}\par
    \caption{Density (top row) and its \textit{Ces\`{a}ro} average (bottom row) for VFV method with $N=8$ at different times.}
    \label{fig:VFV_N9}
\end{figure}

\begin{figure}[ht]
    \centering
    \subfloat[$t=1$]{\label{fig:Kin_N0_t1}%
        \includegraphics[width=.30\linewidth]{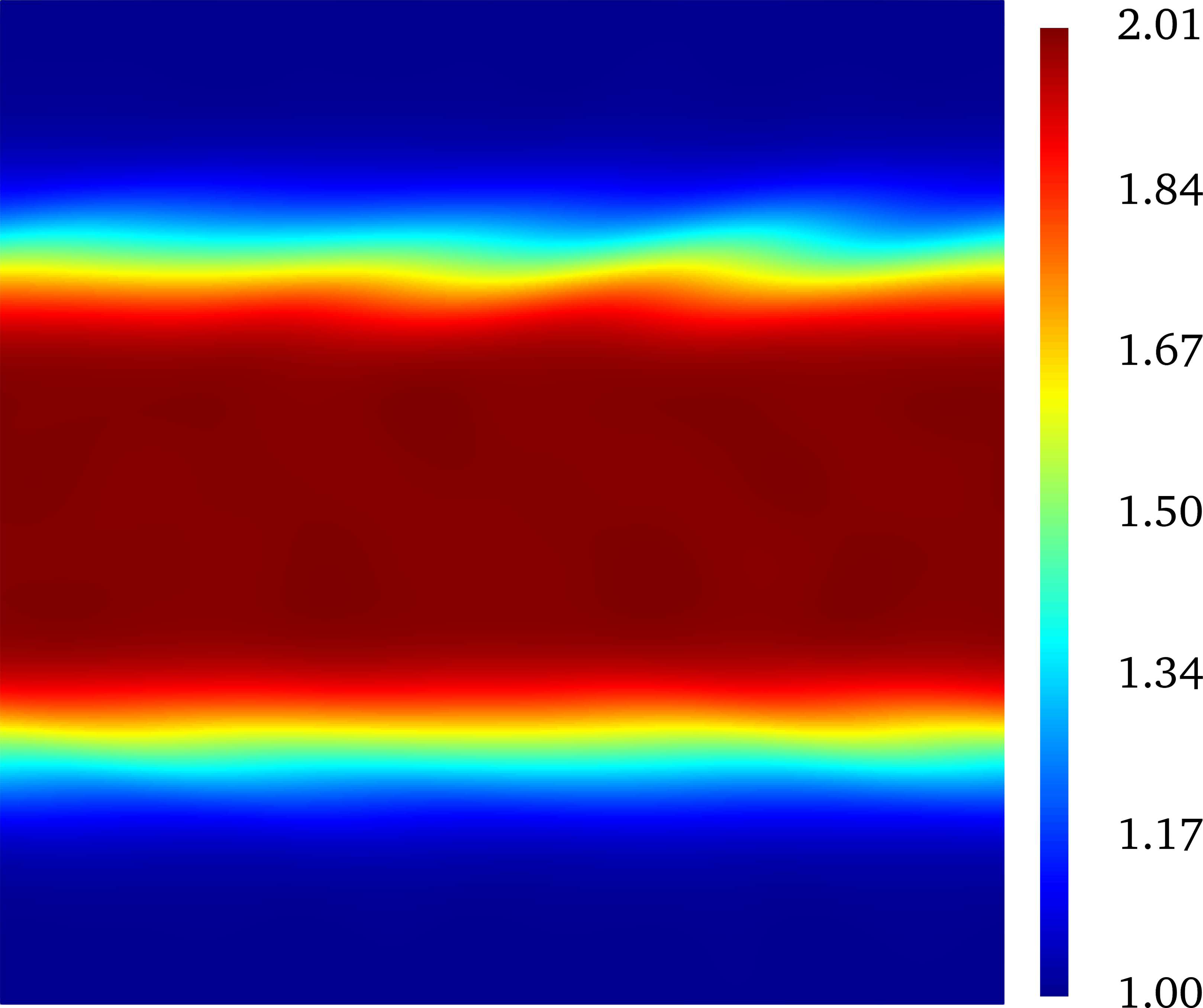}}\hfill
    \subfloat[$t=2$]{\label{fig:Kin_N0_t2}%
        \includegraphics[width=.30\linewidth]{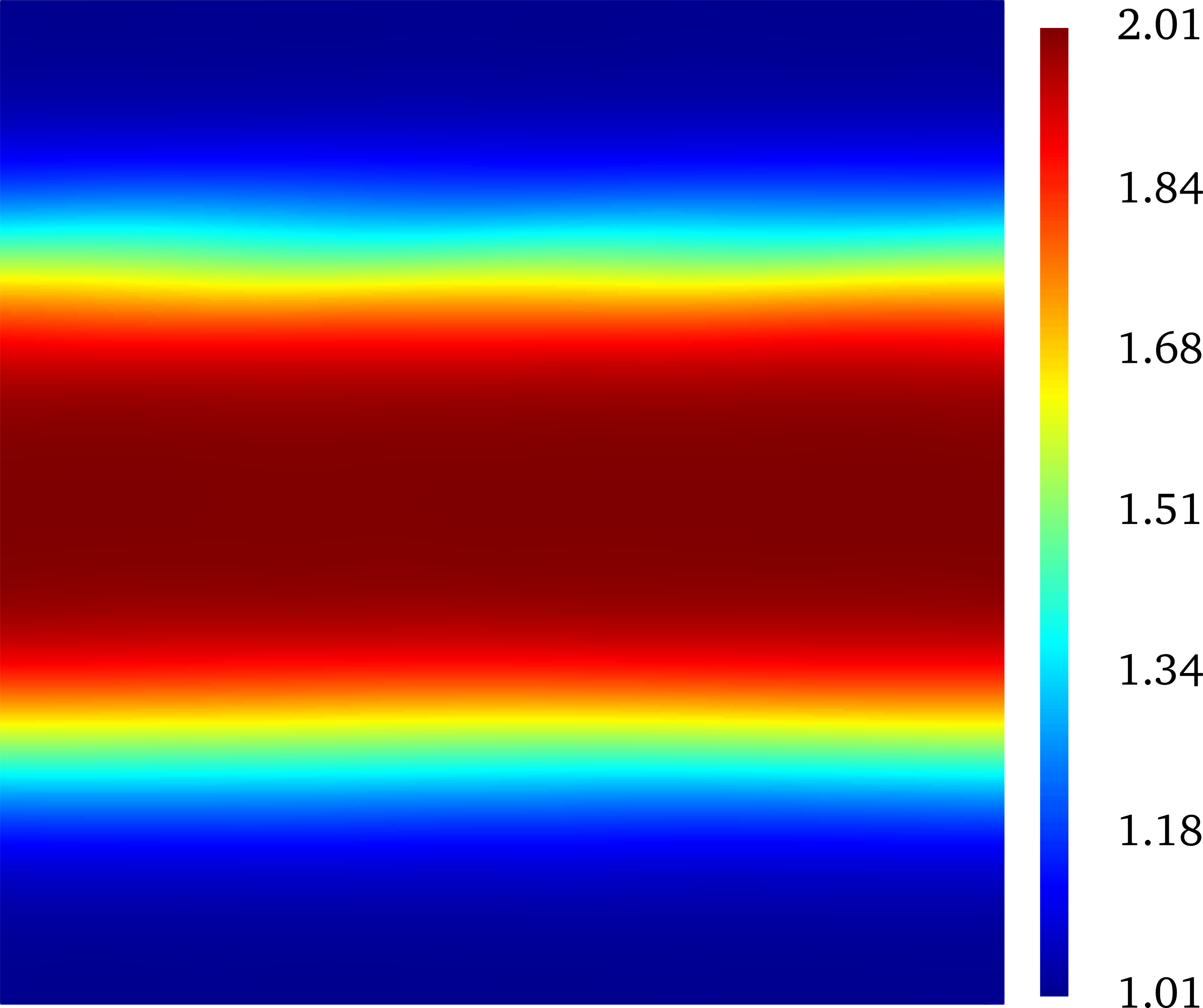}}\hfill
    \subfloat[$t=10$]{\label{fig:Kin_N0_t10}%
        \includegraphics[width=.30\linewidth]{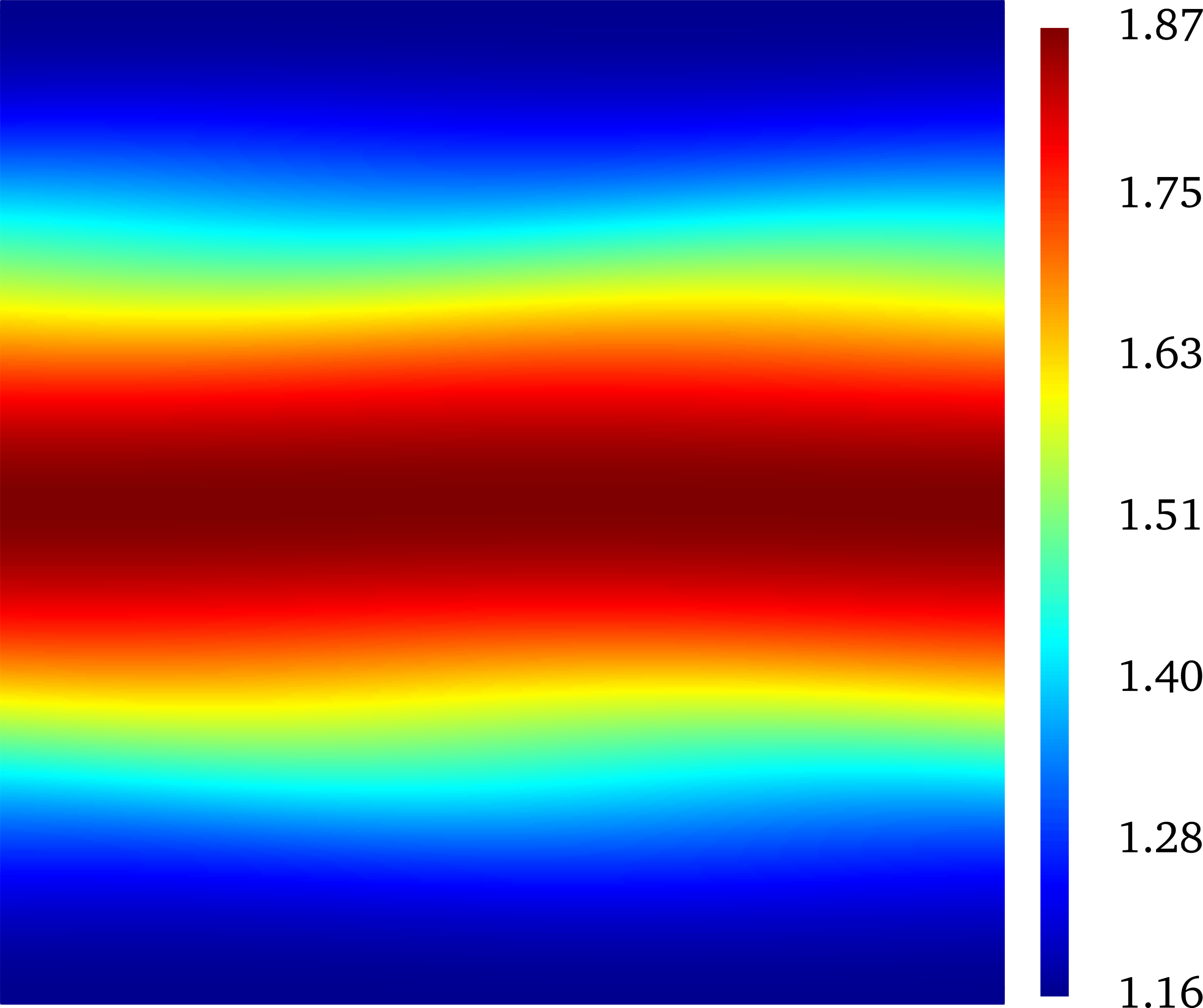}}\par
    \subfloat[$t=1$]{\label{fig:Kin_N0_avg_t1}%
        \includegraphics[width=.30\linewidth]{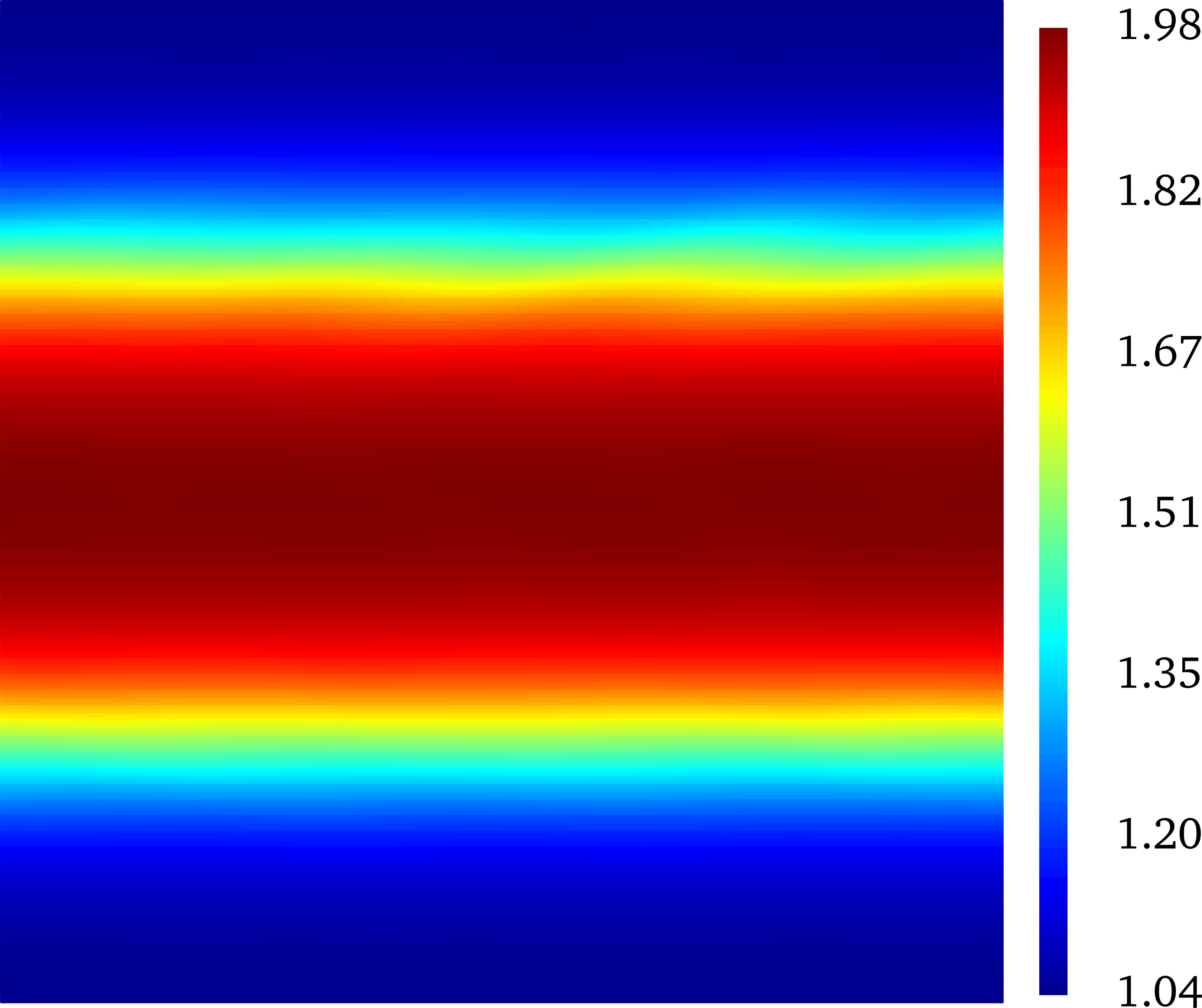}}\hfill
    \subfloat[$t=2$]{\label{fig:Kin_N0_avg_t2}%
        \includegraphics[width=.30\linewidth]{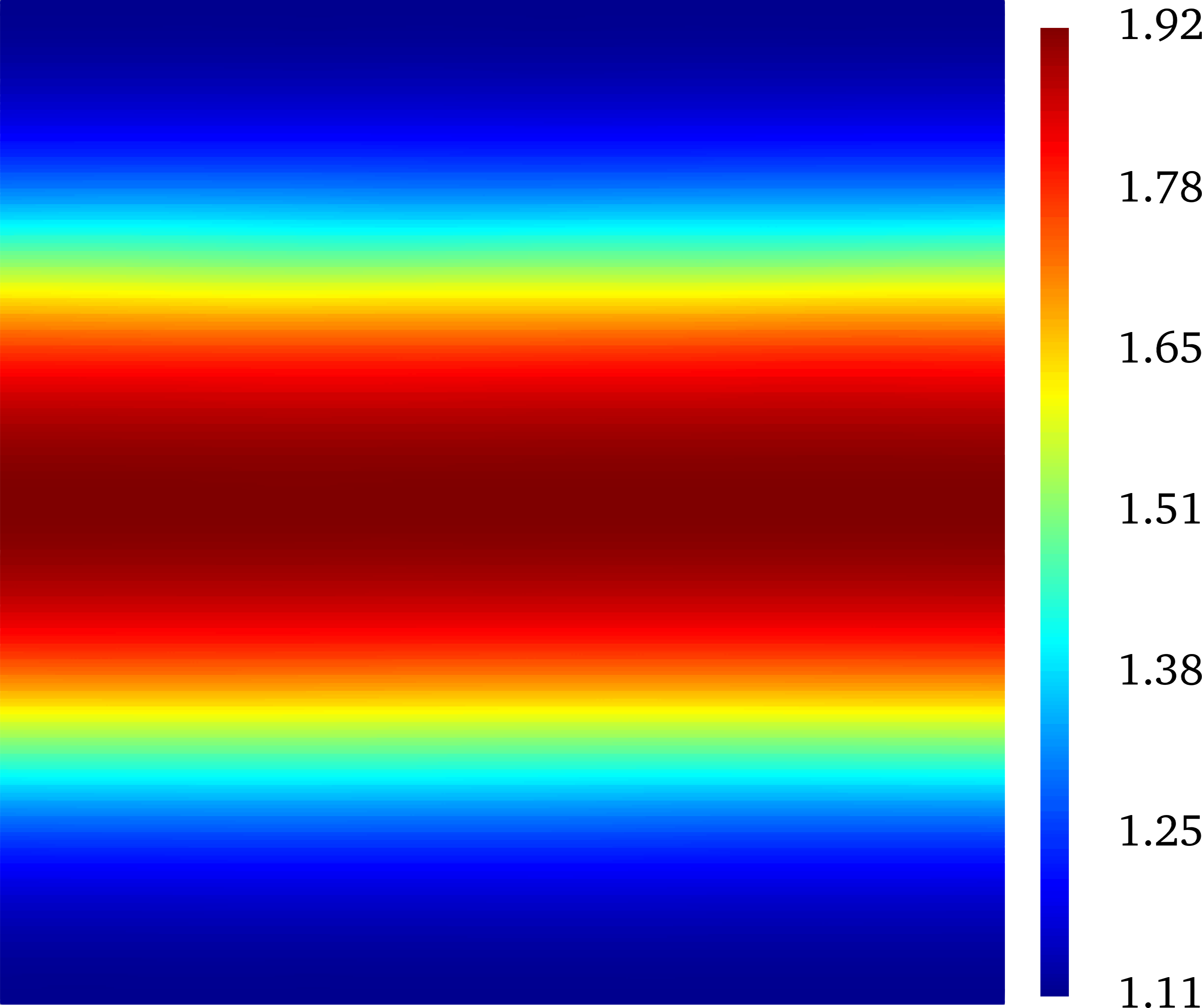}}\hfill
    \subfloat[$t=10$]{\label{fig:Kin_N0_avg_t10}%
        \includegraphics[width=.30\linewidth]{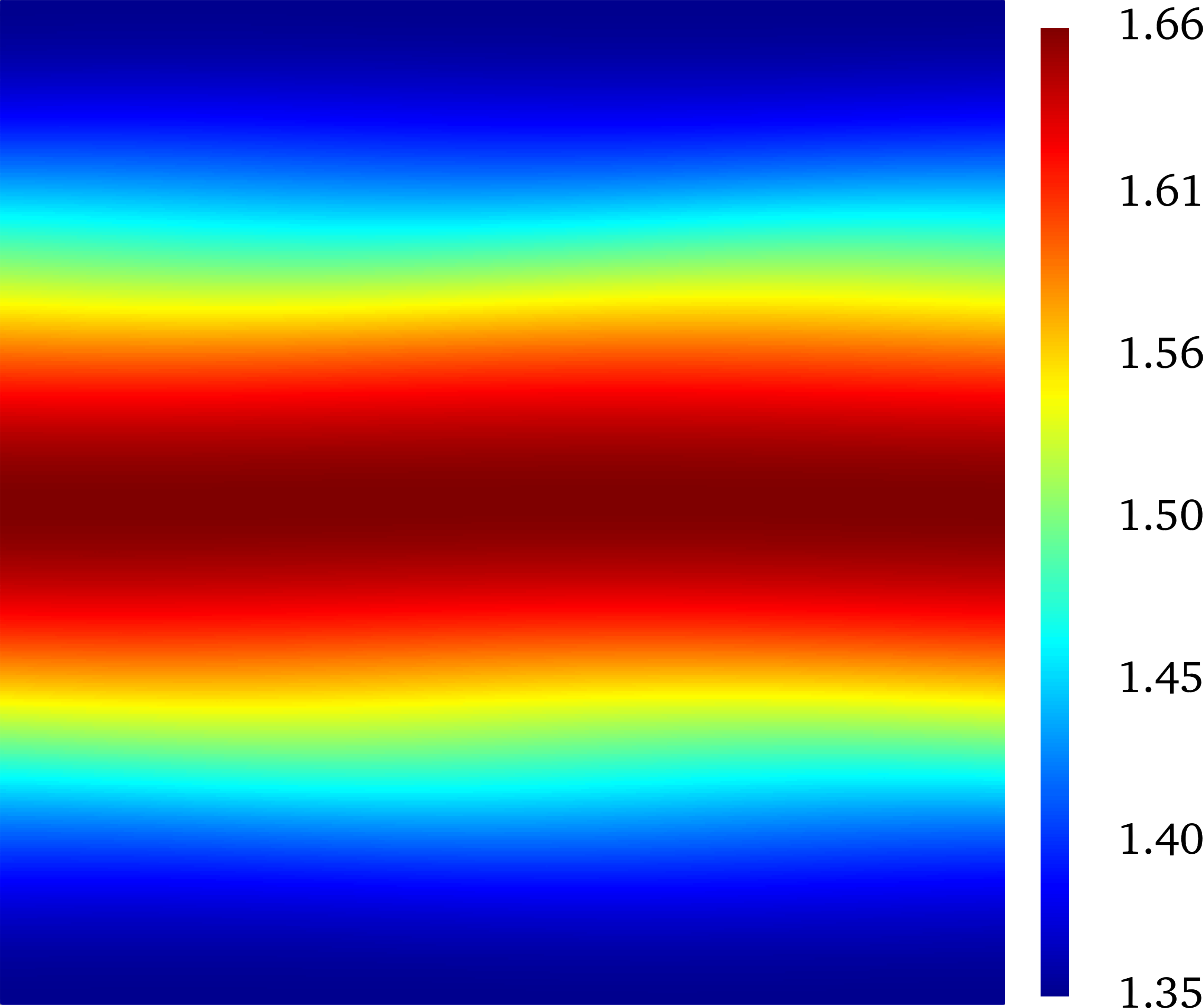}}\par
    \caption{Density (top row) and its \textit{Ces\`{a}ro} average (bottom row) for DVBFV method with $N=0$ at different times.}
    \label{fig:Kin_N0}
\end{figure}

\begin{figure}[ht]
    \centering
    \subfloat[$t=1$]{\label{fig:Kin_N1_t1}%
        \includegraphics[width=.30\linewidth]{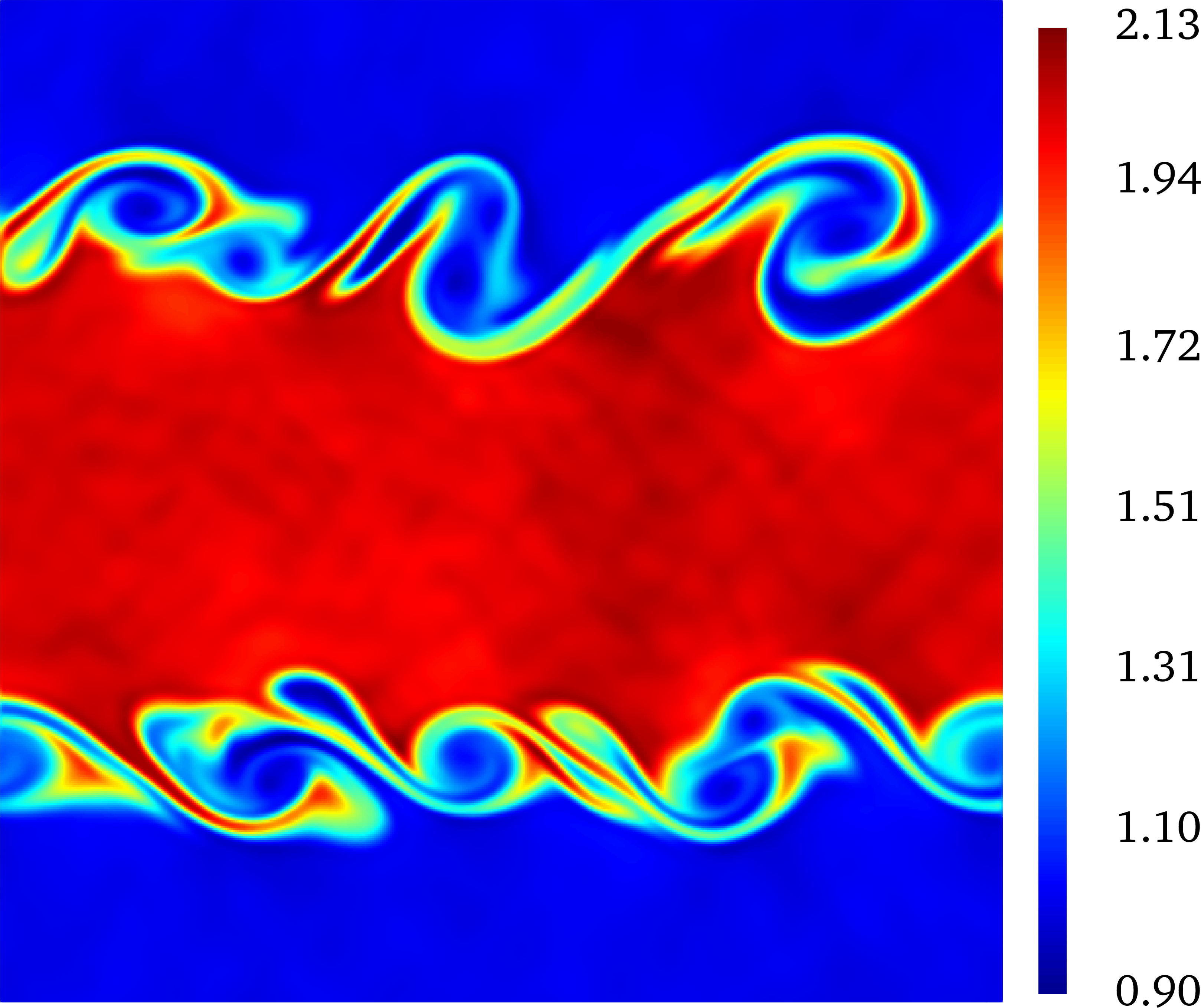}}\hfill
    \subfloat[$t=2$]{\label{fig:Kin_N1_t2}%
        \includegraphics[width=.30\linewidth]{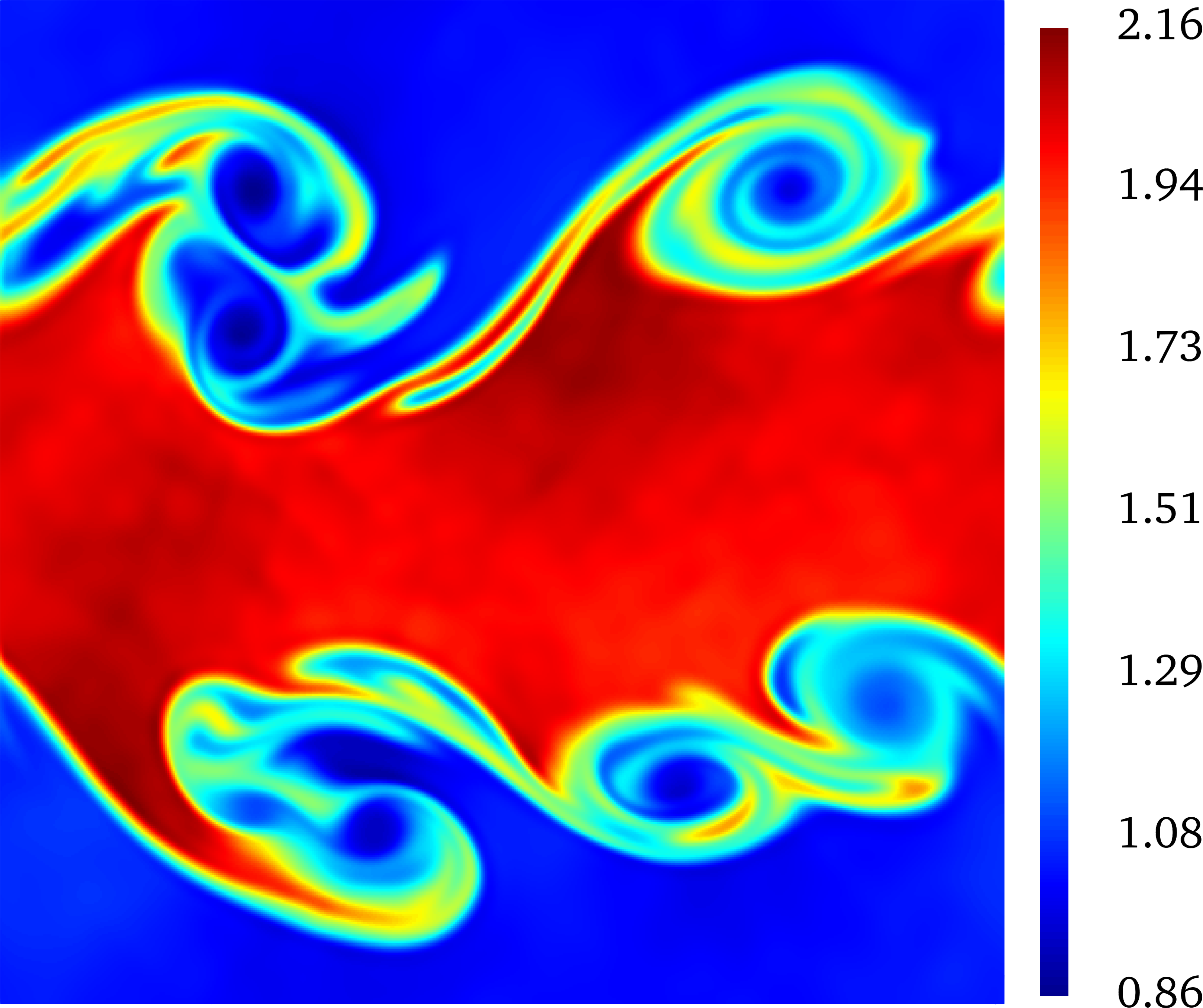}}\hfill
    \subfloat[$t=10$]{\label{fig:Kin_N1_t10}%
        \includegraphics[width=.30\linewidth]{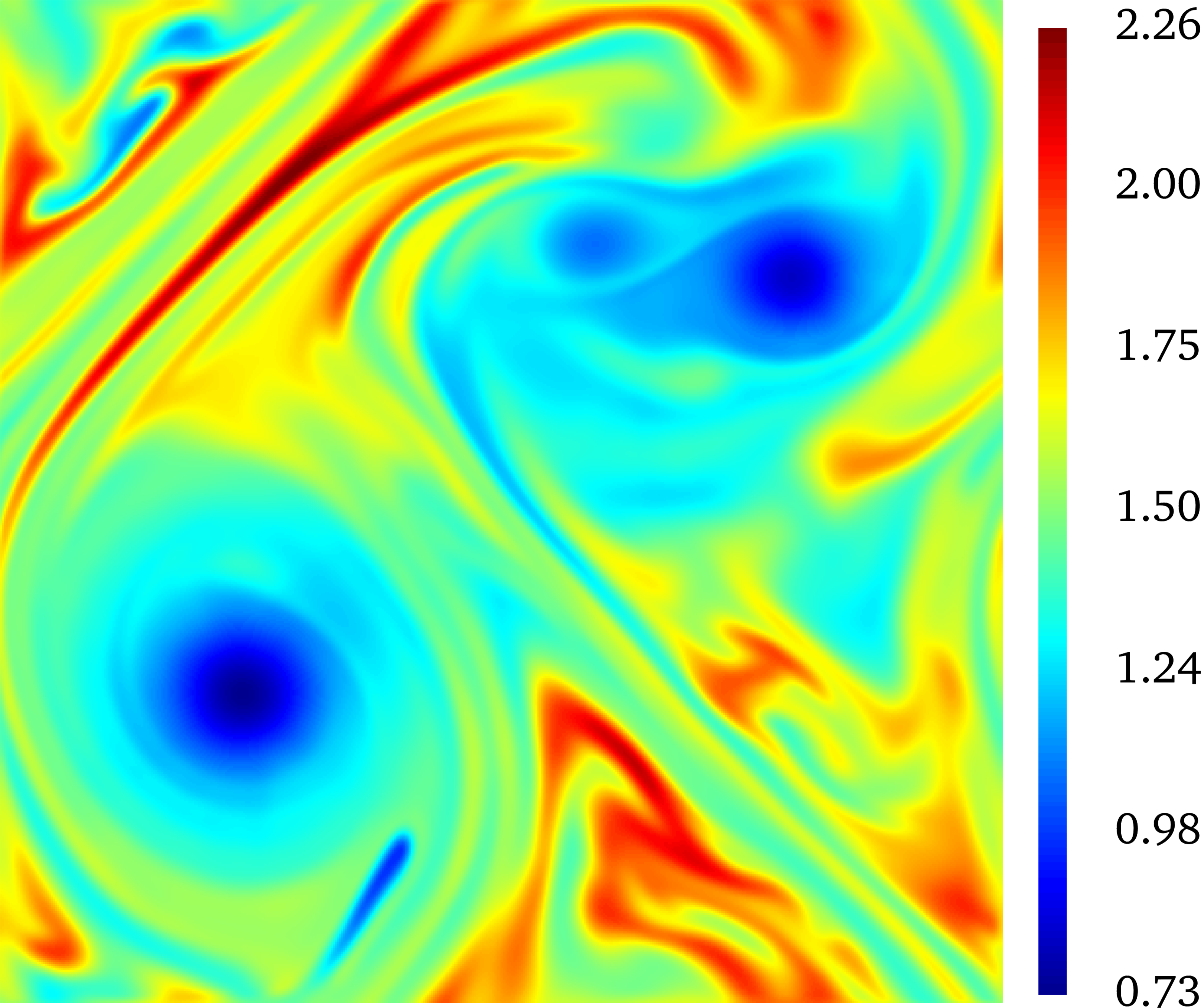}}\par
    \subfloat[$t=1$]{\label{fig:Kin_N1_avg_t1}%
        \includegraphics[width=.30\linewidth]{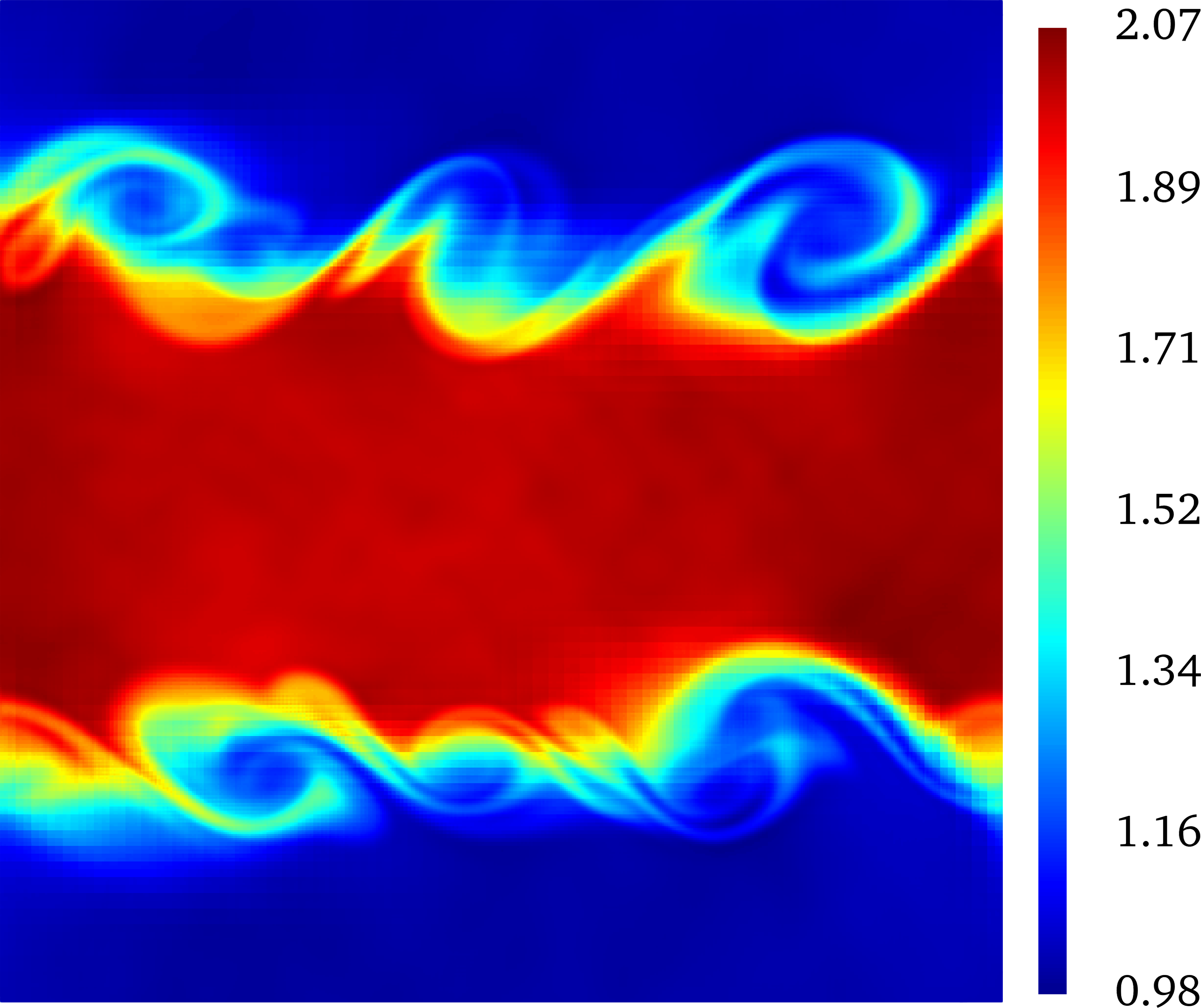}}\hfill
    \subfloat[$t=2$]{\label{fig:Kin_N1_avg_t2}%
        \includegraphics[width=.30\linewidth]{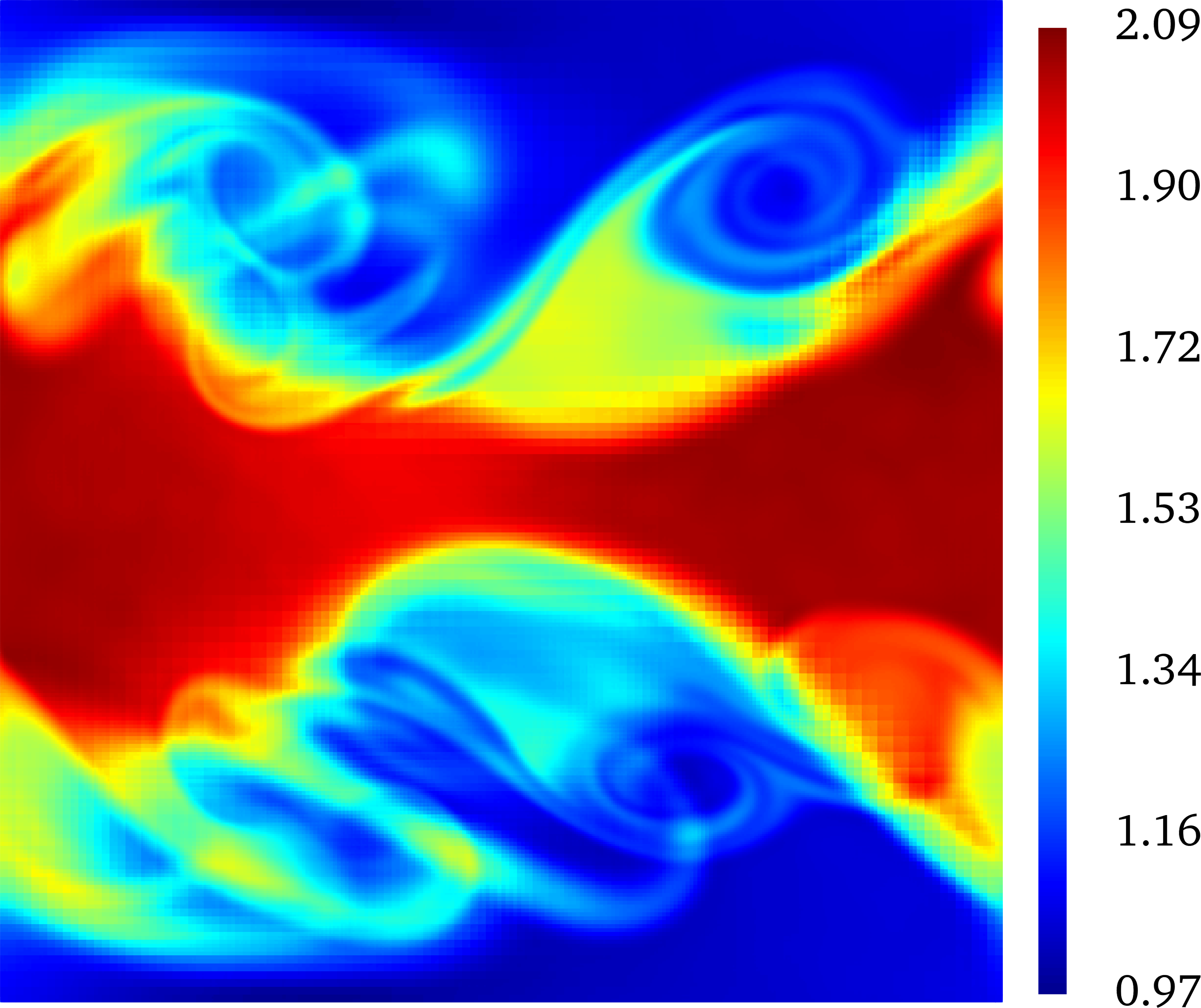}}\hfill
    \subfloat[$t=10$]{\label{fig:Kin_N1_avg_t10}%
        \includegraphics[width=.30\linewidth]{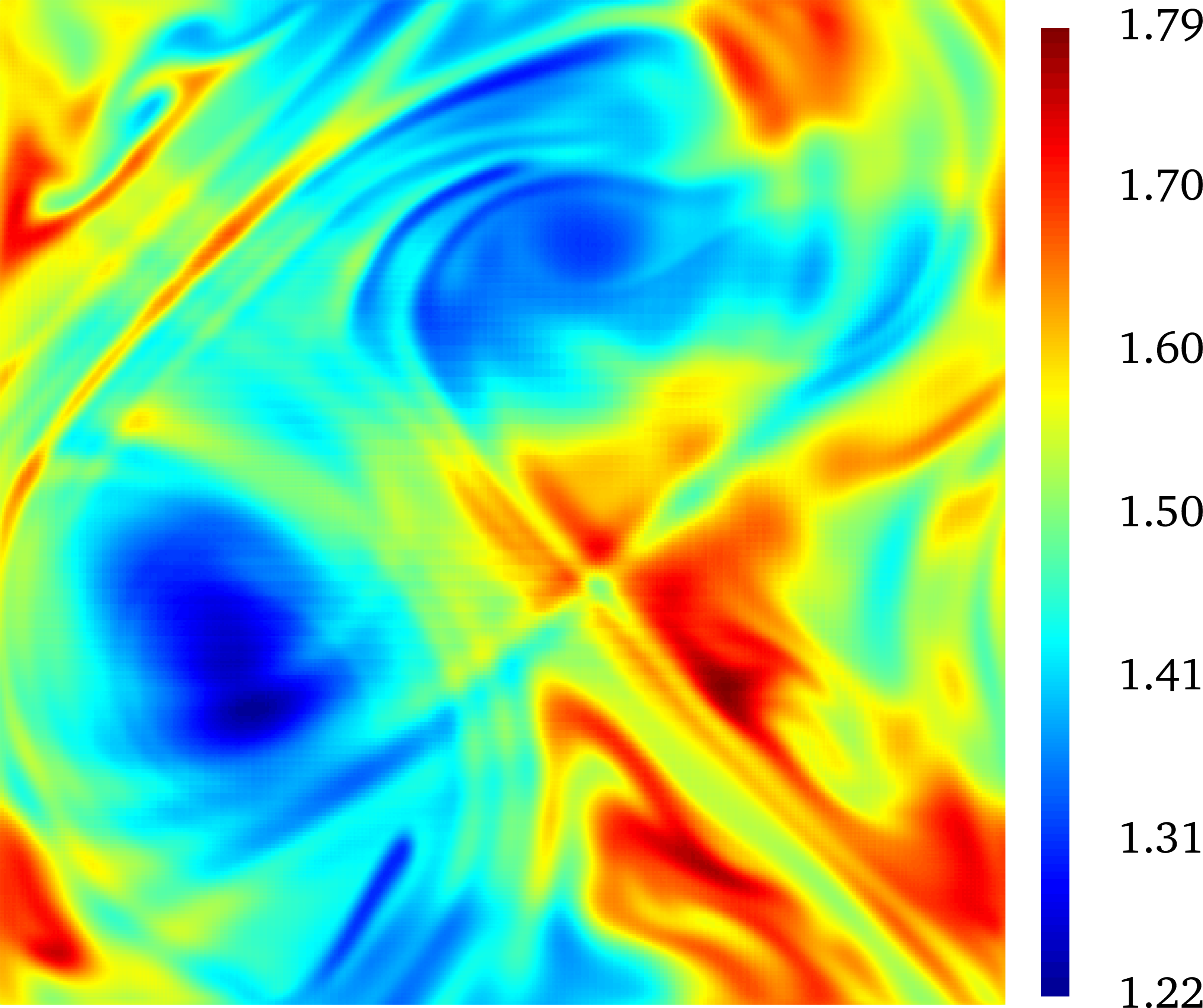}}\par
    \caption{Density (top row) and its \textit{Ces\`{a}ro} average (bottom row) for DVBFV method with $N=1$ at different times.}
    \label{fig:Kin_N1}
\end{figure}

\begin{figure}[ht]
    \centering
    \subfloat[$t=1$]{\label{fig:DG_N0_t1}%
        \includegraphics[width=.333\linewidth]{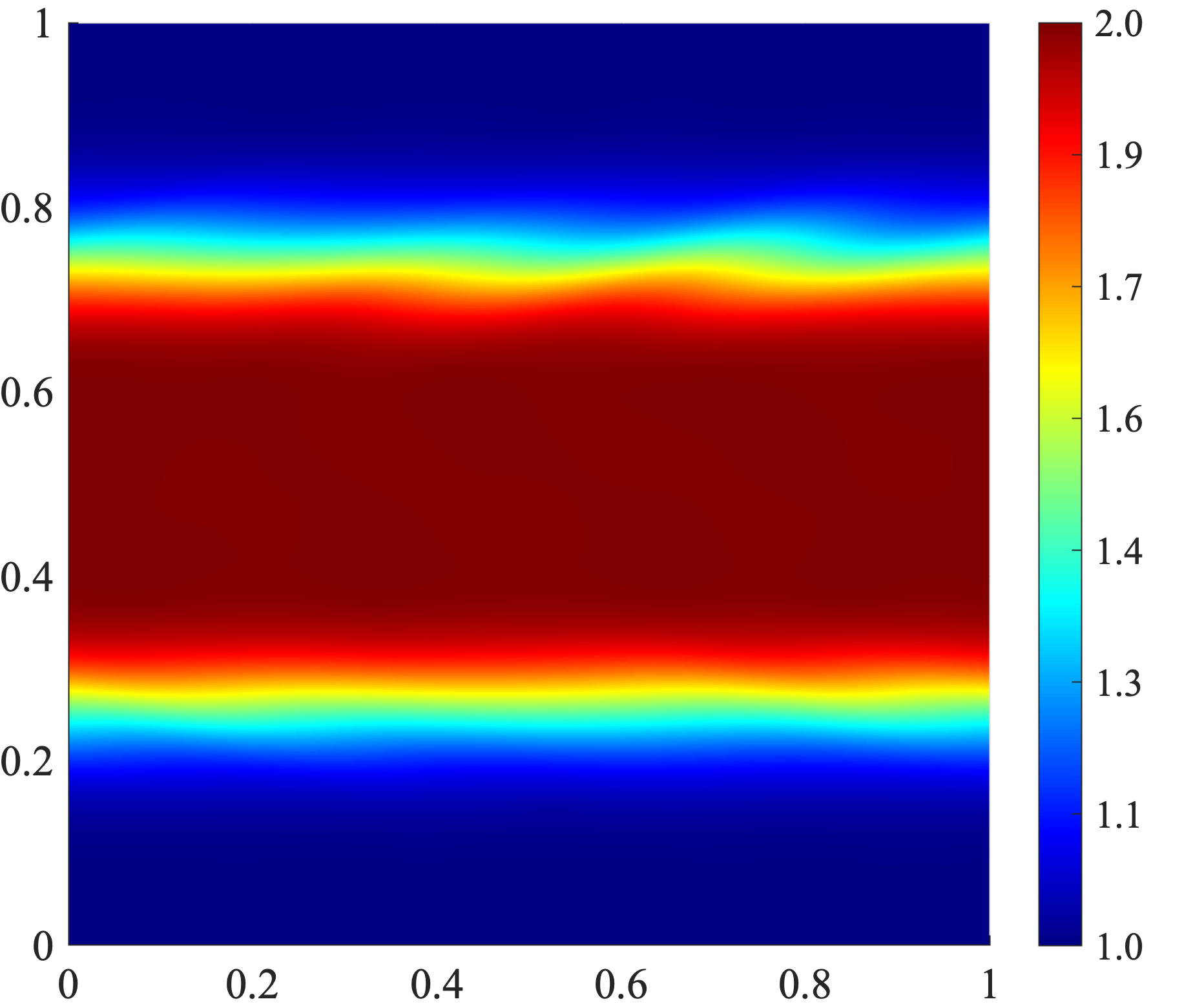}}\hfill
    \subfloat[$t=2$]{\label{fig:DG_N0_t2}%
        \includegraphics[width=.333\linewidth]{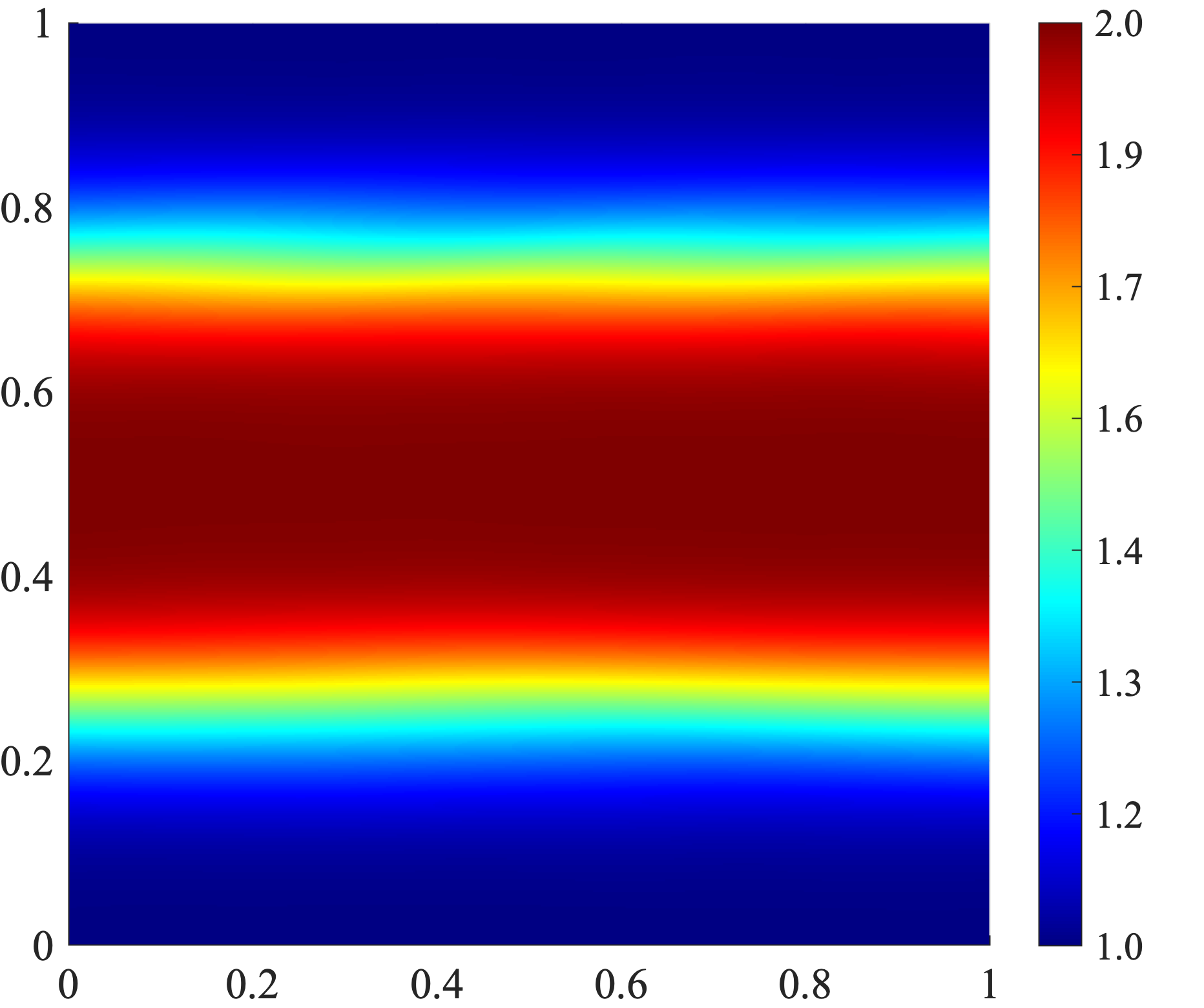}}\hfill
    \subfloat[$t=10$]{\label{fig:DG_N0_t10}%
        \includegraphics[width=.333\linewidth]{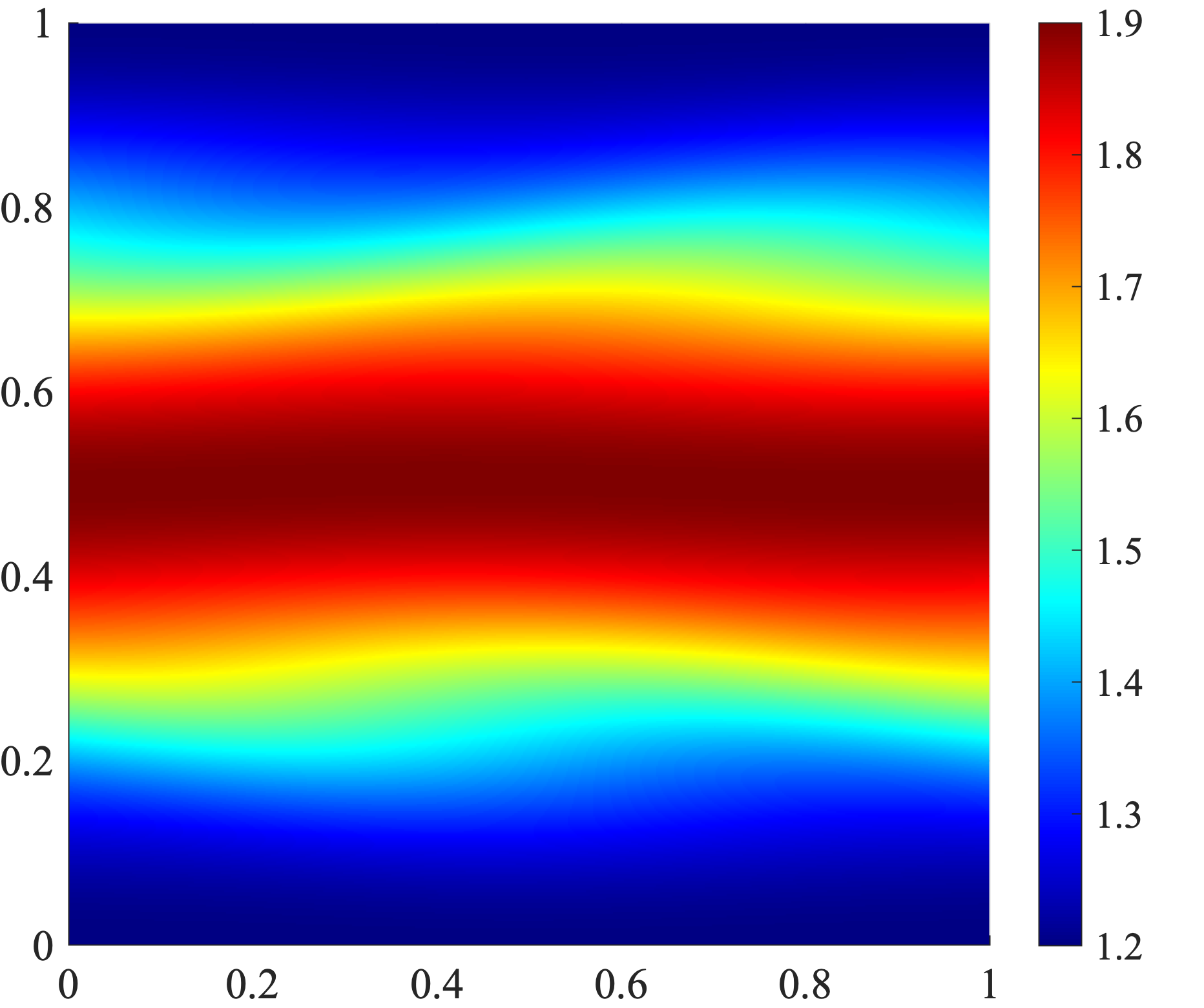}}
    \vspace{-8pt}
    \subfloat[$t=1$]{\label{fig:DG_N0_avg_t1}%
        \includegraphics[width=.333\linewidth]{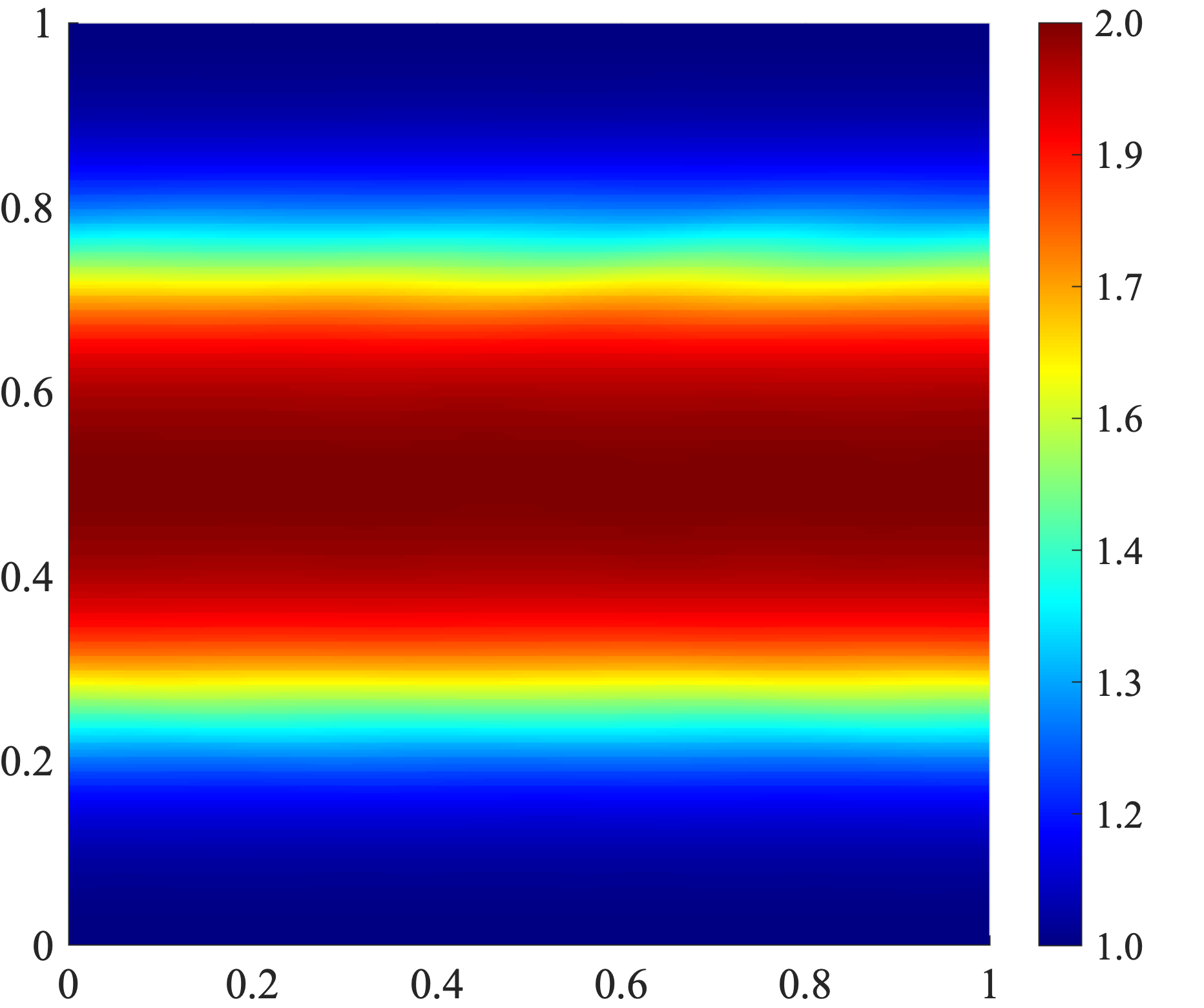}}\hfill
    \subfloat[$t=2$]{\label{fig:DG_N0_avg_t2}%
        \includegraphics[width=.333\linewidth]{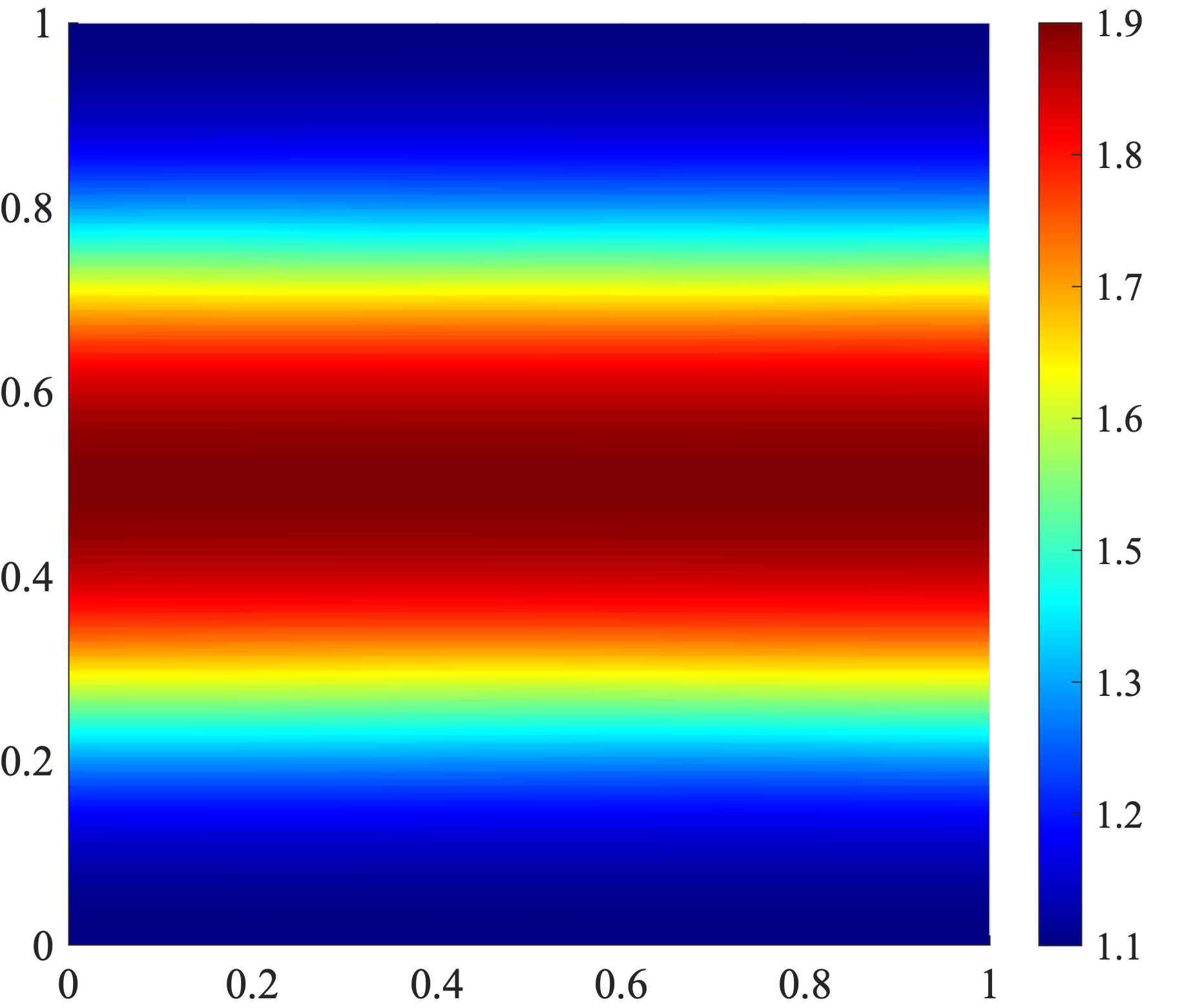}}\hfill
    \subfloat[$t=10$]{\label{fig:DG_N0_avg_t10}%
        \includegraphics[width=.333\linewidth]{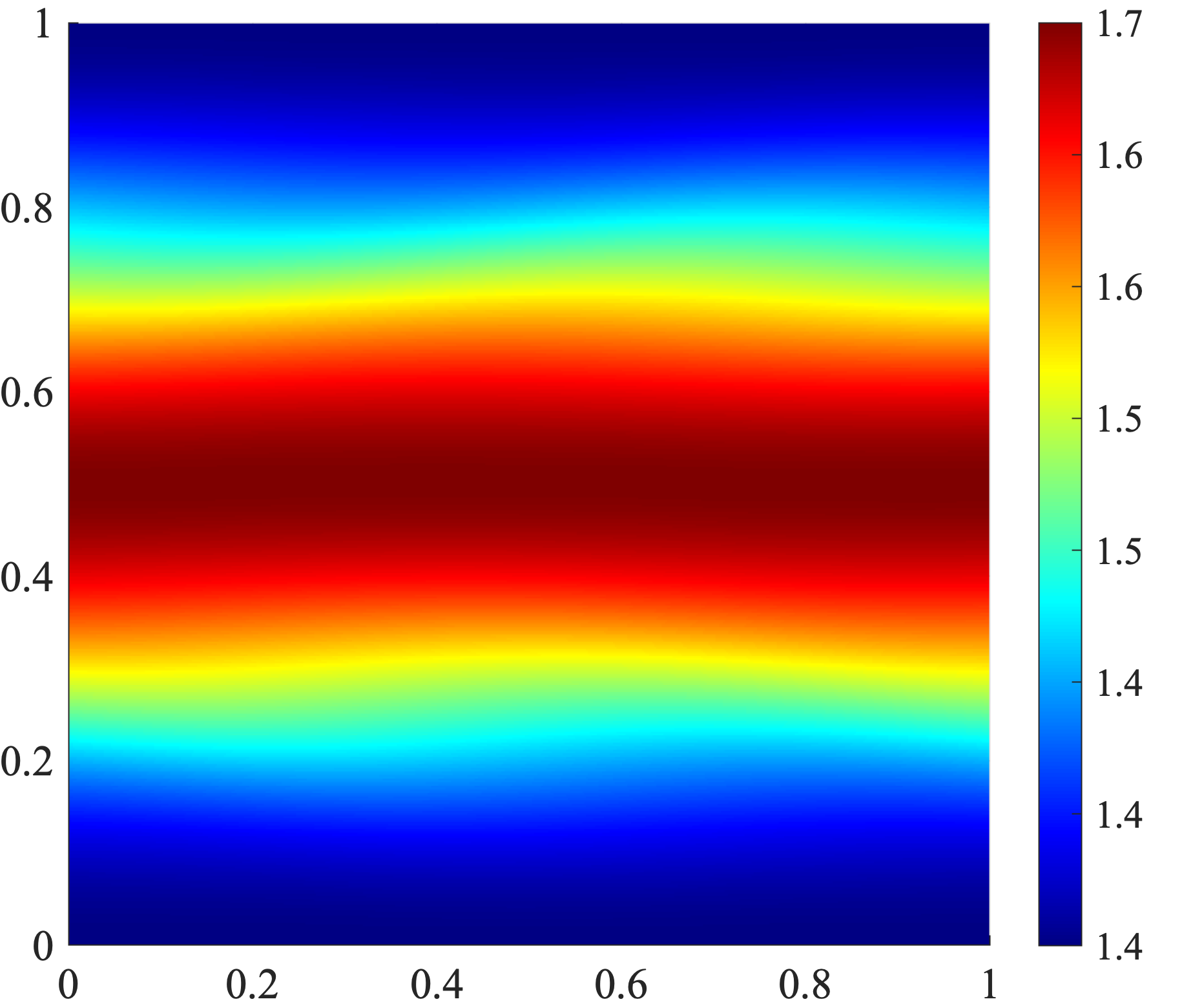}}\par
    \caption{Density (top row) and its \textit{Ces\`{a}ro} average (bottom row) for RKDG method with $N=0$ at different times.}
    \label{fig:DG_N0}
\end{figure}

\begin{figure}[ht]
    \centering
    \subfloat[$t=1$]{\label{fig:DG_N1_t1}%
        \includegraphics[width=.333\linewidth]{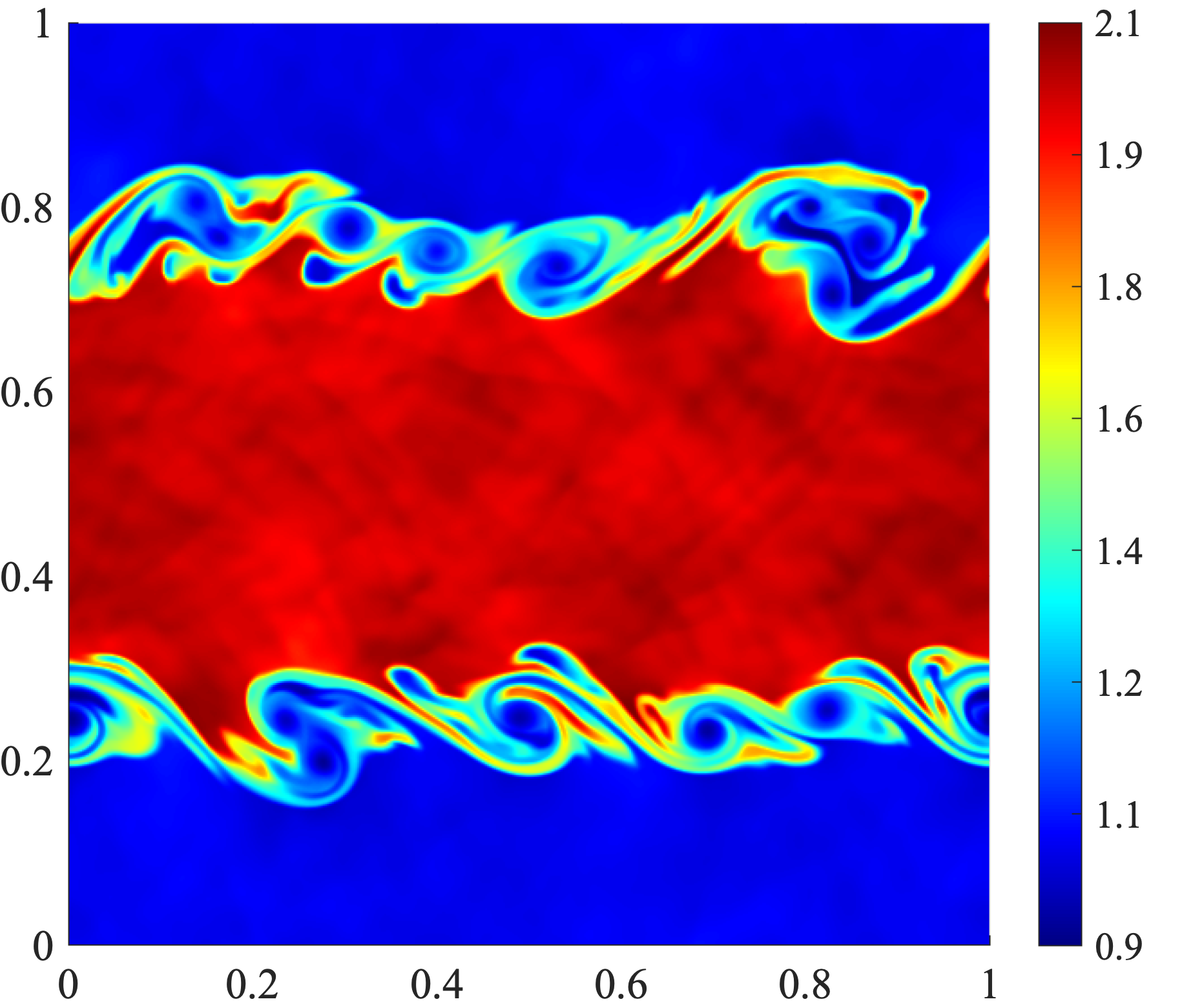}}\hfill
    \subfloat[$t=2$]{\label{fig:DG_N1_t2}%
        \includegraphics[width=.33\linewidth]{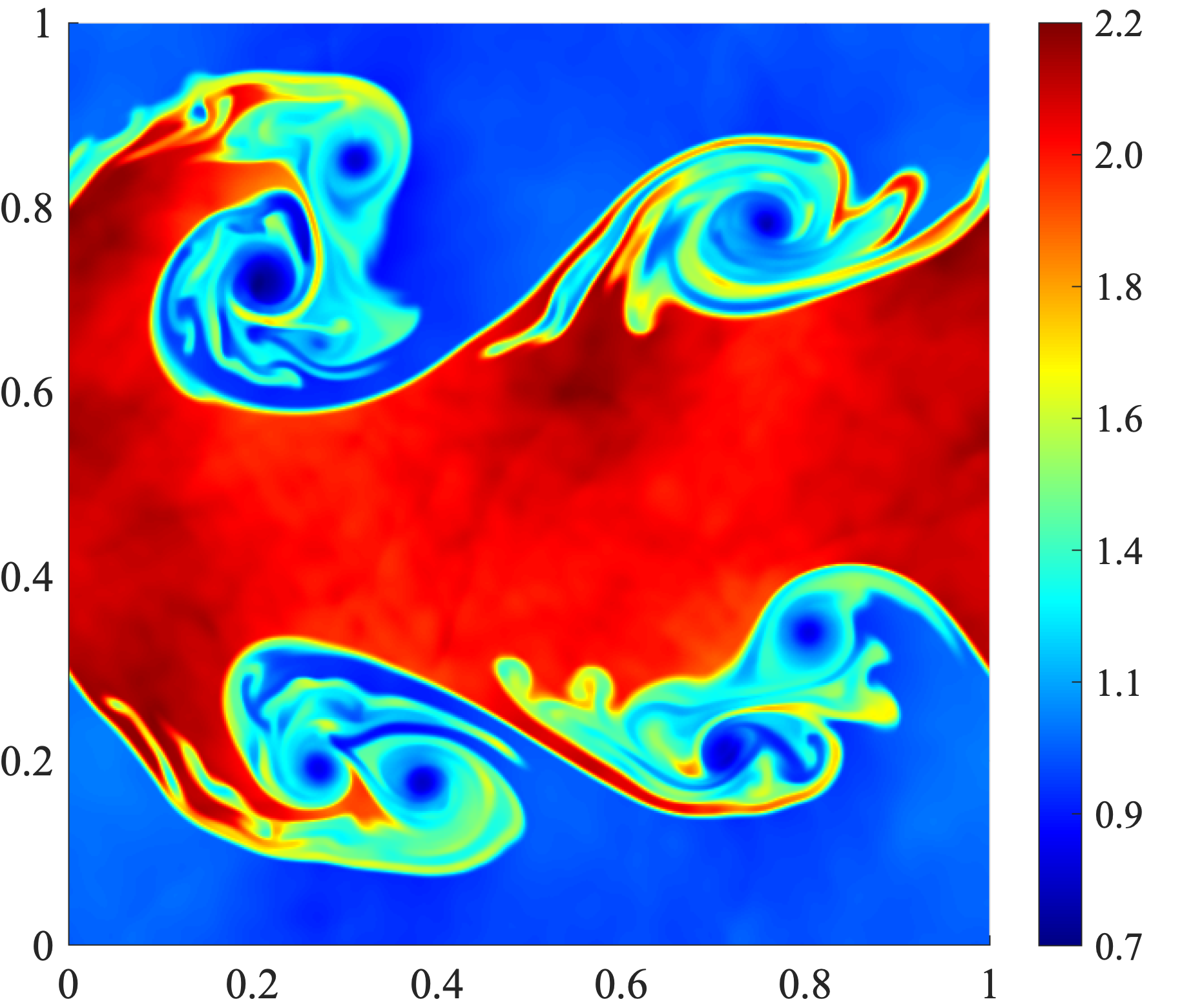}}\hfill
    \subfloat[$t=10$]{\label{fig:DG_N1_t10}%
        \includegraphics[width=.333\linewidth]{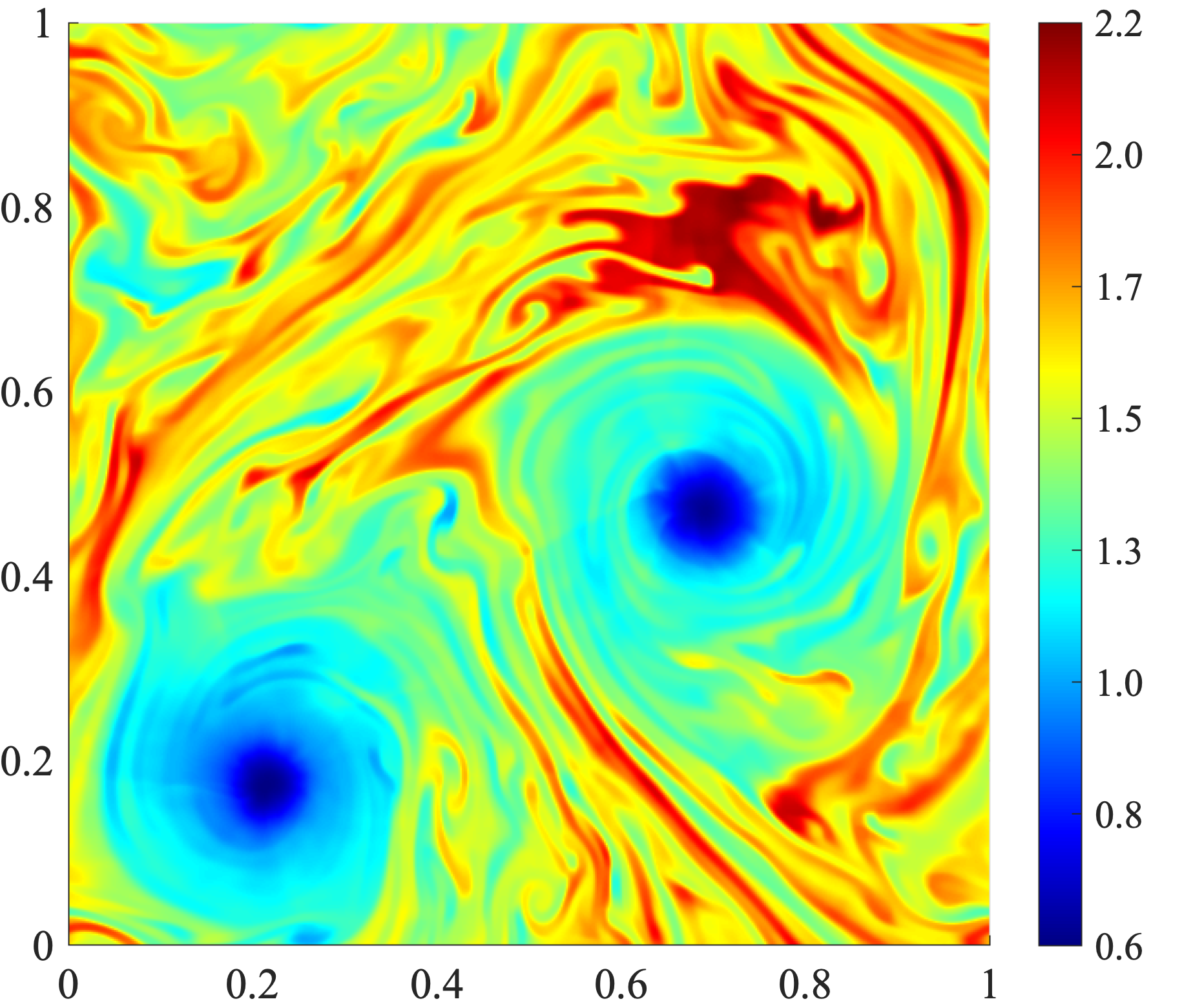}}
    \vspace{-8pt}
    \subfloat[$t=1$]{\label{fig:DG_N1_avg_t1}%
        \includegraphics[width=.333\linewidth]{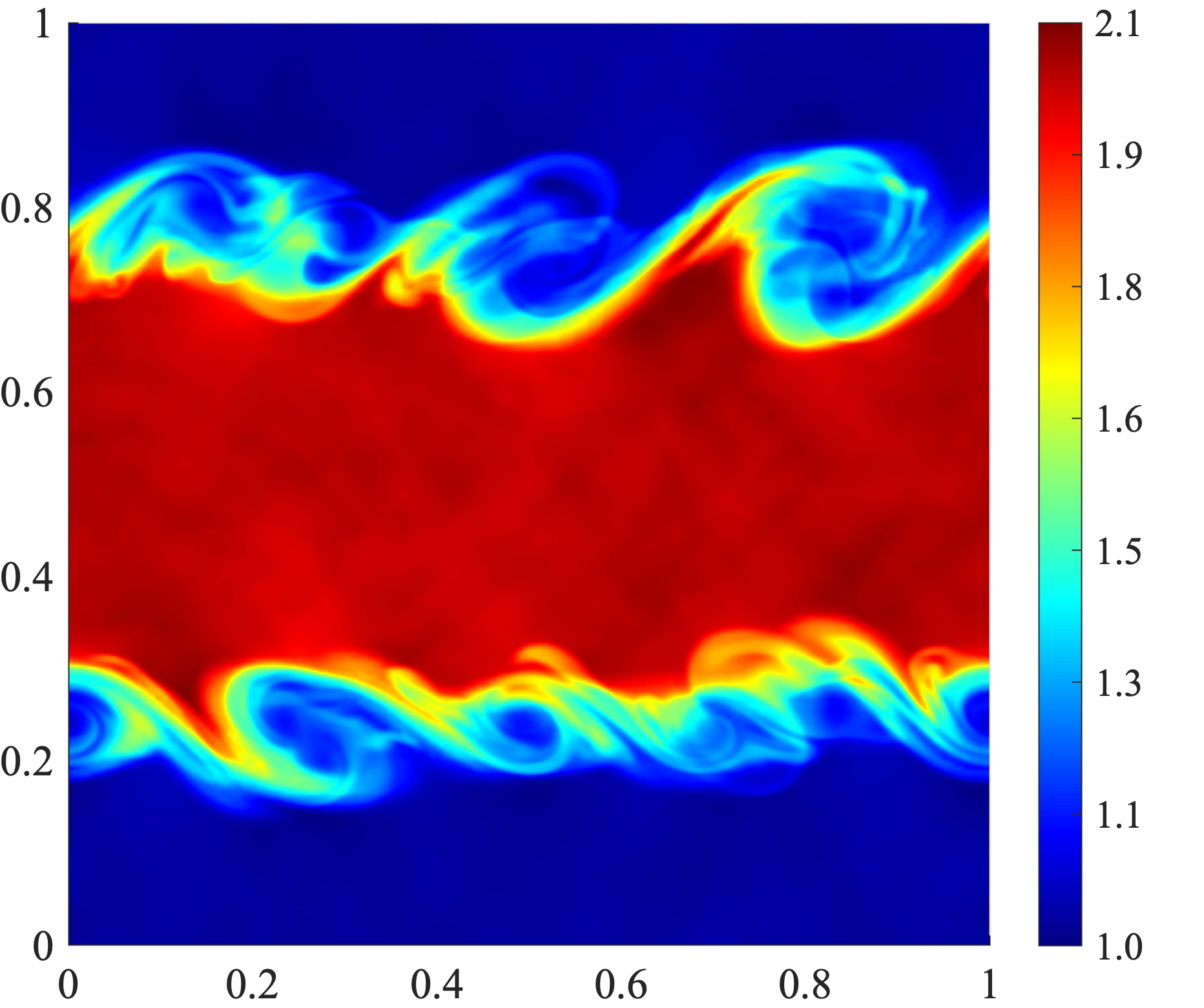}}\hfill
    \subfloat[$t=2$]{\label{fig:DG_N1_avg_t2}%
        \includegraphics[width=.333\linewidth]{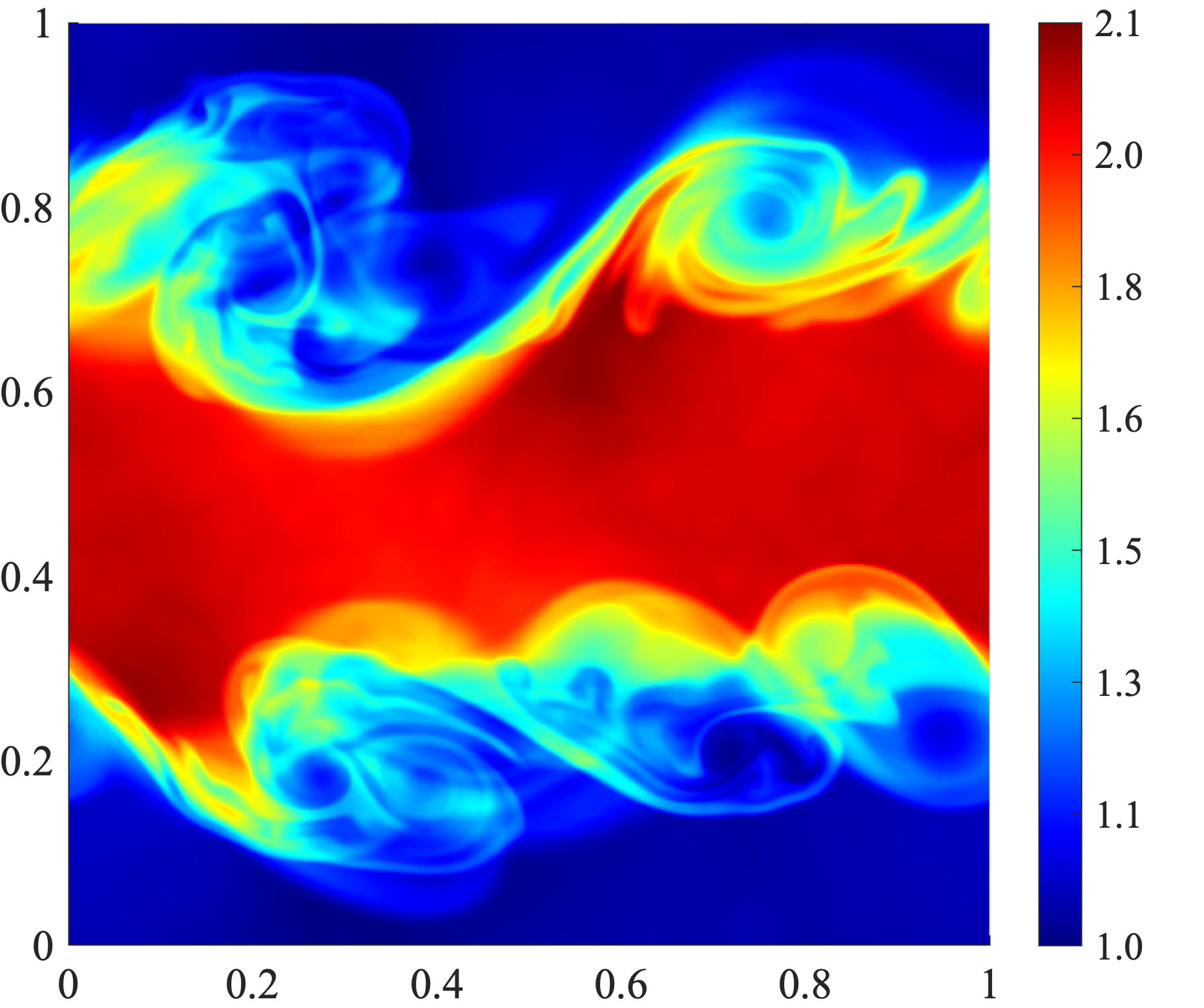}}\hfill
    \subfloat[$t=10$]{\label{fig:DG_N1_avg_t10}%
        \includegraphics[width=.333\linewidth]{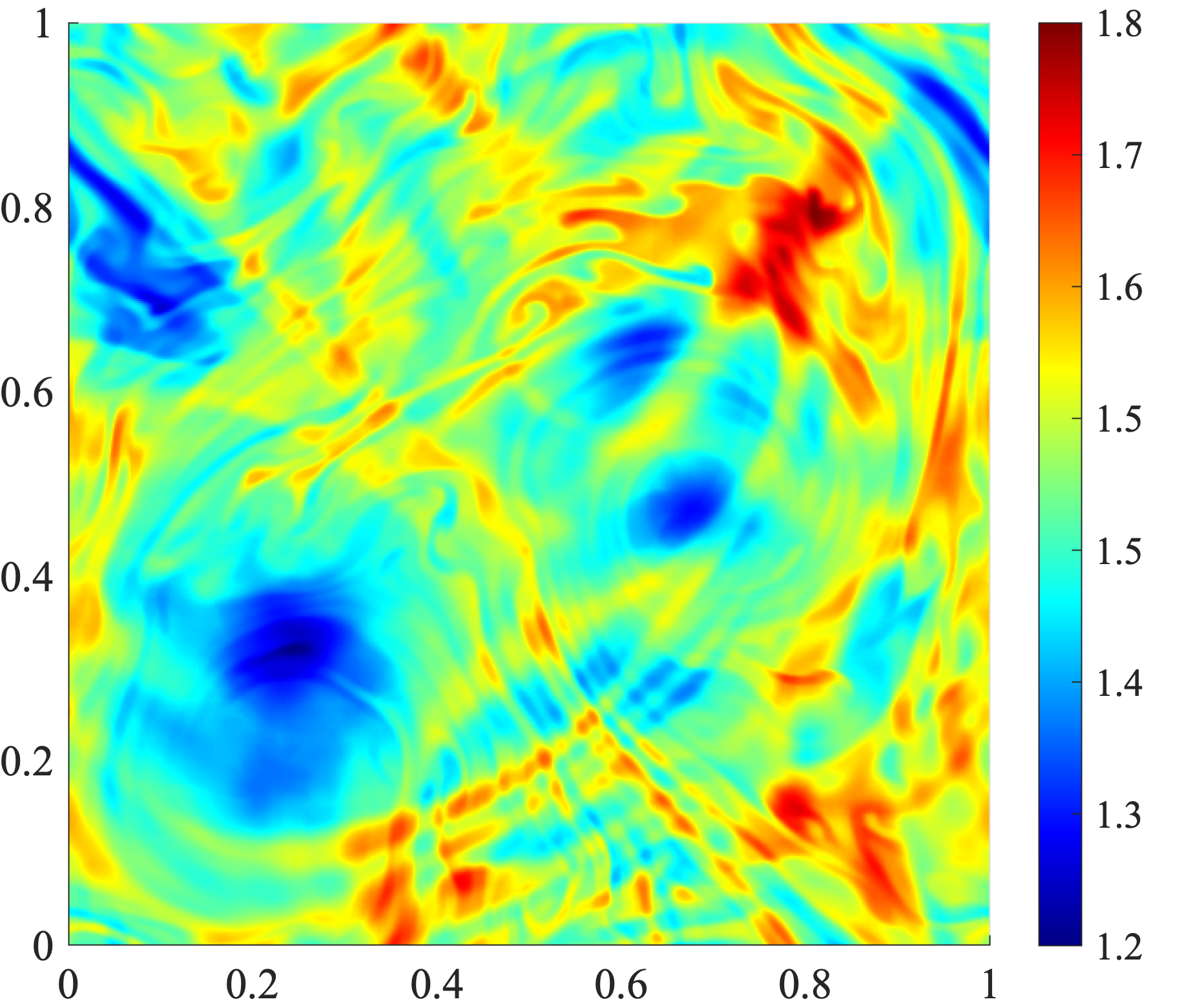}}\par
    \caption{Density (top row) and its \textit{Ces\`{a}ro} average (bottom row) for RKDG method with $N=1$ at different times.}
    \label{fig:DG_N1}
\end{figure}

\begin{figure}[ht]
    \centering
    \subfloat[$t=1$]{\label{fig:DG_N2_t1}%
        \includegraphics[width=.333\linewidth]{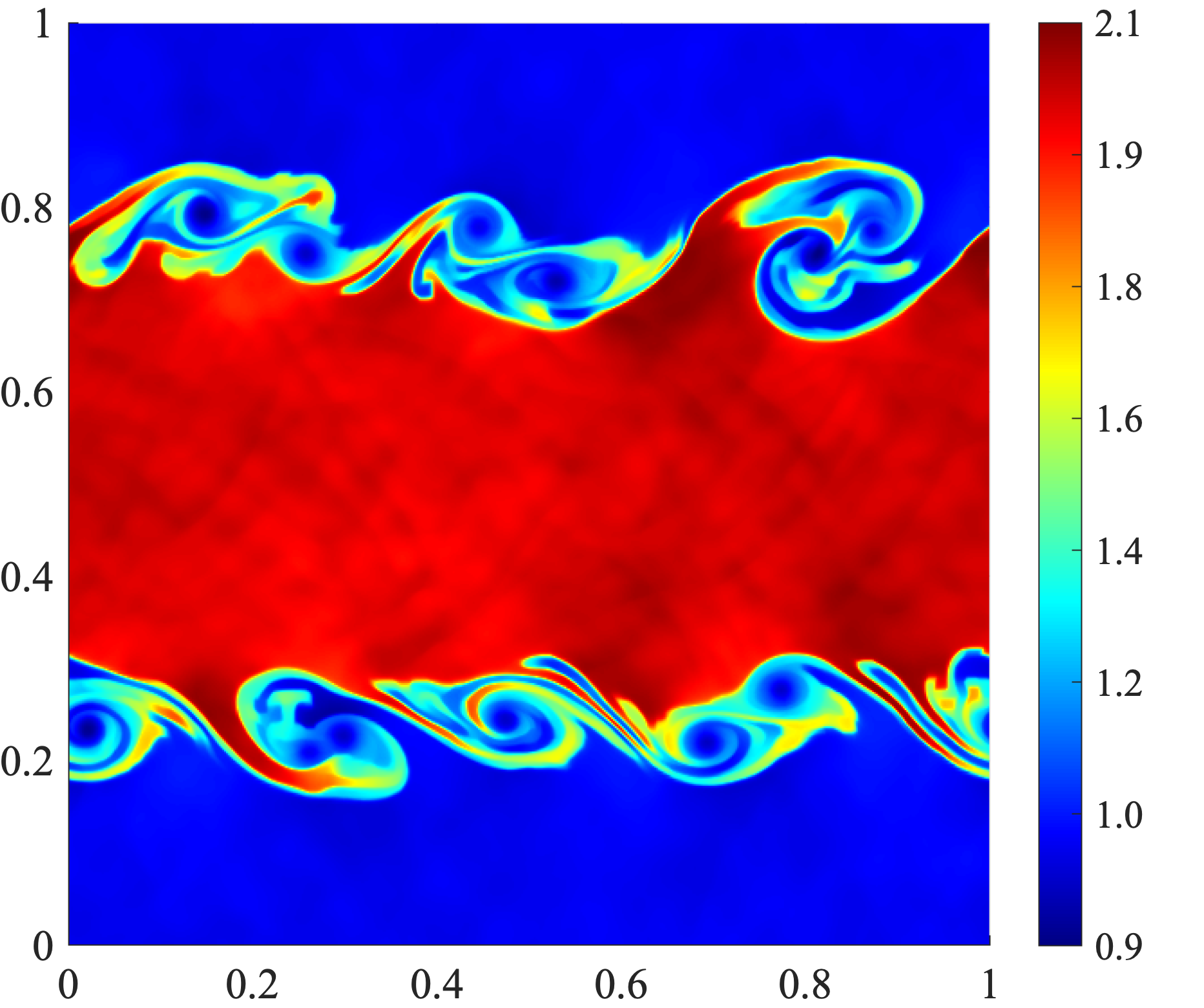}}\hfill
    \subfloat[$t=2$]{\label{fig:DG_N2_t2}%
        \includegraphics[width=.333\linewidth]{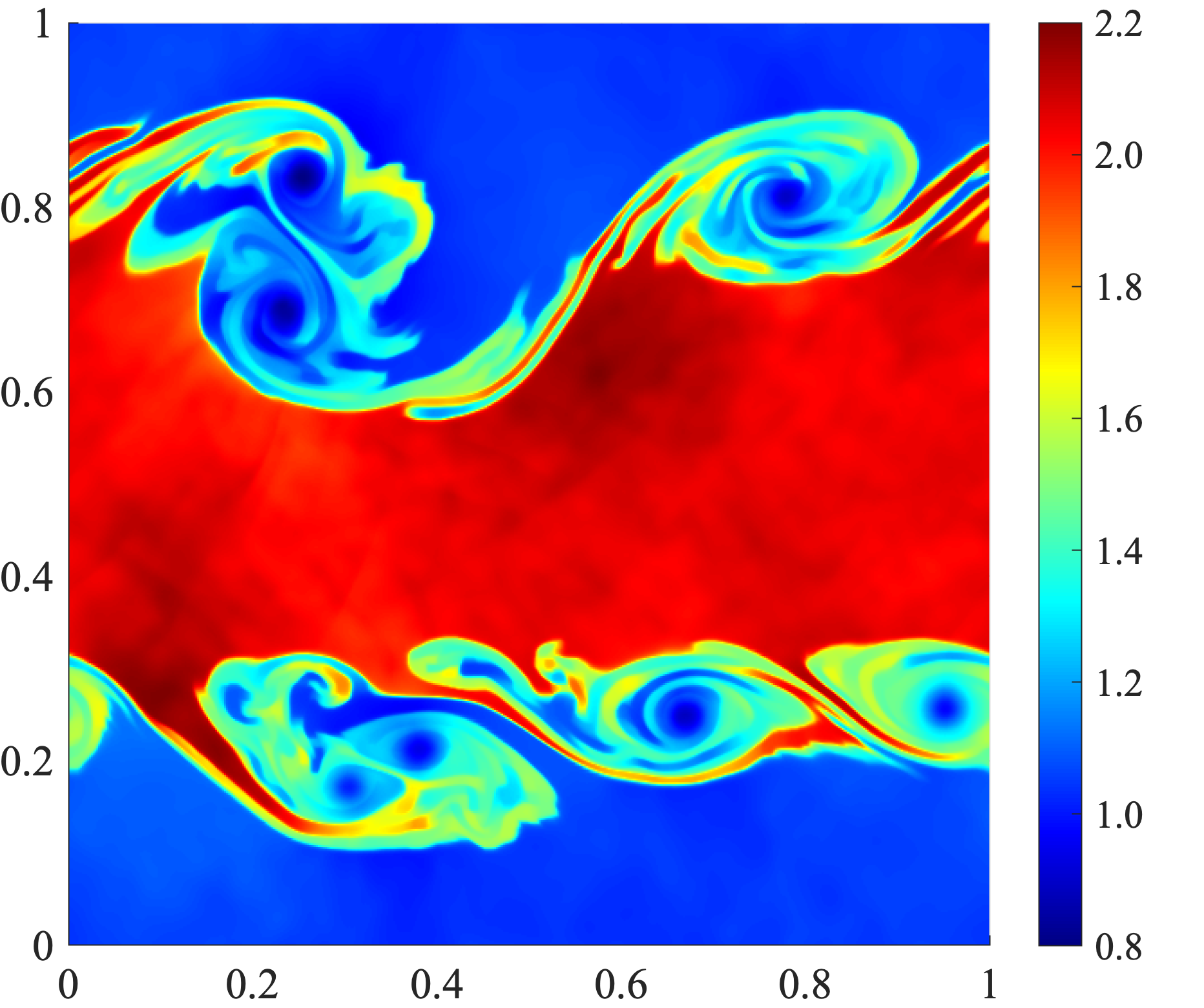}}\hfill
    \subfloat[$t=10$]{\label{fig:DG_N2_t10}%
        \includegraphics[width=.333\linewidth]{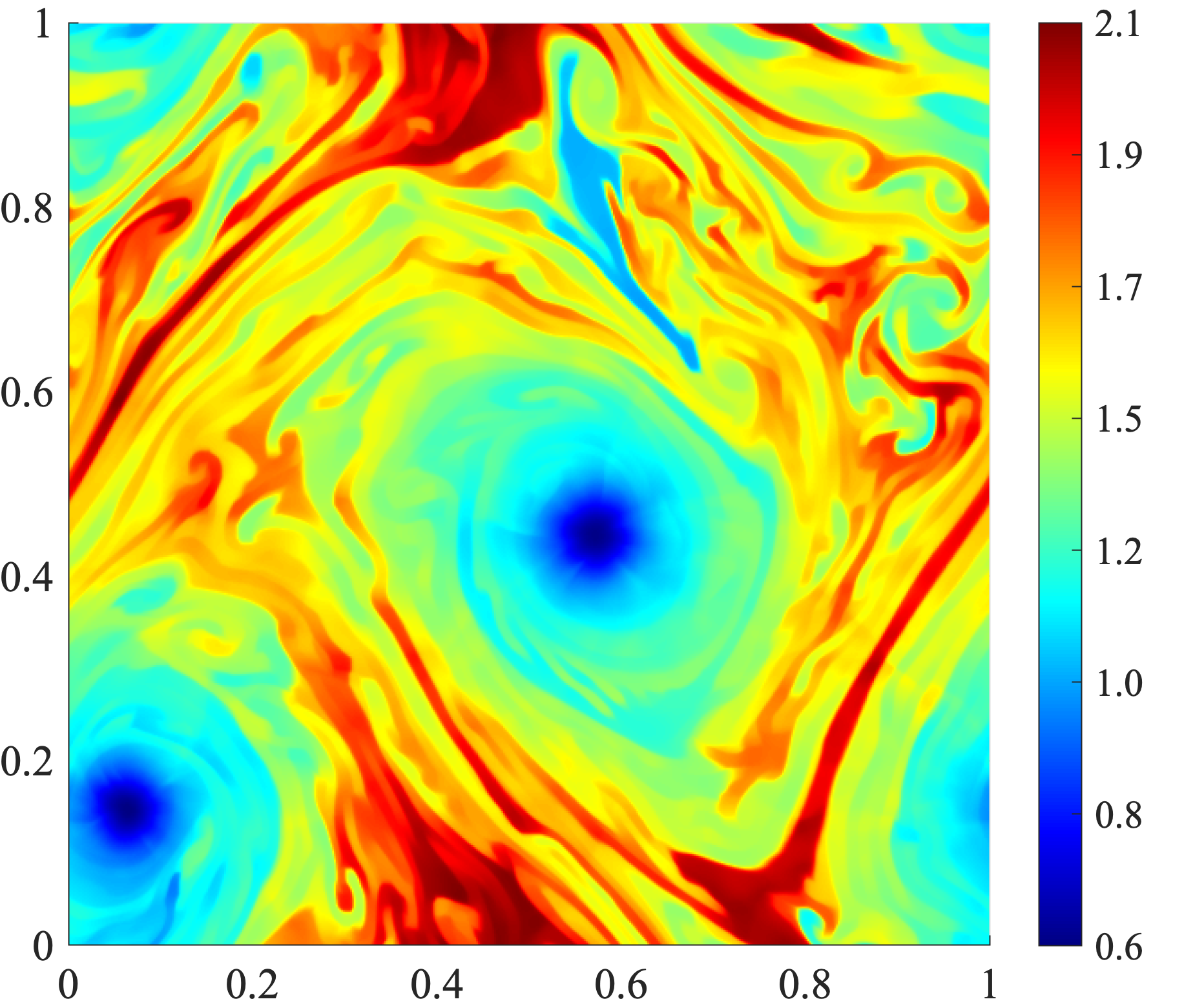}}
    \vspace{-8pt}
    \subfloat[$t=1$]{\label{fig:DG_N2_avg_t1}%
        \includegraphics[width=.333\linewidth]{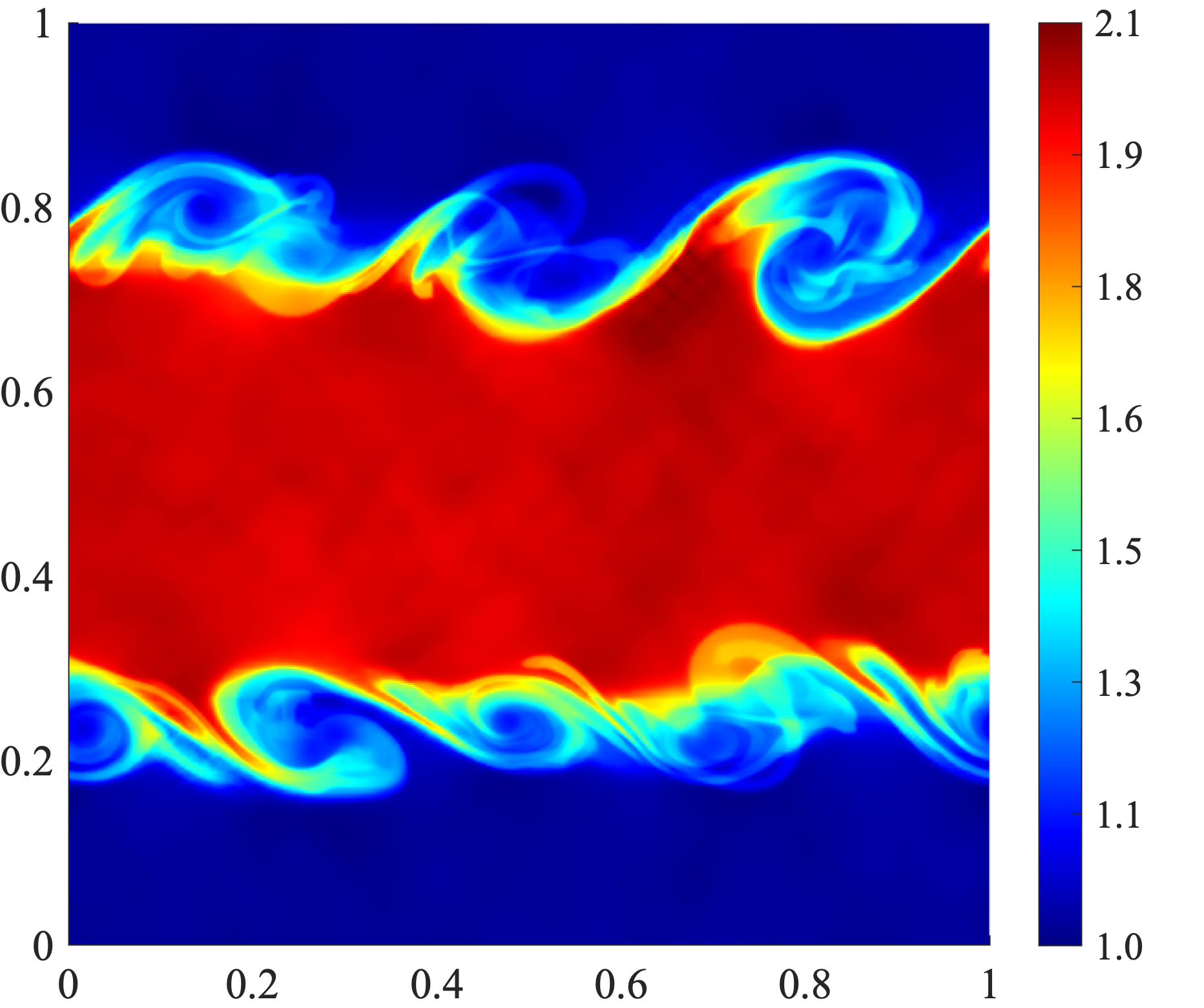}}\hfill
    \subfloat[$t=2$]{\label{fig:DG_N2_avg_t2}%
        \includegraphics[width=.333\linewidth]{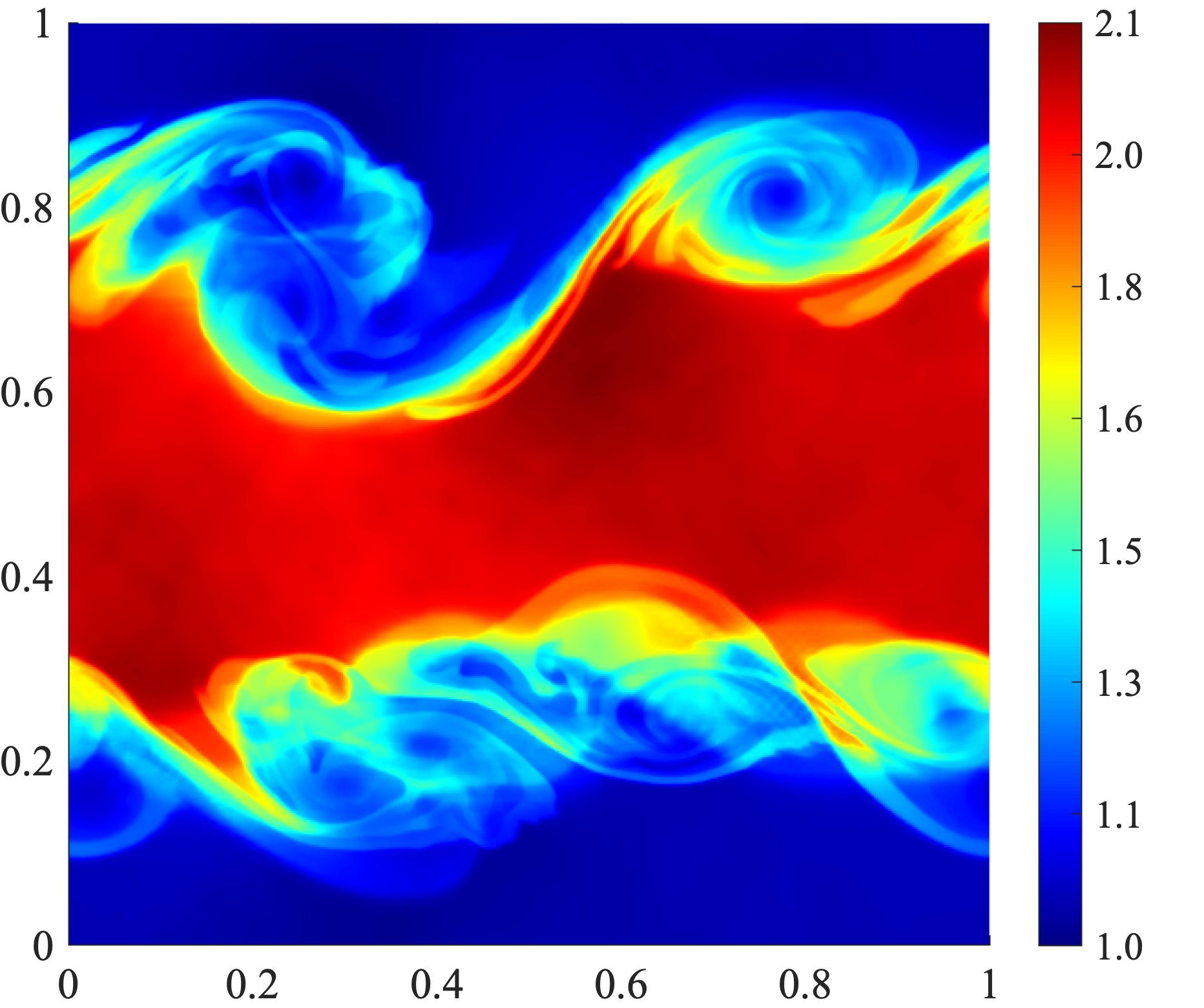}}\hfill
    \subfloat[$t=10$]{\label{fig:DG_N2_avg_t10}%
        \includegraphics[width=.333\linewidth]{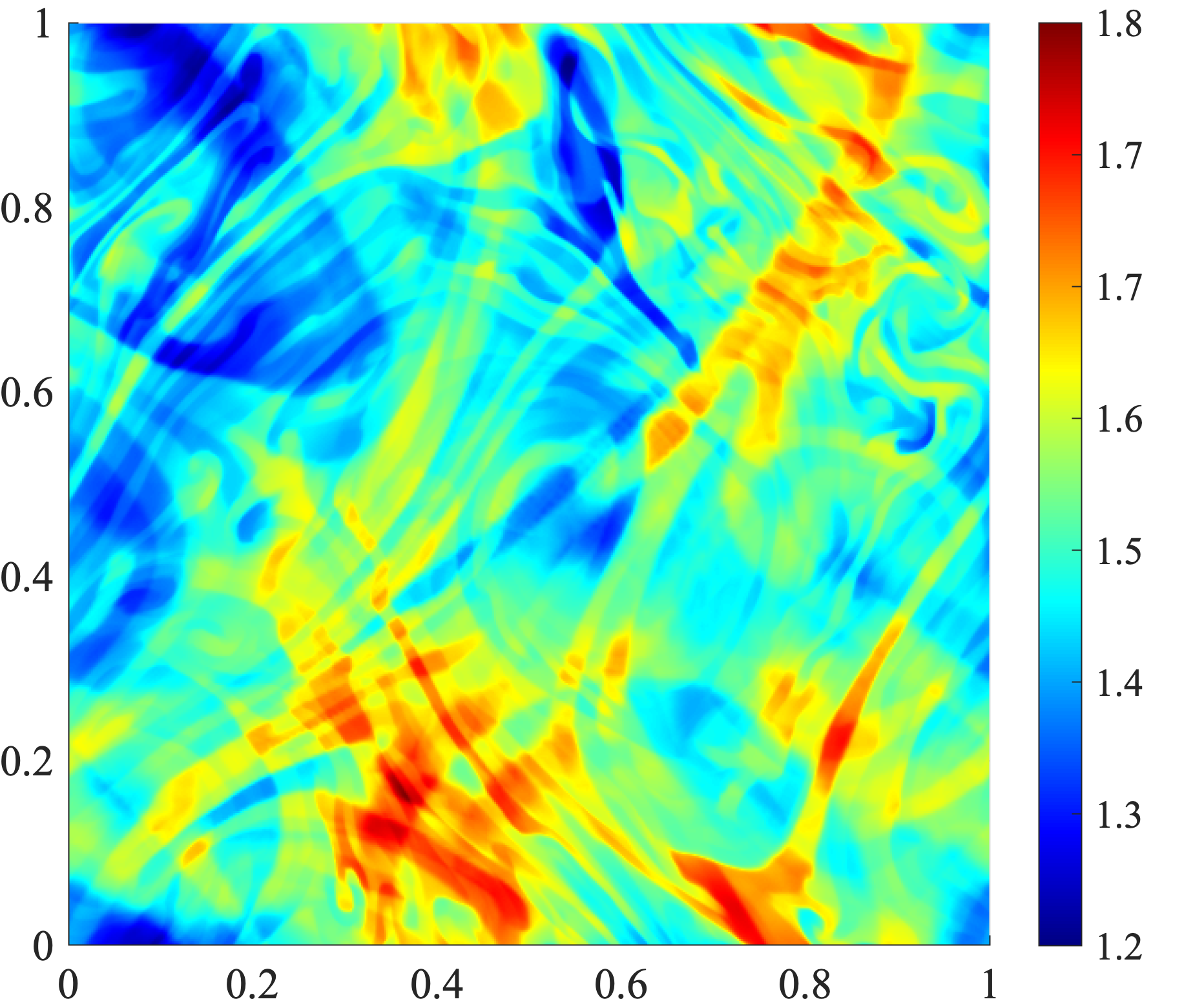}}\par
    \caption{Density (top row) and its \textit{Ces\`{a}ro} average (bottom row) for RKDG method with $N=2$ at different times.}
    \label{fig:DG_N2}
\end{figure}

\begin{figure}[ht]
    \centering
    \subfloat[$t=1$]{\label{fig:DG_N4_t1}%
        \includegraphics[width=.333\linewidth]{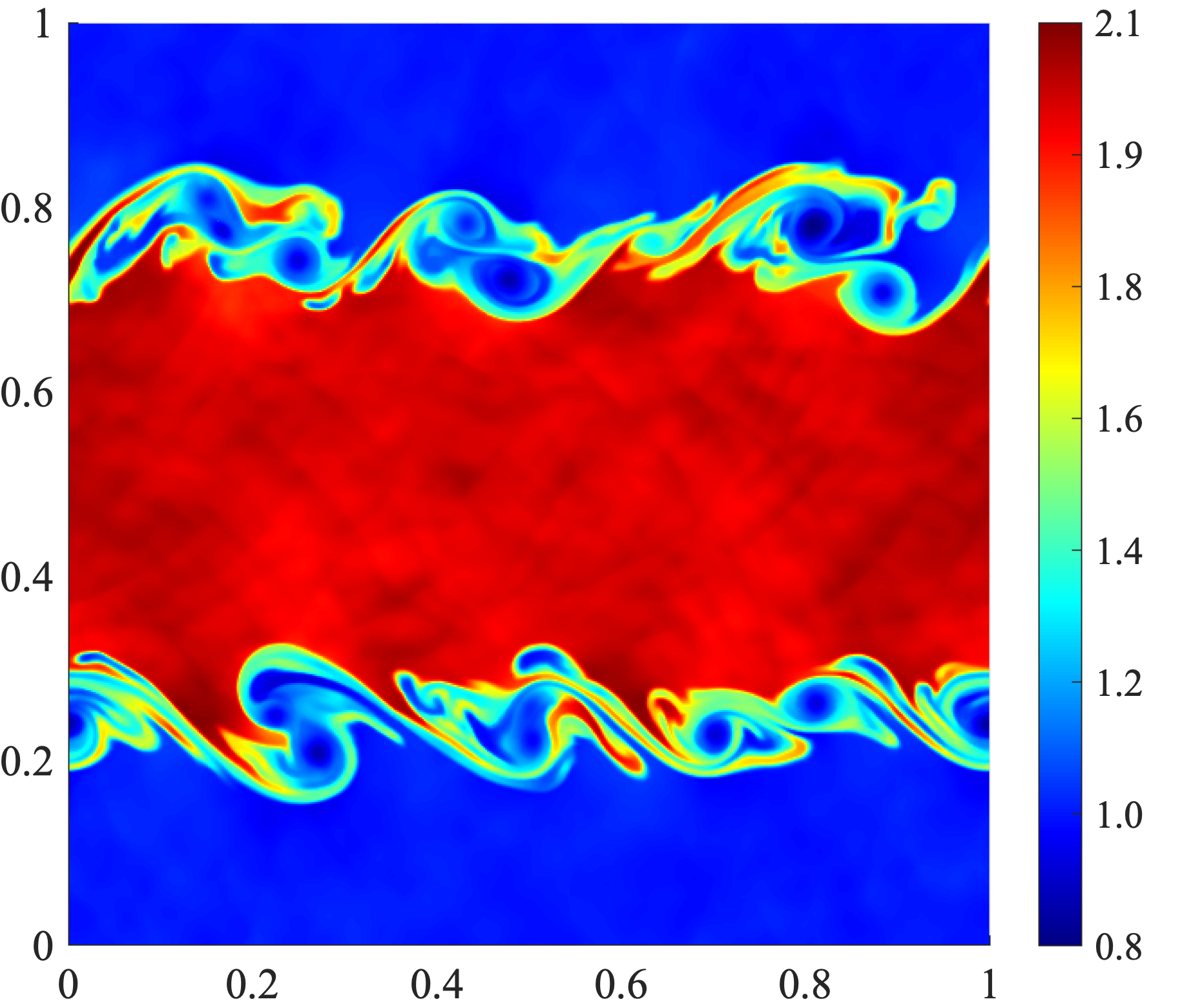}}\hfill
    \subfloat[$t=2$]{\label{fig:DG_N4_t2}%
        \includegraphics[width=.333\linewidth]{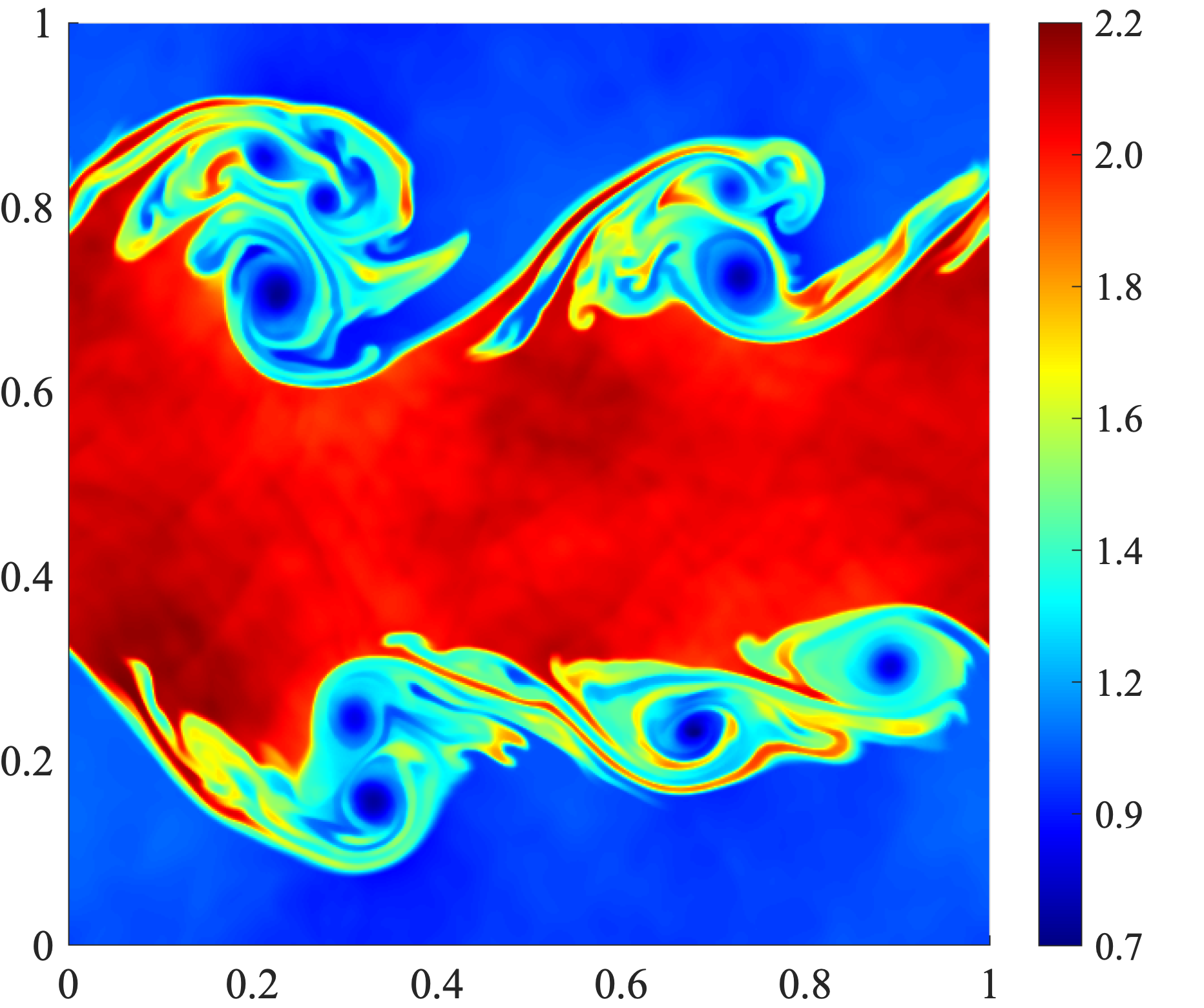}}\hfill
    \subfloat[$t=10$]{\label{fig:DG_N4_t10}%
        \includegraphics[width=.333\linewidth]{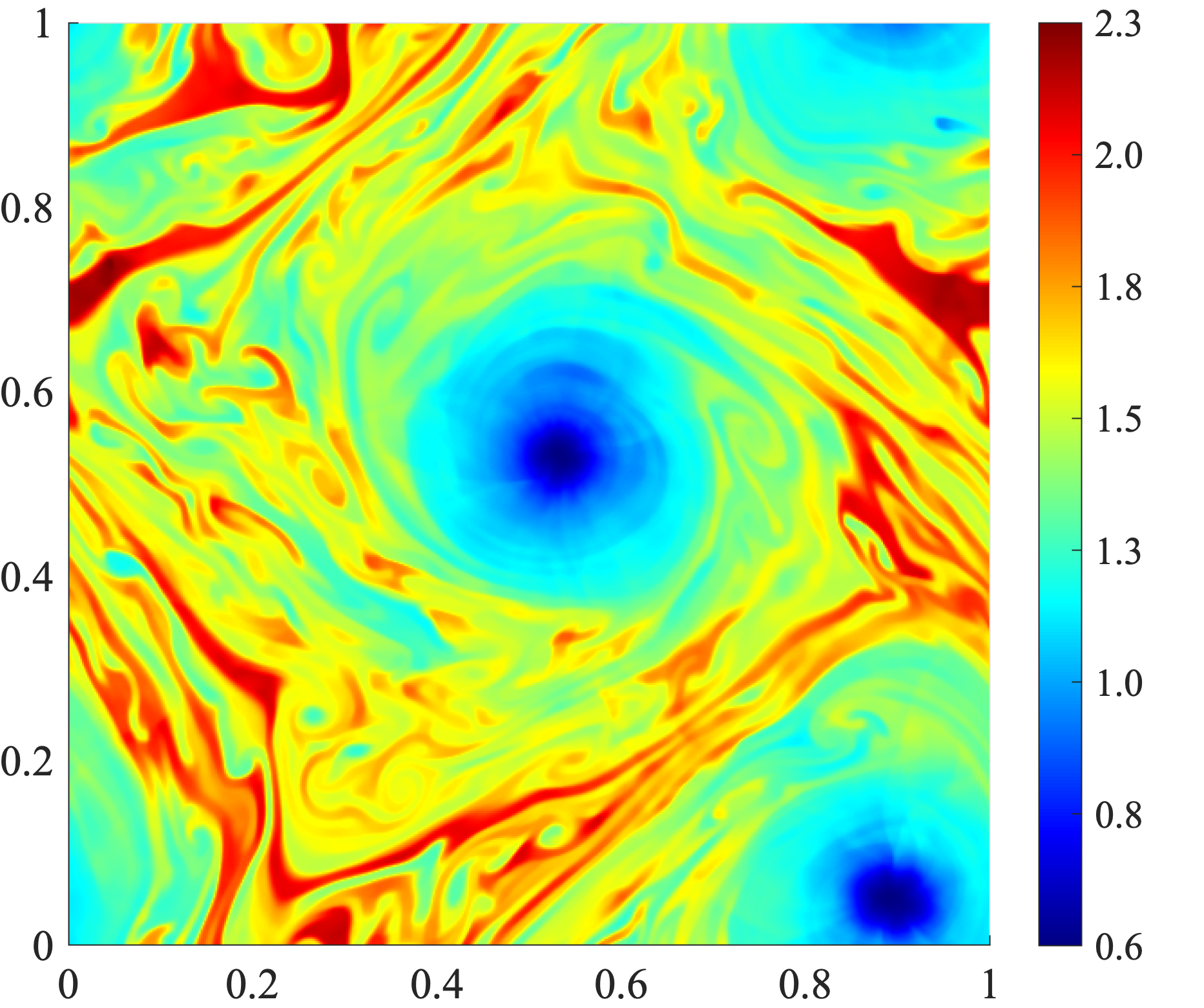}}
    \vspace{-8pt}
    \subfloat[$t=1$]{\label{fig:DG_N4_avg_t1}%
        \includegraphics[width=.333\linewidth]{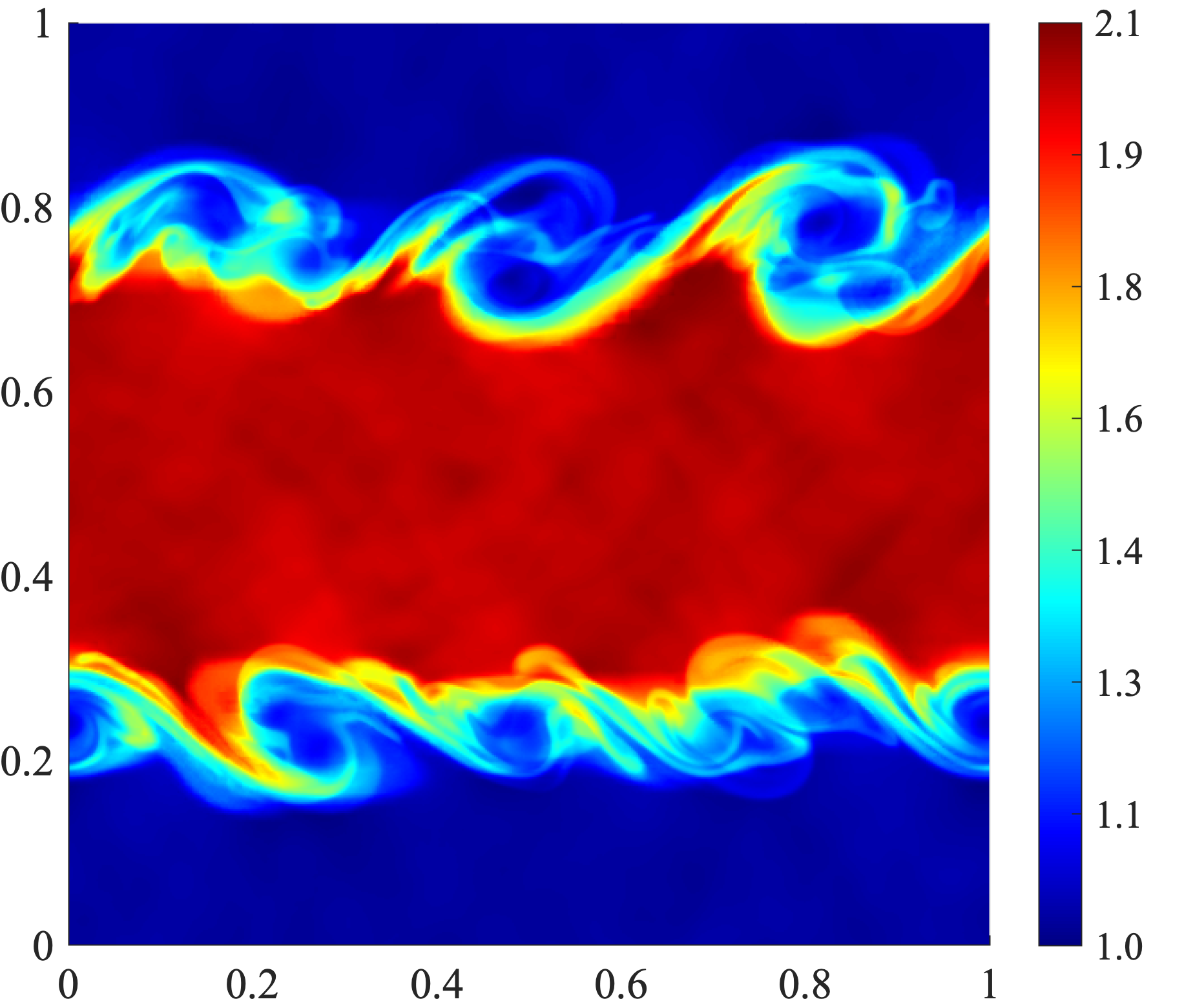}}\hfill
    \subfloat[$t=2$]{\label{fig:DG_N4_avg_t2}%
        \includegraphics[width=.333\linewidth]{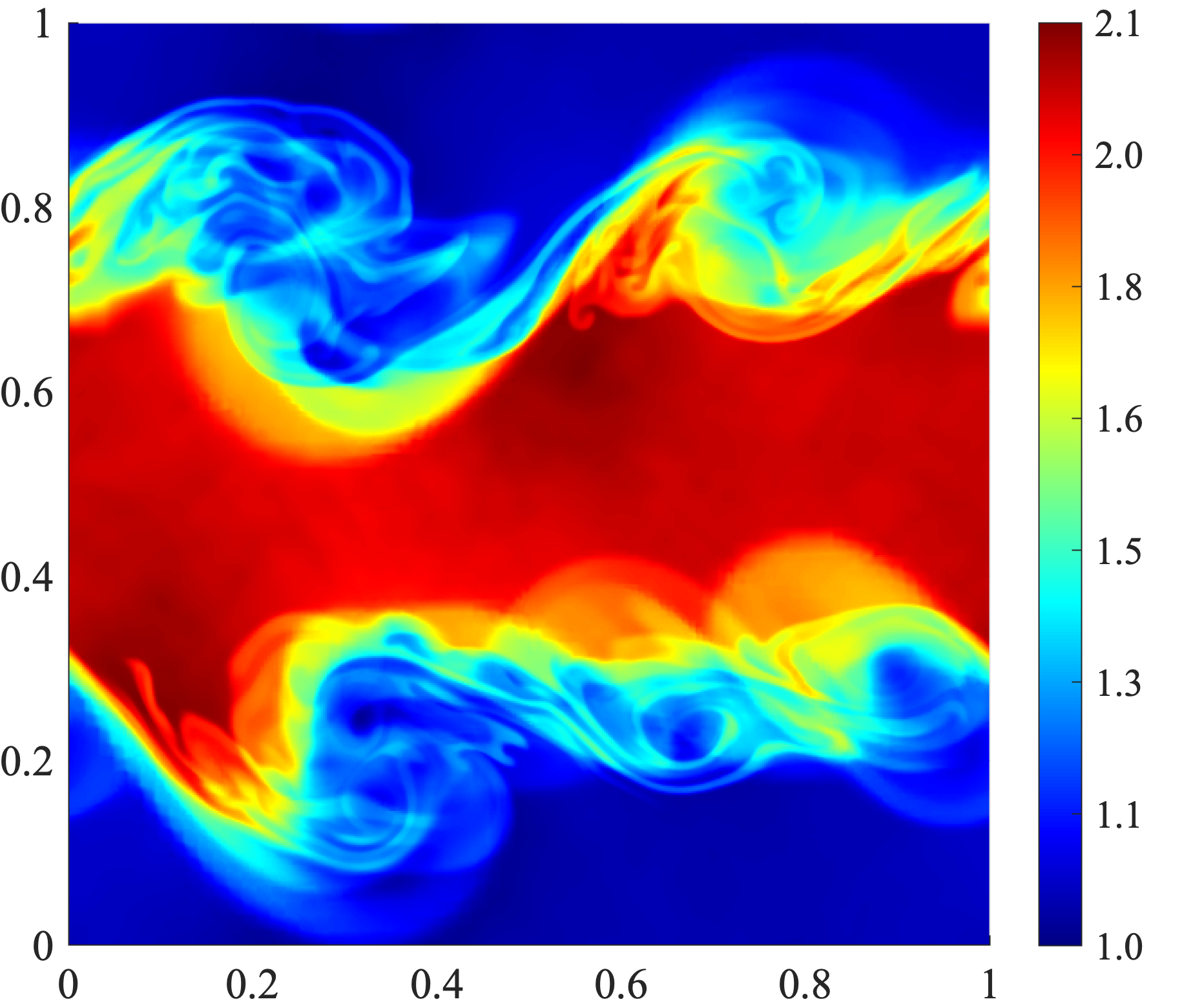}}\hfill
    \subfloat[$t=10$]{\label{fig:DG_N4_avg_t10}%
        \includegraphics[width=.333\linewidth]{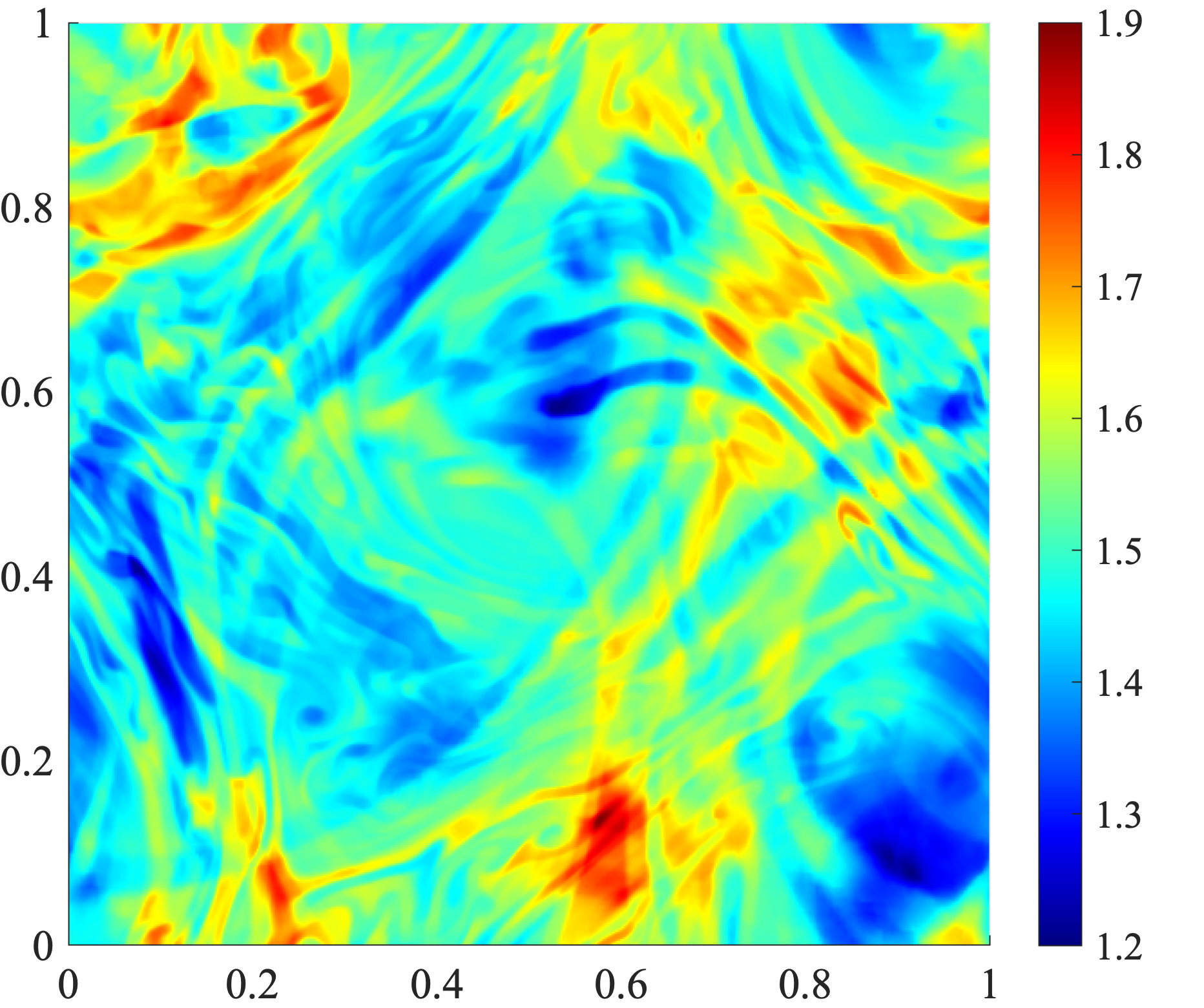}}\par
    \caption{Density (top row) and its \textit{Ces\`{a}ro} average (bottom row) for RKDG method with $N=4$ at different times.}
    \label{fig:DG_N4}
\end{figure}

\begin{figure}[ht]
    \centering
    \subfloat[$t=1$]{\label{fig:DGSEM_N1_t1}%
        \includegraphics[width=.33\linewidth]{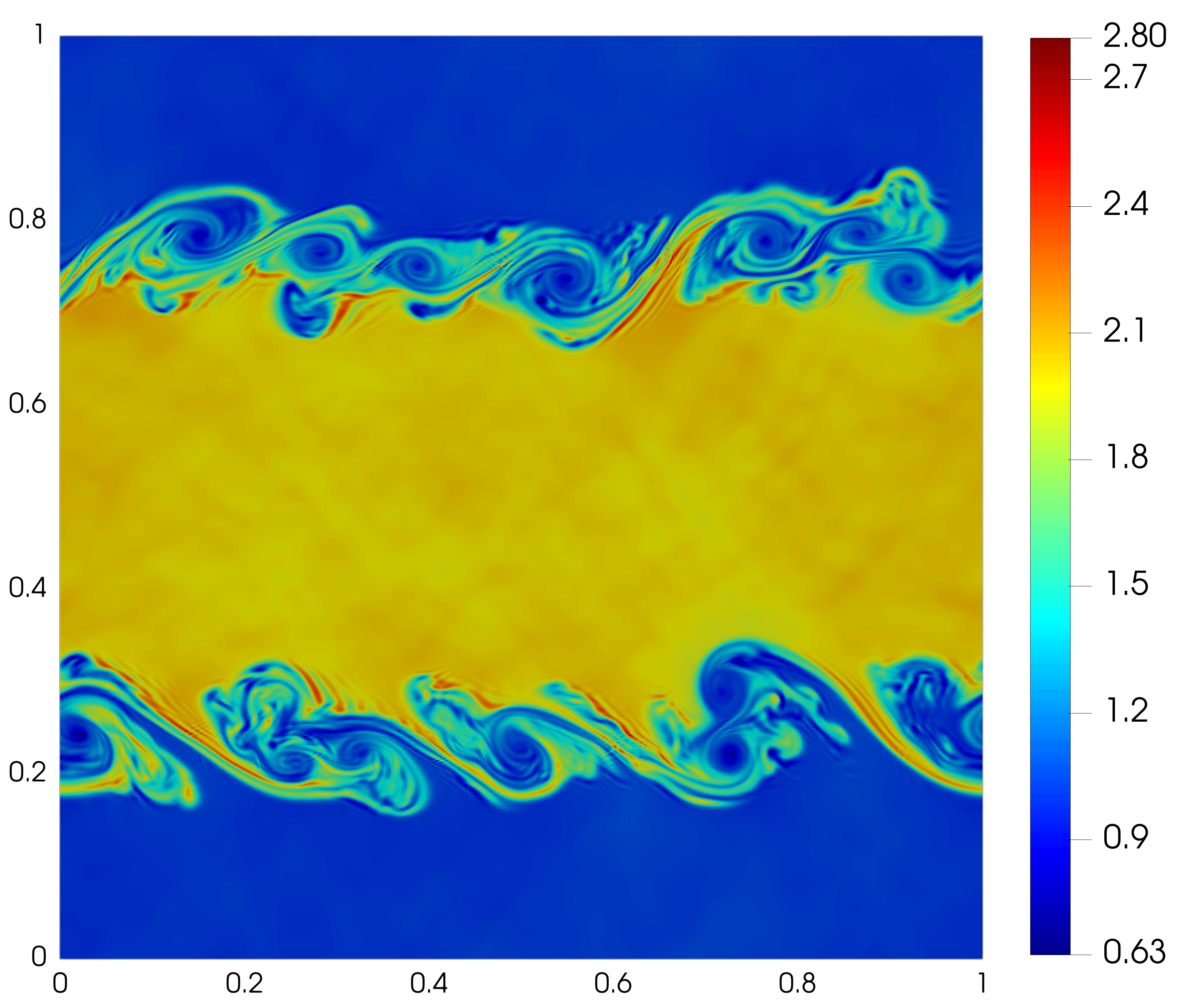}}\hfill
    \subfloat[$t=2$]{\label{fig:DGSEM_N1_t2}%
        \includegraphics[width=.33\linewidth]{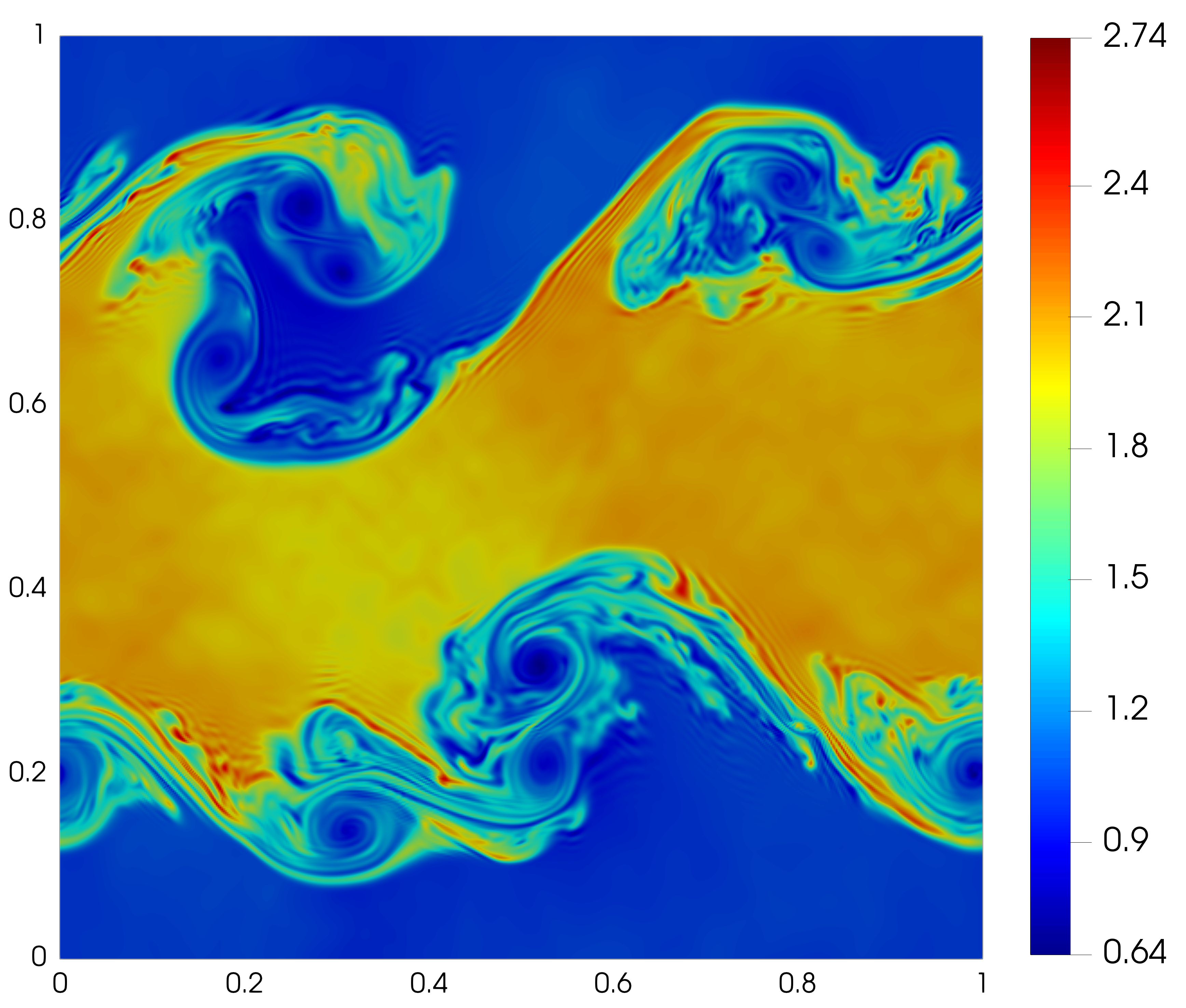}}\hfill
    \subfloat[$t=10$]{\label{fig:DGSEM_N1_t10}%
        \includegraphics[width=.33\linewidth]{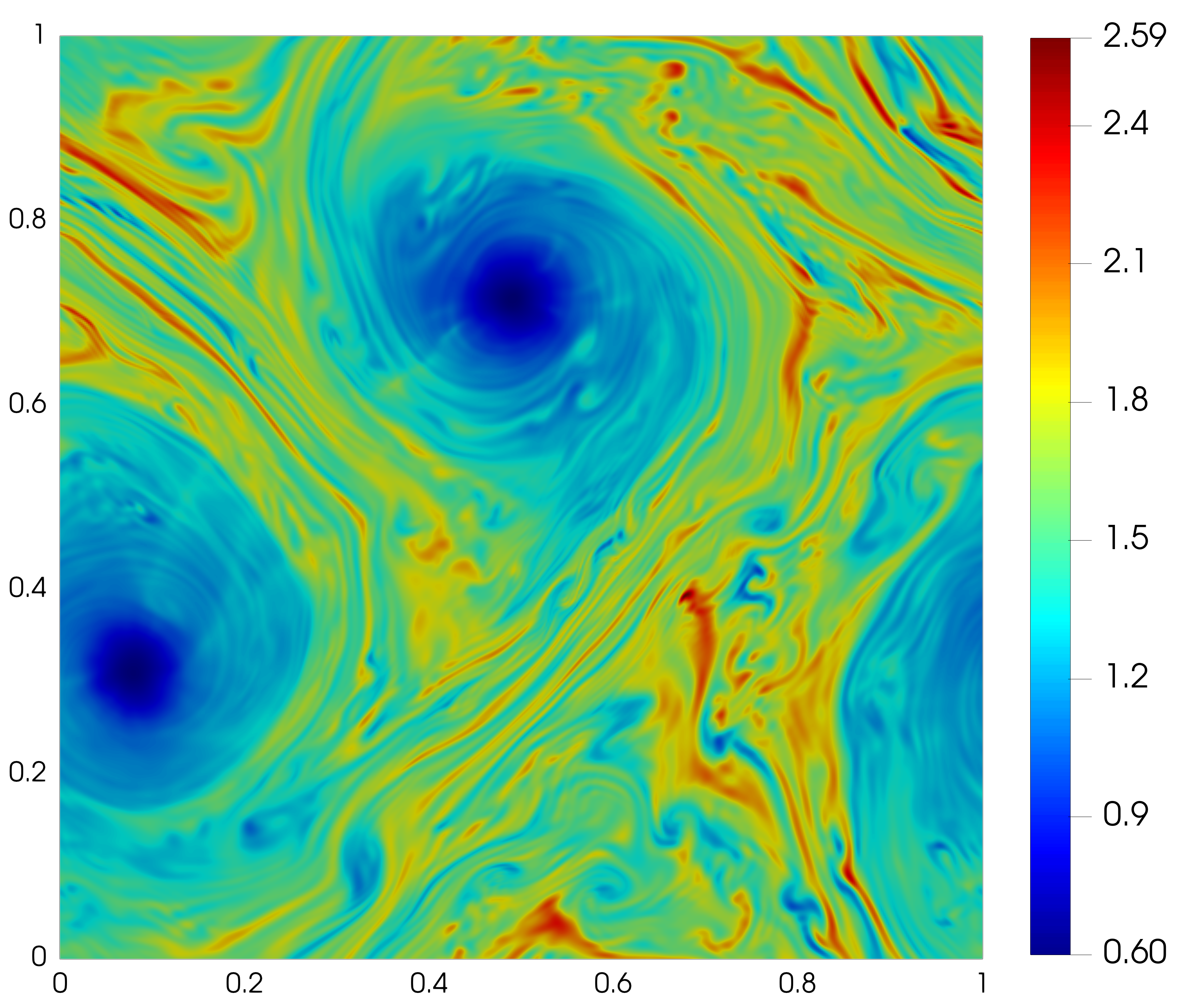}}
    \vspace{-8pt}
    \subfloat[$t=1$]{\label{fig:DGSEM_N1_avg_t1}%
        \includegraphics[width=.33\linewidth]{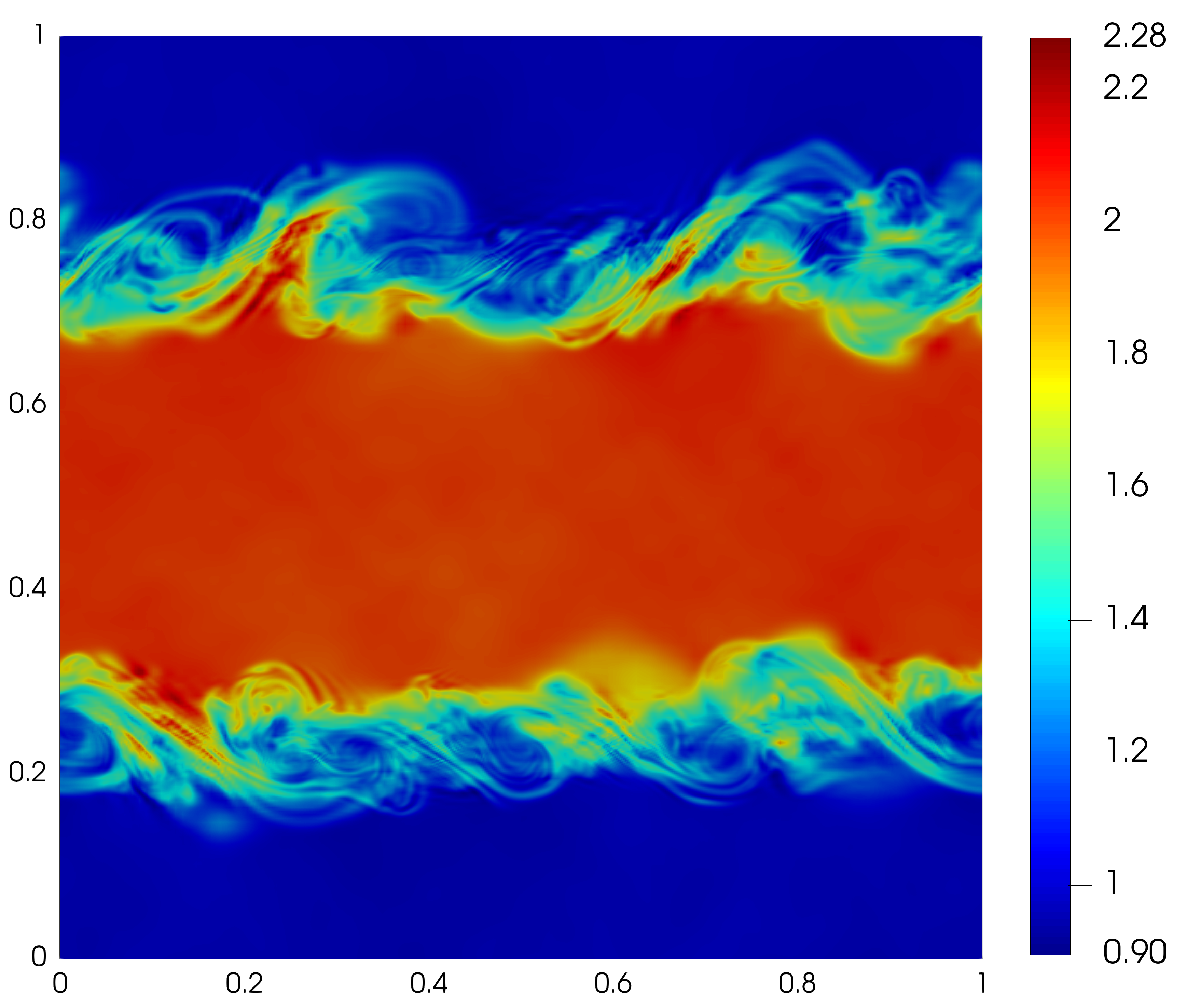}}\hfill
    \subfloat[$t=2$]{\label{fig:DGSEM_N1_avg_t2}%
        \includegraphics[width=.33\linewidth]{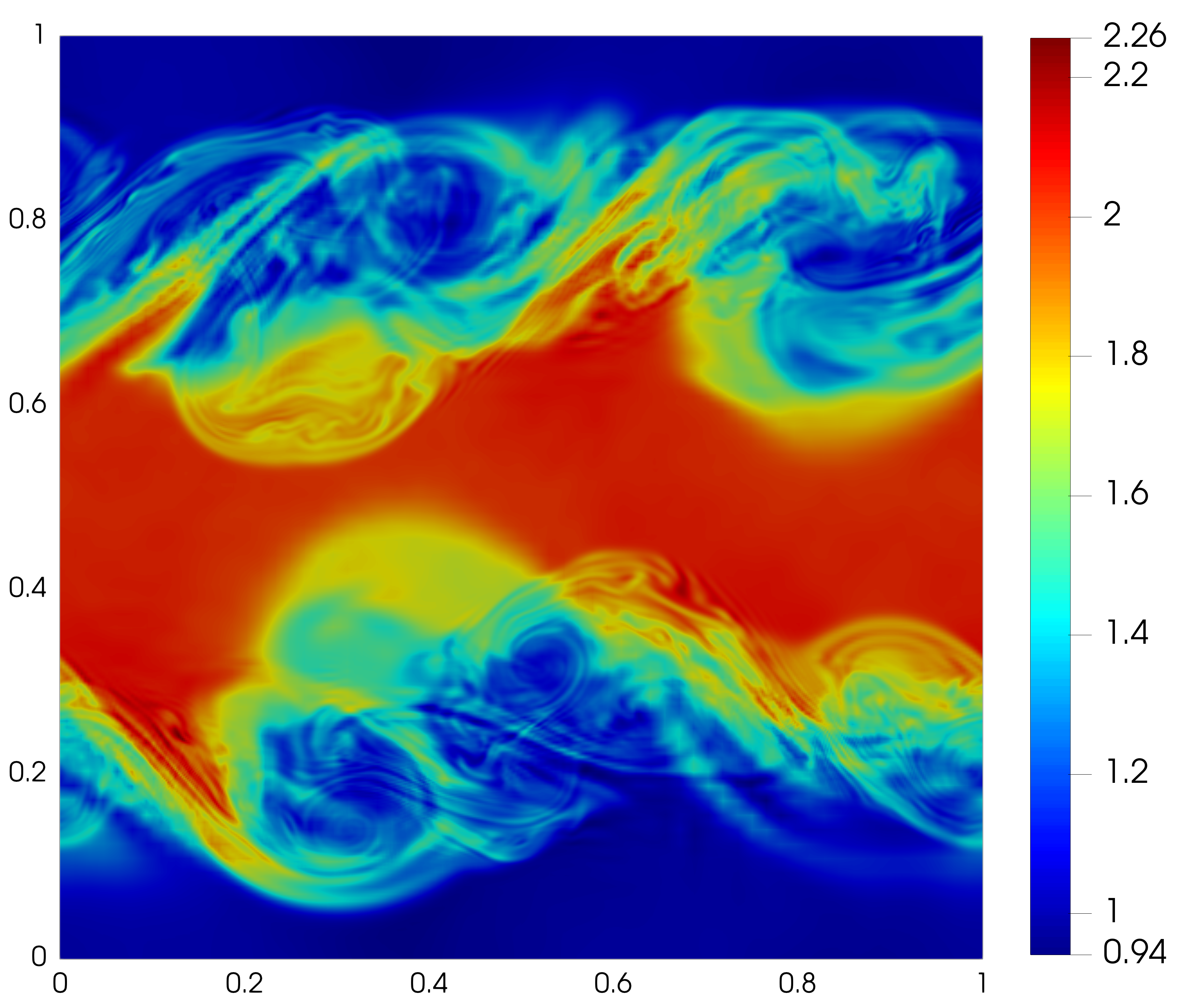}}\hfill
    \subfloat[$t=10$]{\label{fig:DGSEM_N1_avg_t10}%
        \includegraphics[width=.33\linewidth]{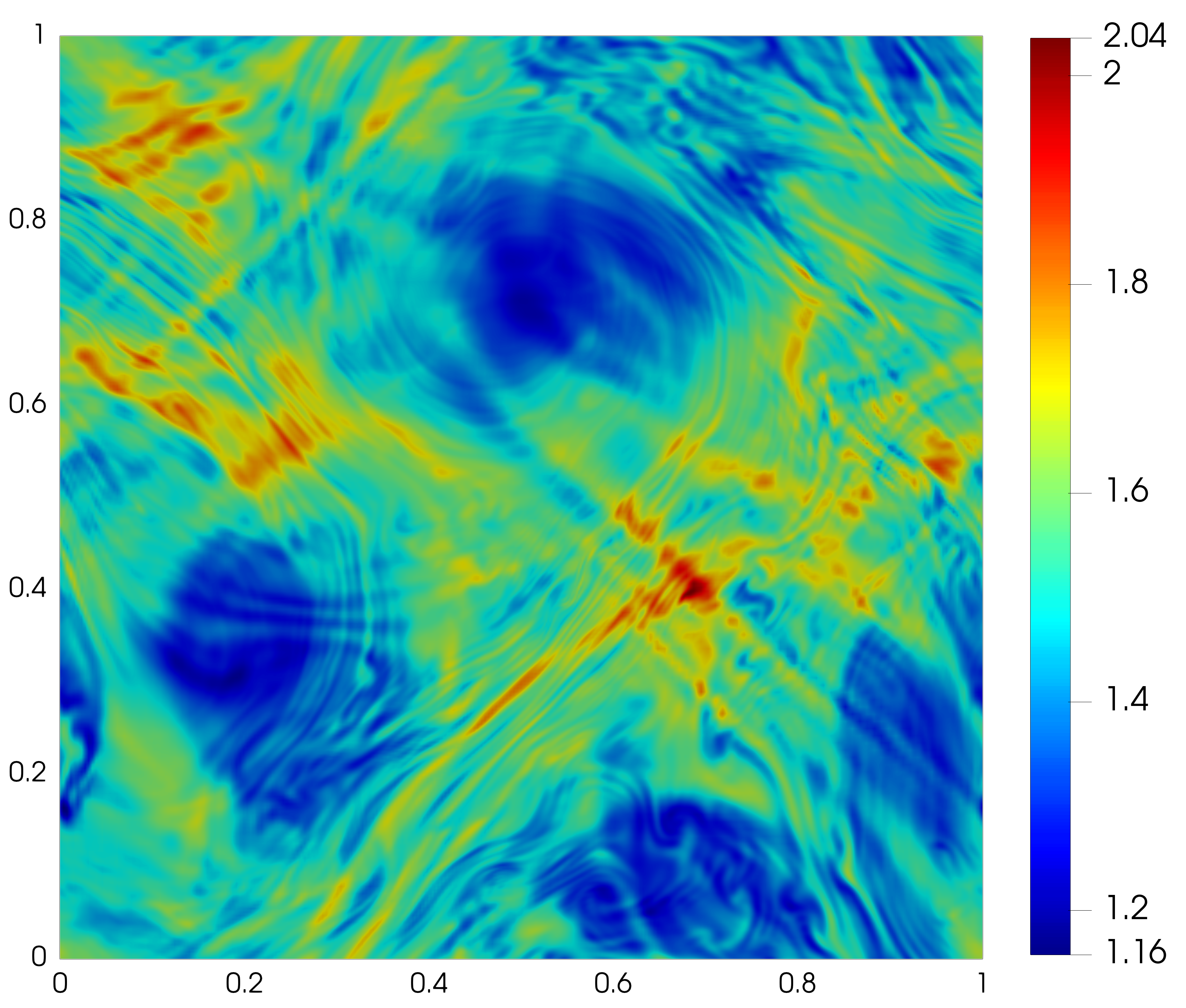}}\par
    \caption{Density (top row) and its \textit{Ces\`{a}ro} average (bottom row) for DGSEM with $N=1$ at different times.}
    \label{fig:DGSEM_N1}
\end{figure}

\begin{figure}[ht]
    \centering
    \subfloat[$t=1$]{\label{fig:DGSEM_N2_t1}%
        \includegraphics[width=.33\linewidth]{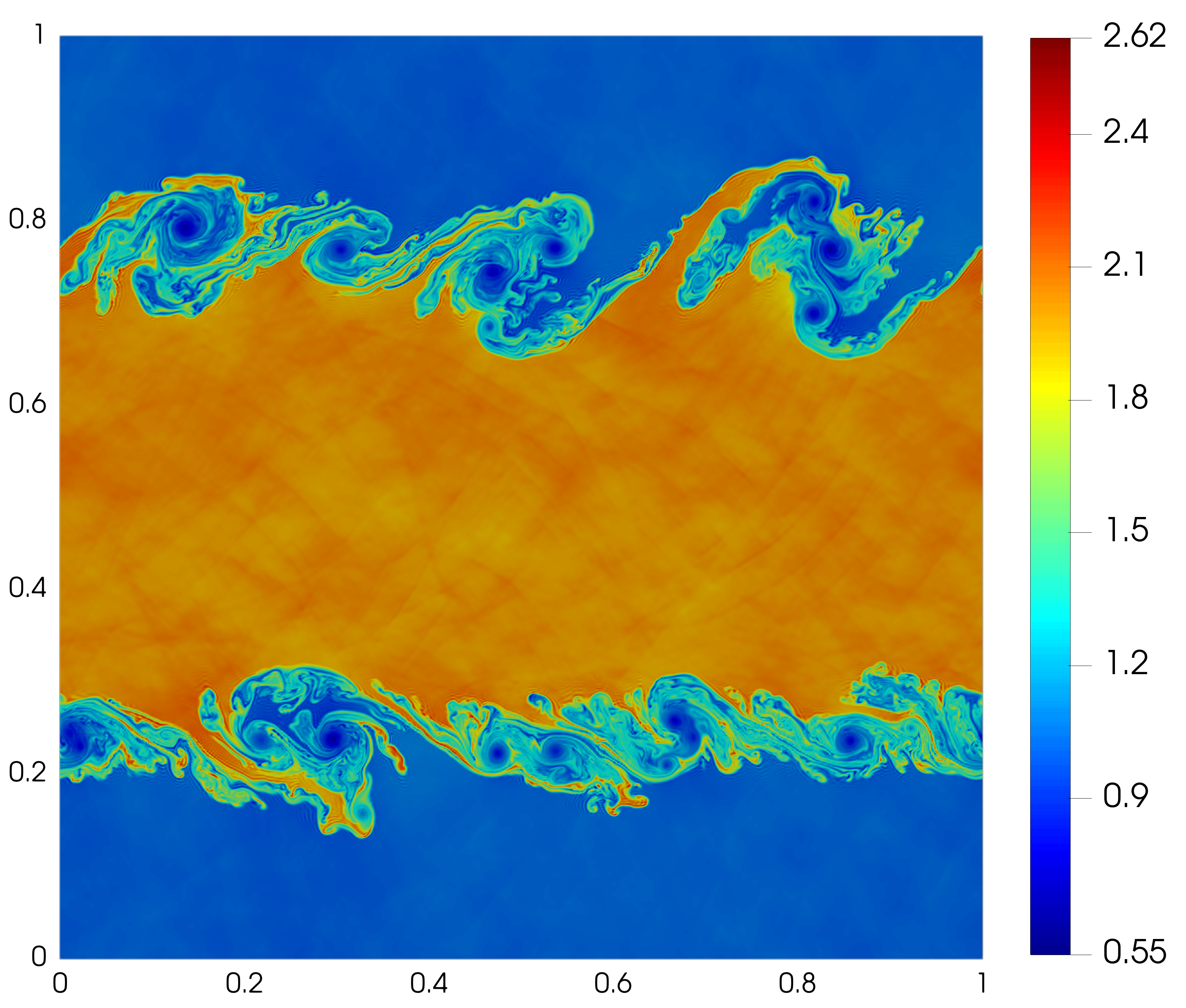}}\hfill
    \subfloat[$t=2$]{\label{fig:DGSEM_N2_t2}%
        \includegraphics[width=.33\linewidth]{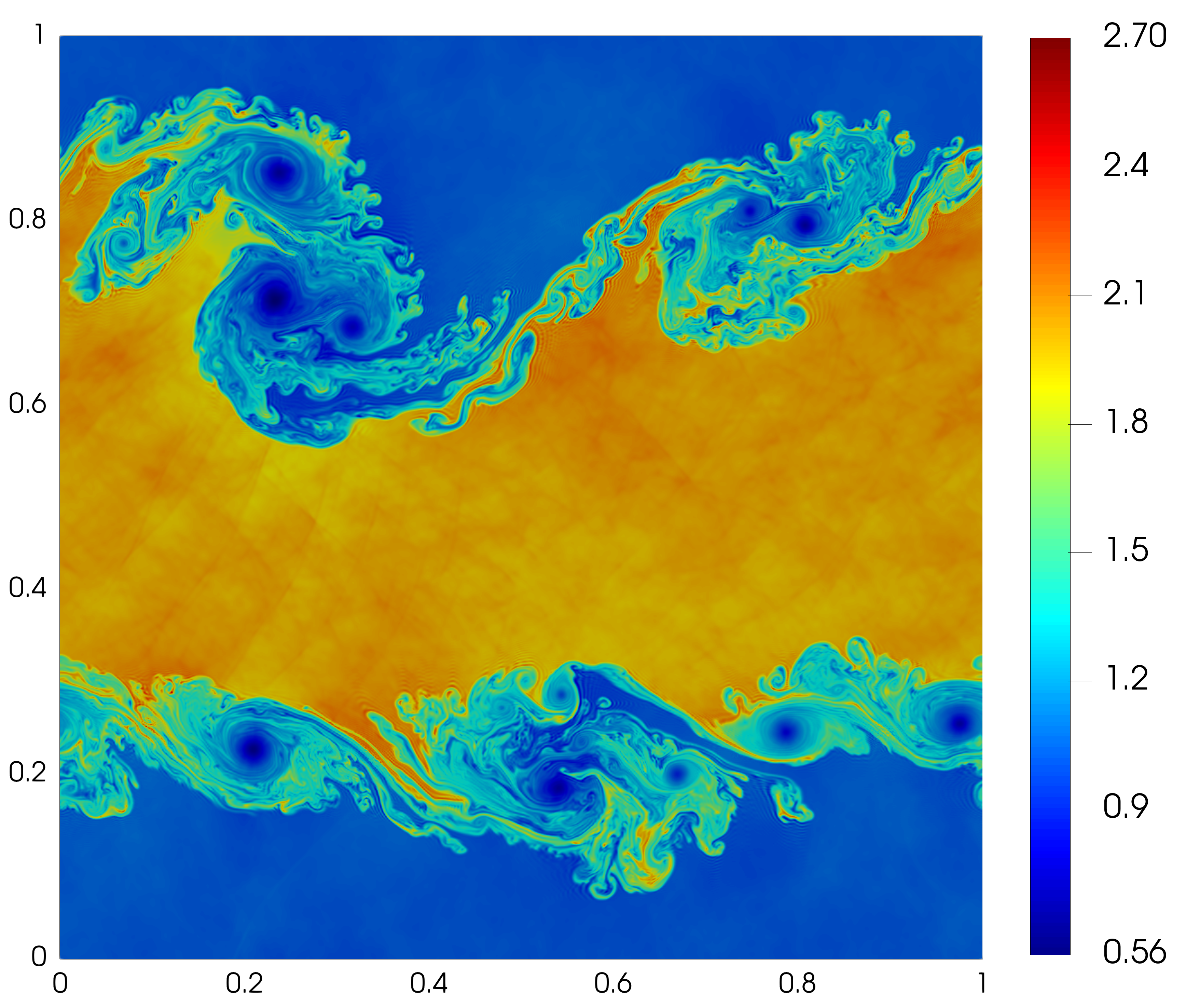}}\hfill
    \subfloat[$t=10$]{\label{fig:DGSEM_N2_t10}%
        \includegraphics[width=.33\linewidth]{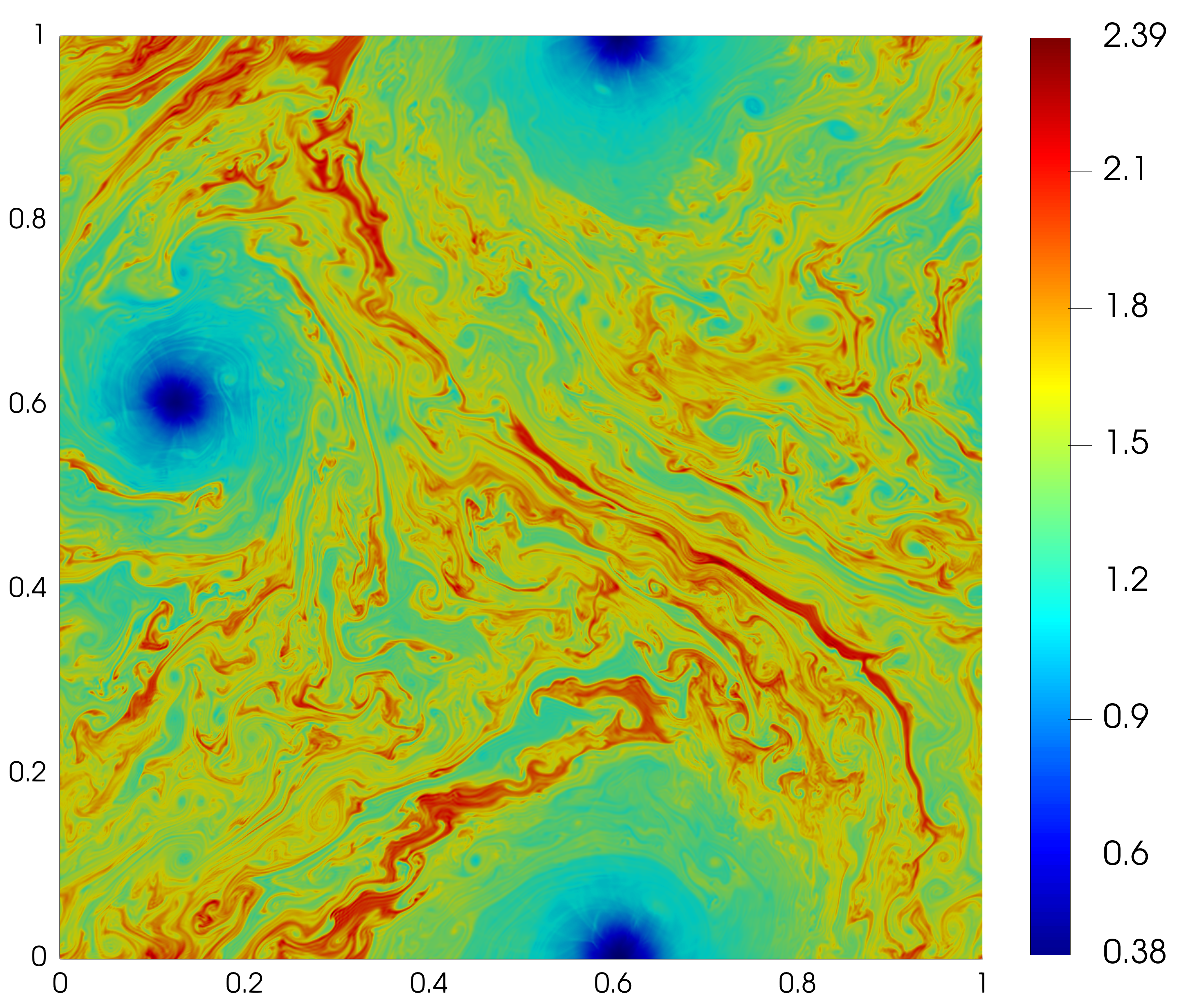}}
    \vspace{-8pt}
    \subfloat[$t=1$]{\label{fig:DGSEM_N2_avg_t1}%
        \includegraphics[width=.33\linewidth]{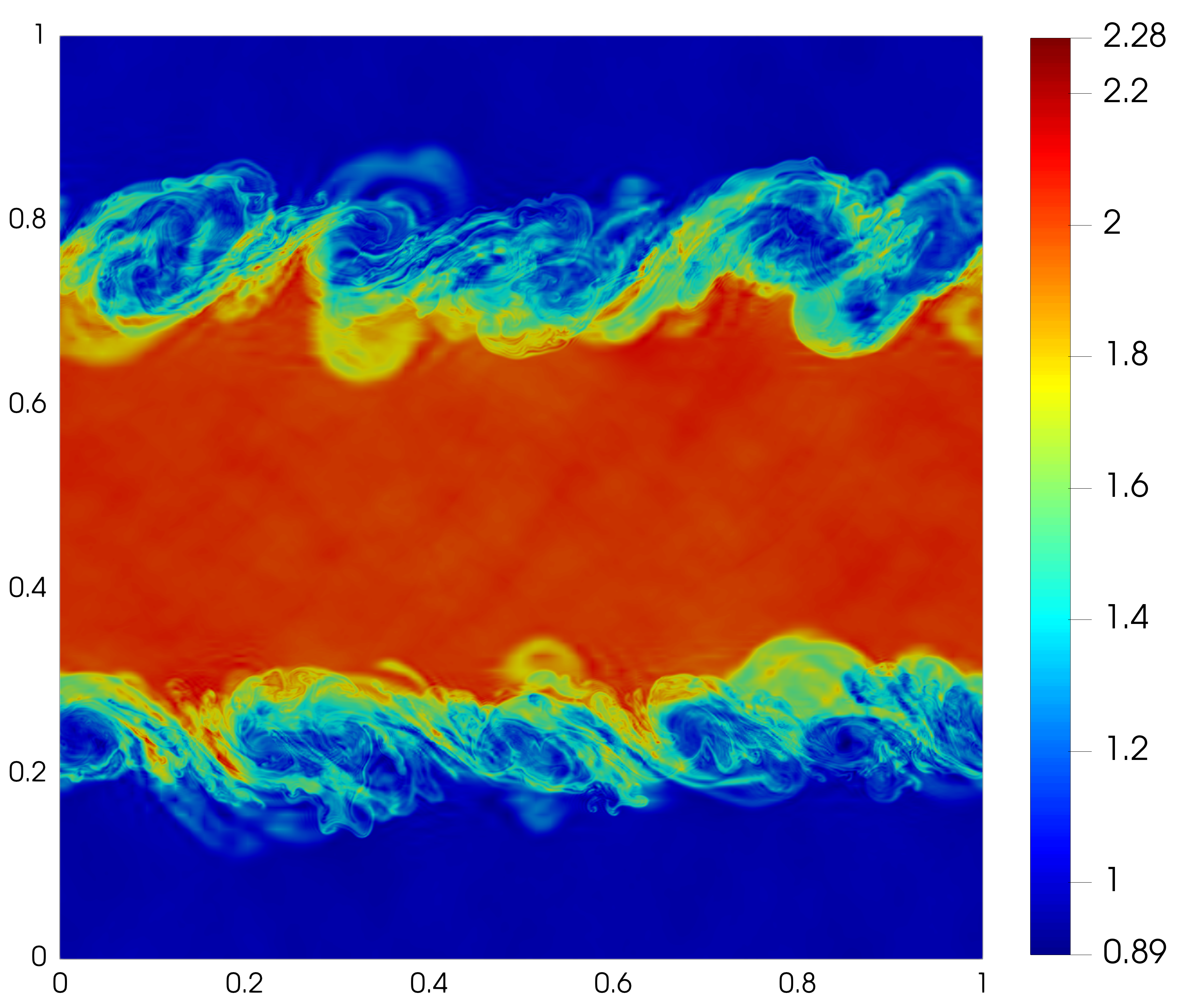}}\hfill
    \subfloat[$t=2$]{\label{fig:DGSEM_N2_avg_t2}%
        \includegraphics[width=.33\linewidth]{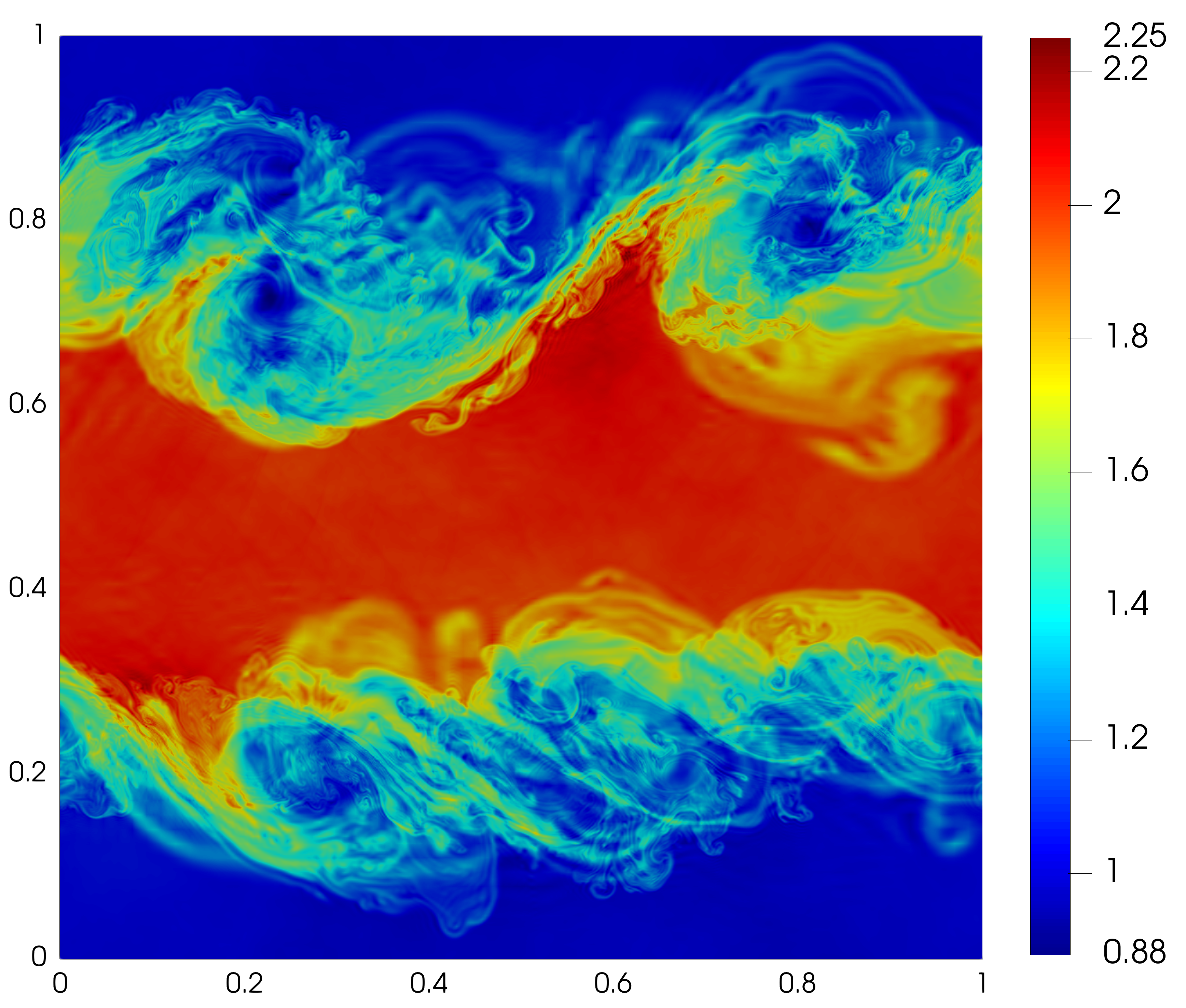}}\hfill
    \subfloat[$t=10$]{\label{fig:DGSEM_N2_avg_t10}%
        \includegraphics[width=.33\linewidth]{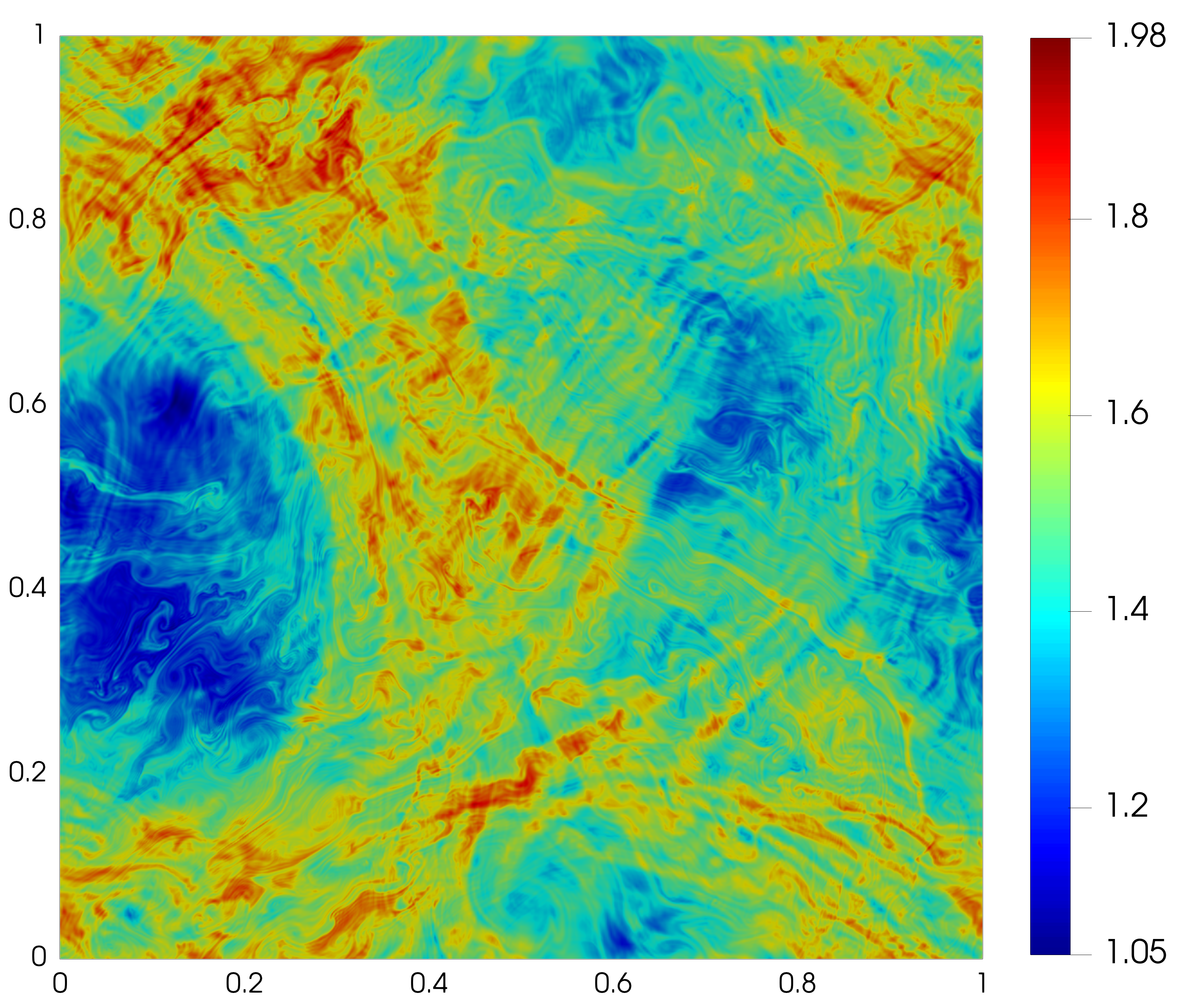}}\par
    \caption{Density (top row) and its \textit{Ces\`{a}ro} average (bottom row) for DGSEM with $N=2$ at different times.}
    \label{fig:DGSEM_N2}
\end{figure}

\begin{figure}[ht]
    \centering
    \subfloat[$t=1$]{\label{fig:DGSEM_N4_t1}%
        \includegraphics[width=.33\linewidth]{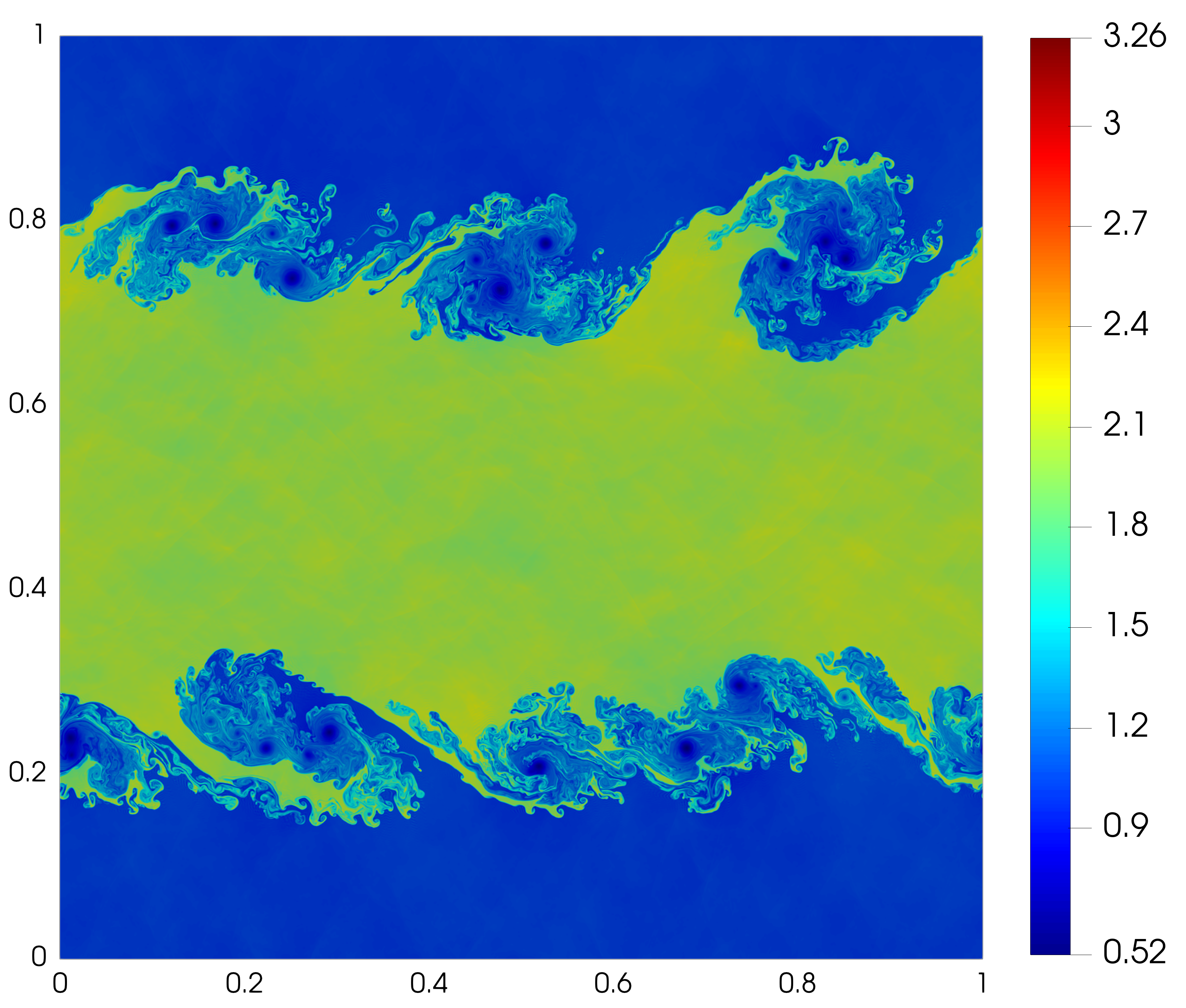}}\hfill
    \subfloat[$t=2$]{\label{fig:DGSEM_N4_t2}%
        \includegraphics[width=.33\linewidth]{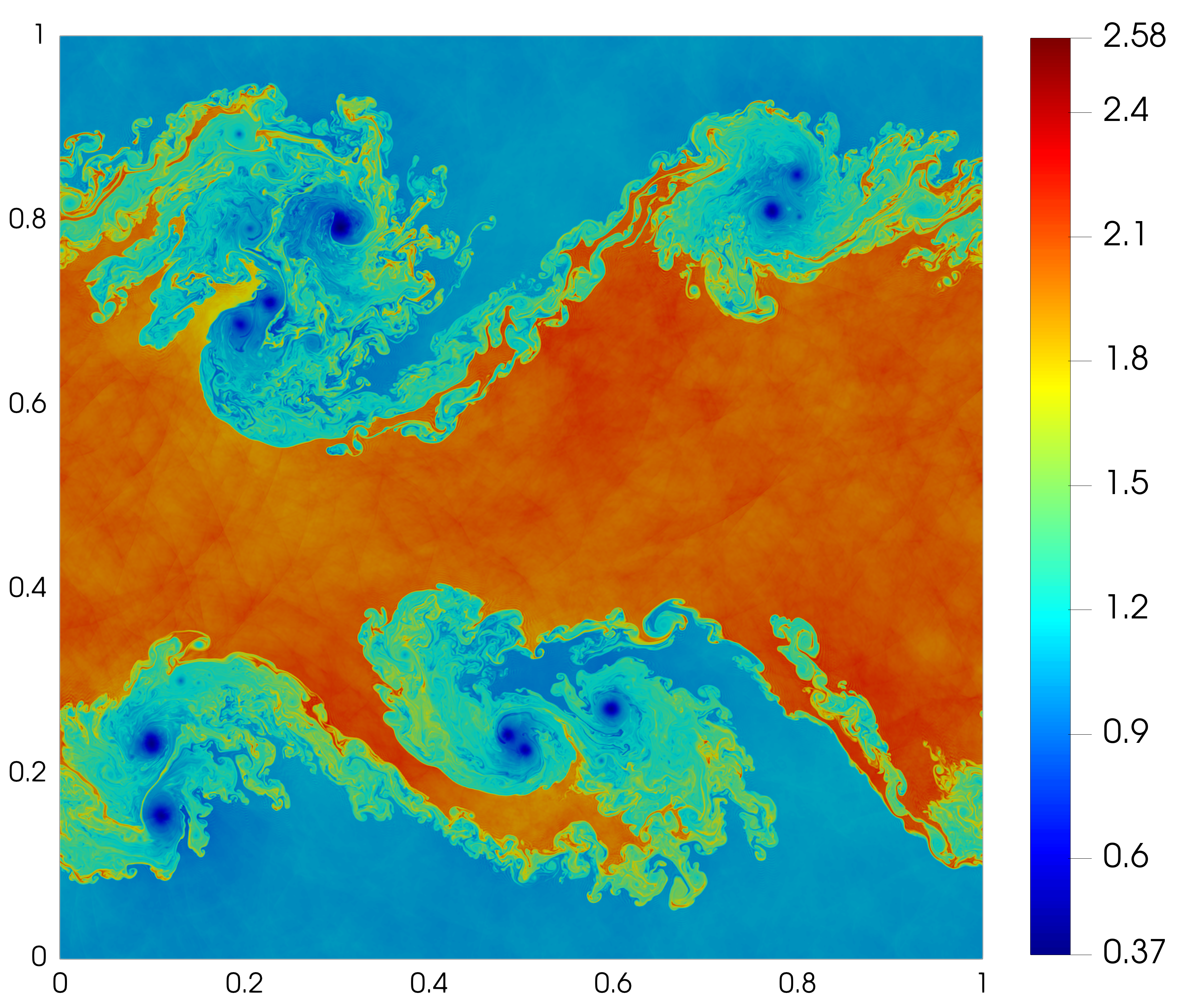}}\hfill
    \subfloat[$t=10$]{\label{fig:DGSEM_N4_t10}%
        \includegraphics[width=.33\linewidth]{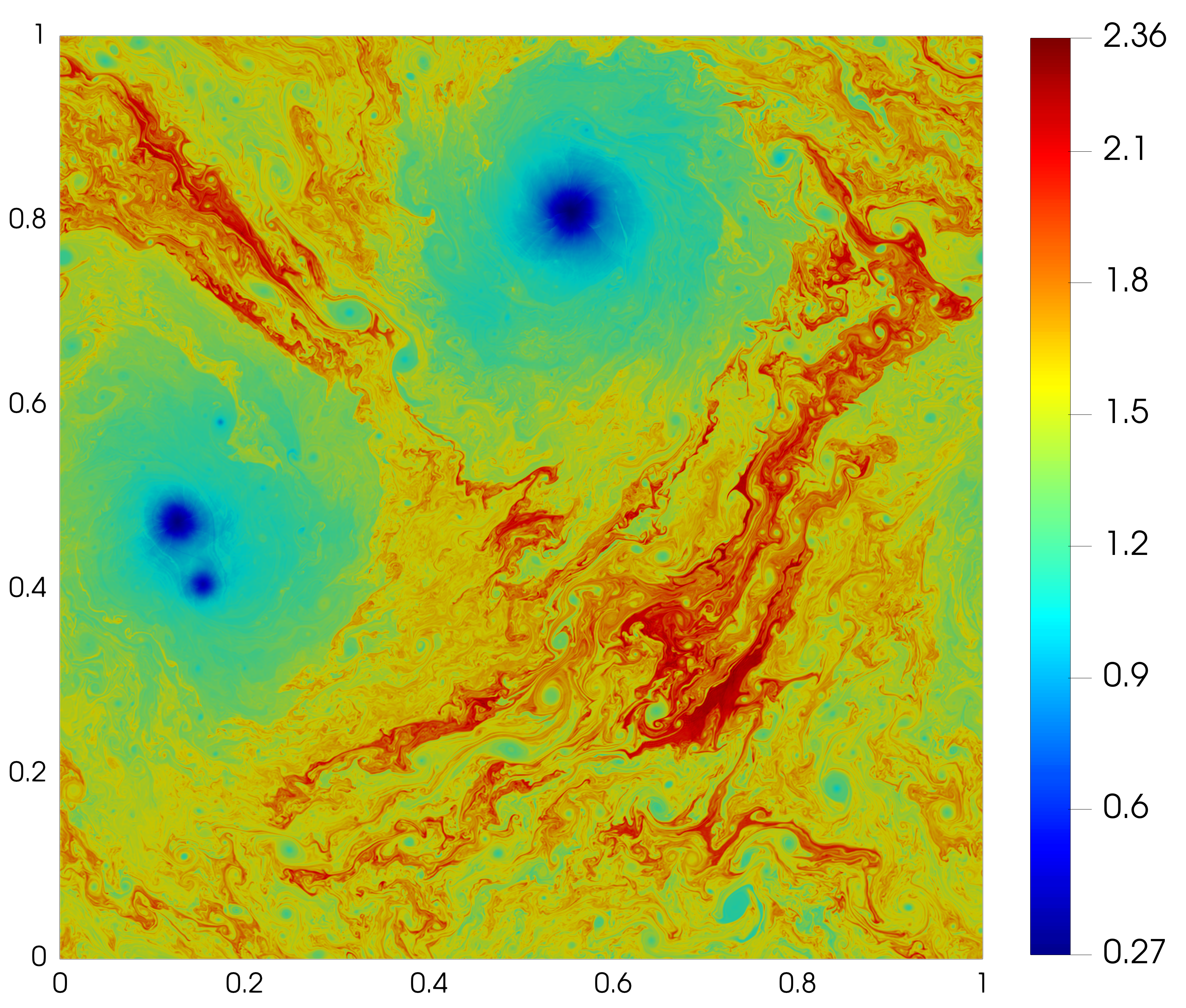}}
    \vspace{-8pt}
    \subfloat[$t=1$]{\label{fig:DGSEM_N4_avg_t1}%
        \includegraphics[width=.33\linewidth]{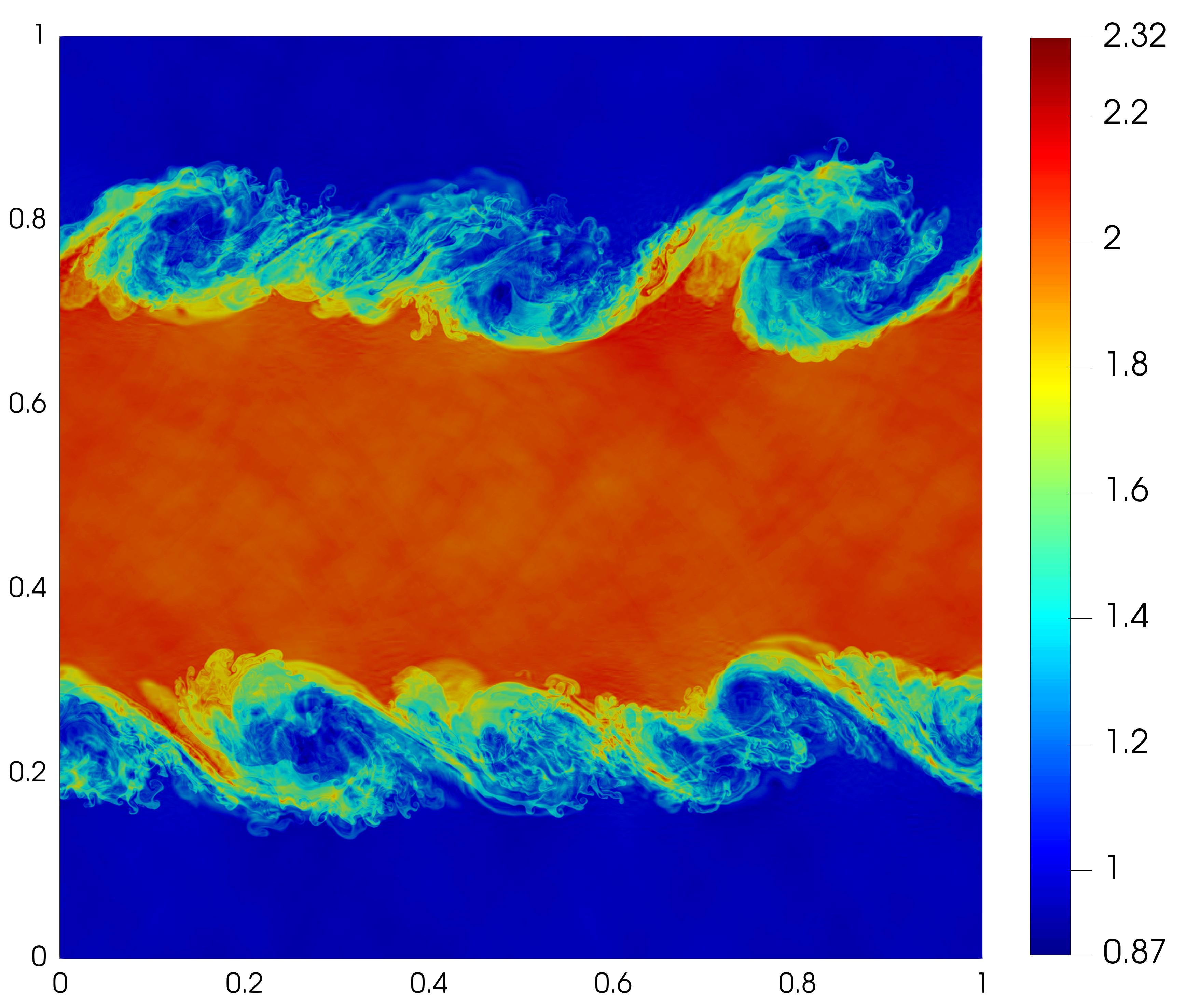}}\hfill
    \subfloat[$t=2$]{\label{fig:DGSEM_N4_avg_t2}%
        \includegraphics[width=.33\linewidth]{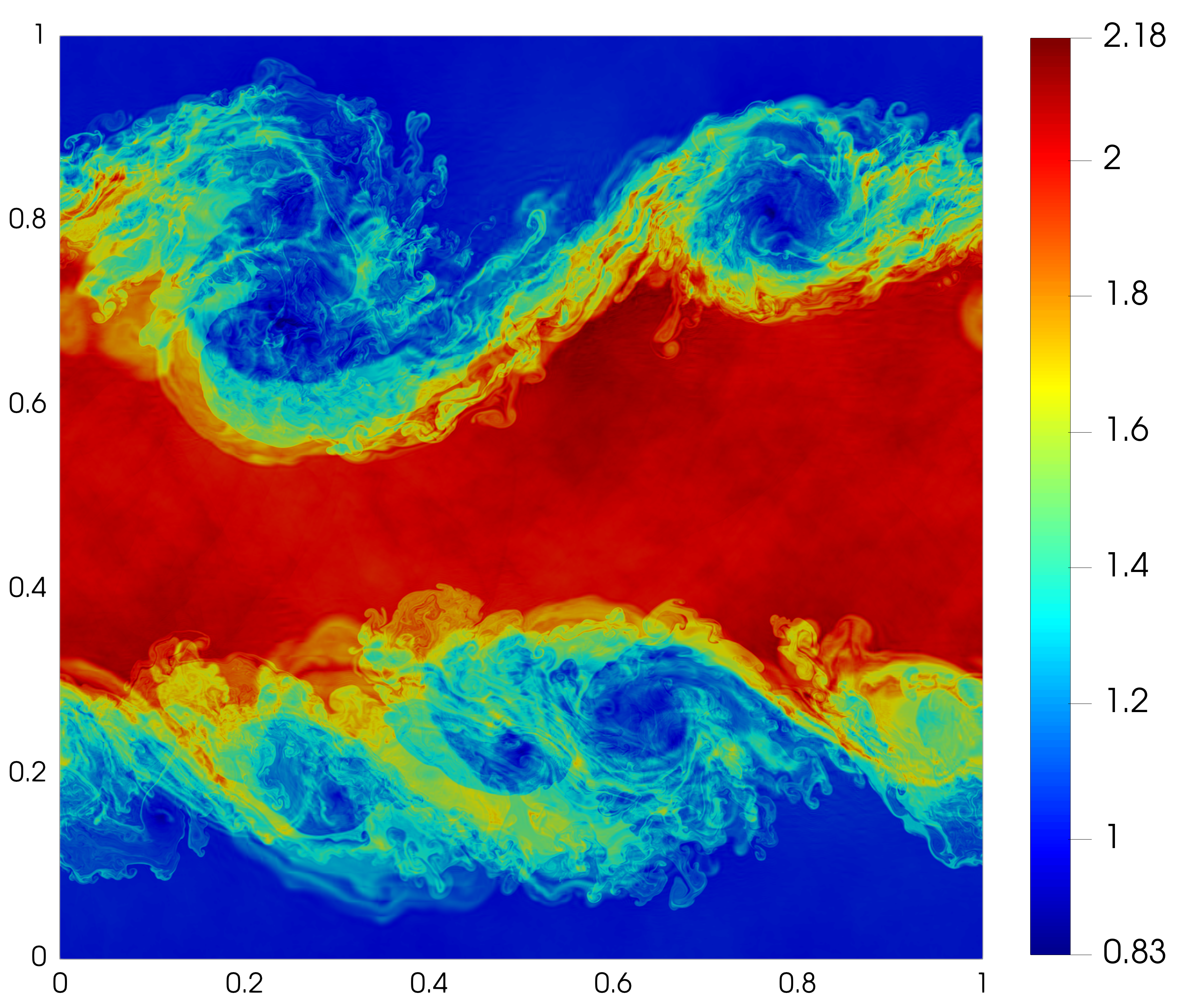}}\hfill
    \subfloat[$t=10$]{\label{fig:DGSEM_N4_avg_t10}%
        \includegraphics[width=.33\linewidth]{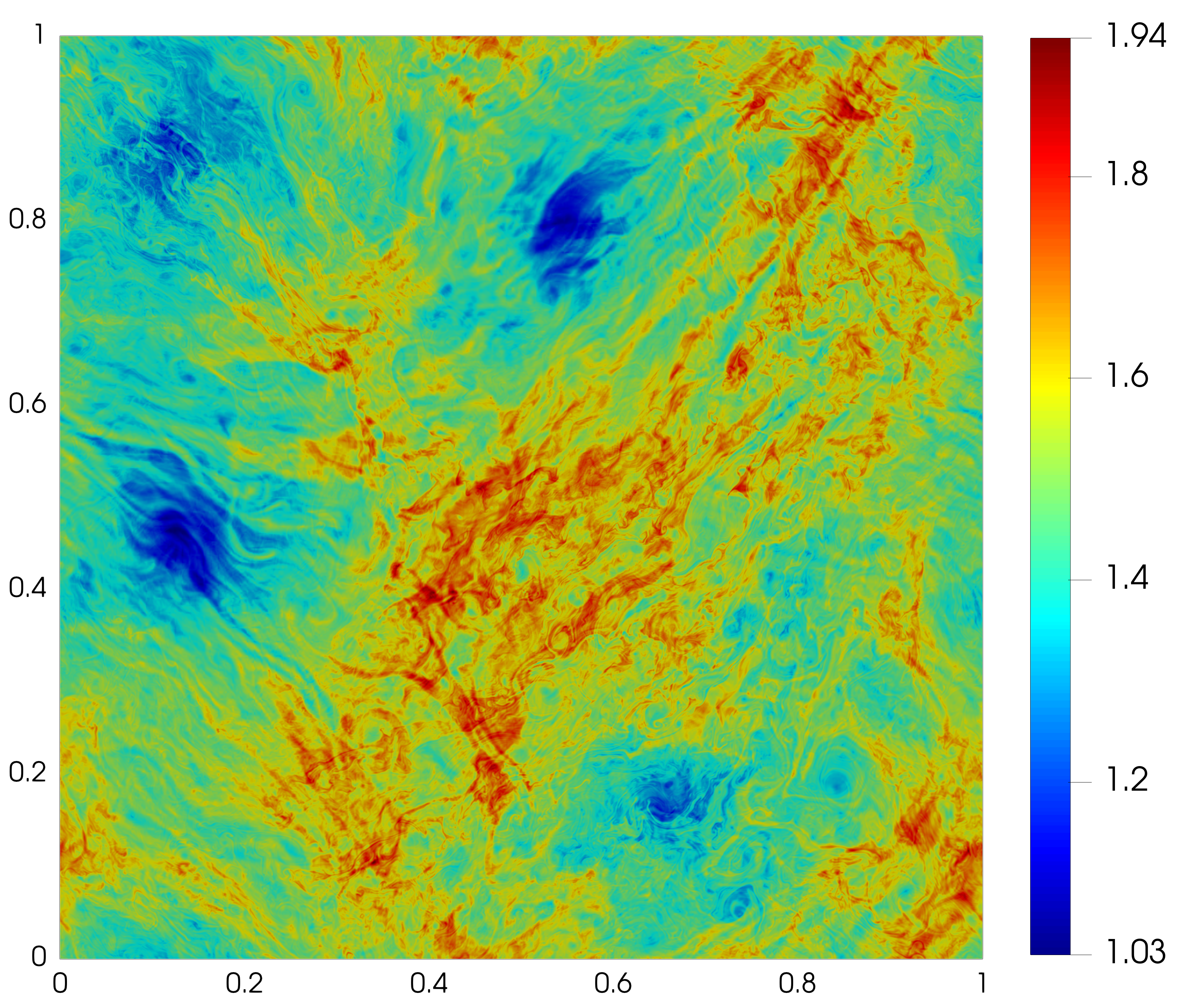}}\par
    \caption{Density (top row) and its \textit{Ces\`{a}ro} average (bottom row) for DGSEM with $N=4$ at different times.}
    \label{fig:DGSEM_N4}
\end{figure}

 We study the numerical approximations of density $\varrho_{h_j} (t,\mathbf{x})$ and its \textit{Ces\`{a}ro} average. The \textit{Ces\`{a}ro} average of any quantity is defined as follows:
\begin{equation}
    \widetilde{(\cdot)}_{h_n} = \frac{1}{n}\sum_{j=1}^n (\cdot)_{hj}.
\end{equation} 
Here, $h_j$ represents the mesh size corresponding to the uniform rectangular mesh with $32 \times 2^j$ cells in each direction, $n=1,2, \dots, N_r$, and ${N_r=4}$ corresponds to the index of the finest mesh size $h_{N_r}$. \Cref{fig:VFV_N1,fig:VFV_N2,fig:VFV_N3,fig:VFV_N5,fig:VFV_N7,fig:VFV_N9,fig:Kin_N0,fig:Kin_N1,fig:DG_N0,fig:DG_N1,fig:DG_N2,fig:DG_N4,fig:DGSEM_N1,fig:DGSEM_N2,fig:DGSEM_N4} show the plots of density and its \textit{Ces\`{a}ro} average for different numerical methods at times $t=1, 2$ and $10$, all with a grid size of $512 \times 512$. 
\par Further, we study the convergence rates for density and its \textit{Ces\`{a}ro} average through the $L_1$ error norms. The $L_1$ error norms, and the experimental order of convergence (EOC) are calculated as
\begin{eqnarray}\label{eq:err}
    \norm{\varepsilon\left( \widehat{\varrho}_{h_n} \right)}_{L_1(\Omega)} &=& \int_{\Omega} \modL{\widehat{\varrho}_{h_n}(\mathbf{x}) - \widehat{\varrho}_{h_{n+1}} (\mathbf{x})}\dx, \\
    EOC(\widehat{\varrho}_{h_n}) &=& \log_2\left(\frac{ \norm{\varepsilon\left(  \widehat{\varrho}_{h_n} \right)}_{L_1(\Omega)}}{\norm{\varepsilon\left(  \widehat{\varrho}_{h_{n+1}} \right)}_{L_1(\Omega)}} \right),  \quad n=1,\dots,N_r-1.
\end{eqnarray}
Here $\widehat{\varrho}$ stands for $\varrho$ and $\widetilde{\varrho}$, respectively. We work with the values of $\varrho_{h_n}(x), n=1,\dots,N_r$ interpolated to the finest grid $h_{N_r}$, as this allows us to reuse the data computed while calculating the \textit{Ces\`{a}ro} averages. Numerical solutions were obtained by VFV, DVBFV, RKDG and DGSEM methods which are described in more detail in \Cref{Sec: Num schemes}.  
For the computation of the integral in~\eqref{eq:err}, the midpoint quadrature rule with quadrature points as the cell centers is used for the VFV method in \Cref{Subsec: VFV} and DVBFV method in \Cref{Subsec: DVB-FVM}, the Gauss Legendre quadrature rule with $(N+1)\times (N+1)$ quadrature points (for polynomial degree $N$) is used for the RKDG method in \Cref{Subsec: DG}, and the Legendre–Gauss–Lobatto quadrature rule (LGL) with $2\times(N+1)$ quadrature points (for polynomial degree $N$) is used for the DGSEM in \Cref{Subsec: ES-DGSEM}. \Cref{tab:all-errors} shows the convergence of density and its \textit{Ces\`{a}ro} average, and it can be observed that the \textit{Ces\`{a}ro} average of density converges for all numerical methods with $N \geq 1$, while the density itself does not show any convergence. These results are in agreement with theoretical findings on weak and strong convergence of consistent and stable approximations, see \cite{FeireislLukacova_book2021}. 
\begin{table}[!ht]
\centering
\resizebox{\textwidth}{!}{
\begin{tabular}{@{}cccccc@{}}
\toprule
$m$ & \multicolumn{2}{c}{$Y_1$} & & \multicolumn{2}{c}{$Y_2$} \\ 
\cmidrule(lr){2-3} \cmidrule(lr){5-6}
& $a_1$ & $b_1$ & & $a_2$ & $b_2$ \\ 
\midrule
1  & 6.848086824246653e$^{-8}$ & -0.973625473853271  & & 0.009373025805955863 &  3.10750325239443 \\
2  & 0.004450348128947341      &  2.33221742979395   & & 0.0197686121934106   &  1.7482985063786 \\
3  & 0.06156955958786613       & -2.57661895600041   & & 0.129928159795144    & -3.01803367486339 \\
4  & 0.116481555805349         &  2.43965931651801   & & 0.215403817045303    & -2.07430785001108 \\
5  & 0.168204784555961         &  1.26278768686501   & & 0.01758678905771403  &  3.10856394704146 \\
6  & 0.146413246162863         &  1.47373734867445   & & 0.211994809652299    &  1.59399689987015 \\
7  & 0.05857224323849688       & -1.25553236484458   & & 0.02134903127162342  & -1.77202310331374 \\
8  & 0.159868678328304         & -2.82920698380582   & & 0.148283301108284    &  1.00086476379714 \\
9  & 0.139003488203723         &  2.56472949845762   & & 0.198161392506709    &  0.245159802027399 \\
10 & 0.145436027507622         & -2.527985588414849  & & 0.02815106156355688  & -0.125265541490568 \\
\bottomrule
\end{tabular}
}
\caption{Initial condition data for $Y_1$ and $Y_2$.}
\label{tab:init-data}
\end{table}


\begin{table}[!ht]
\centering
\begin{minipage}{\textwidth}
\caption{$L_1$ errors of density and \textit{Ces\`{a}ro} averaged density 
for VFV, DVBFV, RKDG and DGSEM methods at time $t=10$.}
\label{tab:all-errors}
\begin{tabular*}{\textwidth}{@{\extracolsep{\fill}}llllll@{}}
\toprule
Method & $h$ 
& $\|\varepsilon(\varrho_h)\|_{L^1}$ & EOC 
& $\|\varepsilon(\widetilde{\varrho}_h)\|_{L^1}$ & EOC \\
\midrule
\multirow{3}{*}{VFV, $N=1$}
& $1/64$  & 0.0180 & --       & 0.1875 & --      \\
& $1/128$ & 0.0694 & -1.9450 & 0.0740 &  1.3412 \\
& $1/256$ & 0.1026 & -0.5624 & 0.0647 &  0.1932 \\
\midrule
\multirow{3}{*}{VFV, $N=2$}
& $1/64$  & 0.4198 & --       & 0.2099 & --      \\
& $1/128$ & 0.2570 &  0.7076  & 0.0779 &  1.4307 \\
& $1/256$ & 0.2588 & -0.0096 & 0.0604 &  0.3658 \\
\midrule
\multirow{3}{*}{VFV, $N=4$}
& $1/64$  & 0.5281 & --       & 0.2641 & --      \\
& $1/128$ & 0.2689 &  0.9736  & 0.0892 &  1.5664 \\
& $1/256$ & 0.2478 &  0.1180  & 0.0491 &  0.8613 \\
\midrule
\multirow{3}{*}{VFV, $N=6$}
& $1/64$  & 0.4905 & --       & 0.2453 & --      \\
& $1/128$ & 0.2822 &  0.7978  & 0.0973 &  1.3342 \\
& $1/256$ & 0.2691 &  0.0685  & 0.0496 &  0.9726 \\
\midrule
\multirow{3}{*}{VFV, $N=8$} & $1/64$ & 0.4205 & -- & 0.2103 & -- \\
 & $1/128$ & 0.3695 & 0.1867 & 0.1255 & 0.7446 \\
 & $1/256$ & 0.3095 & 0.2557 & 0.0513 & 1.2912 \\
\midrule
\multirow{3}{*}{DVBFV, $N=1$}
& $1/64$  & 0.1457 & --      & 0.0729 & --     \\
& $1/128$ & 0.1958 & -0.4264 & 0.0569 & 0.3551 \\
& $1/256$ & 0.2327 & -0.2487 & 0.0482 & 0.2415 \\
\midrule
\multirow{3}{*}{RKDG, $N=1$}
& $1/64$  & 0.2009 & --      & 0.1005 & --     \\
& $1/128$ & 0.2605 & -0.3745 & 0.0852 & 0.2375 \\
& $1/256$ & 0.3182 & -0.2887 & 0.0624 & 0.4502 \\
\midrule
\multirow{3}{*}{RKDG, $N=2$}
& $1/64$  & 0.2210 & --      & 0.1105 & --     \\
& $1/128$ & 0.2889 & -0.3866 & 0.0843 & 0.3903 \\
& $1/256$ & 0.2937 & -0.0235 & 0.0665 & 0.3426 \\
\midrule
\multirow{3}{*}{RKDG, $N=4$}
& $1/64$  & 0.2099 & --      & 0.1049 & --     \\
& $1/128$ & 0.3048 & -0.5385 & 0.0995 & 0.0763 \\
& $1/256$ & 0.3158 & -0.0508 & 0.0642 & 0.6320 \\
\midrule
\multirow{3}{*}{DGSEM, $N=1$}
& $1/64$  & 0.2416 & --      & 0.1208 & --     \\
& $1/128$ & 0.3120 & -0.3688 & 0.0942 & 0.3594 \\
& $1/256$ & 0.2874 &  0.1185 & 0.0575 & 0.7125 \\
\midrule
\multirow{3}{*}{DGSEM, $N=2$}
& $1/64$  & 0.2932 & --      & 0.1466 & --     \\
& $1/128$ & 0.2953 & -0.0104 & 0.0781 & 0.9073 \\
& $1/256$ & 0.2801 &  0.0764 & 0.0534 & 0.5489 \\
\midrule
\multirow{3}{*}{DGSEM, $N=4$}
& $1/64$  & 0.2242 & --      & 0.1121 & --     \\
& $1/128$ & 0.3036 & -0.4370 & 0.0938 & 0.2573 \\
& $1/256$ & 0.1684 &  0.8504 & 0.0520 & 0.8524 \\
\bottomrule
\end{tabular*}
\end{minipage}
\end{table}

\clearpage
\section{Numerical study of the selection criteria}
\label{Sec: criteria}
In this section, we compare the four different numerical schemes with respect to the selection criteria. We provide relevant functionals to serve as a reference for selecting among the numerical solutions.

First, we examine the entropy production of the numerical solutions.
Following the discussion in \Cref{Sec: Introduction}, we approximate the entropy of DW solutions using a \textit{Ces\`{a}ro} average, and consequently define
\begin{align*}
	\mathcal{I}_{\widetilde{S}}(t) &= \int_{\Omega}\widetilde{S}(t,\mathbf{x}) \,\mathrm{d} \mathbf{x}, \nonumber\\
	D_S(t) &=\mathcal{I}_{\widetilde{S}}(t) - \mathcal{I}_{\widetilde{S}}(0), \nonumber
\end{align*}
where $D_S(t)$ represents the entropy production at time $t$.
Next, the total energy and mean energy of the numerical solutions are defined as
\begin{align*}
	\mathcal{I}_{\widetilde{E}}(t) &=  \int_{\Omega}\widetilde{E}(t,\mathbf{x}) \,\mathrm{d} \mathbf{x}, \nonumber\\
	\mathcal{I}_{\widehat{E}}(t) &= \int_{\Omega}E\left( \widetilde{\varrho},\widetilde{\mathbf{m}},\widetilde{S} \right)(t,\mathbf{x}) \,\mathrm{d} \mathbf{x}, \nonumber
\end{align*}
and the energy defect at time $t$ is given by
\begin{align*}
	\mathcal{D}_E(t) &= \mathcal{I}_{\widetilde{E}}(t) - \mathcal{I}_{\widehat{E}}(t). \nonumber
\end{align*}

We next consider the functionals introduced in \eqref{functionals}.
The functions defined above can serve as candidates for the function $\beta$.
Owing to the boundedness of the approximate solutions, $\beta$ can also be chosen as an appropriate norm.
Accordingly, we define
\begin{align*}
	\mathcal{J}_{\widetilde{S}}(t) &= \int_{0}^t \mathcal{I}_{\widetilde{S}}(\tau) \mathrm{d} \tau \nonumber\\
	\mathcal{J}_{\widehat{E}}(t) &= \int_{0}^t \mathcal{I}_{\widehat{E}}(\tau) \mathrm{d} \tau \nonumber\\
	\mathcal{J}_{(\widetilde{\varrho},\widetilde{\mathbf{m}},\widetilde{S})}(t) &= \int_{0}^t \left( \norm{\widetilde{\varrho}}^q_{L^q}  + \norm{\widetilde{\mathbf{m}}}^q_{L^q} + \norm{\widetilde{S}}^q_{L^q} \right) \mathrm{d} \tau, \nonumber
\end{align*}
where $q=1.1$. Correspondingly, we consider the time-weighted integrals in \eqref{functionals}
\begin{align}
	\mathcal{F}_{\widetilde{S}}(t) &= \int_{0}^t e^{-\tau} \mathcal{I}_{\widetilde{S}}(\tau) \mathrm{d} \tau \nonumber\\
    \mathcal{F}_{\widehat{E}}(t) &= \int_{0}^t e^{-\tau} \mathcal{I}_{\widehat{E}}(\tau) \mathrm{d} \tau \nonumber\\
	\mathcal{F}_{(\widetilde{\varrho},\widetilde{\mathbf{m}},\widetilde{S})}(t) &= \int_{0}^t e^{-\tau} \left( \norm{\widetilde{\varrho}}^q_{L^q}  + \norm{\widetilde{\mathbf{m}}}^q_{L^q} + \norm{\widetilde{S}}^q_{L^q} \right) \mathrm{d} \tau \nonumber.
\end{align}
As suggested in \cite{feireisllukacova2025}, we also consider the following functional,
\begin{equation*}
    \mathcal{A}(t) = \int_0^{t} \int_{\Omega} \left( E\left( \widetilde{\varrho},\widetilde{\mathbf{m}},\widetilde{S} \right) - \overline{\vartheta} \widetilde{S} \right) \, \dx \ \dt = \mathcal{J}_{\widehat{E}}(t) - \overline{\vartheta} \mathcal{J}_{\widetilde{S}}(t). 
\end{equation*}
We recall that $\overline{\vartheta} = \frac{\mathcal{E}_0}{c_v M_0}, \mathcal{E}_0 = \int_{\Omega} \widetilde{E}(t=0,\mathbf{x}) \dx, M_0 = \int_{\Omega} \widetilde{\varrho}(t=0,\mathbf{x}) \dx, c_v = 2.5$. 
\definecolor{darkgreen}{rgb}{0,0.5,0}

The time integrals present in the functions are evaluated numerically by using the trapezoidal rule for $t=0,1,\dots,10$. The approximate solutions are computed up to the finest grid of $512 \times 512$, and the respective functionals are evaluated at different times $t=1,2,\dots,10$.  \Cref{tab: I_tilS,tab: I_hatE,tab: J_tilS,tab: J_hatE,tab: J_tilrhomS,tab: F_tilS,tab: F_hatE,tab: F_tilrhomS,tab: D_S,tab: D_E,tab: A} show the comparison of different numerical methods with respect to the above functionals. Depending on whether the functional is maximised or minimised, the first, second and third best values at each time are colored in \textcolor{red}{red}, \textcolor{darkgreen}{green} and \textcolor{blue}{blue}, respectively.

\begin{table}[ht]
\centering
\begin{minipage}{\textwidth}
\caption{Comparison of $\mathcal{I}_{\widetilde{S}}(t)$ for different
numerical methods.}
\label{tab: I_tilS}
\scriptsize
\setlength{\tabcolsep}{3pt}
\renewcommand{\arraystretch}{1.05}
\begin{tabular*}{\textwidth}{@{\extracolsep{\fill}}lcccccc@{}}
\toprule
& $t=0$ & $t=1$ & $t=3$ & $t=5$ & $t=8$ & $t=10$ \\
\midrule
VFV 0
& 1.01007868 & 1.05862868 & 1.08654903 & 1.09972867 & 1.11413028 & 1.12594499 \\
VFV 1
& 1.01007868 & 1.04627195 & 1.09221545 & 1.13991600 & 1.20013155 & 1.22100191 \\
VFV 2
& 1.01007868 & 1.04408313 & 1.09405266 & 1.14059266 & 1.20442312 & 1.22388276 \\
VFV 4
& 1.01007868 & 1.04648572 & 1.09497316 & 1.13572443 & 1.19341192 & 1.21257901 \\
VFV 6
& 1.01007868 & 1.04617724 & 1.09541076 & 1.13603215 & 1.19883008 & 1.22182899 \\
VFV 8
& 1.01007868 & 1.04201593 & 1.08817481 & 1.12567020 & 1.18013418 & 1.20003649 \\
DVBFV 0
& 1.01007511 & \textcolor{red}{1.18790998} & \textcolor{red}{1.28763887} & \textcolor{red}{1.33019702} & \textcolor{red}{1.36200131} & \textcolor{red}{1.37409692} \\
DVBFV 1
& 1.01007511 & \textcolor{blue}{1.07227314} & \textcolor{blue}{1.16167257} & \textcolor{blue}{1.24139944} & \textcolor{blue}{1.29878302} & \textcolor{blue}{1.31518688} \\
RKDG 0
& 1.01287994 & \textcolor{darkgreen}{1.18779301} & \textcolor{darkgreen}{1.28741071} & \textcolor{darkgreen}{1.32994641} & \textcolor{darkgreen}{1.36179903} & \textcolor{darkgreen}{1.37394570} \\
RKDG 1
& 1.01080963 & 1.06421187 & 1.12127941 & 1.19209802 & 1.26616253 & 1.28763923 \\
RKDG 2
& 1.00985814 & 1.06063207 & 1.11838446 & 1.18441618 & 1.26824808 & 1.29043256 \\
RKDG 4
& 1.01017394 & 1.06264146 & 1.11945184 & 1.18039395 & 1.25948572 & 1.28174348 \\
DGSEM 1
& 1.01377577 & 1.05657103 & 1.11638356 & 1.17798664 & 1.25540045 & 1.27613701 \\
DGSEM 2
& 1.01073646 & 1.05416769 & 1.10918702 & 1.16906939 & 1.24741640 & 1.27319154 \\
DGSEM 4
& 1.01019004 & 1.04978581 & 1.10841348 & 1.16928116 & 1.25330492 & 1.27807331 \\
\bottomrule
\end{tabular*}
\end{minipage}
\end{table}

\begin{table}[ht]
\centering
\begin{minipage}{\textwidth}
\caption{Comparison of $\mathcal{J}_{\widetilde{S}}(t)$ for different
numerical methods.}
\label{tab: J_tilS}
\scriptsize
\setlength{\tabcolsep}{3pt}
\renewcommand{\arraystretch}{1.05}
\begin{tabular*}{\textwidth}{@{\extracolsep{\fill}}lcccccc@{}}
\toprule
& $t=0$ & $t=1$ & $t=3$ & $t=5$ & $t=8$ & $t=10$ \\
\midrule
VFV 0
& 0 & 1.03435368 & 3.18270947 & 5.36991461 & 8.69045634 & 10.93027917 \\
VFV 1
& 0 & 1.02817531 & 3.17014300 & 5.40244214 & 8.91918818 & 11.34205754 \\
VFV 2
& 0 & 1.02708091 & 3.16799503 & 5.40203674 & 8.93045323 & 11.36038370 \\
VFV 4
& 0 & 1.02828220 & 3.17333206 & 5.40472445 & 8.90477427 & 11.31229673 \\
VFV 6
& 0 & 1.02812796 & 3.17323243 & 5.40501142 & 8.91156938 & 11.33470098 \\
VFV 8
& 0 & 1.02604731 & 3.15812040 & 5.37194796 & 8.83487464 & 11.21626552 \\
DVBFV 0
& 0 & \textcolor{darkgreen}{1.09899254}
& \textcolor{darkgreen}{3.58707088}
& \textcolor{darkgreen}{6.20845874}
& \textcolor{red}{10.25184635}
& \textcolor{red}{12.98853835} \\
DVBFV 1
& 0 & \textcolor{blue}{1.04117412}
& \textcolor{blue}{3.27301848}
& \textcolor{blue}{5.68346825}
& \textcolor{blue}{9.50518595}
& \textcolor{blue}{12.12078902} \\
RKDG 0
& 0 & \textcolor{red}{1.10033647}
& \textcolor{red}{3.58805037}
& \textcolor{red}{6.20895245}
& \textcolor{darkgreen}{10.25164408}
& \textcolor{darkgreen}{12.98798340} \\
RKDG 1
& 0 & 1.03751075 & 3.22292872 & 5.53380663 & 9.23514675 & 11.79110196 \\
RKDG 2
& 0 & 1.03524511 & 3.21681045 & 5.51855852 & 9.21670351 & 11.77716680 \\
RKDG 4
& 0 & 1.03640770 & 3.21883889 & 5.51971295 & 9.19192226 & 11.73460359 \\
DGSEM 1
& 0 & 1.03517340 & 3.21007154 & 5.50463775 & 9.16748699 & 11.70063093 \\
DGSEM 2
& 0 & 1.03245208 & 3.19818517 & 5.47499809 & 9.10887500 & 11.63134177 \\
DGSEM 4
& 0 & 1.02998793 & 3.19053381 & 5.46803817 & 9.10886014 & 11.64279218 \\
\bottomrule
\end{tabular*}
\end{minipage}
\end{table}

\begin{table}[ht]
\centering
\begin{minipage}{\textwidth}
\caption{Comparison of $\mathcal{F}_{\widetilde{S}}(t)$ for different
numerical methods.}
\label{tab: F_tilS}
\scriptsize
\setlength{\tabcolsep}{3pt}
\renewcommand{\arraystretch}{1.05}
\begin{tabular*}{\textwidth}{@{\extracolsep{\fill}}lcccccc@{}}
\toprule
& $t=0$ & $t=1$ & $t=3$ & $t=5$ & $t=8$ & $t=10$ \\
\midrule
VFV 0
& 0 & 0.69976321 & 1.06712434 & 1.11791586 & 1.12555686 & 1.12590749 \\
VFV 1
& 0 & 0.69749031 & 1.06230779 & 1.11378177 & 1.12178536 & 1.12216399 \\
VFV 2
& 0 & 0.69708770 & 1.06142951 & 1.11296040 & 1.12098240 & 1.12136224 \\
VFV 4
& 0 & 0.69752963 & 1.06267113 & 1.11419616 & 1.12216646 & 1.12254281 \\
VFV 6
& 0 & 0.69747289 & 1.06256720 & 1.11410441 & 1.12208198 & 1.12246047 \\
VFV 8
& 0 & 0.69670746 & 1.05986383 & 1.11101837 & 1.11891003 & 1.11928223 \\
DVBFV 0
& 0 & \textcolor{darkgreen}{0.72354138}
& \textcolor{darkgreen}{1.14330933}
& \textcolor{darkgreen}{1.20388334}
& \textcolor{darkgreen}{1.21315787}
& \textcolor{darkgreen}{1.21358641} \\
DVBFV 1
& 0 & \textcolor{blue}{0.70227118}
& \textcolor{blue}{1.07930438}
& \textcolor{blue}{1.13454679}
& \textcolor{blue}{1.14325822}
& \textcolor{blue}{1.14366742} \\
RKDG 0
& 0 & \textcolor{red}{0.72492228}
& \textcolor{red}{1.14463707}
& \textcolor{red}{1.20520003}
& \textcolor{red}{1.21447288}
& \textcolor{red}{1.21490136} \\
RKDG 1
& 0 & 0.70115565 & 1.07269621 & 1.12576468 & 1.13416479 & 1.13456424 \\
RKDG 2
& 0 & 0.70002144 & 1.07074820 & 1.12364836 & 1.13201985 & 1.13241998 \\
RKDG 4
& 0 & 0.70054894 & 1.07158087 & 1.12450510 & 1.13282922 & 1.13322655 \\
DGSEM 1
& 0 & 0.70123327 & 1.07067112 & 1.12344548 & 1.13175138 & 1.13214732 \\
DGSEM 2
& 0 & 0.69927154 & 1.06749743 & 1.11988500 & 1.12812377 & 1.12851766 \\
DGSEM 4
& 0 & 0.69819233 & 1.06523980 & 1.11762664 & 1.12587171 & 1.12626745 \\
\bottomrule
\end{tabular*}
\end{minipage}
\end{table}

\begin{table}[ht]
\centering
\begin{minipage}{\textwidth}
\caption{Comparison of $D_S(t)$ for different numerical methods.}
\label{tab: D_S}
\scriptsize
\setlength{\tabcolsep}{3pt}
\renewcommand{\arraystretch}{1.05}
\begin{tabular*}{\textwidth}{@{\extracolsep{\fill}}lcccccc@{}}
\toprule
& $t=0$ & $t=1$ & $t=3$ & $t=5$ & $t=8$ & $t=10$ \\
\midrule
VFV 0
& 0 & 0.04855000 & 0.07647035 & 0.08964999 & 0.10405160 & 0.11586630 \\
VFV 1
& 0 & 0.03619326 & 0.08213677 & 0.12983732 & 0.19005287 & 0.21092323 \\
VFV 2
& 0 & 0.03400445 & 0.08397398 & 0.13051398 & 0.19434444 & 0.21380408 \\
VFV 4
& 0 & 0.03640704 & 0.08489448 & 0.12564575 & 0.18333324 & 0.20250032 \\
VFV 6
& 0 & 0.03609856 & 0.08533208 & 0.12595347 & 0.18875140 & 0.21175030 \\
VFV 8
& 0 & 0.03193725 & 0.07809613 & 0.11559152 & 0.17005550 & 0.18995780 \\
DVBFV 0
& 0 & \textcolor{red}{0.17783487}
& \textcolor{red}{0.27756376}
& \textcolor{red}{0.32012191}
& \textcolor{red}{0.35192620}
& \textcolor{red}{0.36402180} \\
DVBFV 1
& 0 & \textcolor{blue}{0.06219802}
& \textcolor{blue}{0.15159746}
& \textcolor{blue}{0.23132433}
& \textcolor{blue}{0.28870791}
& \textcolor{blue}{0.30511177} \\
RKDG 0
& 0 & \textcolor{darkgreen}{0.17491307}
& \textcolor{darkgreen}{0.27453078}
& \textcolor{darkgreen}{0.31706647}
& \textcolor{darkgreen}{0.34891910}
& \textcolor{darkgreen}{0.36106577} \\
RKDG 1
& 0 & 0.05340224 & 0.11046978 & 0.18128839 & 0.25535290 & 0.27682959 \\
RKDG 2
& 0 & 0.05077393 & 0.10852632 & 0.17455803 & 0.25838993 & 0.28057441 \\
RKDG 4
& 0 & 0.05246752 & 0.10927790 & 0.17022001 & 0.24931178 & 0.27156954 \\
DGSEM 1
& 0 & 0.04279526 & 0.10260779 & 0.16421086 & 0.24162467 & 0.26236123 \\
DGSEM 2
& 0 & 0.04343123 & 0.09845056 & 0.15833293 & 0.23667993 & 0.26245508 \\
DGSEM 4
& 0 & 0.03959577 & 0.09822344 & 0.15909112 & 0.24311488 & 0.26788327 \\
\bottomrule
\end{tabular*}
\end{minipage}
\end{table}

\begin{table}[ht]
\centering
\begin{minipage}{\textwidth}
\caption{Comparison of $\mathcal{I}_{\widehat{E}}(t)$ for different
numerical methods.}
\label{tab: I_hatE}
\scriptsize
\setlength{\tabcolsep}{3pt}
\renewcommand{\arraystretch}{1.05}
\begin{tabular*}{\textwidth}{@{\extracolsep{\fill}}lcccccc@{}}
\toprule
& $t=0$ & $t=1$ & $t=3$ & $t=5$ & $t=8$ & $t=10$ \\
\midrule
VFV 0
& 6.43414983 & \textcolor{red}{6.34950925} & 6.30574902 & 6.28765279 & 6.25883512 & 6.24788676 \\
VFV 1
& 6.43414983 & 6.38915860 & 6.29201476 & 6.16193844
& \textcolor{red}{6.16788110} & \textcolor{darkgreen}{6.17371417} \\
VFV 2
& 6.43414983 & 6.38479686 & 6.27331554 & 6.15871171 & 6.20472573 & 6.21238206 \\
VFV 4
& 6.43414983 & 6.37730410 & 6.28384191
& \textcolor{blue}{6.15427350} & 6.19625143 & 6.20168454 \\
VFV 6
& 6.43414983 & \textcolor{blue}{6.37623384} & 6.27888475
& \textcolor{red}{6.13147767}
& \textcolor{darkgreen}{6.18342277} & \textcolor{blue}{6.19247630} \\
VFV 8
& 6.43414983 & 6.37738192 & 6.27184213 & 6.19240200
& \textcolor{blue}{6.18436240} & \textcolor{red}{6.16622644} \\
DVBFV 0
& 6.43413102 & 6.42458411 & 6.40717955 & 6.40332474 & 6.40857755 & 6.41345243 \\
DVBFV 1
& 6.43413102 & 6.40184157 & 6.26295211 & 6.28564231 & 6.31279613 & 6.30603905 \\
RKDG 0
& 6.43552999 & 6.42456309 & 6.40717510 & 6.40335515 & 6.40868288 & 6.41362678 \\
RKDG 1
& 6.43589758 & 6.40470077 & \textcolor{darkgreen}{6.22217964} & 6.23530431 & 6.29431800 & 6.28015184 \\
RKDG 2
& 6.43596358 & 6.41114033 & 6.25161516 & 6.22806996 & 6.30008147 & 6.29131841 \\
RKDG 4
& 6.43661494 & 6.40246809 & \textcolor{blue}{6.22725358} & 6.21056335 & 6.27983673 & 6.30461028 \\
DGSEM 1
& 6.43411456 & 6.38349468
& \textcolor{red}{6.19727696} & \textcolor{darkgreen}{6.14837658} & 6.23210073 & 6.27318222 \\
DGSEM 2
& 6.43579876 & \textcolor{darkgreen}{6.37483274} & 6.23881271 & 6.18673478 & 6.24727107 & 6.25136825 \\
DGSEM 4
& 6.43637700 & 6.38433984 & 6.28054645 & 6.23876559 & 6.30404264 & 6.30351149 \\
\bottomrule
\end{tabular*}
\end{minipage}
\end{table}

\begin{table}[ht]
\centering
\begin{minipage}{\textwidth}
\caption{Comparison of $\mathcal{J}_{\widehat{E}}(t)$ for different
numerical methods.}
\label{tab: J_hatE}
\scriptsize
\setlength{\tabcolsep}{3pt}
\renewcommand{\arraystretch}{1.05}
\begin{tabular*}{\textwidth}{@{\extracolsep{\fill}}lcccccc@{}}
\toprule
& $t=0$ & $t=1$ & $t=3$ & $t=5$ & $t=8$ & $t=10$ \\
\midrule
VFV 0
& 0 & \textcolor{red}{6.39182954}
& \textcolor{blue}{19.04091347} & 31.63411811 & 50.45559111 & 62.95938495 \\
VFV 1
& 0 & 6.41165422 & 19.11210806 & 31.55568900 & 50.07350657
& \textcolor{blue}{62.41715432} \\
VFV 2
& 0 & 6.40947334 & 19.08388940 & 31.48181416 & 50.06909022 & 62.48864095 \\
VFV 4
& 0 & 6.40572697 & 19.07648467 & 31.49603826
& \textcolor{blue}{50.02889212} & 62.43330182 \\
VFV 6
& 0 & \textcolor{darkgreen}{6.40519184} & 19.06763182
& \textcolor{blue}{31.46090741}
& \textcolor{darkgreen}{49.94572603} & \textcolor{red}{62.31648962} \\
VFV 8
& 0 & 6.40576588 & 19.06012434 & 31.51668487 & 50.11657223 & 62.46845807 \\
DVBFV 0
& 0 & 6.42935756 & 19.25945878 & 32.06862016 & 51.28518921 & 64.10723107 \\
DVBFV 1
& 0 & 6.41798629 & 19.07260226 & 31.60107446 & 50.52142849 & 63.13721989 \\
RKDG 0
& 0 & 6.43004654 & 19.26011215 & 32.06930107 & 51.28606145 & 64.10838326 \\
RKDG 1
& 0 & 6.42029918 & 19.08986656 & 31.52782785 & 50.36463196 & 62.93987757 \\
RKDG 2
& 0 & 6.42355195 & 19.13631021 & 31.58697273 & 50.42134277 & 63.01242955 \\
RKDG 4
& 0 & 6.41954151 & 19.09106838 & 31.47496367 & 50.23631126 & 62.82216038 \\
DGSEM 1
& 0 & 6.40880462
& \textcolor{red}{19.00315333}
& \textcolor{red}{31.32447679}
& \textcolor{red}{49.89977358}
& \textcolor{darkgreen}{62.40620929} \\
DGSEM 2
& 0 & \textcolor{blue}{6.40531575}
& \textcolor{darkgreen}{19.02700771}
& \textcolor{darkgreen}{31.41410006} & 50.07282187 & 62.57485654 \\
DGSEM 4
& 0 & 6.41035842 & 19.08270224 & 31.55758836 & 50.37678405 & 62.97880610 \\
\bottomrule
\end{tabular*}
\end{minipage}
\end{table}

\begin{table}[ht]
\centering
\begin{minipage}{\textwidth}
\caption{Comparison of $\mathcal{F}_{\widehat{E}}(t)$ for different
numerical methods.}
\label{tab: F_hatE}
\scriptsize
\setlength{\tabcolsep}{3pt}
\renewcommand{\arraystretch}{1.05}
\begin{tabular*}{\textwidth}{@{\extracolsep{\fill}}lcccccc@{}}
\toprule
& $t=0$ & $t=1$ & $t=3$ & $t=5$ & $t=8$ & $t=10$ \\
\midrule
VFV 0
& 0 & \textcolor{red}{4.38500187} & \textcolor{red}{6.56541708}
& \textcolor{red}{6.85889689} & \textcolor{red}{6.90241098} & \textcolor{red}{6.90437398} \\
VFV 1
& 0 & 4.39229496 & 6.58485992 & 6.87611089 & 6.91885747 & 6.92079395 \\
VFV 2
& 0 & 4.39149267 & 6.58082648 & 6.87096567 & 6.91378458 & 6.91573282 \\
VFV 4
& 0 & \textcolor{blue}{4.39011445} & 6.57763171 & 6.86835837 & 6.91107167 & 6.91301757 \\
VFV 6
& 0 & \textcolor{darkgreen}{4.38991759} & 6.57639676 & 6.86669608 & 6.90926574 & 6.91120648 \\
VFV 8
& 0 & 4.39012876 & 6.57594896 & 6.86694388 & 6.90988554 & 6.91182508 \\
DVBFV 0
& 0 & 4.39880171 & 6.60810546 & 6.90646712 & 6.95083124 & 6.95284293 \\
DVBFV 1
& 0 & 4.39461846 & 6.58369774 & 6.87533012 & 6.91895778 & 6.92093805 \\
RKDG 0
& 0 & 4.39949734 & 6.60879401 & 6.90715593 & 6.95152036 & 6.95353208 \\
RKDG 1
& 0 & 4.39602766 & 6.58920684 & 6.87883128 & 6.92219185 & 6.92416617 \\
RKDG 2
& 0 & 4.39724515 & 6.59575925 & 6.88612138 & 6.92945813 & 6.93143458 \\
RKDG 4
& 0 & 4.39597566 & 6.58894340 & 6.87780064 & 6.92097858 & 6.92295172 \\
DGSEM 1
& 0 & 4.39123551
& \textcolor{darkgreen}{6.57283447} & \textcolor{darkgreen}{6.86043396}
& \textcolor{darkgreen}{6.90317054} & \textcolor{darkgreen}{6.90513004} \\
DGSEM 2
& 0 & 4.39048433
& \textcolor{blue}{6.57300000} & \textcolor{blue}{6.86223563}
& \textcolor{blue}{6.90520795} & \textcolor{blue}{6.90716936} \\
DGSEM 4
& 0 & 4.39252219 & 6.58121312 & 6.87241227 & 6.91573567 & 6.91771341 \\

\bottomrule
\end{tabular*}
\end{minipage}
\end{table}

\begin{table}[ht]
\centering
\begin{minipage}{\textwidth}
\caption{Comparison of $\mathcal{D}_{E}(t)$ for different numerical methods.}
\label{tab: D_E}
\scriptsize
\setlength{\tabcolsep}{3pt}
\renewcommand{\arraystretch}{1.05}
\begin{tabular*}{\textwidth}{@{\extracolsep{\fill}}lcccccc@{}}
\toprule
& $t=0$ & $t=1$ & $t=3$ & $t=5$ & $t=8$ & $t=10$ \\
\midrule
VFV 0
& 0.00334999 & \textcolor{red}{0.08799056} & 0.13175061
& 0.14984742 & 0.17866497 & 0.18961337 \\
VFV 1
& 0.00334999 & 0.04833632 & 0.14548563
& 0.27556111
& \textcolor{red}{0.26961714} & \textcolor{darkgreen}{0.26378460} \\
VFV 2
& 0.00334999 & 0.05270857 & 0.16417959 & 0.27878642 & 0.23277723 & 0.22511498 \\
VFV 4
& 0.00334999 & 0.06003532 & 0.15352655
& \textcolor{blue}{0.28345137} & 0.24130860 & 0.23586913 \\
VFV 6
& 0.00334999 & \textcolor{blue}{0.06140810} & 0.15876044
& \textcolor{red}{0.30602740}
& \textcolor{darkgreen}{0.25410831} & \textcolor{blue}{0.24497004} \\
VFV 8
& 0.00334999 & 0.06008729 & 0.16565759 & 0.24509891
& \textcolor{blue}{0.25313483} & \textcolor{red}{0.27127028} \\
DVBFV 0
& 0.00336898 & 0.01291589 & 0.03032045 & 0.03417526 & 0.02892245 & 0.02404757 \\
DVBFV 1
& 0.00336898 & 0.03565843 & 0.17454789 & 0.15185769 & 0.12470387 & 0.13146095 \\
RKDG 0
& 0.00198628 & 0.01295318 & 0.03034117 & 0.03416112 & 0.02883340 & 0.02388950 \\
RKDG 1
& 0.00161869 & 0.03281550
& \textcolor{darkgreen}{0.21533663} & 0.20221197 & 0.14319828 & 0.15736443 \\
RKDG 2
& 0.00155269 & 0.02637595 & 0.18590111 & 0.20944631 & 0.13743480 & 0.14619786 \\
RKDG 4
& 0.00088586 & 0.03503271
& \textcolor{blue}{0.21024722} & 0.22693745 & 0.15766406 & 0.13289051 \\
DGSEM 1
& 0.00339009 & 0.05400997
& \textcolor{red}{0.24022769}
& \textcolor{darkgreen}{0.28912807} & 0.20540391 & 0.16432243 \\
DGSEM 2
& 0.00170515 & \textcolor{darkgreen}{0.06267116}
& 0.19869119 & 0.25076913 & 0.19023284 & 0.18613566 \\
DGSEM 4
& 0.00112525 & 0.05316240 & 0.15695580 & 0.19873665 & 0.13345961 & 0.13399076 \\
\bottomrule
\end{tabular*}
\end{minipage}
\end{table}

\begin{table}[ht]
\centering
\begin{minipage}{\textwidth}
\caption{Comparison of $\mathcal{J}_{(\widetilde{\varrho},\widetilde{\mathbf{m}},\widetilde{S})}(t)$
for different numerical methods.}
\label{tab: J_tilrhomS}
\scriptsize
\setlength{\tabcolsep}{3pt}
\renewcommand{\arraystretch}{1.05}
\begin{tabular*}{\textwidth}{@{\extracolsep{\fill}}lcccccc@{}}
\toprule
& $t=0$ & $t=1$ & $t=3$ & $t=5$ & $t=8$ & $t=10$ \\
\midrule
VFV 0
& 0
& \textcolor{blue}{3.56139673}
& \textcolor{blue}{10.32734158}
& \textcolor{blue}{16.88759530}
& \textcolor{blue}{26.50055569}
& \textcolor{blue}{32.78955948} \\
VFV 1
& 0 & 3.64003394 & 10.79690295 & 17.69524763 & 27.51410257 & 33.75271237 \\
VFV 2
& 0 & 3.64451081 & 10.79575391 & 17.65726353 & 27.74535774 & 34.25077560 \\
VFV 4
& 0 & 3.63305285 & 10.74348114 & 17.51657512 & 27.42516451 & 33.99341055 \\
VFV 6
& 0 & 3.63237529 & 10.70443026 & 17.35864373 & 26.96435094 & 33.27968985 \\
VFV 8
& 0 & 3.64281462 & 10.78663615 & 17.72758822 & 27.84004279 & 34.35657041 \\
DVBFV 0
& 0 & \textcolor{darkgreen}{3.51904628}
& \textcolor{darkgreen}{10.12930896}
& \textcolor{darkgreen}{16.60944556}
& \textcolor{darkgreen}{26.24062678}
& \textcolor{darkgreen}{32.65102470} \\
DVBFV 1
& 0 & 3.64055699 & 10.68185019 & 17.46921263 & 27.49730924 & 34.01684603 \\
RKDG 0
& 0 & \textcolor{red}{3.51702797}
& \textcolor{red}{10.12642065}
& \textcolor{red}{16.60603054}
& \textcolor{red}{26.23758628}
& \textcolor{red}{32.64886528} \\
RKDG 1
& 0 & 3.65585346 & 10.81461406 & 17.70616061 & 28.14287915 & 34.86957814 \\
RKDG 2
& 0 & 3.65664438 & 10.85866821 & 17.78454692 & 28.18108173 & 34.88430074 \\
RKDG 4
& 0 & 3.65590199 & 10.81239434 & 17.63916183 & 27.91768424 & 34.77913098 \\
DGSEM 1
& 0 & 3.63908001 & 10.67214698 & 17.21971921 & 26.77576538 & 33.28836077 \\
DGSEM 2
& 0 & 3.64080880 & 10.74213912 & 17.57941752 & 27.70042685 & 34.27671861 \\
DGSEM 4
& 0 & 3.65252872 & 10.86427080 & 17.88948825 & 28.38090315 & 35.28975580 \\
\bottomrule
\end{tabular*}
\end{minipage}
\end{table}

\begin{table}[ht]
\centering
\begin{minipage}{\textwidth}
\caption{Comparison of $\mathcal{F}_{(\widetilde{\varrho},\widetilde{\mathbf{m}},\widetilde{S})}(t)$
for different numerical methods.}
\label{tab: F_tilrhomS}
\scriptsize
\setlength{\tabcolsep}{3pt}
\renewcommand{\arraystretch}{1.05}
\begin{tabular*}{\textwidth}{@{\extracolsep{\fill}}lcccccc@{}}
\toprule
& $t=0$ & $t=1$ & $t=3$ & $t=5$ & $t=8$ & $t=10$ \\
\midrule
VFV 0
& 0
& \textcolor{blue}{2.46803039}
& \textcolor{blue}{3.64393818}
& \textcolor{blue}{3.79754366}
& \textcolor{blue}{3.81989121}
& \textcolor{blue}{3.82088093} \\
VFV 1
& 0 & 2.49695940 & 3.73486398 & 3.89737650 & 3.92041559 & 3.92139970 \\
VFV 2
& 0 & 2.49860635 & 3.73738998 & 3.89886097 & 3.92229472 & 3.92332113 \\
VFV 4
& 0 & 2.49439120 & 3.72599339 & 3.88600450 & 3.90887665 & 3.90990848 \\
VFV 6
& 0 & 2.49414194 & 3.72246304 & 3.87997753 & 3.90220772 & 3.90320193 \\
VFV 8
& 0 & 2.49798236 & 3.73487466 & 3.89780207 & 3.92135129 & 3.92237974 \\
DVBFV 0
& 0 & \textcolor{darkgreen}{2.45244418}
& \textcolor{darkgreen}{3.59994885}
& \textcolor{darkgreen}{3.75129192}
& \textcolor{darkgreen}{3.77357242}
& \textcolor{darkgreen}{3.77457800} \\
DVBFV 1
& 0 & 2.49714547 & 3.72399324 & 3.88268165 & 3.90600839 & 3.90703653 \\
RKDG 0
& 0 & \textcolor{red}{2.45058282}
& \textcolor{red}{3.59792896}
& \textcolor{red}{3.74925670}
& \textcolor{red}{3.77153727}
& \textcolor{red}{3.77254298} \\
RKDG 1
& 0 & 2.50313668 & 3.74749512 & 3.90832629 & 3.93241251 & 3.93347823 \\
RKDG 2
& 0 & 2.50394321 & 3.75296677 & 3.91494922 & 3.93895778 & 3.94002006 \\
RKDG 4
& 0 & 2.50393012 & 3.74721425 & 3.90743493 & 3.93105697 & 3.93213411 \\
DGSEM 1
& 0 & 2.49449051 & 3.72268811 & 3.87741450 & 3.89945275 & 3.90046684 \\
DGSEM 2
& 0 & 2.49772903 & 3.73027245 & 3.89062047 & 3.91410319 & 3.91514449 \\
DGSEM 4
& 0 & 2.50252877 & 3.74827332 & 3.91305248 & 3.93730766 & 3.93839442 \\
\bottomrule
\end{tabular*}
\end{minipage}
\end{table}

\begin{table}[ht]
\centering
\begin{minipage}{\textwidth}
\caption{Comparison of $\mathcal{A}(t)$ for different numerical methods.}
\label{tab: A}
\scriptsize
\setlength{\tabcolsep}{3pt}
\renewcommand{\arraystretch}{1.05}
\begin{tabular*}{\textwidth}{@{\extracolsep{\fill}}lcccccc@{}}
\toprule
& $t=0$ & $t=1$ & $t=3$ & $t=5$ & $t=8$ & $t=10$ \\
\midrule
VFV 0 & 0 & 4.61618905 & 13.57726220 & 22.41576468 & 35.53697436 & 44.19573901 \\
VFV 1 & 0 & 4.64661993 & 13.67002923 & 22.28149664 & 34.76223350 & 42.94662217 \\
VFV 2 & 0 & 4.64631777 & 13.64549792 & 22.20831774 & 34.73847881 & 42.98664889 \\
VFV 4 & 0 & 4.64050919 & 13.62893129 & 22.21792794 & 34.74236293 & 43.01385906 \\
VFV 6 & 0 & 4.64023884 & 13.62024947 & 22.18230445 & 34.64753190 & 42.85858623 \\
VFV 8 & 0 & 4.64438466 & 13.63868431 & 22.29484085 & 34.95003740 & 43.21386889 \\
DVBFV 0
& 0 & \textcolor{darkgreen}{4.54275369}
& \textcolor{darkgreen}{13.10165377}
& \textcolor{red}{21.41076600}
& \textcolor{red}{33.68618631}
& \textcolor{red}{41.81024024} \\
DVBFV 1
& 0 & \textcolor{blue}{4.63063738}
& \textcolor{blue}{13.45392054}
& \textcolor{blue}{21.84445397}
& 34.20419262
& 42.32986541 \\
RKDG 0
& 0 & \textcolor{red}{4.54129477}
& \textcolor{red}{13.10114472}
& \textcolor{darkgreen}{21.41149754}
& \textcolor{darkgreen}{33.68888876}
& \textcolor{darkgreen}{41.81422390} \\
RKDG 1
& 0 & 4.63938914 & 13.55763848 & 22.02892698 & 34.51229931 & 42.70019155 \\
RKDG 2
& 0 & 4.64653093 & 13.61458427 & 22.11424557 & 34.60066834 & 42.79666353 \\
RKDG 4
& 0 & 4.64038227 & 13.56541767 & 21.99949539 & 34.45690957 & 42.67784035 \\
DGSEM 1
& 0 & 4.63179973 & 13.49266318 & 21.87507614
& \textcolor{blue}{34.16263309}
& \textcolor{blue}{42.32060971} \\
DGSEM 2
& 0 & 4.63297554 & 13.53690092 & 22.01554349 & 34.43623615 & 42.60812381 \\
DGSEM 4
& 0 & 4.64223304 & 13.60568288 & 22.17089864 & 34.74008927 & 42.99224522 \\
\bottomrule
\end{tabular*}
\end{minipage}
\end{table}

\subsection{Discussion}
We consider two selection strategies: the first involves maximising entropy-related functionals and minimising energy-related functionals, and the second is concerned with minimising $\mathcal{A}(t)$ which involves both energy and entropy. \\ 
\textbf{First selection} \\
\textit{Entropy-related functionals}
\begin{enumerate} 
    \item Maximising $\mathcal{I}_{\widetilde{S}}(t)$: DVBFV~0 and RKDG~0 dominate globally at all times. At t=10, DVBFV~0, RKDG~0, DVBFV~1, RKDG~2 and RKDG~1 are the top 5 methods that maximise $\mathcal{I}_{\widetilde{S}}(t)$, see \Cref{tab: I_tilS}. 
    \item Maximising $\mathcal{J}_{\widetilde{S}}(t)$: DVBFV~0 and RKDG~0 dominate globally at all times. At t=10, DVBFV~0, RKDG~0, DVBFV~1, RKDG~1 and RKDG~2 are the top 5 methods that maximise $\mathcal{J}_{\widetilde{S}}(t)$, see \Cref{tab: J_tilS}. 
    \item Maximising $\mathcal{F}_{\widetilde{S}}(t)$: RKDG~0 and DVBFV~0 dominate globally at all times. At t=10, RKDG~0, DVBFV~0, DVBFV~1, RKDG~1 and RKDG~4 are the top 5 methods that maximise $\mathcal{F}_{\widetilde{S}}(t)$, see \Cref{tab: F_tilS}. 
    \item Maximising $D_{S}(t)$: DVBFV~0 and RKDG~0 dominate globally at all times. At t=10, DVBFV~0, RKDG~0, DVBFV~1, RKDG~2 and RKDG~1 are the top 5 methods that maximise $D_S(t)$, see \Cref{tab: D_S}.   
\end{enumerate}
From the discussion, it is clear that the methods DVBFV~0, RKDG~0, and DVBFV~1 yield the required maximisation of entropy-related functionals. \\
\textit{Energy-related functionals}
\begin{enumerate}
    \item Minimising $\mathcal{I}_{\widehat{E}}(t)$: At t=10, VFV~8, VFV~1, VFV~6, VFV~4 and VFV~2 are the top 5 methods that minimise $\mathcal{I}_{\widehat{E}}(t)$, see \Cref{tab: I_hatE}.  
    \item Minimising $\mathcal{J}_{\widehat{E}}(t)$: At t=10, VFV~6, DGSEM~1, VFV~1, VFV~4 and VFV~8 are the top 5 methods that minimise $\mathcal{J}_{\widehat{E}}(t)$, see \Cref{tab: J_hatE}.  
    \item Minimising $\mathcal{F}_{\widehat{E}}(t)$: At t=10, VFV~0, DGSEM~1, DGSEM~2, VFV~6 and VFV~8 are the top 5 methods that minimise $\mathcal{F}_{\widehat{E}}(t)$, see \Cref{tab: F_hatE}. 
    \item Maximising $\mathcal{D}_{E}(t)$: At t=10, VFV~8, VFV~1, VFV~6, VFV~4 and VFV~2 are the top 5 methods that maximise $\mathcal{D}_{E}(t)$, see \Cref{tab: D_E}.  
\end{enumerate}
From our discussion of the energy-related functionals, we cannot infer which methods yield minimisation, but we can observe that higher-order methods yield smaller mean energy and maximise the energy defect. 

A possible alternative is to minimise the norm of the solution, represented in \Cref{tab: F_tilrhomS,tab: J_tilrhomS}. Here, we infer that the first-order methods RKDG~0, DVBFV~0 and VFV~0 minimise the norm of the solution. 

Now, we consider the second selection strategy consisting of the minimisation of $\mathcal{A}(t)$, which is a functional involving both energy and entropy. \\
\textbf{Second selection} \\
At $t=10$, DVBFV~0, RKDG~0, DGSEM~1, DVBFV~1 and DGSEM~2 are the top 5 methods that minimise $\mathcal{A}(t)$, see \Cref{tab: A}. 

Thus, DVBFV~0, RKDG~0 and DVBFV~1 maximise all the entropy-related functionals and also minimise $\mathcal{A}(t)$ that involves both energy and entropy. Further, although DGSEM~1 and DGSEM~2 do not maximise entropy-related functionals or minimise energy-related functionals, they yield minimisation of $\mathcal{A}(t)$ that involves both energy and entropy. 

\section{Conclusions}
We have investigated numerically the selection criteria for DW 
solutions of the compressible Euler equations using the Kelvin-Helmholtz problem 
as a benchmark. The selection process was carried out in two steps, combining the 
maximisation of entropy-related functionals and the minimisation of energy-related functionals, or in one step using the Bregman distance. 

In the first case, the entropy-related functionals $\mathcal{I}_{\widetilde{S}}$, 
$\mathcal{J}_{\widetilde{S}}$, $\mathcal{F}_{\widetilde{S}}$, and $D_S$ consistently 
identify DVBFV~0, RKDG~0, and DVBFV~1 as the dominant methods across all time steps. 
In contrast, the energy-related functionals $\mathcal{I}_{\widehat{E}}$, 
$\mathcal{J}_{\widehat{E}}$, $\mathcal{F}_{\widehat{E}}$, and $\mathcal{D}_E$ show a trend to the higher-order methods, with different VFV variants and DGSEM methods appearing 
among the top performers depending on the functional considered. 

In the second case, minimisation of the combined functional $\mathcal{A}(t)$, 
which incorporates both energy and entropy, identifies DVBFV~0, RKDG~0, and DVBFV~1 
as the top performers, consistent with the entropy-based selection. DGSEM~1 
and DGSEM~2 also appear among the top five methods minimising $\mathcal{A}(t)$.

Overall, our results suggest that DVBFV~0, RKDG~0, and DVBFV~1 are the most consistent 
candidates for selection across all criteria considered, satisfying both the entropy 
maximisation principle and the minimisation of the combined energy-entropy functional. 
These findings highlight the importance of combining energy/entropy-based selection criteria and 
demonstrate the sensitivity of the selection process to the numerical diffusion 
inherent in each scheme. In future, it will be important to construct PDE-constrained optimisation algorithms that directly approximate a selected solution. 

\begin{appendices}
\section{Description of the numerical methods}
\label{Sec: Num schemes}
In this section, we present the four numerical schemes that we use to compute our functions of interest. We consider the following notations for the conserved variable vector, $\mathbf{U}=\begin{bmatrix} \varrho & \mathbf{m} & E \end{bmatrix}^T$, and the flux vector in $x,y$ directions, $\mathbf{G}^{x,y}(\mathbf{U})
=\begin{bmatrix} m^{x,y} & \mathbf{m}\frac{m^{x,y}}{\varrho} + p\mathbf{e}^{x,y} & (E + p) \frac{m^{x,y}}{\varrho} \end{bmatrix}^T$. Here $m^{x}, m^{y}$ represent the $x,y-$components of $\mathbf{m}$, and $\mathbf{e}^{x}, \mathbf{e}^{y}$ represent the unit vectors $[1,0]^T$ and $[0,1]^T$. 
\subsection{Viscous finite volume (VFV) method}
\label{Subsec: VFV}
In this section, we present the VFV method \cite{feireisl2020finite} for $d=2$. Consider the computational domain to be covered with uniform cells $C_{j,\,k}:=[x_\jmh,x_\jph]\times[y_\kmh,y_\kph]$ centered at
$(x_j,y_k)=\big((x_\jmh+x_\jph)/2$, $(y_\kph+y_\kmh)/2\big)$ with $x_\jph-x_\jmh\equiv\delx$ and $y_\kph-y_\kmh\equiv\dely$ for all $j,k$, and denote by $\xbar\mU_{j,k}(t)$
as the cell averages of $\mU(\cdot,\cdot,t)$ over the corresponding cells $C_{j,k}$: $\xbar\mU_{j,k}(t):\approx\dfrac{1}{\delx \dely}\int\limits_{C_{j,k}}\mU(x,y,t)\dx{}\dy{}.$
Assume that the cell averages $\xbar \mU_{j,k}(t)$ are available at a certain time level $t\ge0$.

\subsubsection{First-order VFV scheme}
Let $\xbar u_x = \frac{\xbar {m^x}}{\xbar {\varrho}}$ and $\xbar u_y=\frac{\xbar {m^y}}{\xbar {\varrho}}$ be the average of velocities, and $\xbar p=(\gamma-1) \left( \xbar E-\frac{\left( \xbar {m^x}\right)^2+\left( \xbar {m^y}\right)^2}{2\xbar \varrho} \right)$ be the average of pressure. The first-order scheme for the approximate density $\xbar \varrho_{j,k}$, momentum $\left(\xbar {m^x}\right)_{j,k},\left(\xbar {m^y}\right)_{j,k}$ and total energy $\xbar E_{j,k}$ in the cell $C_{j,k}$ are given by
\begin{gather*}
		\frac{{\rm d}\xbar \varrho_{j,k}}{{\rm d}t}=-\frac{{{\cal F}}^{\rm up,1}_{\jph,k}-{{\cal F}}^{\rm up,1}_{\jmh,k}}{\delx}-\frac{{{\cal G}}^{\rm up,1}_{j,\kph}-{{\cal G}}^{\rm up,1}_{j,\kmh}}{\dely}, \\ 
		\frac{{\rm d}\left(\xbar {m^x}\right)_{j,k}}{{\rm d}t}=-\frac{{{\cal F}}^{\rm up,2}_{\jph,k}-{{\cal F}}^{\rm up,2}_{\jmh,k}}{\delx}-\frac{{{\cal G}}^{\rm up,2}_{j,\kph}-{{\cal G}}^{\rm up,2}_{j,\kmh}}{\dely}
		-\frac{1}{2\delx} \left(\xbar p_{j+1,k}-\xbar p_{j-1,k} \right) \\
		+h^{\alpha-1} \left( \frac{(\xbar u_1)_{j+1,k}+(\xbar u_1)_{j-1,k}-2(\xbar u_1)_{j,k} }{\delx}+\frac{(\xbar u_1)_{j,k+1}+(\xbar u_1)_{j,k-1}-2(\xbar u_1)_{j,k}}{\dely} \right), \\
		\frac{{\rm d}\left(\xbar {m^y}\right)_{j,k}}{{\rm d}t}=-\frac{{{\cal F}}^{\rm up,3}_{\jph,k}-{{\cal F}}^{\rm up,3}_{\jmh,k}}{\delx}-\frac{{{\cal G}}^{\rm up,3}_{j,\kph}-{{\cal G}}^{\rm up,3}_{j,\kmh}}{\dely}
		-\frac{1}{2\dely} \left(\xbar p_{j,k+1}-\xbar p_{j,k-1} \right) \\
		+h^{\alpha-1} \left( \frac{(\xbar u_2)_{j+1,k}+(\xbar u_2)_{j-1,k}-2(\xbar u_2)_{j,k}}{\delx}+\frac{(\xbar u_2)_{j,k+1}+(\xbar u_2)_{j,k-1}-2(\xbar u_2)_{j,k}}{\dely}\right), \\
		\frac{{\rm d}\xbar E_{j,k}}{{\rm d}t}=-\frac{{{\cal F}}^{\rm up,4}_{\jph,k}-{{\cal F}}^{\rm up,4}_{\jmh,k}}{\delx}-\frac{{{\cal G}}^{\rm up,4}_{j,\kph}-{{\cal G}}^{\rm up,4}_{j,\kmh}}{\dely}\\
		-\frac{1}{2\delx}(\xbar u_1)_{j,k}\left( \xbar{p}_{j+1,k}-\xbar{p}_{j-1,k} \right) -\frac{1}{2\dely}(\xbar u_2)_{j,k} \left( \xbar{p}_{j,k+1}-\xbar{p}_{j,k-1} \right) \\
		-\frac{1}{2\delx}\xbar{p}_{j,k}\left( (\xbar u_1)_{j+1,k}-(\xbar u_1)_{j-1,k} \right) -\frac{1}{2\dely}\xbar{p}_{j,k} \left( (\xbar u_2)_{j,k+1}-(\xbar u_2)_{j,k-1} \right) \\
		+\frac{h^{\alpha-1}}{2\delx} \left((\xbar u_1)_{j+1,k}^2+(\xbar u_2)_{j+1,k}^2+(\xbar u_1)_{j-1,k}^2+(\xbar u_2)_{j-1,k}^2 - 2(\xbar u_1)_{j,k}^2-2(\xbar u_2)_{j,k}^2 \right), \\
		+\frac{h^{\alpha-1}}{2\dely} \left((\xbar u_1)_{j,k+1}^2+(\xbar u_2)_{j,k+1}^2+(\xbar u_1)_{j,k-1}^2+(\xbar u_2)_{j,k-1}^2 - 2(\xbar u_1)_{j,k}^2-2(\xbar u_2)_{j,k}^2 \right).
\end{gather*}
Here, we use the upwind numerical fluxes
\begin{align*}
		\left( {{\cal F}}^{\rm up,1}_{\jph,k},{{\cal F}}^{\rm up,2}_{\jph,k},{{\cal F}}^{\rm up,3}_{\jph,k},{{\cal F}}^{\rm up,4}_{\jph,k} \right) 
		= \frac{1}{4}\left( \xbar \mU_{j,k}+\xbar \mU_{j+1,k}\right) \left((\xbar u_1)_{j,k}+(\xbar u_1)_{j+1,k} \right) \\
		-\frac{1}{4} \modL{(\xbar u_1)_{j,k}+(\xbar u_1)_{j+1,k}} \left( \xbar \mU_{j+1,k}-\xbar \mU_{j,k}\right) 
		-h^\varepsilon \left( \xbar \mU_{j+1,k}-\xbar \mU_{j,k}\right) .
\end{align*}
The upwind numerical fluxes $\left( {{\cal G}}^{\rm up,1}_{j,\kph},{{\cal G}}^{\rm up,2}_{j,\kph},{{\cal G}}^{\rm up,3}_{j,\kph},{{\cal G}}^{\rm up,4}_{j,\kph} \right) $ in the $y$ direction are computed using an analogous procedure..
The two parameters $\alpha$ and $\varepsilon$ control the numerical diffusion. When these parameters are in an appropriate range, the VFV scheme produces consistent approximations, see \cite{feireisl2020finite}.

\subsubsection{Higher-order VFV schemes}
The higher-order VFV scheme based on the A-WENO reconstruction is given in \cite{chu2025numerical}:
\begin{align}\label{2.1}
		\frac{{\rm d}\xbar \mU_{j,k}}{{\rm d}t}=-\frac{\bm{{\cal F}}^{\rm VFV}_{\jph,k}-\bm{{\cal F}}^{\rm VFV}_{\jmh,k}}{\delx}-\frac{\bm{{\cal G}}^{\rm VFV}_{j,\kph}-\bm{{\cal G}}^{\rm VFV}_{j,\kmh}}{\dely}.
\end{align}
The numerical fluxes in the $x$ direction are
\begin{gather*}
		\bm{{\cal F}}^{\rm VFV}_{\jph,k}\big(\bm U_{\jph,k}^-,\bm U_{\jph,k}^+\big) =
		\left( {{\cal F}}^{\rm VFV,1}_{\jph,k},{{\cal F}}^{\rm VFV,2}_{\jph,k},{{\cal F}}^{\rm VFV,3}_{\jph,k},{{\cal F}}^{\rm VFV,4}_{\jph,k} \right) ,\\
		{{\cal F}}^{\rm VFV,1}_{\jph,k}= {{\cal F}}^{\rm up,1}_{\jph,k},\\
		{{\cal F}}^{\rm VFV,2}_{\jph,k}= {{\cal F}}^{\rm up,2}_{\jph,k}+\frac{1}{2}\left( p_{\jph,k}^-+p_{\jph,k}^+\right) -h^{\alpha-1} \left( (u_1)_{\jph,k}^+-(u_1)_{\jph,k}^-\right) ,\\
		{{\cal F}}^{\rm VFV,3}_{\jph,k}= {{\cal F}}^{\rm up,3}_{\jph,k}-h^{\alpha-1} \left( (u_2)_{\jph,k}^+-(u_2)_{\jph,k}^-\right) ,\\
		{{\cal F}}^{\rm VFV,4}_{\jph,k}= {{\cal F}}^{\rm up,4}_{\jph,k}+\frac{1}{2}\left( p_{\jph,k}^-(u_1)_{\jph,k}^-+p_{\jph,k}^+(u_1)_{\jph,k}^+\right)\\
		-h^{\alpha-1} \left( \left( (u_1)_{\jph,k}^+\right) ^2+\left( (u_2)_{\jph,k}^+\right) ^2-\left( (u_1)_{\jph,k}^-\right) ^2-\left( (u_2)_{\jph,k}^-\right) ^2\right) ,
\end{gather*}
where
\begin{multline*}
		\left( {{\cal F}}^{\rm up,1}_{\jph,k},{{\cal F}}^{\rm up,2}_{\jph,k},{{\cal F}}^{\rm up,3}_{\jph,k},{{\cal F}}^{\rm up,4}_{\jph,k} \right) 
		\\ =\frac{1}{4}\left( \bm U_{\jph,k}^-+\bm U_{\jph,k}^+ \right) \left( (u_1)_{\jph,k}^-+ (u_1)_{\jph,k}^+ \right) \\
		-\frac{1}{4} \modL{ (u_1)_{\jph,k}^-+(u_1)_{\jph,k}^+} \left( \bm U_{\jph,k}^--\bm U_{\jph,k}^+\right) 
		- h^\varepsilon \left( \bm U_{\jph,k}^+-\bm U_{\jph,k}^-\right) .
\end{multline*}
Here, $\mU^\pm_{\jph,k}$ are the left/right-sided point values of $\mU$ at the cell interfaces $(x_\jph,y_k)$. $(u_1)_{\jph,k}^\pm$, $(u_2)_{\jph,k}^\pm$ and $p_{\jph,k}^\pm$ are the velocities and pressure of $\bm U_{\jph,k}^\pm$, respectively. For the second-order scheme, these two values are reconstructed using the minmod interpolant; see, e.g., \cite{kurganov2023new}. For higher-order schemes, we employ the WENO interpolation method introduced in \cite{chu2025high,gao2020seventh}. By analogy, the numerical fluxes $\bm{{\cal G}}^{\rm VFV}_{j,\kph}\big(\bm U_{j,\kph}^-,\bm U_{j,\kph}^+\big)$ in the $y$ direction are obtained. We employed the three-stage third-order strong stability preserving Runge-Kutta (SSPRK3) method (see, e.g., \cite{gottlieb2001strong}) for time integration. 

The numerical results corresponding to this method are represented in this paper as VFV $N$, where $N=0,1,2,4,6,8$ correspond to the polynomial degree. 

\subsection{Discrete velocity Boltzmann finite volume (DVBFV) method}
\label{Subsec: DVB-FVM}
In this section, we present a positivity-preserving discrete velocity Boltzmann scheme \cite{aravind2025prep}, which is influenced by entropy-stable schemes for vector-kinetic models \cite{Anandan2024}. The discrete velocity Boltzmann system of equations with the Boltzmann-Gross-Krook (BGK) \cite{BGK_1954} approximation  of the collision term, is given as,
 \begin{equation*}
    \partial_t \mathbf{f}_k+\lambda_k \divx{} \mathbf{f}_k=-\frac{\mathbf{f}_k-\mathbf{f}_{\mathbf{eq},k}
(\mathbf{U})}{\varepsilon} \text{ for } k=1,2,3,\dots,M,
  \end{equation*}
  where $M$ represents the number of discrete velocities.  $\mathbf{f}_{\mathbf{eq},k}
(\mathbf{U})$ satisfies
  \begin{equation}\label{eq:m}
      \sum_{k=1}^N \mathbf{f}_{\mathbf{eq},k}
=\mathbf{U}, \ 
      \sum_{k=1}^N \lambda_k \cdot \mathbf{n}(\theta) \mathbf{f}_{\mathbf{eq},k}
=\mathbf{G}(\mathbf{U})\cdot\mathbf{n}(\theta),
  \end{equation}
  where $\mathbf{U}$ and $\mathbf{G}(\mathbf{U})$ are respectively the vectors of conserved variables and the corresponding fluxes of Euler's system. $\mathbf{n}(\theta)$ is the unit vector $\cos{\theta} \hat{i}+\sin{\theta}\hat{j}$. By taking instantaneous relaxation ($\varepsilon=0$), choosing  $M=2$ and assuming $\mathbf{f}_k(t=0,\mathbf{x})=\mathbf{f}_{\mathbf{eq},k}
(t=0,\mathbf{x})$, we obtain the following system of equations in integral form.

  \begin{equation*}
    \frac{d}{dt}\int_\mathcal{K}\left[\begin{array}{c}
         \mathbf{f_{eq+}} \\
         \mathbf{f_{eq-}}
    \end{array}\right]d\boldsymbol{x} + \int_{\partial \mathcal{K}}\left[\begin{array}{cc}
            \lambda^+\cdot \mathbf{n}&0 \\
            0&\lambda^-\cdot \mathbf{n}
    \end{array}\right]\left[\begin{array}{c}
         \mathbf{f_{eq+}} \\
         \mathbf{f_{eq-}}
    \end{array}\right]ds =\left[\begin{array}{c}
         0 \\
         0
    \end{array}\right].
\end{equation*}
For notational simplicity, $\lambda^\pm \cdot \mathbf{n}$ will henceforth be denoted by $\lambda^\pm$.By decomposing the spatial domain $\Omega$ into $K$ non-overlapping elements as: $ \Omega \simeq \Omega_h = \bigcup_{j=1}^{K}\Omega_{j} $, the first-order accurate finite volume numerical scheme  has the following update formula in cell $j$:
\begin{gather*}
    \mathbf{f}_{\mathbf{eq+},j}^{n+1}=\mathbf{f}_{\mathbf{eq+},j}^n-\frac{\Delta t}{A_j} \sum_{i=1}^{N_e} \Tilde{\mathbf{G}}^+_{i+\frac{1}{2}} \Delta x_{i+\frac{1}{2}},  \   \mathbf{f}_{\mathbf{eq-},j}^{n+1}=\mathbf{f}_{\mathbf{eq-},j}^n-\frac{\Delta t}{A_j}  \sum_{i=1}^{N_e} \Tilde{\mathbf{G}}^-_{i+\frac{1}{2}} \Delta x_{i+\frac{1}{2}},\\
    \mathbf{U}^{n+1}_j=\mathbf{f}_{\mathbf{eq+},j}^{n+1}+\mathbf{f}_{\mathbf{eq-},j}^{n+1},
\end{gather*}
Here $N_e$ and $A_j$ are respectively the number of edges and the area of cell $j$. $\Delta x_{i+\frac{1}{2}}$ is the length of $i^{th}$ edge and $\mathbf{\Tilde{G}}$ is the numerical flux. The update formula can also be written in the form,
\begin{equation*}
   \mathbf{U}_j^{n+1}=\mathbf{U}_j^n- \frac{\Delta t}{A_j} \sum_{i=1}^{N_e} (\Tilde{\mathbf{G}}^+_{i+\frac{1}{2}}+\Tilde{\mathbf{G}}^-_{i+\frac{1}{2}}) \Delta x_{i+\frac{1}{2}}.
\end{equation*}
The numerical flux is defined as,
\begin{equation*}
    \Tilde{\mathbf{G}}^+_{i+\frac{1}{2}}=\lambda^+_{i+\frac{1}{2}}\mathbf{f}_{\mathbf{eq+},L}, \ \Tilde{\mathbf{G}}^-_{i+\frac{1}{2}}=\lambda^-_{i+\frac{1}{2}}\mathbf{f}_{\mathbf{eq+},R}.
\end{equation*}
The subscripts $L$ and $R$ represent the states left and right of the interface $i+\frac{1}{2}$, respectively. The values of $\lambda^\pm$ are chosen such that the numerical scheme maintains the positivity of pressure and density at all time levels. 
\begin{eqnarray*}
    \lambda^+_{i+\frac{1}{2}}=\max\left(\frac{1}{2}(V^\perp_L+a_L+\modL{V^\perp_L+a_L}),\frac{1}{2}(V^\perp_R+a_R+\modL{V^\perp_R+a_R})\right),\\
    \lambda^-_{i+\frac{1}{2}}=\min\left(\frac{1}{2}(V^\perp_L-a_L-\modL{V^\perp_L-a_L}),\frac{1}{2}(V^\perp_R-a_R-\modL{V^\perp_R-a_R})\right),\\
\end{eqnarray*}
where $V^\perp = (u \hat{i}+v\hat{j})\cdot n_{i+\frac{1}{2}}$ is the fluid velocity normal to the edge $i+\frac{1}{2}$ and $a$ is the local speed of sound. The definitions of $\mathbf{f_{eq\pm}}$ can be obtained by using the moment relations in \eqref{eq:m}.
\begin{eqnarray*}
    \mathbf{f}_{\mathbf{eq+},L}=\frac{1}{\Delta \lambda_{i+\frac{1}{2}}}(\mathbf{G}(\mathbf{U}_L)\cdot n_{i+\frac{1}{2}}-\lambda^-_{i+\frac{1}{2}}\mathbf{U}_L),\\ \mathbf{f}_{\mathbf{eq+},R}=\frac{1}{\Delta \lambda_{i+\frac{1}{2}}}(\lambda^+_{i+\frac{1}{2}}\mathbf{U}_R-\mathbf{G}(\mathbf{U}_R)\cdot n_{i+\frac{1}{2}}).
\end{eqnarray*}
where $\Delta \lambda_{i+\frac{1}{2}}= \lambda^+_{i+\frac{1}{2}}-\lambda^-_{i+\frac{1}{2}}$.
\par \noindent \textbf{Higher order numerical scheme} \\
To get second-order accuracy in space, a linear polynomial reconstruction of $\mathbf{U}$ has been utilised in every cell. The gradients are calculated using the Green-Gauss method, and a positivity limiter suggested by Zhang \cite{Zhang_2017} is used to make sure that the reconstructed $\mathbf{U}$ also preserves the positivity of density and pressure. To get the higher order accuracy in time, a strong stability preserving Runge Kutta method of 2$^{\text{nd}}$ order accuracy is employed. Let the first order update formula given above be $\mathscr{T_E}(\mathbf{U}_i^n)$, then the SSPRK2 update is,
\begin{equation*}
    \begin{array}{rcl}
         \mathbf{U}^*_i&=&\mathscr{T_E}({\mathbf{U}}_i^n),  \\
         \mathbf{U}_i^{n+1}&=&\frac{1}{2}\mathbf{U}_i^n+\frac{1}{2}\mathscr{T_E}(\mathbf{U}_i^*).
    \end{array}
\end{equation*}
Both the first-order and second-order numerical schemes preserve the positivity of density and pressure, given that the initial conditions also have the same property. 

The numerical results corresponding to this method are represented in this paper as DVBFV $N$, where $N=0,1$ correspond to the polynomial degree.

\subsection{Runge-Kutta Discontinuous Galerkin (RKDG) method}
\label{Subsec: DG}
In this section, we present a RKDG method for approximating the Euler equations.
Let $\mathcal{T}_h$ denote a tessellation of $\Omega \subset \mathbb{R}^d$ with shape-regular elements $\mathcal {K}$, and the boundary faces of each element $\mathcal{K}\in\mathcal{T}_h$ is denoted as $\partial \mathcal{K}$. In the two-dimensional case,  $\mathcal {K}$ is rectangular for Cartesian meshes.

The finite-dimensional test function space is defined in the following way
 \begin{equation*}
\mathcal{W}_h^N:= \left\{ w: w\textbar_{\mathcal {K}}\in \mathcal{P}^N(\mathcal {K}), \forall \mathcal {K}\in \mathcal{T}_h \right\},
\end{equation*}
where $\mathcal{P}^N(\mathcal {K})$ denotes the space of polynomials in cell $\mathcal {K}$ of degree at most $N$, and
 \begin{equation*}
\boldsymbol{\mathcal{V}}_h^N:= \left\{ \boldsymbol v: \boldsymbol v = (v_1,\dots, v_{l})^T \textbar \ v_1,\dots,v_{l} \in  \mathcal{W}_h^N  \right\}
\end{equation*}
is the piecewise polynomial finite element space with vectors consisting of $l$ components.
In the RKDG methods, for any unknown variable $u$, we project it into the finite element space $\mathcal{W}_h^N$ and still denote the numerical approximation as $u$ with abuse of notation. The values $u^{int_{\mathcal K}},u^{ext_{\mathcal K}}$ denote the approximations at the cell boundary $\partial {\mathcal K}$, and they are defined by
\begin{equation*}
u^{int_{\mathcal K}}(\boldsymbol{x}) := \lim _{\varepsilon \rightarrow 0^+} u(\boldsymbol{x} - \varepsilon\boldsymbol{n}), \
u^{ext_{\mathcal K}}(\boldsymbol{x}) := \lim _{\varepsilon \rightarrow 0^+} u(\boldsymbol{x} + \varepsilon\boldsymbol{n}), \ \forall \boldsymbol{x}\in \partial {\mathcal K}.
\end{equation*}
Here $\boldsymbol{n}$ denotes the outward unit normal vector.

The semi-discrete RKDG scheme reads as: Find $\boldsymbol{U}\in \boldsymbol{\mathcal{V}}_h^N$, s.t. for any test function $\boldsymbol{\varphi} \in \boldsymbol{\mathcal{V}}_h^N$ we have
\begin{equation}\label{scheme}
  \dfrac{d}{dt}\int_{\mathcal K}\boldsymbol{U}\cdot \boldsymbol{\varphi} \ d\boldsymbol{x} = \text{RHS}(\boldsymbol{U},\boldsymbol{\varphi})
\end{equation}
with
\begin{equation}\label{rhs_fg}
 \text{RHS}(\boldsymbol{U},\boldsymbol{\varphi})= 
 \int_{\mathcal K}\boldsymbol{G}(\boldsymbol{U}): \nabla\boldsymbol{\varphi} \ d\boldsymbol{x} -
  \int_{\partial \mathcal K} \widehat{\boldsymbol{G}} \cdot\boldsymbol{\varphi}^{int_{\mathcal K}} \ ds
\end{equation}
Here $"\cdot"$ is the Euclidean dot product, $":"$ is the Frobenius inner product. 
Considering the stability of the scheme, we choose the simplest monotone Lax-Friedrichs numerical flux for  $\widehat{\boldsymbol{G}}$, that is
  \begin{multline}\label{LF_FLUX} 
    \widehat{\boldsymbol{G}}  =\widehat {\boldsymbol {G}}(\boldsymbol U^{int_{\mathcal K}},\boldsymbol U^{ext_{\mathcal K}},\boldsymbol{n}) = \dfrac{1}{2}\left(\boldsymbol  {G}(\boldsymbol U^{int_{\mathcal K}})\cdot \boldsymbol{n}  + \boldsymbol {G}( \boldsymbol U^{ext_{\mathcal K}}) \cdot \boldsymbol{n} \right.  \\ \left. -{\alpha}(\boldsymbol U^{ext_{\mathcal K}} - \boldsymbol U^{int_{\mathcal K}}) \right). 
  \end{multline}
$\alpha$ is taken as
$ \max_{\boldsymbol {U}} \left( \modL{\frac{\partial \boldsymbol G}{\partial\boldsymbol U} \cdot\boldsymbol n}\right)$, and the maximum is taken globally in our computation.

\noindent\textbf{Time discretization} 

For hyperbolic conservation laws, usually high-order strong stability preserving Runge-Kutta (SSP-RK) and multistep time discretisation methods \cite{gottlieb2001strong, shu1988total, spiteri2002new} are adopted to solve the semi-discrete RKDG scheme. We start with presenting the standard $s$-stage SSP-RK time discretisation, for $i=1,\cdots,s$,
\begin{equation}\label{rks}
   \int_{\mathcal K} \boldsymbol{U}^{n,i} \cdot \boldsymbol{\varphi} \ dx 
  = \sum_{j=0}^{i-1} \left(\alpha_{ij}\int_{\mathcal K} \boldsymbol{U}^{n,j} \cdot \boldsymbol{\varphi} \ dx  
  + \Delta t \beta_{ij} \ \text{RHS}(\boldsymbol{U}^{n,j},\boldsymbol{\varphi})\right),  
\end{equation}
where $\text{RHS}(\boldsymbol{U},\boldsymbol{\varphi})$ is the spatial operator in (\ref{scheme}). The constants $\{\alpha_{ij}\}$ and $\{\beta_{ij}\}$ refer to the coefficients of the SSP-RK methods.
The piecewise constant ($P^0$), linear ($P^1$), quadratic ($P^2$), and quartic ($P^4$) polynomial basis are adopted for our spatial discretisation.  We take SSP-RK1, SSP-RK2, SSP-RK3, SSP-RK4 for $P^0,P^1,P^2,P^4$ case respectively.  

\noindent\textbf{Slope limiter}

Slope limiters are used to control spurious numerical oscillations that occur with discontinuous solutions. In our experiments, we adopt the total variation bounded (TVB) limiter \cite{cockburn1989tvb3,cockburn1998runge5}.
We apply the limiter procedure on the local characteristic fields of the conservative variables $\boldsymbol {U}$, by adopting a two-step procedure. First, we check whether the limiter is needed in each cell.  Then, we apply this limiter for those  `troubled' cells on the polynomials $\boldsymbol{U}(\boldsymbol{x})$. 

The numerical results corresponding to this method are represented in this paper as RKDG $N$, where $N=0,1,2,4$ correspond to the polynomial degree. 

\subsection{Discontinuous Galerkin Spectral Element Method (DGSEM) with subcell resolution}
\label{Subsec: ES-DGSEM}
In this section, we present the entropy-stable DGSEM method stabilised by the subcell blending limiter for approximating the system of Euler equations. The formulation is presented for the one-dimensional case. \\
The spatial domain $\Omega$ is decomposed into $K$ non-overlapping elements as: $ \Omega \simeq \Omega_h = \bigcup_{j=1}^{K}\Omega_{j} $. Here, $\Omega_j = \left[x^j_L, x^j_R\right]$, $\Delta x_j = x^j_R - x^j_L$ and $h = \max\limits_{1\leq j \leq K} \,\Delta x_j$.
The DG space of polynomial degree $N$ is given as,
\begin{equation*}
      \mathbf{V}_h^N = \left\{ \mathbf{l}_h \in L^2(\Omega) : \left. \mathbf{l}_h \right\vert_{\Omega_j} \in \mathcal{P}_N(\Omega_j) \, ,\, 1\leq j \leq K \right\}.
\end{equation*}
Here, $\mathcal{P}_N(\Omega_j)$ contains polynomials in element $\Omega_j$ of degree at most $N$. The equations on $\Omega_j$ are transformed to a reference element $I = [-1,1]$ through the linear map $x(\xi) = x^j_L + \frac{1+\xi}{2}\Delta x_j$, $x\in\Omega_j$, $\xi \in I$. The Lagrange polynomial 
\begin{align*}
    &l_i(\xi) = \prod_{j=0,j\neq i}^N \frac{\xi - \xi_j}{\xi_i - \xi_j}, \ l_i(\xi_j) = \delta_{ij}, 
    \ \text{where } \delta_{ij} = \begin{cases}
        1\quad &\text{if} \quad i=j,\\
        0\quad &\text{else}
    \end{cases} \quad 0\leq i,j \leq N,
\end{align*}
is used as the (nodal) basis of $\mathbf{V}_h^N$ to approximate the conserved variable vector $\mathbf{U}_{h} \in \mathbf{V}_h^N $ (Note that the subscript $h$ is dropped hereafter): 
\begin{equation*}
\left. \mathbf{U}(t,\xi)\right\vert_{\Omega_{k}} \approx \mathbf{U}^k(t,\xi) = \sum_{j=0}^N \mathbf{U}_{j}^k (t) l_j(\xi).
\end{equation*}
We collocate the non-linear functions like the flux vector $\mathbf{G(U}(t,x))$ at the nodal points, as follows,
\begin{align*}
\left. \mathbf{G}(\mathbf{U}(t,\xi))\right\vert_{\Omega_{k}} \approx \mathbf{G}^k(\mathbf{U}(t,\xi)) = \sum_{j=0}^N \mathbf{G}^k( \mathbf{U}^k_j(t) ) l_j(\xi) = \sum_{j=0}^N \mathbf{G}_{j}^k(t) l_j(\xi).
\end{align*}
Choosing the Lagrange polynomial $\mathbf{l}_i(\xi)$ as test function and using the Legendre–Gauss–Lobatto quadrature rule (LGL) for numerical integration and nodes, the so-called strong form of DGSEM for node $i$ in any element $\Omega_k$ is given as,
\begin{equation*}
    J_k \frac{d \mathbf{U}^k_i}{d t} +  \sum_{j=0}^N D_{ij} \mathbf{G}^k_j + \frac{\delta_{iN}}{w_N}\left(\widetilde{\mathbf{G}}^k_R -   \mathbf{G}^k_N\right) -  \frac{\delta_{i0}}{w_0}\left(\widetilde{\mathbf{G}}^k_L -   \mathbf{G}^k_0\right)= 0.
\end{equation*}
Here, $w_i$ are the LGL quadrature weights, $D$ is the polynomial derivative matrix with $D_{ij} = l^{'}_j (\xi_i)$  and $J_k = \Delta x_k / 2$ is the Jacobian of transformation. Following this notation, the entropy-stable DGSEM~\cite{chen2017entropy} can then be written as,
\begin{equation}\label{eqn:DGSEM}
    J_k \frac{d \mathbf{U}^k_i}{d t} +  \sum_{j=0}^N 2D_{ij} \mathbf{G}_k^{*}(ij) + \frac{\delta_{iN}}{w_N}\left(\widetilde{\mathbf{G}}^k_R -   \mathbf{G}^k_N\right) -  \frac{\delta_{i0}}{w_0}\left(\widetilde{\mathbf{G}}^k_L -   \mathbf{G}^k_0\right)= 0,
\end{equation}
where $\mathbf{G}^{*}(ij)$ is the two-point symmetric entropy conservative flux (also referred to as the volume flux) and $\widetilde{\mathbf{G}}_I$ is the entropy-stable numerical flux evaluated at the element edges (also referred to as the surface flux). We choose the entropy conservative and kinetic energy preserving flux of Chandrashekar~\cite{chandrashekar2013kinetic} as volume flux and local Lax-Friedrichs flux as the surface flux.\\
The entropy-stable DGSEM scheme~\eqref{eqn:DGSEM} is stabilised following the blending limiter approach of Hennemann et al.~\cite{hennemann2021provably}. In this approach, the higher order DG scheme~\eqref{eqn:DGSEM} is blended with a lower order entropy-stable finite volume (FV) scheme derived on subcells formed within the element. The size of subcells is determined by the quadrature weights in each dimension, and the FV scheme interprets the nodal DG values as the subcell averages. The lower-order scheme for $N+1$ subcells in an element $\Omega_k$ is written as
\begin{equation*}
    J_k  \frac{d \mathbf{U}^{k}_j}{d t} + \frac{1}{w_j}\left( \widetilde{\mathbf{G}}^k_{(j,j+1)} - \widetilde{\mathbf{G}}^k_{(j-1,j)} \right) = 0,\quad j\in \{0,1,\dots,N\}.
\end{equation*}
The numerical flux $\widetilde{\mathbf{G}}$ (also referred to as the blending flux) should be entropy-stable, and we choose the local Lax-Friedrichs flux for our simulations. The element-wise blending of higher order and lower order schemes can be written as
\begin{equation*}
    \frac{d \mathbf{U}}{d t} = (1-\alpha)\frac{d \mathbf{U}^{DG}}{d t} + \alpha \frac{d \mathbf{U}^{FV}}{d t},
\end{equation*}
and the element-wise blending factor $\alpha \in [0,1]$ is computed from the shock indicator. According to the procedure described in~\cite{hennemann2021provably}, the blending factor $\alpha$ is a function of the highest mode energy indicator $ \mathbb{E}$ and polynomial order dependent threshold value $\mathbb{T}(N)$. The energy indicator is defined based on the modal energy of the shock indicator quantity $\phi$ as
\begin{equation*}
    e = \max \left(  \frac{\hat{\phi}_N^2}{\sum_{j=0}^N \hat{\phi}_j^2} , \frac{\hat{\phi}_{N-1}^2}{\sum_{j=0}^{N-1} \hat{\phi}_j^2} \right),
\end{equation*}
where $\hat{\phi}_j$ are the modal coefficients of $\phi$. The threshold value is defined as $f (N) = 0.5\cdot 10^{-1.8(N+1)^{\frac{1}{4}}}$ and the blending factor function with sharpness factor $s = 9.21024$ is given as
\begin{equation*}
    \alpha = \frac{1}{1+\exp(\frac{-s}{f}\left(e - f\right))}.
\end{equation*}
Further, a clipping procedure is performed on $\alpha$ to clear the incidents in the choice of pure lower-order or higher-order configurations as
\begin{equation*}
\Tilde{\alpha} = \begin{cases}
    0, \quad & \text{if}\quad \alpha < \alpha_{min}\\
    \alpha, \quad & \text{if}\quad \alpha_{min} \leq \alpha \leq 1 - \alpha_{min}\\
    1, \quad & \text{if}\quad 1 - \alpha_{min} < \alpha
\end{cases}; \ \alpha = \min(\Tilde{\alpha}, \alpha_{max}),
\end{equation*}
followed by element-wise smoothing of $\alpha$, see~\cite{hennemann2021provably} for more details. The shock indicator quantity is chosen as $\phi = \varrho p$, and the tuning parameters are taken as $\alpha_{min} = 0.0001$ and $\alpha_{max} = 0.001$  for all the simulations. We also use the scaling limiter of Zhang and Shu~\cite{zhang2010positivity} to ensure the positivity of the density and pressure. 
\par \noindent\textbf{Time discretisation} \\
We use an explicit higher-order strong-stability preserving Runge-Kutta (SSP-RK) method to integrate the semi-discrete form of our DGSEM formulations. We follow the SSP-RK methods presented in Spiteri and Ruuth~\cite{spiteri2002new} where the update formula for every RK stage $s$ is given as,
\begin{equation*}
    \mathbf{U}^{s+1} = a_{ss} \mathbf{U}^s + b_{ss} \Delta t \frac{d \mathbf{U}^s}{d t} + \sum_{k=1}^{s-1}\left( a_{sk} \mathbf{U}^k + \Delta t b_{sk} \frac{d \mathbf{U}^k}{d t} \right),
\end{equation*}
where $a_{sk}$ and $b_{sk}$ are the RK coefficients, $\Delta t$ is the time step and $d \mathbf{U}^s/d t$ is the value of RHS at stage $s$.\\
We compute the time step value as
\begin{align*}
    \Delta t^k =& C_{cfl}\,\frac{2}{N+1}\,\min_{\forall i,j}\frac{J^k_{ij}}{\sum_{l=1}^d \left(J\mathbf{a}^l\right)_{ij} \lambda_{l,ij}^{max}} \text{ for } k \in \{ 1,2,\dots,K\}, \\  \Delta t =& \min_{k=1:K} \Delta t^k,
\end{align*}
where $C_{cfl}$ is the tuning parameter, $d$ is the number of spatial dimensions, $N$ is the degree of solution polynomial, $J^k_{ij}$ is the grid Jacobian for element $k$ evaluated at nodes $i,j$, $\left(J\mathbf{a}^l\right)_{ij}$ is the contravariant basis vector in direction $l$ at nodes $i,j$ (see Chapter 6,~\cite{kopriva2009implementing}),  $\lambda_{l,ij}^{max}$ is the maximum wave speed estimate in direction $l$ at nodes $i,j$ for element $k$ and 2 is the size of reference element. We take $C_{cfl}=1.0$ for the simulations. 
\par All the simulations corresponding to the entropy-stable DGSEM are carried out using the open source code \texttt{Trixi.jl}~\cite{schlottkelakemper2020trixi}. The numerical results corresponding to this method are represented in this paper as DGSEM $N$, where $N=1,2,4$ correspond to the polynomial degree. Suitably, we use SSP-RK2 ($2-$stage, $2^{\text{nd}}$ order accurate), SSP-RK3 ($4-$stage, $3^{\text{rd}}$ order accurate) and SSP-RK4 ($5-$stage, $4^{\text{th}}$ order accurate) for $N=1,2$ and $4$ respectively.

\end{appendices}

\section*{End Matter}

\subsection*{Data availability}Data will be made available from the corresponding author on reasonable request.

\subsection*{Conflict of interest}On behalf of all authors, the corresponding author states that there is no conflict of interest.

\subsection*{Ethical approval}Not applicable.

\bibliographystyle{acm}
\bibliography{references}
\end{document}